\documentclass[11pt,letterpaper, envcountsame, envcountsect]{llncs}
\usepackage{soul}
\usepackage[utf8]{inputenc}
\usepackage[table,dvipsnames]{xcolor}
 \usepackage{cancel} 
\usepackage{float}
\usepackage{diagbox}
\usepackage{lineno}

\usepackage[margin=1in]{geometry}
\usepackage{graphicx}
\usepackage{amsmath}
\usepackage{amssymb}
\usepackage[sort,nocompress]{cite}
\usepackage[ruled,vlined,lined,boxed,commentsnumbered]{algorithm2e}
\usepackage{todonotes}
\usepackage{verbatim}
\usepackage{subcaption}
\usepackage[normalem]{ulem}
\usepackage{array}
\usepackage{algorithmic}
\usepackage{mathrsfs}
 \usepackage{multirow}

\usepackage{todonotes}

\usepackage{pgf}
\usepackage{tikz}
\usepackage{imakeidx}

\newtheorem{Problem}{Problem}

\newcommand{\pf}[1]{\text{Pref} (\mathcal #1)}

\newcommand{\nats}{\mathbb N} 

\newcommand{\mt}[1]{\mathcal #1}

\newcommand{\notaG}[1]{\textcolor{cyan}{#1 (G)}}

\SetKwComment{Comment}{$\triangleright$\ }{}

\usetikzlibrary{arrows,automata,positioning}
\makeindex
\title{Co-lexicographically Ordering Automata and Regular Languages - Part I}

\author{Nicola Cotumaccio\inst{1, 2} \and Giovanna D'Agostino\inst{3} \and Alberto Policriti\inst{3}  \and Nicola Prezza\inst{4}} 

\institute{Gran Sasso Science Institute, L'Aquila,  Italy. Email: \email{nicola.cotumaccio@gssi.it} \and Dalhousie University, Halifax, Canada. Email: \email{nicola.cotumaccio@dal.ca} \and University of Udine, Italy.  Email: \email{giovanna.dagostino@uniud.it, alberto.policriti@uniud.it} \and University Ca' Foscari, Venice,  Italy. Email: \email{nicola.prezza@unive.it}}

\date{\today}

\begin{document}
{\def\addcontentsline#1#2#3{}\maketitle}

\thispagestyle{empty}

\begin{abstract}

The states of a finite-state automaton $\mathcal N$ can be identified with collections of words in the prefix closure of the regular language accepted by $\mathcal N$. But words can be ordered, and among the many possible orders a very natural one is the co-lexicographic order. Such naturalness stems from the fact that it suggests a transfer of the order from words to the automaton’s states. This suggestion is, in fact, concrete and in a number of papers  automata admitting a \emph{total} co-lexicographic (\emph{co-lex} for brevity) ordering of states have been proposed and studied. Such class of ordered automata --- \emph{Wheeler automata} --- turned out to require just a constant number of bits per transition to be represented and enable regular expression matching queries in constant time per matched character.

\medskip

Unfortunately, not all automata can be totally ordered as previously outlined. 
In the present work, we lay out a new theory showing that all automata can always be \emph{partially} ordered, and an intrinsic measure of their complexity can be defined and effectively determined, namely, the minimum width $p$ of one of their admissible \emph{co-lex partial orders}—dubbed here the automaton's \emph{co-lex width}.
We first show that this new measure captures \emph{at once} the complexity of several seemingly-unrelated hard problems on automata. 
Any NFA of co-lex width $p$:
(i) has an equivalent powerset DFA whose size is exponential in $p$ rather than (as a classic analysis shows) in the NFA's size;
(ii) can be encoded using just $\Theta(\log p)$ bits per transition;
(iii) admits a linear-space data structure solving regular expression matching queries in time proportional to $p^2$ per matched character. 
Some consequences of this new parameterization of automata are that PSPACE-hard problems such as NFA equivalence 
are FPT in $p$, and quadratic lower bounds for the regular expression matching problem do not hold for sufficiently small $p$.

\medskip

Having established that the co-lex width of an automaton is a fundamental complexity measure, we proceed by (i) determining its computational complexity and (ii) extending this notion from automata to regular languages by studying their smallest-width accepting NFAs and DFAs.
In this work we focus on the deterministic case and prove that  a canonical minimum-width DFA accepting a language $\mathcal L$—dubbed the Hasse automaton $\mathcal H$ of $\mathcal L$—can be exhibited. $\mathcal H$ provides, in a precise sense, the best possible way to (partially) order the states of any DFA accepting $\mathcal L$, as long as we want to maintain an operational link with the (co-lexicographic) order of $\mathcal L$'s prefixes. 
Finally, we  explore the relationship between two  
conflicting objectives: minimizing the width and minimizing the number of states of a DFA. In this context, we provide an analogue of the Myhill-Nerode Theorem for co-lexicographically ordered regular languages. 

\vfill
\hrulefill
\newline
\footnotesize
\emph{A preliminary version of this work appeared in the Proceedings of the 2021 ACM-SIAM Symposium on Discrete Algorithms (SODA) \cite{NicNic2021}}
\normalsize

\end{abstract}


\newpage

\tableofcontents

\newpage

\clearpage
\pagenumbering{arabic}

 \section{Introduction}
 
Equipping the domain of a structure with some kind of order is often a fruitful move performed in both Computer Science and Mathematics. An \emph{order} provides direct access to data or domain elements and sometimes allows to tackle problems otherwise too computationally difficult to cope with. For example, in descriptive complexity it is not known how to logically capture the class $P$ in general, while this can be done on ordered structures (see \cite{Libkin}). 
In general, the price to be paid  when requiring/imposing an order, is a --- sometimes significant --- restriction of the class of structures to which subsequent results refer. If we do not wish to pay such a price, a \emph{partial} order can be a natural alternative. Then, the ``farther'' the partial order is from a total order, the less powerful will be the applications of the established results. In other words, the ``distance'' from a total order of the partial order at hand  becomes a \emph{measure} of the extent to which we have been able to ``tame''  the class of structures under consideration.

In this paper, inspired by the above --- somehow foggy --- considerations, we focus on the class of finite automata.
Partial orders and automata have already met and attracted attention because of their relation with logical, combinatorial, and algebraic characterization of languages.  In the literature (see, among many others,  \cite{DBLP:journals/jcss/BrzozowskiF80, DBLP:conf/dlt/SchwentickTV01,DBLP:journals/lmcs/MasopustK21})  a partially-ordered NFA is an automaton where the transition relation induces a partial order on the set of its states.  Here we pursue a  different approach,  closer to   the one given in \cite{thierrin1974ordered_aut}.  Our starting point  is a work by  Gagie et al. \cite{GAGIE201767}, presenting a simple and unified perspective on several algorithmic techniques related to \emph{suffix sorting} (in particular, to the Burrows-Wheeler transform \cite{burrows1994block}, an ubiquitous string permutation in text compression and indexing --- see also \cite{manzini1999burrows}). 
The general idea is to enforce and exploit a total order among the states of a given automaton, induced by an \emph{a priori} fixed order of its underlying alphabet which propagates through the automaton's transition relation\footnote{The dependence on a fixed   order of the alphabet marks the difference between this approach and the one of \cite{thierrin1974ordered_aut}.}. 
The resulting  automata, called \emph{Wheeler automata}, admit efficient data structures for solving string matching on the automaton's paths and    enable a representation of the automaton in a space proportional to that of the edges' labels --- as well as enabling more advanced compression mechanisms, see \cite{DBLP:conf/dcc/AlankoGNB19,DBLP:conf/soda/Prezza21}. 
This is in contrast with the fact that general graphs require a logarithmic (in the graph's size) number of bits per edge to be represented, as well as with recent results showing that in general, regular expression matching and string matching on labeled graphs can not be solved in subquadratic time, unless the strong exponential time hypothesis is false \cite{backurs2016regular,EquiMT21,EquiGMT19,DBLP:conf/sosa/GibneyHT21, DBLP:journals/ipl/PotechinS20}.

\medskip 

Wheeler languages --- i.e. languages accepted by Wheeler automata --- form an interesting   class of subregular languages, where determinization   becomes computationally easy  (polynomial). 
As was to be expected, however, requiring the existence of a \emph{total} Wheeler order  over an automaton comes with a price. Not all automata enjoy the Wheeler property, and languages recognized by Wheeler automata constitute a relatively small class: a subclass of star-free languages  \cite{Alanko2021Wheeler}.

Our results here show that, as a matter of fact,  Wheeler (automata and) languages can be seen as a first level --- in some sense the optimal level --- of two hierarchies (deterministic  and  non-deterministic) encompassing  all regular languages. A level of each such hierarchy is based on a \emph{partial} order of minimum \emph{width}\footnote{The width of a partial order is the size of its largest antichain.} defined over the states of an automaton accepting the language. More precisely, in Definition \ref{def:colex} we identify a class of \emph{partial} orders --- the \emph{co-lexicographic} ones, or \emph{co-lex} for brevity --- among the states of an automaton $\mathcal N$, reflecting naturally  
the (\emph{total}) co-lexicographic order among the prefixes of the language $\mathcal L(\mathcal N)$ accepted by $\mathcal N$: we require that (1) states with different incoming labels are sorted according to the underlying alphabet's order and (2) the co-lex order propagates through pairs of equally-labeled edges. The minimum \emph{width} of a co-lex order over $\mathcal N$  results in a measure --- dubbed   ${\text{width}}(\mathcal N)$ --- 
that we prove being a fundamental parameter for classifying finite automata as well as regular languages. Letting $\mathcal L$ be a regular language, denote by ${\text{width}^N}(\mathcal L)$  the smallest integer $p$ such that there exists an NFA  $\mathcal N$ such that $\mathcal L=\mathcal L(\mathcal N)$ and ${\text{width}}(\mathcal N)=p$, and by ${\text{width}^D}(\mathcal L)$  the notion similarly defined using DFAs. 
These two  non-deterministic/deterministic widths define two corresponding hierarchies of regular languages. The overall goal of our work is to show that these hierarchies shed new light on the relations existing between the sizes of DFAs and NFAs recognizing a given regular language, as well as classify languages by their propensity to be \emph{searched},  with important applications to regular expression matching and string matching on labeled graphs.

\medskip

In this paper we mostly focus on the deterministic case and study in detail the hierarchy based on $\text{width}^{D}(\mathcal L)$ while, 
in a companion complementary paper (see \cite{parttwo}), we complete the picture by studying the non-deterministic case, proving (among other results) that the two hierarchies are strict and do not collapse, and that in general they are exponentially separated, save for level one --- the Wheeler languages --- where allowing nondeterminism does not add expressive power. The present work and the companion paper \cite{parttwo} are an extension of the SODA'21 paper \cite{NicNic2021}. 

Consider an NFA $\mathcal N$ of co-lex width $p$. As far as motivations for introducing the new concept of co-lex orders are concerned, in this paper we show that: 

\begin{enumerate}
    \item the well-known  explosion in the number of states occurring when computing the \emph{powerset} DFA $ \text{Pow}(\mathcal N)$ is exponential in $p$, rather than (as a classic analysis shows) in the size of $\mathcal N$. We prove that a similar exponential bound holds for $\text{width}(\text{Pow}(\mathcal N))$.
    \item $\mathcal N$ can be encoded using just $\log(p\sigma) + O(1)$ bits per transition on DFAs, where $\sigma$ is the alphabet's size. NFAs can be encoded using additional $\log p$ bits per transition.
    \item String matching on $\mathcal N$'s paths and testing membership in the automaton's accepted language can be solved in $O(p^2 \log\log(p\sigma))$ time per matched character.
\end{enumerate}

Result (1) provides one of the few known parameterizations of NFAs and immediately implies that hard problems such as NFA equivalence and universality are actually easy on small-width NFAs (for example Wheeler NFAs \cite{GAGIE201767}, for which ${\text{width}}(\mathcal N) = 1$ holds). The result allows us to conclude also that the two deterministic and non-deterministic hierarchies of regular languages are exponentially related.
Result (2) provides a new compression paradigm for labeled graphs. Result (3) breaks existing lower bounds for regular expression matching 
\cite{backurs2016regular} and for string matching on labeled graphs. More details on these connections, themes and trends in the literature, are discussed in Section \ref{sec:state art}.

\medskip


Having established that the co-lex width of a language/automaton is a fundamental complexity measure, we address the problem of the effectiveness of such a measure in the deterministic case:
are the width of a DFA and the deterministic width of a language $\mathcal L$ (presented  by an automaton) computable and, if so, at which cost?
We observe that the width of a language  is not in general equal to the width of, say,  its  minimum DFA, since already at level one of the deterministic hierarchy (i.e. Wheeler languages) there are   languages whose minimum DFA has deterministic width larger than one \cite{DBLP:conf/soda/AlankoDPP20}. This makes the language-width problem non-trivial.

\medskip

\begin{table}
\renewcommand{\arraystretch}{2.5}
\centering
\begin{tabular}{|c|c|c|}\hline
\backslashbox{{\hspace{1em}compute}}{{\hspace{-3.5em}given}}
&\makebox[18em]{{ $\mathcal A$ : DFA}}&\makebox[19em]{{ $\mathcal A$ : NFA}}\\\hline\hline
{$\text{width}(\mathcal A)$} & $\tilde O(|\delta|^2)$ \emph{w.h.p.} [Cor. \ref{cor:complexity width dfa}] &NP-HARD \cite[Thm. 2]{gibney2022complexity} \\\hline
{$p = {\text{width}^D}(\mathcal L(\mathcal A))$} & $|\delta|^{O(p)}$ [Thm. \ref{thm:dyn prog}]& PSPACE-HARD \cite[Thm. 10]{dagostino2023complexity}\\\hline
{$p = {\text{width}^N}(\mathcal L(\mathcal A))$} & $|\delta|^{O(1)}$ for $p = 1$ \cite[Thm. 3.4]{Alanko2021Wheeler} & PSPACE-HARD \cite[Thm. 10]{dagostino2023complexity}\\\hline
\end{tabular}
\vspace{1em}
\caption{Lower and upper bounds for computing the width of an automaton (DFA/NFA) and the deterministic/non-deterministic widths of a regular language (encoded as a DFA/NFA). 
In this table, $n = |Q|$ is the number of states and $|\delta|$ is the number of transitions of the automaton $\mathcal A = (Q,s,\delta,F)$.
All hardness bounds follow from recent works dealing with the Wheeler case (automata of co-lex width equal to 1 and languages of deterministic and non-deterministic widths equal to 1). Computing the non-deterministic width in the case $p=1$ is a polynomial problem if the input is a DFA because, only for this level of the two hierarchies, the deterministic and non-deterministic widths coincide \cite{Alanko2021Wheeler} (they are both equal to 1). In this case, an algorithm for recognizing languages of deterministic (thus also non-deterministic) width equal to 1 was already presented in\cite{Alanko2021Wheeler}. }\label{tab:width complexities}
\end{table}

Table \ref{tab:width complexities} reports our complexity results, as well as hardness results that follow from recent works dealing with the Wheeler case.
We  prove that the width of a DFA can be computed in polynomial time. This is in contrast with a recent result showing that Wheeler NFAs are NP-complete to recognize \cite{gibney2022complexity}, which implies that deciding whether the width of a NFA is smaller than a given value is NP-hard.
We then show that, although the deterministic width of a language, $\text{width}^D(\mathcal L)$, differs in general from the width of its (unique) minimum DFA,  $\text{width}^D(\mathcal L)$ can be computed in polynomial time for constant values of  $\text{width}^D(\mathcal L)$ given a DFA for $\mathcal L$, and a canonical  automaton realising this minimum width can be exhibited. 
 Again, this is in contrast with the fact that recognizing Wheeler languages ($\text{width}^D(\mathcal L) = \text{width}^N(\mathcal L) = 1$) is a PSPACE-complete problem when the input language $\mathcal L$ is provided as an NFA \cite{dagostino2023complexity}. 
The key observation for our result is a  combinatorial   property of automata that we called the {\em entanglement} number of a DFA, a quantity measuring the intrinsic co-lex incomparability of the automaton's states.
The entanglement of the minimum DFA for a regular language  turns out to exactly correspond both to the deterministic width of the language and to the width of   the above mentioned canonical automaton, dubbed the \emph{Hasse} automaton for the language. 
As we shall prove, these results imply that 
$\text{width}^D(\mathcal L)$ can be computed from the minimum deterministic automaton recognizing $\mathcal L$. 

\medskip

A further contribution of this paper is to explore the relationship between two (conflicting) objectives on deterministic automata: minimizing the co-lex width and minimizing the number of states. In this context, we provide an analogue of the Myhill-Nerode Theorem for regular languages applied to the concept of co-lex ordered automata. 


\subsection{Our Work in Context}\label{sec:state art}


Our proposal in this paper aims at proving that, by pairing automata with co-lex orders, we can classify regular languages by their  \emph{propensity to be sorted}. 
Our classification represent a useful parameterization \emph{simultaneously} for diverse automata-related measures: (1) the complexity of NFA determinization by the powerset-construction algorithm, (2) the encoding bit-complexity of automata, and (3) the complexity of operations on regular languages (e.g. membership) and on labeled graphs (e.g. pattern matching). 
 As we discuss below, previous works focused on studying the complexity of points (1-3) \emph{separately} and by cases, i.e. by studying notable classes of automata, regular languages, and graphs on which these problems are easy. To the best of our knowledge, ours is the only parameterization of automata/labeled graphs capturing \emph{simultaneously} all these aspects.

\paragraph*{NFA determinization and existing subregular classifications}
An extensive and detailed classification of the complexity of the powerset construction algorithm on families of subregular languages is carried out  in \cite{BORDIHN20093209}. That study proves that for the most popular and studied classes of subregular languages --- including (but not limited to) star-free \cite{mcnaughton1971counter}, ordered \cite{thierrin1974ordered_aut}, comet \cite{brzozowski1969decompositions} and suffix/prefix/infix-closed languages --- the output of powerset construction is exponential in the size of the input NFA: for all the mentioned families, the resulting DFA may have at least $2^{n-1}$ states in the worst case, where $n$ is the number of states of the input NFA. Previously-known families with a sub-exponential upper bound include unary regular languages, with a bound of $e^{\Theta(\sqrt{n\ln n})}$ states \cite{chrobak1986finite} and the family of finite languages over alphabet of size $\sigma$, with a bound of $O(\sigma^{\frac{n}{\log_2\sigma+1}})$ states \cite{salomaa1997nfa}. In this context, our nondeterministic hierarchy of subregular languages --- classifying languages by the width $p$ of their smallest-width accepting NFA and guaranteeing an equivalent DFA of size at most $2^p(n-p+1)-1$ (see Theorem \ref{theor:powersetcrucial}) --- represents a more complete classification than the above-mentioned classes (since it captures all regular languages), even though our deterministic and nondeterministic hierarchies are orthogonal to these classes in some cases (see Section \ref{sec: relation star free} for a study of the relations existing between our proposal and the class of star-free languages).

Interestingly, \cite{BORDIHN20093209} shows that the class of ordered automata \cite{thierrin1974ordered_aut} --- automata admitting a total states' order that must propagate through pairs of equally-labeled edges --- does have a worst-case exponential-output powerset construction. 
Since in our work we show that the powerset construction builds a small-size DFA on bounded-width automata, this fact shows that the small difference between simply imposing an order on the states which is consistent with the transition relation (ordered automata) and linking this property with a fixed order of the underlying alphabet (our framework), does have significant practical consequences in terms of deterministic state complexity. 

As far as other parameterizations of powerset construction are involved, we are aware of only one previous attempt in the literature: the notion of \emph{automata width} introduced in \cite{majumdar2019computing}. Intuitively, given an NFA $\mt N$ the width of $\mt N$ as defined in \cite{majumdar2019computing} is the maximum number of $\mt N$'s states one needs to keep  track of simultaneously while looking for an accepting path for some input word (for the word maximizing such quantity). By its very definition, this quantity is directly linked to the output's size of powerset construction (while for our co-lex width, establishing such a connection will be more involved). 

Further notable classifications of subregular languages include the star-height hierarchy \cite{eggan1963transition} (capturing all regular languages) and the  Straubing-Th\'erien hierarchy \cite{straubing1981generalization,therien1981classification} (capturing the star-free languages).
To the best of our knowledge, these classifications do not lead to useful parameterizations for the automata/graph problems considered in this paper.

\paragraph*{Graph compression}

Graph compression is a vast topic that has been extensively studied in the literature (see for example the survey \cite{besta2019survey}). 
Most solutions discussed below consider the unlabeled and undirected case; in those cases, a compressor for labeled graphs can be obtained by just storing the labels and the edges' directions separately using $\lceil \log\sigma\rceil + 1$ additional bits per edge ($\sigma$ is the alphabet's size). 

Existing results studying worst-case information-theoretic lower bounds of graph encodings can be used as the reference base for the compression methods discussed below.
First of all, note that the worst-case information-theoretic number of bits needed to represent a directed graph with $m$ edges and $n$ vertices is $\log\binom{n^2}{m} = m \log (n^2/m) + \Theta(m)$, that is, $\log (n^2/m) + \Theta(1)$ bits per edge. The same lower bound holds on undirected graphs up to a constant additive number of bits per edge. 
Other useful bounds (on automata) are studied in the recent work of Chakraborty et al. \cite{chakraborty2021succinct}. In that paper, the authors present a succinct encoding for DFAs using $\log\sigma + \log n + 1.45$ bits per transition ($n$ is the number of states) and provide worst-case lower bounds as a function of the number of states: in the worst case, DFAs cannot be encoded using less than  $(\sigma-1)\log n + O(1)$ bits per state and NFAs cannot be encoded in less than $\sigma n + 1$ bits per state. The same paper provides encodings matching these lower bounds up to low-order terms. 

In order to compare our encodings with the state-of-the-art, we anticipate that 
our solutions store DFAs within $\log (p\sigma) + O(1)$ bits per transition (Corollary \ref{cor:encoding DFA}) and NFAs within $\log (p^2\sigma) + O(1)$ bits per transition (Corollary \ref{cor:encoding NFA}). 
Most of the parameterized graph encodings discussed in the literature (read below) provide a space bound \emph{per vertex}. In Corollary \ref{cor:encoding DFA} 
we show that our DFA encoding uses no more than $\sigma \log(p\sigma) + O(\sigma)$ bits per state. In Corollary \ref{cor:encoding NFA} we show that our NFA encoding uses no more than $2p\sigma\log(p^2\sigma) + O(p\sigma)$ bits per state. 
Note that the former bound asymptotically matches  Chakraborty et al.'s lower bound for DFAs (since that $p\leq n$), while the latter   matches  Chakraborty et al.'s lower bound for NFAs up to a logarithmic multiplicative factor.

For our purposes, it is useful to divide graph compression strategies into \emph{general graph compressors} and \emph{compact encodings} for particular graph classes. The former compressors work on arbitrary graphs and exploit sources of redundancy in the graph's topology in order to achieve a compact representation. Compressors falling into this category include (this list is by no means complete, see \cite{besta2019survey} for further references) $K^2$ trees on the graph's adjacency matrix \cite{brisaboa2009k2}, straight-line programs on the graph's adjacency list representation \cite{claude2007fast}, and context-free graph grammars \cite{Engelfriet1997}. A shared feature of these compressors is that, in general, they do not provide guarantees on the number of bits per edge that will be used to encode an arbitrary graph (for example, a guarantee linked with a particular topology or graph parameter such as the ones discussed below); the compression parameter associated with the graph is simply the size of the compressed representation itself. This makes these techniques not directly comparable with our approach (if not experimentally).

Techniques exploiting particular graph topologies or structural parameters of the graph to achieve more compact encodings are closer to our parameterized approach, bearing in mind that also in this case a direct comparison is not always possible in the absence of known relations between our parameter $p$ and the graph parameters mentioned below (the investigation of such relations represents an interesting future research direction). 
A first example of such a parameter (on undirected graphs) is represented by \emph{boxicity} \cite{roberts1969boxicity}, that is, the minimum number $b$ of dimensions such that the graph's edges correspond to the intersections of $b$-dimensional axes-parallel boxes (the case $b=1$ corresponds to interval graphs). Any graph with boxicity $b$ can be represented naively using $O(b)$ words per vertex (that is, storing each vertex as a $b$-dimensional box), regardless of the fact that its number of edges could be quadratic in the number of vertices (even in the interval graph case). Similar results are known for graphs of small clique-width/bandwidth/treedepth/treewidth \cite{kamali2018compact,KAMALI2022104867,farzan2014compact} and bounded genus \cite{deo1998structural}; 
any graph from these graph families can be encoded in $O(k)$ bits per vertex, where $k$ is the graph parameter under consideration. 
Similarly, posets (transitively-closed DAGs) of width $w$ can be encoded succinctly using $2w + o(w)$ bits per vertex \cite{yanagita_et_al:LIPIcs.SWAT.2022.33}. While the above-mentioned methods focus on particular graph parameters, another popular approach is to develop ad-hoc compact encodings for particular graph topologies. Separable graphs (graphs admitting a separator of size $O(n^c)$ breaking the graph into components of size $\alpha n$ for some $c<1$ and $\alpha<1$) allow for an encoding using $O(1)$ bits per vertex \cite{separable_graphs}. This class includes planar graphs --- admitting also an encoding of $4$ bits per vertex \cite{FFSGHNcgta20} --- and trees --- admitting an encoding of $2$ bits per vertex (e.g. a simple balanced-parenthesis representation) and an encoding of $1+o(1)$ bits per vertex when every internal node has exactly two children \cite{jansson2012ultra}. Circular-arc graphs (a class including interval graphs) of maximum degree $\Delta$ can be encoded in $\log \Delta + O(1)$ bits per vertex, and this bound is asymptotically tight \cite{CHAKRABORTY2023156}; in the same paper, the authors show that circular-arc graphs with chromatic number $\chi$ admit an encoding using $\chi + o(\chi)$ bits per vertex. As mentioned above, on NFAs our proposed encoding uses a space per state that can be expressed as a function of $p$ and $\sigma$, thereby fitting with previous research on compact parameterized representations of graphs.

\paragraph*{Regular expression matching and string matching on labeled graphs}
``Regular expression matching'' (REM) refers to the problem of determining whether there exist substrings of an input string that can be derived from an input regular expression. This problem generalizes that of determining membership of a string to a regular language, and it finds important applications which include text processing utilities (where regular expressions are used to define search patterns), computer networks (see \cite{7468531}), and databases (see \cite{cruz1987graphical}).
A closely-related problem is that of ``exact string matching on labeled graphs'' (SMLG): find which paths of an edge-labeled graph match (without edits) a given string (see \cite{EquiGMT19}). 
This problem arises naturally in several fields, such as bioinformatics \cite{baier2015, siren2014}, where the \emph{pan-genome} is a labeled graph capturing the genetic variation within a species (and pattern matching queries are used to match an individual's genome on this graph), and graph databases \cite{ angles2008}.
Since NFAs can be viewed as labeled graphs, it is not surprising that existing lower and upper bounds for both problems have been derived using the same set of techniques.

In order to compare our approach with the state of the art, we anticipate that we describe labeled graph indexes solving SMLG (and thus REM by constructing the index on an NFA derived from the regular expression) in $O(p^2\log\log(p\sigma))$ time per matched character, where $p$ is the co-lex width of the graph/automaton under consideration (Theorem \ref{thm:aBWT index}).
The general idea behind our approach is that co-lexicographically ordering the states of the underlying automaton accepting a language is a way of \emph{structuring} the search-space where  functionalities will eventually operate. In this sense the co-lex width of a labeled graph (or that of a language)  is a measure capturing the intrinsic complexity of the \emph{entire collection} of strings that we aim to represent.

Backurs and Indyk in \cite{backurs2016regular} carry out a detailed study of the complexity of the REM problem as a function of the expression's structure for all regular expressions of depth up to 3. For each case (there are in total 36 ways of combining the regular operators “$|$”, Kleene plus, Kleene star, and concatenation up to depth 3), they either derive a sub-quadratic upper bound (where \emph{quadratic} means the string's length multiplied by the regular expression's size) or a quadratic lower bound conditioned on the Strong Exponential Time Hypothesis \cite{impagliazzo2001complexity} or on the Orthogonal Vectors conjecture \cite{bringmann2015quadratic}. 
Note that this classification does not capture regular expressions of arbitrary depth. 
Similarly, Equi et al. \cite{EquiGMT19} establish lower and upper bounds for the SMLG problem, even in the scenario where one is allowed to pre-process the graph in polynomial time \cite{EquiMT21} (that is, building a graph index); their work represents a \emph{complete} classification of the graph topologies admitting either sub-quadratic pattern matching algorithms or quadratic lower bounds (obtained assuming the Orthogonal Vectors conjecture); in this context, \emph{quadratic} means proportional to the string's length times the graph's size. 
In all these works, as well as in further papers refining these analyses by providing finer lower bounds or better upper bounds for particular cases \cite{bringmann2017dichotomy,gibney2021simple,gibney2020efficient,gibney2021text,bernardini2019even}, the problem's complexity is studied by cases and does not depend on a parameter of the language or the graph.


Techniques parameterizing the problem's complexity on a graph parameter do exist in the literature, and are closer to the spirit of our work. These parameters include (these works consider the SMLG problem) the size of the labeled direct product \cite{rizzo2022solving}, the output size of powerset construction \cite{nellore_et_al:LIPIcs.CPM.2021.20}, and a generalization of DAGs called $k$-funnels \cite{caceres2022parameterized}. Like in our setting, in all these cases quadratic query complexity is obtained in the worst case (on graphs maximizing the parameter under consideration).

\subsection{Organization of the Paper}

The paper is organized as follows.  In Section \ref{sec:preliminary} we give some preliminary definitions, we  formally introduce the width-based deterministic and non-deterministic hierarchies, prove some preliminary results related to them, and state all the problems to be considered  in the rest of the paper. Problems are classified as  \emph{automata}, \emph{language}, and  \emph{compression/indexing} related. In Section \ref{sec:determinization} we prove one of the strongest properties of our notion of co-lex width: this parameter yields a new upper bound to the size of the smallest DFA equivalent to a given NFA and implies new FPT analyses for hard problems on NFAs such as universality, equivalence, and membership.
In the central part of the paper  we  discuss and propose solutions to the problems defined in Section \ref{sec:preliminary}. In the first paragraphs of Section \ref{sec:det} we give a simple polynomial solution to the DFA-width problem, thus solving Problem \ref{pr:width_complexity} (NFA width problem) when the input is a DFA. In the following three subsections, we introduce tools that allow us to compute the deterministic width of a language: in particular, Section \ref{sec:entanglement} introduces the concept of \emph{entanglement} of a DFA, a measure able to capture the language's deterministic width on the minimum DFA; Section \ref{sec:Hasse automaton of L} exhibits a canonical DFA (dubbed the \emph{Hasse automaton} of a language) of minimum width, thus solving  Problem \ref{pr:hasse} (Minimum-width DFA), and gives an automata-free characterization for each level of the deterministic hierarchy, thus solving Problem \ref{pr:aut_free} (Automata-free characterization); Section \ref{sec:computing} puts together the notions developed in the two previous subsections to derive an algorithm computing the deterministic width of a language, which solves the deterministic side of Problem \ref{problem: language width} (Language width problem). 
In Section \ref{sec: relation star free} we compare our notion of deterministic width with an important class of subregular languages: the star-free languages. 
In Section \ref{sec:minimal}  we prove a Myhill-Nerode theorem for each   level  of the deterministic hierarchy, thereby providing an alternative solution to Problem \ref{pr:aut_free} (Automata-free characterization). 
In Section \ref{subsec:ind_dfa}, we consider compression and indexing problems over finite state automata. More in detail, in Section \ref{sec:path_coherence} we establish the necessary tools at the core of our data structures (in particular, the \emph{path coherence} property); in Section \ref{sec:aBWT} we present a space-efficient representation for automata (the \emph{automata BWT}, or \emph{aBWT}) preserving the accepted regular language (for any input NFA) and the automaton's topology for a strict superclass of the DFAs; in Section \ref{subsubsec:index} we augment the aBWT to obtain an index solving subpath queries on languages and NFAs; finally, in Section \ref{sec:aaBWT} we augment the aBWT to make it faithful (i.e. preserving the automaton's topology) also on NFAs.
These last contributions solve  Problems \ref{problem:encoding} (Compressing automata) and \ref{problem:indexing} (Indexing automata).
  In Section \ref{sec:conc} we draw our conclusions.

After the bibliography, an index gathers the main mathematical symbols and definitions used throughout the paper, linking them to the   location where they are defined. 
The paper is concluded with Appendix \ref{app:minimal_convex}, where we prove general results related to partitions and orders which are needed in some of our results in Section \ref{sec:Hasse automaton of L} and could be of independent interest.

Notice that in Section \ref{sec:preliminary} we present the main idea of the paper: how to (partially) order the sets of states of an automaton by lifting the co-lex order on strings. Then, Sections \ref{sec:determinization}, \ref{sec:det} and \ref{subsec:ind_dfa} can be read independently from one another: the reader interested in automata theory can focus on Sections \ref{sec:determinization} and \ref{sec:det}, and the reader interested in graph compression can focus on Section \ref{subsec:ind_dfa}.

\section{Definitions,  First Results, and  Problems}\label{sec:preliminary}

In this section we first present all basic definitions required in order to follow the paper: sequences, finite-state automata, orders, and model of computation used in our data-structures results. After giving all the necessary definitions, in Section \ref{sec:width of automaton} we introduce the core concept of our paper: the co-lex width of a finite-state automaton.
Section \ref{sub:language_width} extends the notion of co-lex width to regular languages. 
These two sections also formally introduce the problems, tackled in the next sections, related to the notion of co-lex width, and prove some preliminary results. 
To conclude, Section \ref{sec:definition indexing and compression} formally defines the problems of automata compression and indexing.

\subsection{Basics}\label{sec:basics}

\subsubsection*{Sequences}

Let $\Sigma$  \index{Sigma@$\Sigma$, alphabet}  be a finite alphabet and let  $\Sigma^*$ \index{Sigma*@$\Sigma^*$, set of finite strings} be the set of  all finite sequences (also called \emph{words} or \emph{strings}) on $\Sigma$, with $\varepsilon$ being the empty sequence. We write $\beta \dashv  \alpha$  \index{beta is a suffix of alpha@$\beta \dashv  \alpha$} if $\alpha, \beta\in \Sigma^*$ and $\beta$ is a suffix of $\alpha$.

Throughout the paper, we assume that  there is a fixed total order $ \preceq $ \index{$ \preceq $, co-lex order} on $ \Sigma $ (in our examples, the alphabetical order). When using an integer alphabet in Section  \ref{subsec:ind_dfa}, the total order will be the standard order on integers. The special symbol \index{$\#$, special symbol}  $\#\not \in \Sigma$ is considered smaller than any element in $\Sigma$. We extend  $\preceq$ to words in $\Sigma^*$ \emph{co-lexicographically}, that is, for $ \alpha, \beta \in \Sigma^* $, we have \index{alpha is smaller than beta@ $ \alpha \preceq \beta $} $ \alpha \preceq \beta $ if and only if either $ \alpha $ is a suffix of $ \beta $, or there exist $ \alpha', \beta', \gamma \in \Sigma^* $ and $ a, b \in \Sigma $, such that $ \alpha = \alpha' a \gamma $ and $ \beta = \beta' b \gamma $ and $ a \prec b $.

\subsubsection*{Finite-state automata}

A  \emph{non-deterministic} finite  automaton  (an NFA\index{NFA}) accepting strings in $ \Sigma^{*} $ is a tuple  $\mathcal N=(Q, s, \delta,F)$  where $ Q $ is a finite set of states, $ s $  is a \emph{unique} initial state, $ \delta(\cdot, \cdot): Q \times \Sigma \rightarrow \mathcal Pow(Q) $ is the  transition function  (where $\mathcal Pow(Q)$ is the set of  all subsets of $Q$), and $ F\subseteq Q $ is the set of  final states. We write $Q_{\mt N}, s_{\mt N}, \delta_{\mt N}, F_{\mt N}$ when the automaton $ \mathcal{N} $ is not clear from the context and, conversely, if the automaton is clear from the context we will not explicitly say that $ s $ is the initial state, $ \delta $ is the transition function and $ F $ is the set of final states.

  
 With $|\delta|$ we denote the cardinality of $\delta$ when seen as a set of triples over $Q \times Q \times \Sigma$. In other words, $|\delta| = |\{(u,v,a)\ |\ v\in \delta(u,a),\ u,v\in Q, a\in \Sigma\}$.   In fact, in our results (especially the data-structure related ones) we will often treat NFAs as edge-labeled graphs having as set of nodes $Q$ and set of edges $\{(u,v,a)\ |\ v\in \delta(u,a),\ u,v\in Q, a\in \Sigma\}$. 

 As customary, we extend $ \delta $ to operate on strings as follows: for all $ u \in Q, a\in \Sigma, $ and $ \alpha \in \Sigma^{*} $:
\begin{align*}
 {\delta}(u,\varepsilon)=\{u\},    \hspace{.5cm}
 {\delta}(u,\alpha a) = \bigcup_{v\in{\delta}(u,\alpha)} \delta(v,a).
\end{align*}   
 We denote by   \index{L(N)@$ \mathcal L(\mathcal N)$, language recognized by $ \mathcal{N} $} 
 $ \mathcal L(\mathcal N)=\{\alpha\in \Sigma^*: \delta(s, \alpha)\cap F\neq \emptyset\}$   the language accepted by the automaton $ \mathcal N $.
 We say that two automata are {\em equivalent} if they accept the same language. 
 
 We assume, without loss of generality,  that all states in our automata are \emph{useful}, that is, from each state one can reach a final state (possibly, the state itself). We assume also that every state is reachable from the (unique) initial state. 
Hence,  the collection of prefixes of words accepted by $ \mathcal N $, $ \text{Pref}(\mathcal L(\mathcal N))$, \index{Pref(L(N))@$ \text{Pref}(\mathcal L(\mathcal N))$, set of all prefixes of $ \mathcal{L(N)}$}  will consist of the set of  words that can be read on $ \mathcal{N} $ starting from the initial state.

 Following \cite{Alanko2021Wheeler}, we  adopt a specific notation to denote the set of words reaching a given state and to denote the set of states reached by a given word:
\begin{itemize}
\item if $u\in Q$, let $I_u$ \index{Iu@$I_u$} be the set of words \emph{reaching} $u$ from the initial state: 
\begin{align*}
    I_u &=\{\alpha \in\text{Pref}(\mathcal L(\mathcal N)): u\in \delta(s, \alpha)\};
\end{align*}
\item if $\alpha \in \text{Pref}(\mathcal L(\mathcal N))$, let $I_\alpha$ \index{Ialpha@$I_\alpha$} be the set $ \delta(s, \alpha) $ of all states   \emph{reached}   from the initial state by $\alpha$.
\end{itemize}

A {\em deterministic}  finite  automaton (a DFA), \index{DFA} is an NFA $\mt D$  where $ |\delta(u,a)|\leq 1 $,  for any $ u \in Q $ and $ a \in \Sigma $.
If the automaton is deterministic we  write  ${\delta}(u,\alpha)=v$  for  the unique $v$  such that ${\delta}(u,\alpha)=\{v\}$  (if defined: we are not assuming a DFA to be complete).

\medskip

Let $ \mathcal{L} \subseteq \Sigma^* $ be a language. An equivalence relation $ \sim $ on $ \pf L $ is \emph{right-invariant} if for every $ \alpha, \beta \in \pf L $ such that $ \alpha \sim \beta $ and for every $ a \in \Sigma $ it holds $ \alpha a \in \pf L $ iff $ \beta a \in \pf L $ and, if so, $ \alpha a \sim \beta a $. We will extensively use the \emph{Myhill-Nerode equivalence} induced by $\mathcal{L} $, namely, the right-invariant equivalence relation $\equiv_\mathcal{L}$   on $ \pf L $ such that for every $ \alpha, \beta \in \pf L $ it holds:

\begin{equation*} 
    \alpha \equiv_\mathcal{L} \beta \iff \{\gamma \in \Sigma^*\ |\  \alpha \gamma \in \mathcal{L}  \} = \{\gamma \in \Sigma^*\ |\ \beta \gamma \in \mathcal{L}  \}.
\end{equation*}

We denote by  \index{D(L)@ $\mathcal D_\mathcal{L}$, minimum DFA recognizing $ \mathcal{L} $} $\mathcal D_\mathcal{L}$ the minimum (with respect to state-cardinality) deterministic automaton recognizing a regular language $\mathcal L$. 

If $u\in Q$, then $\lambda(u)$ \index{lambda(u)@$\lambda(u)$} denotes the set of labels of edges entering $u$,
 except when $u=s$ when we also add  $\#\not \in \Sigma$ to $\lambda(s)$, with $\# \prec e$ for all $e\in \Sigma$ (see e.g. Figure \ref{fig:1-colex} where $s=0$, $\lambda(s)=\{\#, a\}, \lambda(1)=\{a\}$, and $\lambda(5)=\{b,c\}$). If $u\in Q$, by  \index{min(lambda(u))@$\text{min}_{\lambda(u)}$} $\text{min}_{\lambda(u)}$,  \index{max(lambda(u))@$\text{max}_{\lambda(u)}$} $\text{max}_{\lambda(u)}$ we denote the minimum and the maximum, with respect to the order $\preceq$,   among the elements in $\lambda(u)$.

\subsubsection*{Orders}
A  {\em (non-strict) partial order} \index{partial order} is a pair $ (Z, \le) $,   where     $ \le $ is a reflexive ($u \le u$ for all $u\in Z$), transitive ($u \le v$ and $v \le w$ implies $u \le w$, for all $u,v,w \in Z$), and antisymmetric ($u \le v$ and $v \le u$ implies $u = v$) binary relation on $ Z $. 

Two elements $  u, v \in Z $ are said to be \index{comparable@$ \le $-{comparable}} $ \le $-{\em comparable} if either $ u \le v $ or $ v \le u $. A  partial order  can also be described using the relation  $ u < v $ which holds when  $ u \le v $ and $ u \not = v $.  
We write     $u ~\|~ v$    if $ u $ and $ v $ are not  $\leq$-comparable.
 
If $(Z, \le)$ is a partial order and $Z'\subseteq Z$  we denote by  \index{Z'@$\le_{Z'}$, restriction of $ \le $ to $ Z' $}
$(Z', \le_{Z'})$ the restriction of the partial order $\le$ to the set $Z'$. To simplify notation, we will also  use 
$(Z', \leq)$ when clear from the context.

A partial order $ (Z, \le) $ is a {\em total order} if for all $ y, z \in Z $, $ y $ and $ z $  are $  \le $ -comparable.

A subset $ Z' \subseteq Z $ is a \index{chain@$\le$-{chain}} $\le$-{\em chain} if $ (Z', \le) $ is a total order. A family $ \{Z_i \; |\; 1\leq i \leq p\} $ is a \index{chain partition@$\le$-{chain partition}}  $\le$-{\em chain partition}  of $Z$ if $ \{Z_i \; |\; 1\leq i \leq p\} $ is a partition of $ Z $ and each $ Z_i $ is a $\le$-chain. 

The {\em  width} of a partial order  $ (Z, \le) $, denoted by \index{width@$\text{width} (\le)$ }  $ \text{width} (\le) $, is the smallest cardinality of a $ \le$-chain partition.  A subset  $ U \subseteq Z $ is a \index{antichain@$\le$-{antichain}}  $\le$-{\em  antichain} if any two distinct elements in $ U $ are not $ \le $-comparable. 
Dilworth's Theorem \cite{dilworth} states that the width of $ (Z, \le) $ is  equal to the cardinality of a largest $\le$-antichain.

 If $A,B$ are \emph{disjoint } subsets of a partial order $ (Z, \le) $, then  $A<B$ \index{A<B@$A<B$} denotes: \[(\forall a \in A)(\forall b \in B) (a<b).\]  
 
A  \index{monotone sequence}{\em  monotone sequence} in (a partial order) $(Z, \le)$ is a sequence $(v_n)_{n\in \mathbb N}$ with $v_n\in Z$ and either $v_i\le v_{i+1}$, for all $i\in \nats$, or  $v_i\ge v_{i+1}$, for all $i\in \nats$. 

 A   subset  $C $ of a partial order $(V, \le)$  is   \index{convex@$\le$-{convex}}  $\le$-{\em  convex}   if     for every $ u, v, z \in V $ with  $ u, z \in C $ and $ u < v < z $ we have  $ v \in C $.
We drop  $\leq$ when clear from the context.

If $\alpha\preceq  \alpha'\in \Sigma^*$, we define \index{alphaalpha'@$[\alpha, \alpha']$}
$[\alpha, \alpha']=\{\beta: \alpha\preceq \beta \preceq \alpha'\}$;  if the relative order between $\alpha, \alpha'$ is not known,  we set 
\index{alphaalpha'+-@$[\alpha, \alpha']^{\pm} $}
$[\alpha, \alpha']^{\pm} =
[\alpha, \alpha']$, if $\alpha \preceq  \alpha'$, while  $[\alpha, \alpha']^{\pm}= 
[\alpha', \alpha]$,  if $\alpha' \preceq  \alpha$.

\subsubsection*{Other assumptions}

Throughout the paper, $[a,b]$ with $a, b\in \mathbb N$ and 
$a\leq b$ denotes the integer set $\{a,a+1,\dots, b\}$. If $b<a$, then $[a,b] = \emptyset$. All logarithms used in the paper are in base 2.

 \subsection{The {Co-lex Width} of an Automaton}\label{sec:width of automaton}
 
The notion of ordering stands at the core of the most successful compression and pre-processing algorithmic techniques: integer sorting and suffix sorting are two illuminating examples. 
This concept is well-understood in the case of strings (where the co-lexicographic order of the string's prefixes or the lexicographic order of the string's suffixes are typically used)  and has been generalized to a special class of subregular languages (the Wheeler languages)  in \cite{GAGIE201767}.  The goal of this section is to provide a generalization of the co-lexicographic (\emph{co-lex} for brevity) order  among prefixes  of a word    to an order among  the states of an  NFA recognizing     \emph{any} regular language, by imposing axioms on the accepting NFA. This will allow us to generalize powerful compression and indexing algorithmic techniques from strings to arbitrary regular languages and NFAs, as well as proving relations between the sizes of NFAs and DFAs recognizing a given language. 
 
 We capture co-lex orders on an NFA by means of two axioms which ensure a \emph{local} comparability between pairs of states.  Given two states $u$ and $v$ such that $u<v$, Axiom 1 of Definition \ref{def:colex} imposes that all words in $ I_u $ end with letters being smaller than or equal to letters ending words in $ I_v $;   Axiom 2, instead, requires that the order among states $u$ and $v$ propagates backwards when following pairs of equally-labelled transitions. 
 These two axioms generalize to NFAs the familiar notion of prefix sorting: if the NFA is a simple path --- i.e., a string ---, this order reduces to the well-known co-lexicographic order among the string's prefixes.

\begin{definition}\label{def:colex}
Let $\mathcal N = (Q, s, \delta, F) $ be an NFA. A \index{co-lex order} \emph{ co-lex order} on $ \mathcal N $ is a partial order $ \le $ on $ Q $ that satisfies the following two axioms:
\begin{enumerate}
    \item (Axiom 1) For every $ u, v \in Q $, if $ u < v $, then $ \text{max}_{\lambda(u)} \preceq \text{min}_{\lambda(v)} $;
    \item (Axiom 2) For every $ a \in \Sigma $ and $ u, v, u', v' \in Q $, if $ u \in \delta (u', a) $, $ v \in \delta (v', a) $ and $ u < v $, then $ u' \le v' $.
\end{enumerate}
\end{definition}

\begin{remark}\label{rem:initial} 

\

\vspace{-.3cm}

\begin{enumerate} 
\item    Since $ \# \in \lambda (s) $ and $ \# \not \in \lambda (u) $ for $u\neq s$, then from Axiom 1 it follows that for every $ u \in Q $ it holds $ u \not < s $.
\item  If $ \mathcal{D} $ is a DFA, then we can restate Axiom 2 as follows: for every $ a \in \Sigma $, if $  u= \delta (u', a) $, $ v = \delta (v', a) $, and $ u < v $, then $ u' < v' $ (it must be $ u' \not = v' $ because $ u $ and $ v $ are distinct).
    
\end{enumerate} 

\end{remark}

When $ \le $ is a total order, we say that the co-lex order $\le$ is a \index{Wheeler order} \emph{Wheeler order}.  Wheeler orders were first introduced in \cite{GAGIE201767} in a slightly less general setting \footnote{More in detail: they required the set $\lambda(u)$ of labels entering in state $u$ to be a singleton for all $u\in Q$ (\emph{input consistency} property). In this paper we drop this requirement and work with arbitrary NFAs.}. The class of \index{Wheeler languages} Wheeler languages --- that is, the class of all regular languages recognized by some Wheeler NFA --- is rather small: for example, unary languages are Wheeler only if they are finite or co-finite, and all Wheeler languages are star-free (see \cite{Alanko2021Wheeler}). Moreover, Wheeler languages are not closed under union, complement, concatenation, and Kleene star \cite{Alanko2021Wheeler}. In contrast, as observed in the following remark, any regular automaton admits a co-lex order. 

\begin{remark}
Every NFA $ \mathcal{N} $ admits some co-lex order. For example, the   order $ \{(u, u)\ |\ u \in Q \} $ and  the order   $ \{(u, v)\ |\ \text{max}_{\lambda(u)} \prec \text{min}_{\lambda(v)}   \} \cup \{(u, u)\ |\ u \in Q \}  $ are co-lex orders on $ \mathcal{N} $.
\end{remark}

We note that Axiom 2 implies that the order between two states is  not defined whenever their predecessors cannot be unambiguously compared, as observed in the following remark.

\begin{remark} 
Let $\mathcal N = (Q, s, \delta, F) $ be an NFA and let $ \leq $ be a co-lex order on $ \mathcal N $. Let $ u, v \in Q $ be two distinct states. Then, $ u~\|~v $ if at least one of the following holds:
\begin{enumerate}
    \item There exist $ u', v' \in Q $ and $ a \in \Sigma $ such that $ u \in \delta (u', a) $, $ v \in \delta (v', a) $ and $ u'~\|~v' $.
    \item There exist $ u', v', u'', v'' \in Q $ and $ a, b \in \Sigma $ such that $ u \in \delta (u', a) \cap \delta (u'', b) $, $ v \in \delta (v', a) \cap \delta (v'', b) $, $ u' < v' $ and $ v'' < u'' $.
\end{enumerate}
Indeed, if e.g. it were $ u < v $, then Axiom 2 would imply that in case 1 it should hold $ u' \le v' $ and in case 2 it should hold $ u'' \le v'' $ (which is forbidden by the antisymmetry of $\leq$).
\end{remark}

\begin{figure*}[h!]
\centering
\begin{subfigure}{.30\textwidth}
	\centering
	\begin{tikzpicture}[shorten >=1pt,node distance=1.6cm,on grid,auto]
	\tikzstyle{every state}=[fill={rgb:black,1;white,10}]
	
	\node[state,initial]   (q_0)                    {$0$};
	\node[state ] (q_2)  [right of=q_0]    {$2$};
		\node[state]           (q_1)  [above   of=q_2]    {$1$};
	\node[state]           (q_3)  [below   of=q_2]    {$3$};
	\node[state,accepting] (q_4)  [right of=q_2]    {$4$};	
	\node[state,accepting]           (q_5)  [right of=q_3]    {$5$};
 
	\path[->]
	(q_0) edge node {a}    (q_1)
	(q_0) edge  [loop below] node {a} (q_0)
	(q_0) edge node {b}    (q_2)
		(q_0) edge node {b}    (q_3)
	(q_1) edge node {b}    (q_4)
	(q_2) edge node {b}    (q_4)
		(q_3) edge node {b,c}    (q_5);
	\end{tikzpicture}
\end{subfigure}
\begin{subfigure}{.20\textwidth}
	\centering
	\begin{tikzpicture}[shorten >=1pt,node distance=0.7cm,on grid,auto]
	\tikzstyle{every state}=[fill={rgb:black,1;white,10}]
	
	\node[state,color=white,text=black,inner sep=1pt,minimum size=0pt] (q_0)                   {$0$};
	\node[state,color=white,text=black,inner sep=1pt,minimum size=0pt] (q_1)  [above of=q_0]   {$1$};
	\node[state,color=white,text=black,inner sep=1pt,minimum size=0pt] (q_2)  [above of=q_1]   {$2$};
	\node[state,color=white,text=black,inner sep=1pt,minimum size=0pt] (q_3)  [above of=q_2]   {$3$};
	\node[state,color=white,text=black,inner sep=1pt,minimum size=0pt] (q_4)  [above of=q_3]   {$4$};	
	\node[state,color=white,text=black,inner sep=1pt,minimum size=0pt] (q_5)  [above   of=q_4]   {$5$};

	\path[-]
	(q_0) edge node {}    (q_1)
	(q_1) edge node {}    (q_2)
	(q_2) edge node {}    (q_3)
	(q_3) edge node {}    (q_4)
(q_4) edge node {}    (q_5);
	\end{tikzpicture}
\end{subfigure}
\begin{subfigure}{.20\textwidth}
	\centering
	\begin{tikzpicture}[shorten >=1pt,node distance=0.7cm,on grid,auto]
	\tikzstyle{every state}=[fill={rgb:black,1;white,10}]
	
	\node[state,color=white,text=black,inner sep=1pt,minimum size=0pt] (r_0)                   {$0$};
	\node[state,color=white,text=black,inner sep=1pt,minimum size=0pt] (r_1)  [above of=r_0]   {$1$};
	\node[state,color=white,text=black,inner sep=1pt,minimum size=0pt] (r_3)  [above of=r_1]   {$3$};
	\node[state,color=white,text=black,inner sep=1pt,minimum size=0pt] (r_2)  [above of=r_3]   {$2$};
	\node[state,color=white,text=black,inner sep=1pt,minimum size=0pt]
	(r_5)  [above   of=r_2]   {$5$};
	\node[state,color=white,text=black,inner sep=1pt,minimum size=0pt] 	(r_4)  [right  of=r_5]   {$4$};

	\path[-]
	(r_0) edge node {}    (r_1)
	(r_1) edge node {}    (r_3)
	(r_3) edge node {}    (r_2)
	(r_2) edge node {}    (r_5)
	(r_2) edge node {}    (r_4);
	\end{tikzpicture}
\end{subfigure}

 \caption{An NFA and the Hasse diagrams --- that is, graphs depicting the \emph{transitive reductions} of the partial orders --- of two of its  (maximal)   co-lex orders. Characters are sorted according to the standard alphabetical order.} \label{fig:1-colex}
\end{figure*}
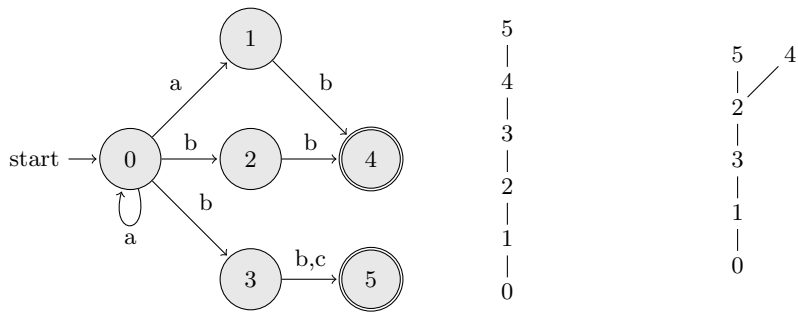

More than one non-trivial co-lex order can be given on the same automaton. As an example, consider Figure \ref{fig:1-colex}: the automaton on the left admits the two co-lex orders whose Hasse diagrams are depicted on the right.  The first, $\leq_1$,  is total and 
states that $2<_1 3$, while the width of the second one,  $\leq_2$, is equal to $2$ and   
$3<_2 2$ holds. As a matter of fact, in any co-lex order $\leq$ for this automaton in which   $3< 2$ holds,  nodes $4$ and $5$ must be incomparable.

In  Section \ref{subsec:ind_dfa} we will prove that a co-lex order over an automaton  enables compression and indexing mechanisms whose efficiency is parameterized by the width of the co-lex order  (the smaller, the better): this  justifies introducing the  \emph{co-lex width} of an NFA (Definition \ref{def:width}) as a meaningful measure for compression and indexing. In fact, the co-lex width can  also be used for further,  interesting, language-theoretic consequences --- more on this in Sections \ref{sec:determinization} and \ref{sec:det} for the deterministic case and in \cite{parttwo} for the non-deterministic case. 
 
\begin{definition} \label{def:width} The \emph{co-lex  width}   of an    NFA $\mt N$    is the  minimum  width of a co-lex order on $\mt N$: \index{width(N)@$\text{width}(\mt N)$} 
\[ \text{width}(\mt N)=min \{\text{width}(\leq)\mid \leq  \text{is a co-lex order on $\mt N$}\}\]
\end{definition}

In Example \ref{ex:depedent} below  we shall see that the value $ \text{width}(\mt N) $ may depend  on the choice of the total order $ \preceq $ on $ \Sigma $. 

As a matter of fact, string sorting stands at the core of the most popular  string compression and indexing paradigms, which for this reason also suffer from a sharp dependence on the total alphabet order. For example, the number $ r $ of equal-letter runs of the Burrows-Wheeler transform (BWT) of a string \cite{burrows1994block} is an important string compressibility parameter (see \cite{gagieJACM}) and its value depends on the choice of the total order on the alphabet; deciding whether there exists an ordering of the alphabet of a string such that $ r $ is bounded by a given value, however, is an NP-complete problem \cite{bentley2020}. Despite this limitation, the BWT and the data structures based on it --- such as the FM-index \cite{ferragina2000opportunistic} and the r-index \cite{gagieJACM}  --- are widely used in applications with a fixed (often sub-optimal) alphabet order.

Similarly, in  our scenario  it is natural to wonder whether it is possible to determine an ordering of the alphabet that minimizes the width of an automaton. Unfortunately, also this problem is not tractable: deciding whether there exists a total alphabet order under which a given DFA is Wheeler (that is, it has co-lex width equal to one) is already an NP-complete problem \cite{dagostino2021ordering}. 
In such situations, one possible way to tame the problem's complexity is to study a more constrained version of the problem, with the goal of shedding new light on the more general (unconstrained) scenario. 
For this reason, in this paper we start by fixing a total order on the alphabet and 
investigating the implications of this choice. In particular we will prove that, if we fix an order on the alphabet, then 
the width of a DFA with respect to that order can be determined in polynomial time (see Corollary \ref{cor:complexity width dfa}). This finding can already be used, for example, as a black-box to test candidate alphabet orderings in search for the one minimizing the automaton's width (more advanced search strategies will be the subject of forthcoming works, bearing in mind that the optimal solution is NP-hard to find).

In the following, we establish preliminary useful properties of the new measure $\text{width}(\mt N)$. Our first observation is that this measure is linked with the graph's \emph{sparsity}.

\begin{lemma}\label{lem: p sparsity}
Let $\mt N$ be an NFA on an alphabet of cardinality $\sigma$ with $n$ states and $|\delta|$ transitions, and let $p=\text{width}(\mt N)$. Then:
$$
|\delta| \leq (2n - p) p \sigma.
$$
\end{lemma}

\begin{proof}
    Let $\mt N$ be an NFA on an alphabet of cardinality $\sigma$ with $n$ states and $|\delta|$ transitions, and let $p=\text{width}(\mt N)$. Let $ \le $ be a co-lex order of width $p$, $ \{Q_i \;|\; 1 \le i \le p \} $ be a $ \le $-chain partition of the set of states, and  for all $1\leq i \leq p$ let $Q_i = \{v_{i,1},\dots, v_{i,n_i}\}$, where $n_i = |Q_i|$ and $v_{i,k} < v_{i,k'}$ whenever $k<k'$.
    Fix $1\leq i,j \leq p$ and $a\in \Sigma$, and consider all the transitions $(v_{i, k}, v_{j, l}, a)$ labeled with $a$ that leave $Q_i$ and reach $Q_j$. We denote with $e_{i,j,a}$ the number of such transitions; our goal is to establish an upper bound to this quantity for all $i,j,a$. 
    Sort these $e_{i,j,a}$ edges $(v_{i,k}, v_{j, l}, a)$ by the index $ l $ of their destination state, breaking ties by the index $ k $ of their source state. Now, let us prove that the value $ k  + l $ is strictly increasing with respect to this order. In order words, we want to prove that if we pick two edge $(v_{i, k}, v_{j, l}, a)$ and $ (v_{i, k'}, v_{j, l'}, a) $ being consecutive with respect to the edge order, then $ k + l < k' + l' $. By the definition of the edge order, we have $ l \le l' $. If $ l = l' $, again by the definition of the edge order we have $ k < k' $ and so $ k + l < k' + l' $. If $ l < l' $, by Axiom 2 of co-lex orders we have $ k \le k' $, and again we conclude $ k + l < k' + l' $. Since we have proved that $ k  + l $ is strictly increasing with respect to the edge order, then from $2\leq k + l \leq n_i+n_j$ we obtain $e_{i,j,a} \leq n_i+n_j-1$. Observing that $\sum_{i=1}^p n_i = n$, we conclude:
    $$|\delta| = \sum_{a\in \Sigma}\sum_{i=1}^p\sum_{j=1}^p e_{i,j,a}
    \leq \sum_{a\in \Sigma}\sum_{i=1}^p\sum_{j=1}^p (n_i+n_j-1) = 2\sigma p n - \sigma p^2 = (2n - p) p \sigma. $$ \qed
\end{proof}

The above lemma will be useful later, when measuring the size of our NFA encodings as a function of the number of states. Note that Wheeler automata ($p=1$) have a number of transitions proportional to $O(\sigma n)$. This relation was already noted in the literature \cite[Thm. 4]{gibney2022complexity}.

\medskip
   
Next, we move on to studying some preliminary properties of the smallest-width co-lex order.
We say that $ \le^* $ is a   \emph{refinement}   of $ \le $ if, for all  $ u, v \in Q$, $u \le v $ implies $ u \le^* v$. Since there are only finitely many co-lex orders over  an automaton,   every co-lex order $ \le $ is maximally refined by a   co-lex order.
Moreover, 
if $ \le^* $ is a refinement of $ \le $, then it must be  that $ \text{width} (\le^*)$ is less than or equal to  $\text{width} (\le) $, since every $ \le $-chain partition is also a $ \le^* $-chain partition. This implies that there is always a \emph{maximal} co-lex order $ \le $ on  an NFA $ \mathcal{N} $ such that $ \text{width} (\mathcal{N}) = \text{width} (\le) $. In general  an NFA admits several maximal co-lex orders of  different widths. For example, the two co-lex orders presented  in Figure \ref{fig:1-colex} are both maximal and have different widths.   
This cannot happen over DFAs: in the following lemma we prove that a   DFA    always admits  a unique maximal co-lex order (the \emph{maximum} co-lex order) so that  this order realizes the width of the DFA. In particular, the maximum co-lex order refines every co-lex order on the DFA. This simplifies the search for a   co-lex order realizing the width of the automaton and, indeed, in Lemma \ref{lem:complexity prec DFA}  and  Corollary  \ref{cor:complexity width dfa} we prove that such a co-lex order can be determined in polynomial time.

   \begin{definition}\label{def:prec_DFA} Let $\mathcal D$ be a DFA. The relation $<_{\mathcal D}$ over $Q$  is  defined by:
\[u <_{\mathcal D} v \text{ if and only if } (\forall \alpha \in I_u) ( \forall \beta \in I_v) ~(\alpha \prec \beta).    \]
\end{definition}

One can easily prove that $\leq_{\mathcal D}$ (that is, ${<_{\mathcal D}}\cup\{(u,u) \; |\; u\in Q\}$)   is a partial order over $Q$. Moreover:

\begin{lemma}\label{lem:maximumorderDFAs}
If  $\mathcal D$ is a DFA then   $(Q,\leq_{\mathcal D} )$ is the maximum co-lex order on $\mathcal D$.
\end{lemma}
\begin{proof}
First, let us prove that $ \leq_\mathcal{D} $ is a co-lex order on $ \mathcal{D} $.     

To see that  Axiom 1 holds assume that $ u <_\mathcal{D} v $: we must prove that $ e=\text{max}_{\lambda (u)} \preceq e'=\text{min}_{\lambda (v)} $. Notice that it must be $ v \not = s $ because the empty string $\varepsilon$ is in $ I_s $ and $\varepsilon$ is co-lexicographically smaller than any other string. Hence, $ e'\succ\#$ and if $ e= \#$ we are done. Otherwise, there are  $\alpha e \in I_u $ and $ \alpha' e' \in I_{v} $, so that $ u <_\mathcal{D} v $ implies $ \alpha e \prec \alpha' e' $ and therefore $ e \preceq e' $. 
As for Axiom 2, assume that $ u \in \delta (u', a) $, $ v \in \delta (v', a) $, and $ u <_\mathcal{D} v $. We must prove that $ u' <_\mathcal{D} v' $. Fixing $ \alpha \in I_{u'} $ and $ \beta \in I_{v'} $, we must prove that $ \alpha \prec \beta $. We have $ \alpha a \in I_u $ and $ \beta a \in I_{v} $, hence from $ u <_\mathcal{D} v $ it follows $ \alpha a \prec \beta a $, and therefore $ \alpha \prec \beta $.

Let us now prove that $ \leq_\mathcal{D} $ is the maximum co-lex order.  

Suppose, reasoning for contradiction, that  $ \leq  $ is a co-lex order on $ \mathcal{D} $ and for some distinct $u,v \in Q$, $ u < v $, and   $ u \not <_\mathcal{D} v $. Then, there exist $\alpha \in I_u, \beta\in I_v$ with $\beta \prec \alpha$. Let us fix $ u $, $ v $, $ \alpha $ and $ \beta $ with the above properties such that $ \beta $ has the minimum possible length. Notice that $\beta$ cannot be the empty word, otherwise $v$ would be the initial state $s$, while $z\not < s$ for all $z\in Q$ (see Remark \ref{rem:initial}). Hence, $\beta=\beta' e$ for some $e\in \Sigma$  and $\beta\prec \alpha$ implies $\alpha=\alpha'f$ for some $f\in \Sigma$. We then have $e\in \lambda(v), f \in \lambda(u)$, and by    Axiom 1 of  co-lex orders   we get $f\preceq e$.  From  $\beta=\beta'e\prec \alpha=\alpha'f$ we conclude $f=e$ and $\beta'\prec \alpha'$.   If  $u',v'$ are  such that $\delta(u',e)=u, \delta(v', e)=v$  and $\alpha'\in I_{u'}, \beta'\in I_{v'}$, then by   Axiom 2 of  co-lex orders and Remark \ref{rem:initial} we get $u'<v'$; however,   the pair $\alpha',\beta'$  witnesses $u'\not \leq_\mathcal{D}v'$, 
 contradicting the minimality of $ \beta $. \qed
\end{proof}

Having proved that $\leq_{\mt D}$ is the  maximum co-lex order over  $\mt D$,   we immediately deduce that  its characterizing property is satisfied by \emph{any} co-lex order. 

\begin{corollary}\label{cor:axiom1DFA} If $\leq$ is a
co-lex order over a DFA    and  $u<v$   then $ (\forall \alpha \in I_u) ( \forall \beta \in I_v) ~(\alpha \prec \beta)$.
\end{corollary} 

\begin{remark}The previous corollary shows that 
Axiom 1 of  co-lex orders   \emph{propagates} from a
\emph{local} level (i.e. letters in $\lambda(u), \lambda(v)$, for which it holds $(\forall e
\in \lambda(u))(\forall f \in \lambda(v)) (e\preceq f)$) to a \emph{global} one (i.e. words in
$I_u, I_v$, for which it holds $(\forall \alpha \in I_u)(\forall \beta \in I_v)
(\alpha\preceq \beta)$).
This works for DFAs  because different states are reached by disjoint sets of words: if $u\neq v$ then $I_u\cap I_v= \emptyset$.  On NFAs things become more complicated   and  the existence of  a  maximum co-lex order is no longer guaranteed.   
\end{remark}

Lemma \ref{lem:maximumorderDFAs} established that a DFA $\mt D$ always  has a maximum co-lex order $\leq_{\mathcal D}$. This lemma will be used in  Section \ref{sec:determinization} to  prove that  
both the   cardinality and the   width of the automaton resulting from the powerset construction applied to any NFA $\mt N $  are fixed-parameter linear in $|\mt N|$ with parameter $\text{width}(\mathcal N)$. Since $\leq_{\mathcal D}$ extends any possible co-lex order on $Q$, it realizes the width of the automaton $\mt D$, as stated in the following lemma.

\begin{lemma}\label{lem:widthDFA} If $\mt D$ is a DFA then 
$\text{width}(\leq_{\mathcal D}) = \text{width}({\mathcal D})$.
\end{lemma}

With the next example we show that the 
co-lex width of an automaton
may depend on the total order on the alphabet.

\begin{example}\label{ex:depedent}
Let $ \mathcal{D} $ be a DFA. 
Let us show that, in general, the value $ \text{width}(\mathcal{D})=\text{width}(\leq_{\mathcal D})  $ (Lemma \ref{lem:widthDFA}) may depend on the total order $ \preceq $ on the alphabet.
Let $ \mathcal{D}$ be the DFA in Figure \ref{dependent}.

First, assume that $ \preceq $ is the standard alphabetical order such that $ a \prec b \prec c \prec d $. We have $ q_1 <_\mathcal{D} q_2 <_\mathcal{D} q_4 $ and $ q_1 <_\mathcal{D} q_3 <_\mathcal{D} q_4 $, so $\text{width}(\leq_{\mathcal D}) $ is at most two. Notice that $ q_2 $ and $ q_3 $ are not $ \le_\mathcal{D} $-comparable because $ acc, ac \in I_{q_2} $, $ bc \in I_{q_3} $ and $ ac \prec bc \prec acc $, so $\text{width}(\leq_{\mathcal D}) $ is equal to two.

Next, assume that $ \preceq $ is the total order such that $a\prec c \prec b\prec d$. Then, $ q_1 <_\mathcal{D} q_2 <_\mathcal{D} q_3 <_\mathcal{D} q_4 $, hence $\text{width}(\leq_{\mathcal D}) $ is equal to one.
\end{example}
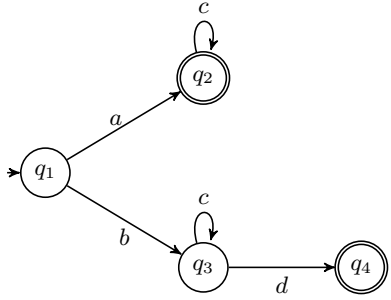
\begin{figure}[ht]%
    \centering
\begin{tikzpicture}[->,>=stealth', semithick, initial text={}, auto, scale=.42]
 \node[state, minimum size=2pt, initial] (0) at (0,0) {$q_1$};
 \node[state, label=above:{}, minimum size=2pt, accepting] (1) at (5,3) {$q_2$};
 \node[state, label=above:{}, minimum size=2pt] (2) at (5,-3) {$q_3$};
 \node[state, label=above:{}, minimum size=2pt, accepting] (3) at (10,-3) {$q_4$};
\draw (0) edge [above] node [above, xshift=-3, yshift=-3] {$a$} (1);
\draw (0) edge [below] node [below] {$b$} (2);
\draw (1) edge [loop above] node {$c$} (1);
\draw (2) edge [loop above] node {$c$} (2);
\draw (2) edge [above] node [below] {$d$} (3);
\end{tikzpicture}
 \caption{A DFA $ \mathcal{D} $ where the value $ \text{width}(\mathcal{D}) $ depends on the total order $ \preceq $ on the alphabet.}%
    \label{dependent}
\end{figure}

We now     generalize   Corollary \ref{cor:axiom1DFA} to NFAs, coping with the fact that, on NFAs, sets $I_u, I_v$ may intersect for $u\neq v$ (we shall use this result in Sections \ref{sec:determinization} and  \ref{subsec:ind_dfa}).      

 \begin{lemma}\label{lem:monot}
 Let $\mathcal N = (Q, s, \delta, F) $ be an NFA and let $ \le $ be a co-lex order on $ \mathcal{N} $. If $ u < v $, then $ (\forall \alpha \in I_u)(\forall \beta \in I_v)(\{\alpha, \beta \} \not \subseteq I_u \cap I_v \implies \alpha \prec \beta) $.
 \end{lemma}

 \begin{proof}
       Let $ \alpha \in I_u $ and $ \beta \in I_v $ such that $ \{\alpha, \beta \} \not \subseteq I_u \cap I_v $. We must prove that $ \alpha \prec \beta $. Let $ \gamma \in \Sigma^* $ be the longest string such that  $ \alpha = \alpha' \gamma $ and $ \beta = \beta' \gamma $, for some $ \alpha', \beta' \in \text{Pref}(\mathcal{L(N)}) $. If $ \alpha' = \varepsilon $ the claim follows, therefore we can assume $ |\alpha' | \ge 1 $.
        
      Let $ \gamma = c_p \dots c_1 $, with $ c_i \in \Sigma $ for $ i \in \{1, \dots, p \} $ ($ p \ge 0 $), $ \alpha' = a_q \dots a_1 $, with $ a_i \in \Sigma $ for $ i \in \{1, \dots, q \} $ ($ q \ge 1 $), and $ \beta' = b_r \dots b_1 $, with $ b_i \in \Sigma $ for $ i \in \{1, \dots, r \} $ ($ r \ge 0 $).
        
      Assume $ |\gamma| > 0 $. Since $ \alpha \in I_u $ and $ \beta \in I_v $, then there exist $ u_1, v_1 \in Q $ such that $ \alpha' c_p \dots c_2 \in I_{u_1} $, $ \beta' c_p \dots c_2 \in I_{v_1} $, $ u \in \delta (u_1, c_1) $ and $ v \in \delta (v_1, c_1) $. By Axiom 2, we obtain $ u_1 \le v_1 $. However, it cannot be $ u_1 = v_1 $ because this would imply $ \{\alpha, \beta \} \subseteq I_u \cap I_v $, so it must be $ u_1 < v_1 $. By iterating this argument, we conclude that there exist $ u', v' \in Q $ such that $ \alpha' \in I_{u'} $, $ \beta' \in I_{v'} $ and $ u' < v' $, and the same conclusion holds  if $\gamma=\varepsilon$ as well.
        
      Now, it cannot be $ r = 0 $ because this would imply $ v' = s $, and $ u' < s $ contradicts Remark \ref{rem:initial}. Hence, it must be $ |\beta| \ge 1 $. By Axiom 1, it must be $ a_1 \preceq b_1 $. At the same time, the definition of $ \gamma $ implies that it cannot be $ a_1 = b_1 $, so we obtain  $ a_1 \prec b_1 $ and we can conclude $ \alpha \prec \beta $.
      \qed
 \end{proof}

Lemma \ref{lem:monot} has an important implication (which we will use later in Theorem \ref{theor:powersetcrucial} and will stand at the core of the encoding and indexing results in Section \ref{subsec:ind_dfa}): given a co-lex order on an NFA, then the sets  $I_\alpha$'s are convex w.r.t this order.

\begin{corollary}\label{cor:intervals}
 Let $\mathcal N = (Q, s, \delta, F) $ be an NFA and let $ \le $ be a co-lex order on $ \mathcal{N} $. 
 If $\alpha \in  \text{Pref}(\mathcal L(\mathcal N))$ then $I_\alpha$ is convex in $(Q,\leq)$.
\end{corollary}
\begin{proof}
Suppose $u,z \in I_\alpha$    and let $v\in Q$ be  such that $u<v<z$. We have to prove that $v\in I_\alpha$. If this were not true, we would have $\alpha   \in (I_u \cap I_z) \setminus I_v $. Consider any $\beta\in I_v $. By Lemma \ref{lem:monot} we would have $\alpha \prec \beta \prec \alpha$, a contradiction.\qed
\end{proof}

 \medskip
 
The notion of co-lex width of an automaton naturally calls into play the problem of determining the complexity of computing this measure. More precisely, we  define the  \emph{NFA-width problem}   as follows:

\begin{Problem}[NFA width problem] \label{pr:width_complexity}   Given an NFA $ \mathcal{N} $ and an integer $ p $, determine whether: \[\text{width} (\mathcal{N}) \leq  p. \]
\end{Problem}
 
Problem \ref{pr:width_complexity} will be tackled for the particular case of DFAs in Section  \ref{sec:det}.

A natural question is whether there is a connection between the notion of width and the complexity of regular expressions. The case $\text{width}(\mt N) = 1$ corresponds to the class of Wheeler automata \cite{GAGIE201767}.  In \cite{Alanko2021Wheeler}, it was shown that \emph{Wheeler languages}, that is, regular languages recognized by Wheeler automata, are closed essentially only under intersection and under concatenation with a finite language. On the other hand, with the next two remarks we point out that our notion of width can easily be used to capture complementation, intersection and union.

\begin{remark}\label{rem:negation}
    Let $\mathcal D$ be a DFA. Then, there exists a DFA $\mt{ D'}$ such that $\mt L(\mt D') = \Sigma^* \setminus \mt L(\mt D)$ and $\text{width}(\mt D') \le \text{width}(\mt D) + 1 $. The DFA $\mathcal D'$ is obtained by first transforming $ \mathcal{D} $ into a complete DFA by adding a non-final sink state that is reached by all transitions not defined in $ \mt D $ (including the ones leaving the sink itself), then switching final and non-final states, and finally removing all states that neither are final, nor allow to reach a final state. In the worst case, the new sink state is not $\le_{\mt D'}$-comparable with any other state and therefore the width cannot increase by more than one.
\end{remark}

\begin{remark}\label{rem:intersection}
    Let $\mt D_1= (Q_1,s_1, \delta_1, F_1),\mt D_2  = (Q_2,s_2, \delta_2, F_2)$ be DFAs, with $\text{width}(\mt D_1) = p_1$ and $\text{width}(\mt D_2) = p_2$. Let us prove that there exists a DFA $\mathcal D$ such that $\mt L(\mt D) = \mt L(\mt D_1) \cap \mt L(\mt D_2)$ and $\text{width}(\mt D) \leq p_1\cdot p_2$ and a DFA  $\mathcal D'$   such that $\mt L(\mt D') = \mt L(\mt D_1) \cup \mt L(\mt D_2)$ and $\text{width}(\mt D') \leq p_1\cdot p_2+ p_1+p_2$. In the following, we will implicitly use Lemma \ref{lem:widthDFA}. Let $\{Q_1, \dots, Q_{p_1}\}$ and $\{Q^*_1, \dots, Q^*_{p_2}\}$ be a $ \le_{\mt D_1}$-chain partition of $ Q_1 $ and a $ \le_{\mt D_2}$-chain partition of $ Q_2 $, respectively. 
    In order to build both $ \mt D $ and $ \mt D' $, we first turn $ \mathcal{D}_1 $ and $ \mathcal{D}_2 $ into complete DFAs by adding non-final sinks $\star_1, \star_2$ to $Q_1$ and $Q_2$, respectively (like in Remark \ref{rem:negation}), then we build the standard product automaton: the set of states is $(Q_1\cup\{\star_1\})\times (Q_2\cup \{\star_2\})$, the initial state is $(s_1,s_2)$, and the transition function is defined by $\delta((u,v),a)=( \delta_1(u,a), \delta_2(v,a))$, for every state $ (u, v) $ and for every $ a \in \Sigma $. The difference between $\mathcal D$ and $ \mathcal{D}'$ lies in how the set of final states is defined.

 We define $ \mathcal{D} $ by letting $F=F_1\times F_2$ being the set of all final states, and then removing (i) all states that are not reachable from the initial state and (ii) all states that neither are final, nor allow to reach a final state. Let $ \mathcal{D} = (Q, s, \delta, F) $ the resulting DFA at the end of the construction. Notice that $ Q \subseteq Q_1 \times Q_2 $ (that is, the sinks play no role) because in $ \mathcal{D}_1 $ and $ \mathcal{D}_2 $ the sinks are not final and do not allow to reach a final state. Moreover we have $I_{(u,v)} = I_{u } \cap I_{v}$ for every $(u, v)\in Q $, because for every $ \alpha \in \Sigma^* $ there exists a path labeled $ \alpha $ from $ (s_1, s_2) $ to $ (u, v) $ on $ \mathcal{D} $ if and only there exist a path labeled $ \alpha $ from $ s_1 $ to $ u $ on $ \mathcal{D}_1 $ and a path labeled $ \alpha $ from $ s_2 $ to $ v $ on $ \mathcal{D}_2 $. As a consequence, $\mt L(\mt D) = \bigcup_{(u, v) \in F} I_{(u, v)} = \bigcup_{u \in F_1, v \in F_2} I_u \cap I_v = \mt L(\mt D_1) \cap \mt L(\mt D_2)$. Now, let us prove that $\text{width}(\mt D) \leq p_1\cdot p_2$. For every $ i = 1, \dots, p_1 $ and for every $ j = 1, \dots, p_2 $, define:
\[Q_{i,j}=\{(u,v) \in Q \;|\; u\in Q_i,  v\in Q^*_j\}. \]
Since $ \{Q_{i, j} \;|\; 1 \le i \le p_1, 1 \le j \le p_2 \} $ is a partition of $ Q $, we only have to show that every $ Q_{i, j} $ is a $ \le_\mathcal{D}$-chain. Let $ (u_1, v_1), (u_2, v_2) $ be distinct elements in $ Q_{i, j} $. Hence, at least one between $ u_1 \not = u_2 $ and $ v_1 \not = v_2 $ holds true. Assume that $ u_1 \not = u_2 $ (the other case is analogous). Since $ u_1, u_2 \in Q_i $, then $ u_1 <_{\mathcal{D}_1} u_2 $ or $ u_2 <_{\mathcal{D}_1} u_1 $. Assuming without loss of generality that $ u_1 <_{\mathcal{D}_1} u_2 $, then from $ I_{(u_1, v_1)} \subseteq I_{u_1} $ and $ I_{(u_2, v_2)} \subseteq I_{u_2} $ we conclude  $ (u_1, v_1) <_{\mathcal{D}} (u_2, v_2) $.

Next, we define $ \mathcal{D'} $ by letting $F'=(F_1\times Q_2) \cup (Q_1\times F_2)$ be the set of all final states, and then removing (i) all states that are not reachable from the initial state and removing (ii) all states that neither are final, nor allow to reach a final state. Let $ \mathcal{D}' = (Q', (s_1,s_2), \delta, F') $ the resulting DFA at the end of the construction. Again, we have $I_{(u,v)} = I_{u } \cap I_{v}$ for every $(u, v)\in Q' $ such that $ u \in Q_1 $ and $ v \in Q_2 $, so $\mt L(\mt D) = \bigcup_{(u, v) \in F} I_{(u, v)} = \bigcup_{u \in F_1 \lor v \in F_2} I_u \cap I_v = \mt L(\mt D_1) \cup \mt L(\mt D_2)$. Let us prove that $\text{width}(\mt D') \leq p_1\cdot p_2+ p_1+p_2$. Notice that this time if $ (u, v) \in Q' $, then it may happen that $ u = \star_1 $ or $ v = \star_2 $ (but not both). Moreover, if $ (u, \star_2) \in Q' $, then $ I_{(u, \star_2)} = I_u \setminus \text{Pref} (\mathcal{L}_2) $, and if $ (\star_1, v) \in Q' $, then $ I_{(\star_1, v)} = I_v \setminus \text{Pref} (\mathcal{L}_1) $. For every $ i = 1, \dots, p_1 $ and for every $ j = 1, \dots, p_2 $, define:
\begin{equation*}
    \begin{split}
        Q'_{i,j}&=\{(u,v) \in Q' \;|\; u\in Q_i,  v\in Q^*_j \} \\
        Q'_{i, \star_2}&=\{(u,\star_2) \in Q' \;|\;  u \in Q_i \}  \\
        Q'_{\star_1,j}&=\{(\star_1,v) \in Q' \;|\;  v \in Q^*_j \}. 
    \end{split}
\end{equation*}
These sets identify a partition of $ Q' $, and like before one can show that each set is a $ \le_\mathcal{D'} $-chain. We conclude that $\text{width}(\mt D') \leq p_1\cdot p_2+p_1+p_2$.
\end{remark}

The above two remarks suggest that our notion of width is related with the structural complexity of the regular expressions accepting a given regular language. 
A more precise and complete analysis (improving these bounds, describing the behavior on NFAs and including the other regular operators, namely, concatenation and Kleene star)  will be the subject of forthcoming works (see also Remark \ref{rem:booleans width} for the consequences of the above two remarks on the smallest-width DFA recognizing a regular language).

  \subsection{The Co-lex Width of a Regular Language}\label{sub:language_width}

On the grounds of the definition of automata's  width, we also study  some implications on the theory of regular languages.
We start by defining  the width of a regular language based on the co-lex orders of the automata recognizing it.

\begin{definition}\label{def:psrt_languages}
Let $ \mathcal{L} $ be a regular language.
\begin{enumerate}
    \item The \index{non-deterministic co-lex  width} \emph{non-deterministic co-lex  width}   of $ \mathcal{L} $, denoted by \index{widthN(L)@${\text{width}}^N(\mathcal L) $, non-deterministic width of $ \mathcal{L} $} ${\text{width}}^N(\mathcal L) $, is the smallest integer $ p $ for which there exists an NFA $ \mathcal{N} $ such that $ \mathcal{L(N)} = \mathcal{L} $ and ${\text{width}}(\mathcal N) = p $.
    \item The  \index{deterministic co-lex width} \emph{deterministic co-lex 
 width}   of $ \mathcal{L} $, denoted by \index{widthD(L)@${\text{width}}^D(\mathcal L) $, deterministic width of $ \mathcal{L} $}  ${\text{width}}^D(\mathcal L) $, is the smallest integer $ p $ for which there exists a DFA $ \mathcal{D} $ such that $ \mathcal{L(D)} = \mathcal{L} $ and ${\text{width}}(\mathcal D) = p $. 
\end{enumerate}
\end{definition}

In Example \ref{ex:depedent} we showed that the width of an automaton may depend on the total order $ \preceq $ on the alphabet. In Example \ref{ex:lang_depend} below, we will show that the deterministic and nondeterministic widths of a language may also depend on the order $ \preceq $ on the alphabet.

On the grounds of Remarks \ref{rem:negation} and \ref{rem:intersection}, we observe the following relations, which allow us to conclude that already constant-width regular languages form an interesting class:
\begin{remark}\label{rem:booleans width}
    Let $\mt L$, $\mt L_1$, and $\mt L_2$ be any regular languages. Then:
    \begin{enumerate}
        \item ${\text{width}}^D(\Sigma^* \setminus \mt L) \leq  {\text{width}}^D({\mt L}) + 1$
        \item ${\text{width}}^D(\mt L_1 \cap \mt L_2) \leq  {\text{width}}^D({\mt L_1}) \cdot {\text{width}}^D({\mt L_2})$
        \item ${\text{width}}^D(\mt L_1 \cup \mt L_2) \leq  {\text{width}}^D({\mt L_1)}  \cdot  {\text{width}}^D({\mt L_2}) + {\text{width}}^D({\mt L_1})+ {\text{width}}^D({\mt L_2}) $.
    \end{enumerate}
     These inequalities  are a direct consequence of Remarks \ref{rem:negation} and \ref{rem:intersection} by starting from smallest-width DFAs recognizing $\mt L$, $\mt L_1$, and $\mt L_2$.
     \end{remark}

By Remark \ref{rem:booleans width}, if $\mathcal L$ can be written as the boolean combination of a constant number of Wheeler languages (for example, of a constant number of finite languages), then ${\text{width}}^N({\mt L}) \leq {\text{width}}^D({\mt L}) \in O(1)$. Furthermore, this bound holds for any total order $\preceq$ on the alphabet if the starting languages are finite (because finite languages are Wheeler independently on the alphabet order).

Definition \ref{def:psrt_languages} introduces two hierarchies of regular languages. 
As shown in Section \ref{subsec:ind_dfa}, languages in the first levels of these hierarchies are much easier to index and compress (see Section \ref{sec:definition indexing and compression} for a definition of the indexing and compression problems for finite-state automata).
Hence, it is interesting to determine the correct position of a given language in the above hierarchies, that is, to solve the \index{language width problem} \emph{language width} problem. 

\begin{Problem}[Language width problem]\label{problem: language width}
Given a regular language $\mathcal L$ (by means of a DFA or an NFA recognizing it) and an integer $p$, determine whether: 
\begin{itemize}
    \item ${\text{width}}^D(\mathcal L) \leq p$.
    \item ${\text{width}}^N(\mathcal L) \leq p$.
\end{itemize}
\end{Problem}

The deterministic case will be tackled in Section \ref{sec:det}. In
the companion paper \cite{parttwo} we  also study the relationships between the deterministic and the non-deterministic hierarchies,  proving   that every level of both hierarchies is non-empty and  that,  apart from level 1, the levels of the two hierarchies do not coincide.


\medskip

The notion of width of a language naturally calls into play the problem of giving an automaton realizing the width of the language. \index{minimum-width DFA problem}

\begin{Problem}[Minimum-width DFA]\label{pr:hasse}
Given a regular language $\mathcal L$, can we define  a canonical DFA $\mathcal D$ such that $\mt L(\mt D)=\mt L$ and  $\text{width}(\mathcal D) = \text{width}^{D}(\mathcal L)$?
\end{Problem}

A solution to the above problem is provided in Section \ref{sec:Hasse automaton of L}, Theorem \ref{thm:hasse}.  We shall prove that  the minimum-width DFA problem is \emph{orthogonal}   with respect to the well-studied minimum-size DFA problem: it could be necessary to split states in order to minimize the width of a DFA. 
Hence,  reducing the size of a given DFA  does not necessarily bring us to the minimum-width DFA  for the same language.

This side of the topic is complemented considering   the following problem.
\index{automata-free characterization problem}

\begin{Problem}[Automata-free characterization]\label{pr:aut_free}
 Is there an automata-free characterization of languages with deterministic width less than or equal to $ p$? 
\end{Problem}

Two solutions to Problem  \ref{pr:aut_free} will be given: on the one hand, we will prove that we can characterize width-$p$ languages by their co-lex monotone sequences of prefixes. On the other hand, a further solution will be given by proving  a generalization of the Myhill-Nerode Theorem for the class of DFAs of width $p$ (for any fixed $p$).

\subsection{Compression and Indexing}\label{sec:definition indexing and compression}

Another problem that we will consider is that of improving the space usage of the existing encoding for automata (such as the one of Chakraborty et al. \cite{chakraborty2021succinct} discussed in Section \ref{sec:state art}), in the case that the automaton's width is small:

\index{compressing automata problem}

\begin{Problem}[Compressing automata]\label{problem:encoding}
Can we represent a finite state automaton using $\log\sigma + o(\log n)$ bits per edge, provided its co-lex width $p$ satisfies $\log p \in o(\log n) $?
\end{Problem}

The motivation behind Problem \ref{problem:encoding} is that Wheeler automata, that is, automata of co-lex width equal to 1, can be represented using $\log\sigma + O(1)$ bits per edge \cite{GAGIE201767}.
Indeed, 
in Sections \ref{sec:aBWT} and \ref{sec:aaBWT}
we will show that our notion of co-lex order will enable us to solve Problem \ref{problem:encoding} through a generalization of the powerful \emph{Burrows-Wheeler transform} (BWT) \cite{burrows1994block}.
The BWT of a text is a permutation that re-arranges the text's characters according to the co-lex order of the prefixes that precede them\footnote{The original definition considers the lexicographic order of the text's suffixes. We choose a symmetric definition because it can be generalized to NFAs via co-lex orders.}. 
A well-known fact is that the BWT boosts compression and enables efficient indexing (read Problem \ref{problem:indexing} below) in compressed space \cite{ferragina2000opportunistic}.
Previous works generalized this transform to trees \cite{Ferragina2005}, string sets \cite{mantaci2007extension}, de Bruijn graphs \cite{BOSS,GCSA} and Wheeler graphs \cite{GAGIE201767}. In Section \ref{sec:aBWT} we complete the picture by generalizing those BWT-based indexes to NFAs (equivalently, to arbitrary labeled graphs) and languages and showing that this index indeed solves Problem \ref{problem:indexing}: 


\begin{Problem}[Indexing automata]\label{problem:indexing}\index{indexing automata problem}
Let $\mathcal N = (Q, s, \delta,F)$ be a finite-state automaton. Let moreover $T(\alpha)$ be the set of states of $\mathcal N$ which are reached by a path labeled with a given word $\alpha$, i.e. $T(\alpha) = \{u\in Q\ |\ (\exists \beta \in I_u)(\alpha \dashv  \beta)\}$. Pre-process $\mathcal N$ into a small data structure supporting efficient \emph{subpath queries}, i.e. given a query word $\alpha$ solve:
\begin{itemize}
	\item (Existential queries) Determine whether $T(\alpha) \neq \emptyset$, i.e. whether $\alpha$ matches a substring of some string in the language of $\mathcal N$.
	\item (Count queries) Compute the cardinality of $T(\alpha)$.
	\item (Locate queries) Return a representation for all the states in $T(\alpha)$. 
\end{itemize} 
\end{Problem}


Recent works (see Section \ref{sec:state art}) show that, as opposed to the membership problem on DFAs, all the three above-listed subpath queries are hard even on acyclic DFAs: unless the Strong Exponential Time Hypothesis fails, such queries require quadratic time to be solved off-line \cite{EquiGMT19} and even on-line using any index constructible in polynomial time \cite{EquiMT21}. In Section \ref{subsubsec:index} we provide a linear-space index solving subpath (and thus membership) queries on NFAs in time proportional to $p^2$ per character in the input query, $p$ being the automaton's co-lex width. 
The index can be built in polynomial time on DFAs and exponential time on NFAs (the latter complexity is due to the hardness of computing the co-lex width of NFAs). For small $p$, our index for DFAs breaks the indexing lower bound of Equi et al. \cite{EquiMT21}. 

\section{Co-lex Width and NFA Determinization}\label{sec:determinization}

In this  section we show that the notion of width can be used to prove some crucial relationships between an NFA $ \mathcal{N} $ and the powerset automaton $ \text{Pow}(\mathcal{ N}) $   obtained from $ \mathcal{N} $. First, we bound the width of $ \text{Pow}(\mathcal{ N}) $ in terms of the width of $ \mathcal{N} $ and  prove that the number of $ \text{Pow}(\mathcal{ N}) $'s states is exponential in $ \text{width} (\mathcal{N}) $ rather than in the number of $ \mathcal{N} $'s states. This implies that several problems easy on DFAs but difficult on NFAs are in fact fixed-parameter tractable with respect to the width.

\medskip

Recall that, given an NFA $ \mathcal{N} = (Q, s, \delta, F) $, the powerset construction algorithm builds an equivalent  DFA $ \text{Pow}(\mathcal{ N}) = (Q^*, s^*, \delta^*, F^*) $  defined as: 
\begin{itemize}
\item  $ Q^* = \{ I_{\alpha}\ |\ \alpha\in \text{Pref}(\mathcal{L(\mathcal{N})})\}$; 
\item    $ s^* = \{s \} $; 
  \item     $ \delta^{*} (I_\alpha, a) = I_{\alpha a} $ for all $ \alpha \in \Sigma^*$ and $ a \in \Sigma $ such that $ \alpha a \in \text{Pref}(\mathcal{L(N)}) $; 
\item   $ F^* = \{I_\alpha\ |\ \alpha \in \mathcal{L(\mathcal{N})} \} $.
\end{itemize}

For $ \alpha, \alpha' \in \text{Pref}(\mathcal{L(N)}) $ we have:
\begin{equation*}
    \delta^* (s^*, \alpha') = I_{\alpha} \iff I_{\alpha'} = I_\alpha
\end{equation*}

and defining as usual $ I_{u^*} = \{\alpha \in \text{Pref}(\mathcal{L}(\text{Pow}(\mathcal{ N})))\ |\ u^* \in \delta^{*} (s^*, \alpha)  \}  $ for $ u^* \in Q^* $, we have that for $ \alpha \in \text{Pref}(\mathcal{L(N)}) $:
\begin{equation}\label{eq:powersetstates}
    I_{I_\alpha} = \{\alpha' \in \text{Pref}(\mathcal{L(N)})\ |\ I_{\alpha'} = I_\alpha \}.
\end{equation}

We start with a characterization of the maximum co-lex order on $ \text{Pow}(\mathcal{ N}) $ (which exists by Lemma \ref{lem:maximumorderDFAs}).

\begin{lemma}\label{lem:maximalpowerset}
Let $ \mathcal{N} = (Q, s, \delta, F) $ be an NFA  and let  $ \text{Pow}(\mathcal{ N}) = (Q^*, s^*, \delta^*, F^*) $ be the powerset automaton obtained from $ \mathcal{N} $. Let $ \le_{\text{Pow}(\mathcal{ N})} $ be the maximum co-lex order on $\text{Pow}(\mathcal{ N}) $. Then, for $ I_\alpha \not = I_\beta $:
\begin{equation*}
    (I_\alpha <_{\text{Pow}(\mathcal{ N})} I_\beta) \iff (\forall \alpha', \beta' \in \text{Pref}(\mathcal{L(\mt N)}))((I_{\alpha'} = I_{\alpha}) \land (I_{\beta'} = I_\beta) \to \alpha' \prec \beta')
\end{equation*}
Moreover, let $ \le $ be a co-lex order on $ \mathcal{N} $, and fix $ \alpha, \beta \in \text{Pref}(\mathcal{L(N)}) $. Then:
\begin{equation*}
    (\exists u \in I_\alpha)(\exists v \in I_\beta)(\{u, v\} \not \subseteq I_\alpha \cap I_\beta \land u < v) \implies (I_\alpha <_{\text{Pow}(\mathcal{ N})} I_\beta).
\end{equation*}

\end{lemma}

\begin{proof}
              The first part follows immediately from the characterization of the maximum co-lex order over a DFA (Lemma \ref{lem:maximumorderDFAs}) and Equation \ref{eq:powersetstates}.   Let us prove the second part. Consider $ u \in I_\alpha $ and $ v \in I_\beta $ such that $ \{u, v\} \not \subseteq I_\alpha \cap I_\beta $ and $ u < v $. We prove that $I_\alpha <_{\text{Pow}(\mathcal{ N})} I_\beta$ using the characterization of $<_{\text{Pow}(\mathcal{ N})}$  given in the first part of the proof. Fix $ \alpha', \beta' \in \text{Pref}(\mathcal{L(N)}) $ such that $ I_{\alpha'} = I_\alpha $ and $ I_{\beta'} = I_\beta $. We must prove that $ \alpha' \prec \beta' $.  From the hypothesis it follows  $u\in I_{\alpha'}, v\in I_{\beta'} $, and $\{u,v\} \not \subseteq I_{\alpha'}\cap I_{\beta'}$ so that  $\alpha'\in I_u, \beta'\in I_v$, and    $ \{\alpha', \beta' \} \not \subseteq I_{u} \cap I_{v} $ hold. Hence,   $ \alpha' \prec \beta ' $   follows from  $u<v$ and 
               Lemma \ref{lem:monot}. \qed
\end{proof}

We can now prove the main  result of this  section.

\begin{theorem}\label{theor:powersetcrucial}
Let $ \mathcal{N} = (Q, s, \delta, F) $ be an NFA  and let  $ \text{Pow}(\mathcal{ N}) = (Q^*, s^*, \delta^*, F^*) $  be the powerset automaton obtained from $ \mathcal{N} $. Let $ n = |Q| $ and $ p = {\text{width}} (\mathcal{N}) $. Then:
\begin{enumerate}
\item ${\text{width}} (\text{Pow}(\mathcal{ N})) \le 2^{p} - 1 $;
\item $ |Q^*| \le 2^{p}(n - p + 1) - 1 $.
\end{enumerate}
\end{theorem}

\begin{proof}
Let $ \le $ be a co-lex order on $ \mathcal{N} $ such that $ width (\le) = p $, and let $ \{Q_i  \; |\; 1\leq i \leq p\} $ be a $ \le $-chain partition. Let $ \le_{\text{Pow}(\mathcal{ N})} $  be the maximum co-lex order on $\text{Pow}(\mathcal{ N})$. For every nonempty $ K \subseteq \{1, \dots, p \} $, define:
\begin{equation*}
    \mathcal{I}_K = \{I_\alpha \mid(\forall i \in \{1, \dots, p \})(I_\alpha \cap Q_i \not = \emptyset \iff i \in K) \}.
\end{equation*}
Notice that $ Q^* $ is the disjoint union of all $ \mathcal{I}_K $. More precisely:
\begin{equation}\label{eq:2expp-1}
    Q^* = \bigsqcup_{\substack{
  \emptyset \not = K \subseteq \{1, \dots, p \}}}
\mathcal{I}_K.
\end{equation}
Let us prove that each $ \mathcal{I}_K $ is a $\le_{\text{Pow}(\mathcal{ N})}$-chain. Fix $ I_\alpha, I_\beta \in \mathcal{I}_K $, with $ I_\alpha \not = I_\beta $. We must prove that $ I_\alpha $ and $ I_\beta $ are  $\le_{\text{Pow}(\mathcal{ N})}$-comparable. Since $ I_\alpha \not = I_\beta $, there exists either $ u \in I_\alpha \setminus I_\beta $ or $ v \in I_\beta \setminus I_\alpha $. Assume that there exists $ u \in I_\alpha \setminus I_\beta $ (the other case is analogous). In particular, let $ i \in \{1, \dots, p \} $ be the unique integer such that $ u \in Q_i $. Since $ I_\alpha, I_\beta \in \mathcal{I}_K $, from the definition of $ \mathcal{I}_K $ it follows that there exists $ v \in I_\beta \cap Q_i $. Notice that $ \{u, v \} \not \subseteq I_\alpha \cap I_\beta $ (so in particular $ u \not = v $), and since $ u, v \in Q_i $ we conclude that $ u $ and $ v $ are $ \le $-comparable. By Lemma \ref{lem:maximalpowerset} we conclude that $ I_\alpha $ and $ I_\beta $ are $\le_{\text{Pow}(\mathcal{ N})}$-comparable.

\begin{enumerate}
    \item The first part of the theorem follows from equation \ref{eq:2expp-1}, because each $ \mathcal{I}_K $ is a $\le_{\text{Pow}(\mathcal{ N})}$-chain and there are $ 2^p - 1 $ choices for $ K $.
    \item Let us prove the second part of the theorem. Fix $ \emptyset \not = K \subseteq \{1, \dots, p \} $. For every $ I_\alpha \in \mathcal{I}_K $ and for every $ i \in K $, let $ m_\alpha^i $ be the smallest element of $ I_\alpha \cap Q_i $ (this makes sense because $ (Q_i, \le) $ is totally ordered), and let $ M_\alpha^i $ be the largest element of $ I_\alpha \cap Q_i $. Fix $ I_\alpha, I_\beta \in \mathcal{I}_K $, and note the following:
\begin{enumerate}
    \item Assume that for some $ i \in K $ it holds $ m_{\alpha}^i < m_{\beta}^i \lor M_{\alpha}^i < M_{\beta}^i $. Then, it must be $ I_\alpha <_{\text{Pow}(\mathcal{ N})} I_\beta $. Indeed, assume that $ m_{\alpha}^i < m_{\beta}^i $ (the other case is analogous). We have $ m_{\alpha}^i \in I_{\alpha} $, $ m_{\beta}^i \in I_{\beta} $, $ \{m_{\alpha}^i, m_{\beta}^i \} \not \subseteq I_{\alpha} \cap I_{\beta} $ and $ m_{\alpha}^i < m_{\beta}^i $, so the conclusion follows from Lemma \ref{lem:maximalpowerset}. Equivalently, we can state that if $ I_\alpha <_{\text{Pow}(\mathcal{ N})} I_\beta $ then $ (\forall i \in K)(m_{\alpha}^i \le m_{\beta}^i \land M_{\alpha}^i \le M_{\beta}^i) $.
    \item Assume that for some $ i \in K $ it holds $ m_{\alpha}^i = m_{\beta}^i \land M_{\alpha}^i = M_{\beta}^i $. By Corollary \ref{cor:intervals}, the sets  $I_\alpha$ and $I_\beta$ are convex in $(Q, \leq)$. This implies that     $ I_\alpha \cap Q_i $ and $I_\beta \cap Q_i $ are  $ \le_{Q_i} $-convex, and  having the same minimum and maximum  they must be equal, that is, $ I_\alpha \cap Q_i = I_\beta \cap Q_i $.
    \item Assume that $ (\forall i \in K)(m_{\alpha}^i = m_{\beta}^i \land M_{\alpha}^i = M_{\beta}^i) $. Then, it must be $ I_\alpha = I_\beta $. Indeed, from point (b) we obtain $ (\forall i \in K)(I_\alpha \cap Q_i = I_\beta \cap Q_i) $, so $ I_\alpha = \bigcup_{i \in K}(I_\alpha \cap Q_i) = \bigcup_{i \in K}(I_\beta \cap Q_i) = I_\beta $. Notice that we can equivalently state that if $ I_\alpha \not = I_\beta $, then $ (\exists i \in K)(m_{\alpha}^i \not = m_{\beta}^i \lor M_{\alpha}^i \not = M_{\beta}^i) $.
\end{enumerate}
\end{enumerate}

Fix $ I_\alpha, I_\beta \in \mathcal{I}_K $. Now it is easy to show that:
\begin{equation}\label{eq:powersetminmax}
\begin{split}
    I_\alpha <_{\text{Pow}(\mathcal{ N})} I_\beta  & \iff (\forall i \in K)(m_{\alpha}^i \le m_{\beta}^i \land M_{\alpha}^i \le M_{\beta}^i) \land \\
    & \land (\exists i \in K)(m_{\alpha}^i < m_{\beta}^i \lor M_{\alpha}^i < M_{\beta}^i).
\end{split}
\end{equation}
Indeed, ($ \Leftarrow $) follows from point (a). As for ($ \Rightarrow $), notice that $ (\forall i \in K)(m_\alpha^i \le m_{\beta}^i \land M_{\alpha}^i \le M_{\beta}^i) $ again follows from point (a), whereas $ (\exists i \in K)(m_{\alpha}^i < m_{\beta}^i \lor M_{\alpha}^i < M_{\beta}^i) $ follows from point (c).

Let $ |m_{\alpha}^i| $ and $ |M_{\alpha}^i| $ be the positions of $ m_{\alpha}^i $ and $ M_{\alpha}^i $ in the total order $ (Q_i, \le) $ (so $ |m_{\alpha}^i|, |M_{\alpha}^i| \in \{1, \dots, |Q_i| \} $). For every $ I_\alpha \in \mathcal{I}_K $, define:
\begin{equation*}
    T(I_\alpha) = \sum_{i \in K}(|m_\alpha^i| + |M_\alpha^i|).
\end{equation*}
By equation \ref{eq:powersetminmax}, we have that $I_\alpha <_{\text{Pow}(\mathcal{ N})} I_\beta $  implies  $ T(I_\alpha) < T(I_\beta) $, so since $ \mathcal{I}_K $ is a $ <_{\text{Pow}(\mathcal{ N})}$-chain  we have that $ |\mathcal{I}_K| $ is bounded by the values that $ T(I_\alpha) $ can take. For every $ I_\alpha \in \mathcal{I}_K $ we have $ 2 |K| \le T(I_\alpha) \le 2 \sum_{i \in K} |Q_i| $ (because $ |m_{\alpha}^i|, |M_{\alpha}^i| \in \{1, \dots, |Q_i| \} $), so:
\begin{equation}\label{eq:boundIk}
|\mathcal{I}_K| \le 2 \sum_{i \in K} |Q_i| - 2 |K| + 1.
\end{equation}
From Equations \ref{eq:2expp-1} and \ref{eq:boundIk}, we obtain:
\begin{equation*}
\begin{split}
    |\mathcal{Q^*}| & = \sum_{\emptyset \subsetneqq K \subseteq \{1, \dots, p \}} |\mathcal{I}_K| \le \sum_{\emptyset \subsetneqq K \subseteq \{1, \dots, p \}} (2 \sum_{i \in K} |Q_i| - 2 |K| + 1) \\
    & = 2 \sum_{\emptyset \subsetneqq K \subseteq \{1, \dots, p \}} \sum_{i \in K} |Q_i| - 2 \sum_{\emptyset \subsetneqq K \subseteq \{1, \dots, p \}} |K| + \sum_{\emptyset \subsetneqq K \subseteq \{1, \dots, p \}} 1.
\end{split}
\end{equation*}
Notice that $ \sum_{\emptyset \subsetneqq K \subseteq \{1, \dots, p \}} \sum_{i \in K} |Q_i| = 2^{p - 1} \sum_{i \in \{1, \dots, p \}} |Q_i| = 2^{p - 1} n $ because every $ i \in \{1, \dots, p \} $ occurs in exactly $ 2^{p - 1} $ subsets of $ \{1, \dots, p \} $. Similarly, we obtain $ \sum_{\emptyset \subsetneqq K \subseteq \{1, \dots, p \}} |K| = 2^{p - 1} p $ and $ \sum_{\emptyset \subsetneqq K \subseteq \{1, \dots, p \}} 1 = 2^p - 1 $, We conclude:
\begin{equation*}
    |\mathcal{Q^*}| \le 2^p n - 2^p p + 2^p - 1 = 2^p (n - p + 1) - 1.
\end{equation*} \qed
\end{proof}

 As a first  consequence of  Theorem \ref{theor:powersetcrucial} we start  
 comparing  the non-deterministic and deterministic width  hierarchies of  regular languages. 
Clearly, for every regular language $\mathcal L$ we have ${\text{width}}^N(\mathcal L)\leq {\text{width}}^D(\mathcal L)$ since DFAs are particular cases of NFAs.  Moreover, 
for languages  with ${\text{width}}^N(\mathcal L)=1$, the so-called Wheeler languages, it is known that 
the non-deterministic and deterministic widths coincide \cite{DBLP:conf/soda/AlankoDPP20}. Nonetheless,   in the  companion paper \cite{parttwo}  we will prove that this property is truly peculiar of Wheeler languages, because the gap between the deterministic and non-deterministic hierarchies is, in general, exponential.
Here we prove that Theorem \ref{theor:powersetcrucial} provides an \emph{upper} bound for  the deterministic width in terms of the non-deterministic width. 
 
\begin{corollary}\label{lem:widthupperbound}
Let $ \mathcal{L} $ be a regular language. Then, $ {\text{width}}^D(\mathcal L) \le 2^{{\text{width}}^N(\mathcal L)} - 1 $.
\end{corollary}

\begin{proof}
Let $ \mathcal{N} $ be an NFA such that $ \mathcal{L(N)} = \mathcal{L} $ and $ {\text{width}}(\mt N) = { {\text{width}}}^N(\mathcal L) $. By Theorem \ref{theor:powersetcrucial}, we have $ {\text{width}}^D(\mathcal L) \le {\text{width}}(Pow(\mathcal N)) \le 2^{{\text{width}} (\mathcal{N})} - 1 = 2^{{\text{width}}^N(\mathcal L)} - 1 $. \qed
\end{proof}

In a forthcoming work we will prove that the above bound is tight.
Notice that the above corollary  shows once again that ${\text{width}}^N(\mathcal L)=1$ implies ${\text{width}}^D(\mathcal L)=1$, that is, the non-deterministic and deterministic widths are equal for Wheeler languages.

Theorem \ref{theor:powersetcrucial} has another  intriguing consequence: the PSPACE-complete NFA equivalence problem \cite{stockmeyer1973word} is fixed-parameter tractable with respect to the widths of the automata. In order to prove this result, we first update the analysis of Hopcroft et al. \cite{hopcroft2006introduction} of the powerset construction algorithm.

\begin{lemma}[Adapted from \cite{hopcroft2006introduction}]\label{lem:powerset complexity}
Let $ \mathcal{N} = (Q, s, \delta, F) $ be an NFA  and let  $ \text{Pow}(\mathcal{ N}) = (Q^*, s^*, \delta^*, F^*) $ be the powerset automaton obtained from $ \mathcal{N} $. Let $ n = |Q| $ and $ p = {\text{width}} (\mathcal{N}) $. Then, the powerset construction algorithm runs in $O(2^p(n-p+1)n^2\sigma)$ time.
\end{lemma}
\begin{proof}
By Theorem \ref{theor:powersetcrucial}, we know that $N = 2^p(n-p+1)$ is an upper bound to the number of states of the equivalent DFA. Each state in $ Q^* $ consists of $ k \le n $ states $ u_1, \dots, u_k $ of $ Q $. For each character $a \in \Sigma$,   we need to follow all edges labeled $ a $ leaving $u_1, \dots, u_k$. In the worst case (a complete transition function), this leads to traversing $O(k\cdot n) \subseteq O(n^2)$ edges of the NFA. The final complexity is thus  $O(N\cdot n^2\cdot \sigma)$.\qed
\end{proof}

\begin{corollary}\label{cor:NFA equiv}
We can check the equivalence between two NFAs over an alphabet of size $\sigma$, both with number of states at most $ n $ and width at most $ p $, in $O(2^p(n-p+1)n^2\sigma)$ time.
\end{corollary}
\begin{proof}
First, build the powerset automata,  both having at most $N = 2^p(n-p+1)$ states by Theorem \ref{theor:powersetcrucial}. This takes $O(N n^2\sigma)$ time by Lemma \ref{lem:powerset complexity}.
Finally, DFA equivalence can be tested in $O(N \sigma \log N)$ time by DFA minimization using Hopcroft's algorithm. \qed
\end{proof}

Similarly, the powerset construction can be used to test membership of a word of length $m$ in a regular language expressed as an NFA. When $m$ is much larger than $n$ and $2^p$, this simple analysis of a classical method yields a faster algorithm than the state-of-the-art solution by Thorup and Bille, running in time $O(m\cdot e\cdot \log\log m / (\log m)^{3/2} + m + e)$ \cite{BilleThorup09} where $e$ is the NFA's (equivalently, the regular expression's) size:

\begin{corollary}
We can test membership of a word of length $m$ in the language recognized by an NFA with $n$ states and co-lex width $ p $ on alphabet of size $\sigma$ in $O(2^p(n-p+1)n^2\sigma + m)$ time. 
\end{corollary}

\section{The Deterministic Width of a Regular Language}\label{sec:det}

In this section we consider problems related to the notion of width in the deterministic case.
In  Corollary \ref{cor:complexity width dfa} we shall prove  that (the NFA width) Problem \ref{pr:width_complexity} is polynomial for the case of DFAs. 
This motivates the search for an efficient algorithm for  (the language width) Problem \ref{problem: language width}.
Surprisingly,  we show that also this problem admits an efficient (polynomial) solution for any constant width.
Given a DFA, we then consider (the Minimum-width DFA) Problem \ref{pr:hasse}. In particular, we prove that the requests of minimizing the width and the number of states are conflicting: the minimum DFA recognizing a regular language will not, in general, have minimum width, and among the automata of minimum width recognizing a language there can be non-isomorphic automata with minimum number of states.  
We then  propose two different solutions to Problem \ref{pr:aut_free} (Automata-free characterization for languages of deterministic width $p$): in Corollary \ref{cor:monotone_sequences} we consider the behavior of monotone sequences in $(\pf L, \preceq)$, while in Theorem \ref{thm:MN} we  
  present     a Myhill-Nerode   result for languages of fixed deterministic width. 


\medskip

The notion of width of an automaton is simpler if we restrict our attention to the class of DFAs.
We already proved in Section \ref{sec:width of automaton} that any DFA admits a (unique) maximum co-lex order
  $\leq_{\mathcal D}$  realizing  the width of the automaton $\mt D$.
We shall now prove that the order $\leq_{\mathcal D}$ is polynomially computable 
 and so is its width.
We start with a 
 characterization of $\leq_{\mathcal D}$ in terms of graph reachability. 

\begin{definition}
We say that a pair $(u',v')\in Q\times Q$    precedes a pair $(u ,v )\in Q\times Q$  if $u'\neq v', u\neq v$ and there exists $\alpha\in \Sigma^*$ such that $\delta(u',\alpha)=u, \delta(v',\alpha)=v$.
\end{definition}

  \begin{lemma}\label{lem:prec_dfa_char}Let $ \mathcal{D}$ be a DFA  and let $ u, v \in Q $, with $ u \not = v $. Then:
\[u<_{\mathcal D}v  ~ \Leftrightarrow \text{for all  pairs $ (u',v') $ preceding $ (u,v) $ it holds $ \text{max}_{\lambda (u')} \preceq \text{min}_{\lambda (v')}$}.
\]
\end{lemma}

\begin{proof}
($ \Rightarrow $) Suppose $u <_{\mathcal D}v$.  Let $ (u', v') $ be a pair preceding $ (u, v) $  and let $ \gamma \in \Sigma^* $ be such that $ \delta (u', \gamma) = u $ and $ \delta (v', \gamma) = v $. We must prove that $\text{max}_{\lambda (u')} \preceq \text{min}_{\lambda (v')} $. First, notice that it cannot be $ v' = s $, otherwise, given $ \alpha \in I_{u'}  $, we would have $ \alpha \gamma \in I_u $ and $ \gamma \in I_v $, which contradicts $ u <_\mathcal{D} v $. So we are only left with proving that if $ u' = \delta (u'', e) $ and $ v' = \delta (v'', e') $, with $ e,e' \in \Sigma $, then $ e \preceq e' $. Let $ \alpha'' \in I_{u''} $ and $ \beta'' \in I_{v''} $. Then $\alpha'' e \gamma \in I_u $, $\beta'' e' \gamma \in I_v$ and from $ u <_{\mathcal D}v $ it follows that $ \alpha'' e \gamma \prec \beta'' e' \gamma $, which implies  $ e \preceq e' $. 

($ \Leftarrow $) Suppose that for all pairs $(u',v') $ preceding $ (u,v) $ it holds $ \text{max}_{\lambda (u')} \preceq \text{min}_{\lambda (v')}$. Let $ \alpha \in I_u $ and $ \beta \in I_v $. We must prove that $ \alpha \prec  \beta $. Since $ u \not = v $, then $ \alpha \not = \beta $. Write $ \alpha = \alpha' \gamma $ and $ \beta = \beta' \gamma $, where $ \alpha' $ and $ \beta' $ end with a distinct letter (or, possibly, exactly one of them is equal to the empty string). Let $ u', v' \in Q $ be such that $ \alpha' \in I_{u'} $ and $ \beta' \in I_{v'} $. Then, $ (u', v') $ precedes $ (u, v) $, so it must be $ \text{max}_{\lambda (u')} \preceq \text{min}_{\lambda (v')} $. This implies that $ v' \not = s $, so $ \beta $ is not a suffix of $ \alpha $. If $ \alpha $ is a suffix of $ \beta $, we are done. Otherwise, it must be $ \alpha' = \alpha'' a $ and $ \beta' = \beta'' b $, with $ a, b \in \Sigma $, $ a \not = b $;    from $ \text{max}_{\lambda (u')} \preceq \text{min}_{\lambda (v')} $ it then follows  $ a \prec b $, which implies $ \alpha \prec \beta $.
\qed
\end{proof}

Using the previous lemma we are now able to describe a polynomial time algorithm for  computing  $<_{\mathcal D}$.
Let $G=(V,F)$ where
$ V  = \{(u, v)\in Q\times Q\ | \ u \neq v \} $ 
and $ F = \{((u',v'),(u,v)) \in V\times V \ |\ (\exists e \in \Sigma)( \delta(u',e)=u \land \delta(v',e)=v) \}$, where $|F|\leq |\delta|^2$. Intuitively, we will use $G$ to propagate the complement     $ \not <_{\mathcal D}  $ of $<_{\mathcal D}$.  First, 
  mark all nodes $ (u, v) $ of $G$ for which $\text{max}_{\lambda (u)} \preceq \text{min}_{\lambda (v)}$ does not hold. This process takes $O(|Q|^2)$ time:  for any state $u$ we find the minimum and the maximum of    $\lambda(u)$  by scanning the transitions of the automaton (total time $O(|\delta|))$; then  we   decide  in constant time when $\text{max}_{\lambda (u)} \preceq \text{min}_{\lambda (v)}$ does not hold.    Then, mark all nodes reachable on $ G $ from marked nodes.   This can be done with a simple DFS visit of $G$, initiating the stack with all marked nodes. This process takes $O(|\delta|^2)$ time.  By Lemma \ref{lem:prec_dfa_char},  the set of  unmarked pairs is $<_{\mathcal D}$. Hence, we proved:

\begin{lemma}\label{lem:complexity prec DFA} 
Let $ \mathcal{D} = (Q, s, \delta, F) $   be a DFA. We can find the   order  $\leq_ {\mathcal{D}} $  in $ O(|\delta|^2) $ time.
\end{lemma}

It follows   that also the width of a DFA is computable in polynomial time:
  by the following  lemma from \cite{KoganParterICALP}, the width of a partial order $\le$  (and an associated $ \le $-chain partition) can always  be found in    polynomial time from $\le$. In the following results, \emph{with high probability} means with success probability at least $1-N^{-c}$, where $N$ is the size of the input and $c$ is a user-defined constant.

  \begin{lemma}[\hspace{1sp}\cite{KoganParterICALP}]
  \label{lem:complexity min chain partition}
 Let $(V,\leq)$ be a partial order. A smallest $ \le $-chain partition of $(V,\leq)$ and its width can be found in $\tilde O(|V|^2)$ time, with high probability.
 \end{lemma}
  \begin{proof}
  In \cite[Thm 1.6]{KoganParterICALP}, it is shown how to compute a minimum chain partition of a DAG with $n$ vertices and $m$ edges in $\tilde O(m + n^{3/2})$ time with high probability. The claim follows running this algorithm on the DAG corresponding to $(V,\leq)$, having $|V|$ nodes and $O(|V|^2)$ edges.
  \qed
 \end{proof}

\begin{corollary} \label{cor:complexity width dfa}  Let $ \mathcal{D} = (Q, s, \delta, F) $ be a DFA. A co-lex order $ \le $ of width equal to $ \text{width}({\mathcal D}) $ and a corresponding $ \le $-chain partition of  cardinality equal to $ \text{width}({\mathcal D}) $ can be computed in  $ \tilde O(|\delta|^2) $ time with high probability.
\end{corollary}

The previous corollary  solves Problem \ref{pr:width_complexity} for DFAs.  As for the    width-complexity over NFAs,  it is known that the problem is NP-hard, since    already   deciding whether the width of an NFA is equal to $1$ (i.e. deciding whether the NFA is Wheeler)  is an NP-complete problem (see \cite{gibney2022complexity}).
  

We have proved that computing the width of a DFA is an ``easy'' problem. 
We now move to the natural problem of computing the deterministic width  of a regular language, presented by means of a DFA  recognizing it.

It would have been nice to have the deterministic width of a language equal to the width of its minimum DFA: if this were true, we could use Corollary \ref{cor:complexity width dfa}  to  determine $\text{width}^D(\mt L)$ in polynomial time with high probability by inspecting its minimum-size accepting automaton. As we show in Example \ref{ex:no_minimal},  unfortunately this is not the case.  Moreover, there is, in general, no unique (up to isomorphism) minimum DFA of minimum width recognizing a given regular language.

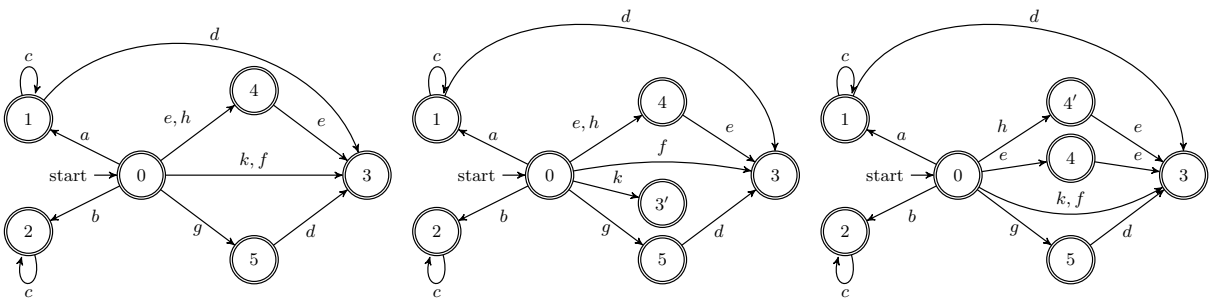
\begin{figure}[h!] 
 \begin{center}
 \scalebox{0.75}{ 
\begin{tikzpicture}[->,>=stealth', semithick, auto, scale=1]
\node[state,accepting,initial] (s)    at (0,0)	{$0$};

\node[state, accepting,label=above:{}] (1)    at (-2,1)	{$1$};
\node[state,accepting, label=above:{}] (2)    at (-2,-1)	{$2$};

\node[state, accepting,label=above:{}] (3)    at (4,0)	{$3$};
\node[state,accepting, label=above:{}] (4)    at (2,1.5)	{$4$};
\node[state,accepting, label=above:{}] (5)    at (2,-1.5)	{$5$};

\draw (s) edge [] node [above] {$a$} (1);
\draw (s) edge [] node [] {$b$} (2);

\draw (s) edge [] node [] {$k,f$} (3);
\draw (s) edge [] node [] {$e,h$} (4);
\draw (s) edge [] node [below] {$g$} (5);
\draw (5) edge [] node [below] {$d$} (3);
 \draw (4) edge [] node [] {$e$} (3);
 \draw (1) edge  [bend left=60, above] node [] {$d$} (3);
 \draw (1) edge  [loop above] node {$c$} (1);
 \draw (2) edge  [loop below] node {$c$} (2);
\end{tikzpicture}
}
 \scalebox{0.75}{ 
\begin{tikzpicture}[->,>=stealth', semithick, auto, scale=1]
\node[state,accepting,initial] (s)    at (-3,0)	{$0$};

\node[state, accepting,label=above:{}] (1)    at (-5,1)	{$1$};
\node[state,accepting, label=above:{}] (2)    at (-5,-1)	{$2$};

\node[state, accepting,label=above:{}] (3)    at (1,0)	{$3$};
 \node[state, accepting,label=above:{}] (3')    at (-1,-0.5)	{$3'$};
\node[state,accepting, label=above:{}] (4)    at (-1,1.3)	{$4$};
\node[state,accepting, label=above:{}] (5)    at (-1,-1.5)	{$5$};

\draw (s) edge [] node [above] {$a$} (1);
\draw (s) edge [] node [] {$b$} (2);

\draw (s) edge [] node [] {$k$} (3');
\draw (s) edge [bend left=10] node [] {$f$} (3);
\draw (s) edge [] node [] {$e,h$} (4);
\draw (s) edge [] node [below] {$g$} (5);
\draw (5) edge [] node [below] {$d$} (3);
 \draw (4) edge [] node [] {$e$} (3);
 \draw (1) edge  [bend left=80, above] node [] {$d$} (3);
 \draw (1) edge  [loop above] node {$c$} (1);
 \draw (2) edge  [loop below] node {$c$} (2);
\end{tikzpicture}
}
 \scalebox{0.75}{ 
\begin{tikzpicture}[->,>=stealth', semithick, auto, scale=1] 
 
 \node[state,accepting,initial] (sb)    at (4,0)	{$0$};

\node[state, accepting,label=above:{}] (1b)    at (2,1)	{$1$};
\node[state,accepting, label=above:{}] (2b)    at (2,-1)	{$2$};

\node[state, accepting,label=above:{}] (3b)    at (8,0)	{$3$};
 \node[state,accepting, label=above:{}] (4b')    at (6,1.3)	{$4'$};
\node[state,accepting, label=above:{}] (4b)    at (6,0.3)	{$4$};
\node[state,accepting, label=above:{}] (5b)    at (6,-1.5)	{$5$};

\draw (sb) edge [] node [above] {$a$} (1b);
\draw (sb) edge [] node [] {$b$} (2b);

\draw (sb) edge [bend right=30] node [] {$k, f$} (3b);
\draw (sb) edge [] node [] {$h$} (4b');
\draw (sb) edge [] node [] {$e$} (4b);
\draw (sb) edge [] node [below] {$g$} (5b);
\draw (5b) edge [] node [below] {$d$} (3b);
 \draw (4b) edge [] node [] {$e$} (3b);
  \draw (4b') edge [] node [] {$e$} (3b);
 \draw (1b) edge  [bend left=80, above] node [] {$d$} (3b);
 \draw (1b) edge  [loop above] node {$c$} (1b);
 \draw (2b) edge  [loop below] node {$c$} (2b);
\end{tikzpicture}
}
 \end{center}
 \caption{Three DFAs recognizing the same language.}
 \label{fig:2minimumDFA}
\end{figure}
  
\begin{example}\label{ex:no_minimal}
In Figure \ref{fig:2minimumDFA}, three   DFAs recognizing the same language $\mathcal L$ are shown. We prove that  $ \text{width}^D(\mathcal{L}) = 2 $, the width of the minimum DFA for $\mathcal L$ is 3, and there is not a  unique minimum automaton among all DFAs recognising $\mathcal L$ of   width $2$. 
 Consider the DFA $ \mathcal{D}_1 $ on the left of Figure  \ref{fig:2minimumDFA} and let $ \mathcal{L} = \mathcal{L}(\mathcal{D}_1) $.  The automaton  $ \mathcal{D}_1 $ is a minimum DFA  for $\mt L$. The   states $0, \dots,5$ are such that: $I_0=\{\varepsilon\}$, $I_1= ac^* $, $I_2= bc^*$, $I_3=ac^*d \cup \{gd,  ee,he, f,k \}$, $I_4=\{e,h\}$, $I_5=\{g\}$. States $1 $ and $ 2 $ are $\leq_{\mt D_1}$-incomparable because $a\in I_1, b\in I_2, ac\in I_1$ and $a\prec b\prec ac$.
Similarly one checks that states $3,4,5$ are pairwise  $\leq_{\mt D_1}$-incomparable. On the other hand, $0$ is the minimum and states $1,2$ precede states $3,4,5 $ in the order $\leq_{\mt D_1}$. We conclude that the Hasse diagram of the partial order $\leq_{\mt D_1}$ is the one depicted in Figure \ref{fig:hasse}.

\begin{figure}
\begin{center}
\begin{tikzpicture}[shorten >=1pt,node distance=1cm,on grid,auto]
	\tikzstyle{every state}=[fill={rgb:black,1;white,10}]
	
	\node[state,color=white,text=black,inner sep=1pt,minimum size=0pt] (r_0)                   {$0$};
	\node[state,color=white,text=black,inner sep=1pt,minimum size=0pt] (r_1)  [above left of=r_0]   {$1$};
	\node[state,color=white,text=black,inner sep=1pt,minimum size=0pt] (r_3)  [above left  of=r_1]   {$3$};
	\node[state,color=white,text=black,inner sep=1pt,minimum size=0pt] (r_2)  [above right of=r_0]   {$2$};
	\node[state,color=white,text=black,inner sep=1pt,minimum size=0pt]
	(r_5)  [above right  of=r_2]   {$5$};
	\node[state,color=white,text=black,inner sep=1pt,minimum size=0pt] 	(r_4)  [above left  of=r_2]   {$4$};

	\path[-]
	(r_0) edge node {}    (r_1)
	(r_1) edge node {}    (r_3)
	(r_1) edge node {}    (r_4)
	(r_1) edge node {}    (r_5)
		(r_0) edge node {}    (r_2)
	(r_2) edge node {}    (r_3)
	(r_2) edge node {}    (r_4)
	(r_2) edge node {}    (r_5);
	\end{tikzpicture}
	\caption{The Hasse diagram of $\leq_{\mt D_1}$} \label{fig:hasse}
	\end{center}
\end{figure}
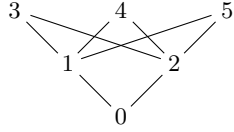

The width of the DFA $\mt D_1$ is $ 3 $  because $ \{3, 4, 5 \} $ is a largest $ \leq_{{\mathcal D}_1} $-antichain. A $ \leq_{{\mathcal D}_1} $-chain partition of cardinality $ 3 $ is, for example, $ \{\{0, 1, 3 \}, \{2, 4 \}, \{5 \} \} $.

Let us prove that $ \text{width}^D(\mathcal{L}) \ge 2 $. Suppose by contradiction that there exists a   DFA $\mt D$ of width $1$    recognizing  $\mt L $. Then, the order $\leq_{\mt D}$ is  total. Moreover,  there   exists a state $u$ such that two words of the infinite set 
   $ac^*\in \pf L$, say $ac^i, ac^j$  with $i<j$,    belong to $I_u$. Consider the word $bc^i \in \pf L$.
   Since   $bc^i\not \equiv_{\mt L} ac^{i}$,   it follows that $bc^i\not \in I_u$.  If $u'$ is such that $bc^i\in I_{q'}$, from $ac^i\prec bc^i\prec ac^{j}$ 
  we have
   that $u$ and $u'$ are $\leq_{\mt D}$-incomparable, a contradiction.

Finally, let $ \mathcal{D}_2 $ be the DFA in the center of Figure  \ref{fig:2minimumDFA} and let $ \mathcal{D}_3 $ be the DFA on the right of Figure  \ref{fig:2minimumDFA}. Notice that $ \mathcal{L}(\mathcal{D}_2) = \mathcal{L}(\mathcal{D}_3) = \mathcal{L} $, and $ \mathcal{D}_2 $ and $ \mathcal{D}_3 $ have just one more state than $ \mathcal{D}_1 $ and are non-isomorphic. We know that $ \mathcal{D}_2 $ and $ \mathcal{D}_3 $ cannot have width equal to 1. On the other hand, they both have width 2, as witnessed by the chains $\{ \{0,1,4\}, \{2,3 ,5, 3' \} \}$ (for $ \mathcal{D}_2 $) and $ \{ \{0,1,3\}, \{2,4, 5, 4' \}\} $ (for $ \mathcal{D}_3 $).
\end{example}

Motivated by the fact that computing the deterministic width of a regular language is not a trivial problem, in the next subsections we develop a set of tools that will ultimately allow us to derive an algorithm solving the problem. 

Example \ref{ex:no_minimal} triggers a further natural and important observation.
It is known  that    languages with  deterministic width equal to $1$  (that is, Wheeler languages)     admit a  (unique)  minimum-size Wheeler DFA  \cite{DBLP:conf/soda/AlankoDPP20}.
Note that Example \ref{ex:no_minimal}  implies that no such minimality result holds true for higher levels of the deterministic width hierarchy. In Section \ref{sec:minimal} we will explain why Example \ref{ex:no_minimal} is not the end of the story and we will derive an adequate notion of minimality.

\subsection{The Entanglement of a Regular Language}\label{sec:entanglement}

We now exhibit a measure that on the minimum DFA will capture exactly the width of the accepted language: the  {\em entanglement number} of a DFA. 
  We shall use the following terminology: if  $\mathcal D$ is a DFA  and $V\subseteq  \pf{L(\mathcal D)}$, then a state $u$ \emph{occurs} in $V$ if there exists $\alpha \in V$ such that $u=\delta(s,\alpha)$.  
 \begin{definition}\label{def:ent}
 Let $\mathcal D$ be a DFA with  set of  states $Q$. 
 \begin{enumerate}
     \item A subset $Q'\subseteq Q$ is \index{entangled set}  {\em entangled} if there exists a monotone sequence  $(\alpha_i)_{i\in \mathbb N}$ in $ \pf{L(\mathcal D)} $ such that for all $u' \in Q'$ it holds $\delta(s,\alpha_i)=u'$ for infinitely many $i$'s. In this case  the sequence $(\alpha_i)_{i\in \mathbb N}$ is said to be a {\em witness} for (the entanglement of) $Q'$. 
     \item A set $V\subseteq  \pf{L(\mathcal D)}$ is \emph{entangled  in $\mathcal D$ }   if there exists a monotone sequence $(\alpha_i)_{i\in \mathbb N}$, with $ \alpha_i \in V $ for every $ i $, witnessing that the set  $\{ \delta(s,\alpha)  ~|~ \alpha \in V\}$, consisting of all states occurring in $V$, is entangled.
     
 \end{enumerate}
 Moreover, define \index{ent(D)@$\text{ent}(\mathcal D)$, $\text{ent}(\mathcal L)$, entanglement}: \begin{align*}
& \text{ent}(\mathcal D) = max\{|Q'| \;|\; Q'\subseteq Q \text{~and~} Q'\text{is entangled } \}&\\
&  \text{ent}(\mathcal L) = {\text min} \{ \text{ent}(\mathcal D) \;|\; \mt D ~\text{is a DFA} ~ \land~\mt L(\mt D)=\mt L\}. &\\
\end{align*}
 \end{definition}

Notice that any singleton $\{u\} \subseteq Q$ is  entangled,  as  witnessed by the trivially monotone sequence $(\alpha_i)_{i\in \mathbb N}$ where all the $\alpha_i$'s are equal and $\delta(s,\alpha_i) = u$.

 As an example consider the entanglement of all  DFAs  in Figure  \ref{fig:2minimumDFA}. For any of them the entanglement is two,   because the only entangled subset of states is $\{1,2\}$, as witnessed by the sequence $a\prec b\prec ac \prec bc \prec acc\prec bcc \prec  \cdots  $. 
 
 \medskip
 
When two states $u\neq u'$ of a DFA  $ \mathcal{D} $  belong to an entangled set,   there are words $\alpha\prec \beta\prec \alpha'$ such that $\alpha,\alpha'\in I_u, \beta\in I_{u'}$,  so that neither $u<_{\mathcal D}  u' $ nor  $u'<_{\mathcal D}  u  $ can hold. In other words,  two distinct states $u, u'$  belonging to an entangled set are always $\leq_{\mt D}$-incomparable.  Since by Lemma \ref{lem:widthDFA} we have $\text{width}(\leq_{\mt D})=\text{width}(\mathcal D)$, it easily follows that   the entanglement of a DFA is always smaller than  or equal to its width.

 \begin{lemma}\label{lem:ent_leq_width} 
  Let $ \mathcal D$ be a DFA. Then
  $\text{ent}(\mathcal D)\leq \text{width}(\mathcal D).$
 \end{lemma}
 
 The converse of the above inequality is not always true:   for the (minimum) DFA $\mt D_1 $ on the left of  Figure \ref{fig:2minimumDFA} we have  $\text{ent}(\mathcal D_1)=2$ and $\text{width}(\mathcal D_1)=3$.
 
 \medskip
 
Contrary to what happens with the width, we  now prove that the    entanglement of a regular  language is realized by the minimum-size automaton accepting it. 
 
 \begin{lemma}\label{lem:minent}
If  $\mathcal D_{\mt L} $ is the minimum-size   DFA recognizing $\mathcal L$, then
 $\text{ent}(\mathcal D_{\mt L})= \text{ent}(\mathcal L)$.
 \end{lemma}
 \begin{proof}
It is enough to  prove that $\text{ent}(\mathcal D_{\mt L})\leq \text{ent}(\mathcal D)$, for any DFA  $\mathcal D$ such that $\mathcal L(\mathcal D_{\mt L})= \mathcal L(\mathcal D)$. Suppose $u_1, \dots, u_k$ are pairwise distinct states which are    entangled in $\mathcal D_{\mt L} $, witnessed by the monotone sequence $(\alpha_i)_{i\in \mathbb N}$. Since $\mathcal D_{\mt L}$ is minimum, each  $I_{u_j}$ is a union of a finite number of  $I_v$, with $v\in Q_{\mt D}$. 
 The monotone sequence $(\alpha_i)_{i\in \mathbb N}$
 goes through $u_j$ infinitely often,  so there must be a state $v_j\in Q_{\mt D}$ such that $I_{v_j}\subseteq I_{u_j}$ and  $(\alpha_i)_{i\in \mathbb N}$
 goes through $v_j$ infinitely often.
Then  $(\alpha_i)_{i\in \mathbb N}$ goes through the pairwise distinct states $v_1,\dots, v_k$ infinitely often  and $v_1, \dots, v_k$  are entangled in  $\mathcal D$. \qed
 
 \end{proof}
 
\subsection{The Hasse Automaton of a Regular Language}\label{sec:Hasse automaton of L}

  Our aim is to prove  that the entanglement measure over the minimum DFA $\mt D_{\mt L}$ captures the deterministic width of a language $\mt L$:
 \[\text{width}^D (\mt L)= \text{ent} (\mt D_{\mt L})\]   In order to prove the previous equality we shall  describe an automaton, the {\em Hasse automaton} of $\mathcal L$,    realizing the width and the entanglement of the language as its width
  (Theorem \ref{thm:hasse}). 
As a first step,  given a DFA $\mt D$  we prove that   there exists an equivalent DFA  $\mt D'$  that realizes the entanglement of $\mt D$ as its width:  $\text{ent} (\mt D)= \text{width} (\mt D')$ (Theorem \ref{thm:equivalentholeproof}). 
  
To give an intuition on the construction of the automaton $\mt D'$  we  use the {\em trace} of the  DFA $\mt D$,  that is, the (in general) transfinite   sequence: $(\delta(s, \alpha))_{\alpha \in \pf{ L}}$, indexed over the totally ordered set $(\pf L,\preceq)$, where $\mt L=\mt L(\mt D)$.  We depict below a hypothetical $(\pf L,\preceq)$,  together with the  trace left by a  DFA $\mt D$ with set of states $\{u_1, u_2, u_3 \}$ and  $\delta(s, \alpha_i)=\delta(s, \alpha')=u_1$, 
  $\delta(s, \beta_i)= \delta(s, \beta_i')=u_2$, and  $\delta(s, \gamma_i)=u_3$: 
  \begin{center}
  \small
   \begin{tabular}{ccccccccccccccccccccccccccccccccccccccccccc}
&$\alpha_1$& $ \prec $ &$\beta_1$& $ \prec $ &$\alpha_2$& $ \prec $ &$\beta_2$& $ \prec $ & $ \cdots $ & $ \prec $ & $\alpha_i$& $ \prec $ &$\beta_i$& $ \prec $ & $ \cdots $ & $ \prec $ &$\beta_1'$ & $ \prec $ &$\gamma_1$& $ \prec $ &$\beta_2'$& $ \prec $ &$\gamma_2$& $ \prec $ & $ \cdots $ & $ \prec $ &$\beta_i'$& $ \prec $ &$\gamma_i$& $ \prec $ & $ \cdots $ & $ \prec $ & $\alpha'$&\\
 &$\color{red}{ u_1}$&&$\color{blue}{ u_2}$&&$\color{red}{ u_1}$&&$\color{blue}{ u_2}$&&$\dots$&&$\color{red}{ u_1}$&&$\color{blue}{ u_2}$&&$\dots $&&$\color{blue}{ u_2}$&&$\color{orange}{ u_3}$&&$\color{blue}{ u_2}$&&$\color{orange}{ u_3}$&&$\dots$&& $\color{blue}{ u_2}$&&$\color{orange}{ u_3}$&&$\dots $&&$\color{red}{ u_1}$
\end{tabular}
\end{center}
   
Consider  the entanglement and  width of $\mt D$.  Notice that the sets $\{u_1, u_2\}$ and $\{u_2,u_3\}$ are entangled. The set  $\{u_1, u_3\}$ is not entangled  and therefore the set $\{u_1, u_2, u_3\}$ is not entangled.  However, $\{u_1, u_2, u_3\}$ contains pairwise incomparable states. 
 Hence the whole triplet $ \{u_1, u_2, u_3 \} $ does not contribute  to the entanglement but does contribute to the width  so that    $\text{ent}(\mt D)=2< \text{width}(\mt D)=3$. \\ 
In general,   an automaton where     incomparability  and  entanglement coincide  would have $\text{ent}(\mt D)=  \text{width}(\mt D)$. Hence, we would like   to force \emph{all} sets of incomparable states in the new automaton $\mt D'$ to be entangled.  
 To this end,   
   we will first prove that there always exists a finite, ordered partition $\mt V=\{V_1, \dots, V_r\}$ of 
$\pf L$ composed of convex sets which are entangled in $\mt D$. In the    example above  we can write $\pf L=V_1\cup V_2\cup V_3 $, where: 
\[V_1=\{\alpha_1,\beta_1, \dots, \alpha_i, \beta_i, \dots\}, V_2=\{\beta_1',\gamma_1, \dots, \beta_i', \gamma_i\dots \},V_3=\{\alpha'\}\]
and the states occurring (and entangled) in $V_1,V_2,V_3$ are, respectively: $\{u_1,u_2\}$, $\{u_2,u_3\}$, and $\{u_1\}$. 
In order to construct an equivalent automaton $\mt D'$ in which  the pairwise incomparability  of the three states  $u_1,u_2,u_3$ is eliminated and $\text{width}(\mt D')=2$,  we  could try  to \emph{duplicate} some of the original states, as it would be   the case if  the states occurring in $V_1,V_2,V_3$ where, respectively, 
  $\{u_1,u_2\}$, $\{u_2,u_3\}$, and $\{u_1'\}$.    To this end, we will consider a refinement $\sim$ of the Myhill-Nerode equivalence on $\pf L$ stating that   two strings are equivalent if and only if they are in the same $ I_u $ and all   $ V \in \mt V $ laying between the two strings intersect $I_u $.
   In the above example we have $\beta_i \sim   \beta'_j$  for all  integers $i,j$, because   no $V \in \mt V $ is contained in $[\beta_i,\beta'_j]$, while    $\alpha_1\not \sim  \alpha'$,  since $V_2\subseteq [\alpha_1,  \alpha'] $ but $V_2\cap I_{u_1}= \emptyset$.

We will prove that the equivalence $\sim $    decomposes  the  set  of   words reaching $u$ into a {\em finite} number of $\sim $-classes  and  induces a well-defined quotient automaton $ \mathcal{D'} $ equivalent to $ \mathcal{D} $.

By construction,  in  the automaton $ \mathcal{D'} $    any  set of $\prec_{\mt {D'}}$-incomparable states $\{u_1, \dots, u_k\}$ will occur  in at least an element   $V\in \mt V$, so that, $V$ being entangled,  they will contribute  to the  entanglement number of $\mt D$ and   $\text{width}({\mt D'})= \text{ent}({\mt D}) $ will follow.

In our example, the new automaton $\mt D'$  will leave  the following trace:
  \begin{center}
  \small
   \begin{tabular}{ccccccccccccccccccccccccccccccccccccccccccc}
&$\alpha_1$& $ \prec $ &$\beta_1$& $ \prec $ &$\alpha_2$& $ \prec $ &$\beta_2$& $ \prec $ & $ \cdots $ & $ \prec $ & $\alpha_i$& $ \prec $ &$\beta_i$& $ \prec $ & $ \cdots $ & $ \prec $ &$\beta_1'$ & $ \prec $ &$\gamma_1$& $ \prec $ &$\beta_2'$& $ \prec $ &$\gamma_2$& $ \prec $ & $ \cdots $ & $ \prec $ &$\beta_i'$& $ \prec $ &$\gamma_i$& $ \prec $ & $ \cdots $ & $ \prec $ & $\alpha'$&\\
 &$\color{red}{ u_1}$&&$\color{blue}{ u_2}$&&$\color{red}{ u_1}$&&$\color{blue}{ u_2}$&&$\dots$&&$\color{red}{ u_1}$&&$\color{blue}{ u_2}$&&$\dots $&&$\color{blue}{ u_2}$&&$\color{orange}{ u_3}$&&$\color{blue}{ u_2}$&&$\color{orange}{ u_3}$&&$\dots$&& $\color{blue}{ u_2}$&&$\color{orange}{ u_3}$&&$\dots $&&$\color{green}{ u_1'}$
\end{tabular}
\end{center}
and     $\text{ent}(\mt D )=\text{width}(\mt D')=2$. 

\medskip
 
In order to give a formal definition of  $\mt D'$ we  need some   properties of entangled sets. Since these properties hold true  with respect to a partition of  a generic total order,  we state  and prove  them in a more general setting in  Appendix \ref{app:minimal_convex},   while we state them  without proof in this section. 
In particular, most  of the properties of $\mt D'$ will be proved using a   finite partition of  $ (\text{Pref}(\mathcal{L(D)}), \preceq)$, composed by entangled, convex sets.

\begin{definition}\label{def:ec decomposition_automaton_case}
If $\mt D$ is a DFA, we say that a  partition $\mt V$  of $\pf {L(\mathcal{D})}$ is an \emph{entangled, convex  decomposition  of $ \mt D$}  (\emph{e.c. decomposition}, for short) if
all the elements of $ \mathcal V  $ are convex in $(\pf{L(\mathcal{D})}, \preceq)$ and entangled in $\mt D$.
\end{definition}

\begin{theorem}\label{thm:fdt} If $\mt D$ is a DFA, then there exists a finite  partition $\mt V$  of $\pf {L(\mathcal{D})}$ which is an e.c. decomposition of $\mt D$. 
\end{theorem}

\begin{proof}The existence of such a decomposition  is guaranteed  by Theorem \ref{thm:generalfdt} of Appendix \ref{app:minimal_convex}, applied to  $(Z,\leq)=(\pf {L(\mathcal{D})}, \preceq)$ and $\mt P=  \{I_u \;|\; u\in Q\}$. \qed
 \end{proof}

Using an    e.c. decomposition of an automaton  $\mt D$ we can  express  a condition implying the equality  $\text{width}(\mathcal D)=\text{ent}(\mathcal D) $.

 \begin{lemma}\label{lem:width=ent}  Let $\mt D$ be a DFA. Suppose $ V_1 \prec V_2 \prec \dots \prec V_m$ is  an e.c. decomposition   of  $\mt D$ such that for every $u\in Q$ there are $1\leq i\leq j\leq m$ with: 
  \[I_u\subseteq V_{i}\cup V_{i+1}\cup \dots \cup V_{j}, ~~~ \text{and } ~~  V_h\cap I_u\neq \emptyset,  ~~ \text{for all } ~~  i\leq h\leq j.\]
Then, $\text{width}(\mathcal D)=\text{ent}(\mathcal D) $.
 \end{lemma}
 
 \begin{proof} We already proved that     $\text{ent}(\mathcal D)\leq \text{width}(\mathcal D) $ in Lemma \ref{lem:ent_leq_width}. In order to prove the reverse inequality, 
 let $\text{width}(\mathcal D) =p$ and let $u_1, \dots, u_p$ be $p$ $\preceq_{\mt D}$-incomparable states. If we prove that $u_1, \dots, u_p$ are   entangled, we are done. 
Fix $i\in \{1, \dots,p\}$ and denote by  $W_i$ the convex set $V_{h}\cup  V_{h+1}\cup  \dots\cup V_{k}$ where   $I_{u_i}\subseteq V_{h}\cup V_{h+1}\cup \dots \cup V_{k}$ and  $V_j\cap I_{u_i}\neq \emptyset$,  for all $h\leq j\leq k$.   
From the incomparability of the $u_i$'s it follows that  $W_i\cap W_j\neq \emptyset$, for all pairs  $i,j$,   so that   
$\bigcap_i W_i\neq \emptyset$   by Lemma \ref{lem:con-int} in Appendix \ref{app:minimal_convex}. Since all $W_i$'s are unions of consecutive elements in the partition $\mt V$,  
 there must be an element  $V\in \mt V$  with 
$V\subseteq \bigcap_i W_i$. Such a  $V$ must contain  an occurrence of every $u_i$,  and since $V$ is an entangled set, we conclude that  $u_1, \dots, u_p$  are entangled. \qed
 \end{proof}

\begin{figure}[H]
 \begin{center}
\begin{tikzpicture}[scale=1.2]
\draw[black, thick] (-5,0) -- (5,0);
\draw[black, thick] (-5,-0.2) -- (-5,0.2);
 \draw node at (-6, 0){$\pf L$};
\draw node at (-4, -0.5){$V_1$};
 \draw node at (6, 0.5){$\mt D$-states};
  \draw node at (-2, -0.5){$V_2$};
   \draw node at (0, -0.5){$V_3$};
 \draw node at (2, -0.5){$V_4$};
  \draw node at (4, -0.5){$V_5$};
  \draw node at (-4, -0.5){$V_1$};
  \draw node at (-2, -0.5){$V_2$};
   \draw node at (0, -0.5){$V_3$};
 \draw node at (2, -0.5){$V_4$};
  \draw node at (4, -0.5){$V_5$};
\draw[black, thick](-3,-0.2) -- (-3,0.2);
\draw[black, thick] (-1,-0.2) -- (-1,0.2);
\draw[black, thick](1,-0.2) -- (1,0.2);
\draw[black, thick](3,-0.2) -- (3,0.2);
\draw[black, thick](5,-0.2) -- (5,0.2);
\filldraw[red] (-3.5,0) circle (2pt) node[anchor=south]{$u_2$}node[anchor=north]{$\alpha\phantom{'}$};
\filldraw[blue] (-4.5,0) circle (2pt) node[anchor=south]{$u_1$};
\filldraw[red] (-1.5,0) circle (2pt) node[anchor=south]{$u_2$};
\filldraw[green] (-2.5,0) circle (2pt) node[anchor=south]{$u_3$};
\filldraw[blue] (0.5,0) circle (2pt) node[anchor=south]{$u_1$};
\filldraw[green] (-0.5,0) circle (2pt) node[anchor=south]{$u_3$};
\filldraw[blue] (2.5,0) circle (2pt) node[anchor=south]{$u_1$};
\filldraw[red] (1.5,0) circle (2pt) node[anchor=south]{$u_2$}node[anchor=north]{$\alpha'$};
\filldraw[blue] (4.5,0) circle (2pt) node[anchor=south]{$u_1$};
\filldraw[red] (3.5,0) circle (2pt) node[anchor=south]{$u_2$}node[anchor=north]{$\alpha''$};
\end{tikzpicture}
\end{center}
  \begin{center}
\begin{tikzpicture}[scale=1.2]
\draw[black, thick] (-5,0) -- (5,0);
\draw[black, thick] (-5,-0.2) -- (-5,0.2);
 \draw node at (-6, 0){$\pf L$};
\draw node at (-4, -0.5){$V_1$};
 \draw node at (6, 0.5){$\mt D'$-states};
  \draw node at (-2, -0.5){$V_2$};
   \draw node at (0, -0.5){$V_3$};
 \draw node at (2, -0.5){$V_4$};
  \draw node at (4, -0.5){$V_5$};
  \draw node at (-4, -0.5){$V_1$};
  \draw node at (-2, -0.5){$V_2$};
   \draw node at (0, -0.5){$V_3$};
 \draw node at (2, -0.5){$V_4$};
  \draw node at (4, -0.5){$V_5$};
\draw[black, thick](-3,-0.2) -- (-3,0.2);
\draw[black, thick] (-1,-0.2) -- (-1,0.2);
\draw[black, thick](1,-0.2) -- (1,0.2);
\draw[black, thick](3,-0.2) -- (3,0.2);
\draw[black, thick](5,-0.2) -- (5,0.2);
\filldraw[red] (-3.5,0) circle (2pt) node[anchor=south]{$u_2$}node[anchor=south]{$u_2$}node[anchor=north]{$\alpha\phantom{'}$};
\filldraw[blue] (-4.5,0) circle (2pt) node[anchor=south]{$u_1$};
\filldraw[red] (-1.5,0) circle (2pt) node[anchor=south]{$u_2$};
\filldraw[green] (-2.5,0) circle (2pt) node[anchor=south]{$u_3$};
\filldraw[violet] (0.5,0) circle (2pt) node[anchor=south]{$u_1'$};
\filldraw[green] (-0.5,0) circle (2pt) node[anchor=south]{$u_3$};
\filldraw[violet] (2.5,0) circle (2pt) node[anchor=south]{$u_1'$};
\filldraw[orange] (1.5,0) circle (2pt) node[anchor=south]{$u_2'$}node[anchor=north]{$\alpha'$};
\filldraw[violet] (4.5,0) circle (2pt) node[anchor=south]{$u_1'$};
\filldraw[orange] (3.5,0) circle (2pt) node[anchor=south]{$u_2'$}node[anchor=north]{$\alpha''$};
\end{tikzpicture}
\end{center}
\caption{The upper line  represents the words in $\pf L$, partitioned by the e.c. decomposition $\mt V$, where  some words  are highlighted and labelled by the state they reach in $\mt D$. 
The lower line still represents  $\pf L$, with the same e.c. decomposition, but now the highlighted words are labelled by states of $\mt D'$.
Note that the strings $\alpha, \alpha'$ reach the same state in $\mt D$, but in different states in $\mt D'$,  because   $V_3\subseteq [\alpha, \alpha']$, and $V_3 \cap I_{u_2}=\emptyset$.  However, the words $\alpha', \alpha''$ reach the same state in both automata. } 
\label{fig:duplication}
\end{figure}
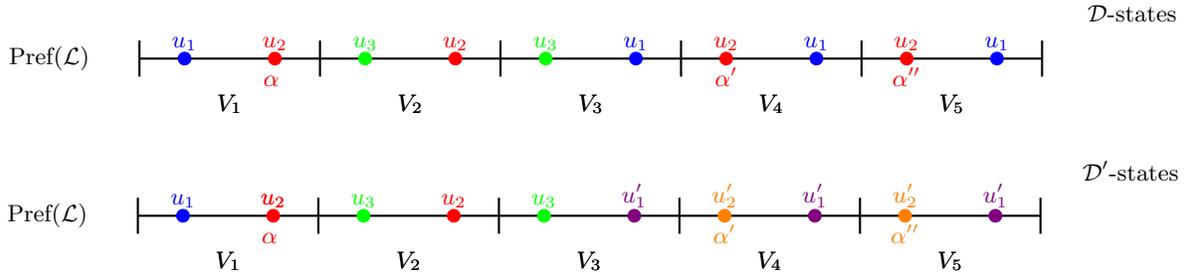

 The previous lemma suggests that  in order to construct an automaton $\mt D'$ recognizing the same language as $\mt D$ and satisfying    $\text{width}(\mathcal D')=\text{ent}(\mathcal D') =\text{ent}(\mathcal  D)$,  we  might \emph{duplicate} some states in $\mt D$ in order to ensure that the new automaton      $\mt D'$    satisfies  the condition of the previous Lemma.   Consider two words 
 $\alpha \prec  \alpha'$ reaching the same state $u$ of $\mt D$:
 if the convex  $[\alpha, \alpha']$  is not contained in a union of consecutive elements of the partition $\mt V$, all having an occurrence of $u$, then in $\mt D'$ we  duplicate the state $u$ into    $u$ and $u'$, with $\alpha$ reaching $u$ and $\alpha'$ reaching $u'$  (see Fig. \ref{fig:duplication}).  
As usual, an  equivalence relation $\sim_{\mt D}$ over $\pf L$ is  used to introduce the new states of the automaton $\mt D'$. In order to maintain  the definition of $\mt D'$ independent from any particular 
e.c. decomposition $\mt V$, in the definition below we use generic   entangled convex sets instead of elements of an e.c. decomposition.  In Lemma \ref{lem:equivalent_sim} we prove that this is equivalent to using elements of an e.c. decomposition of minimum cardinality. 

 \begin{definition}\label{def:easy_sim}
 Let $\mt D$ be a DFA and let $\sim_{\mt D} $ be the equivalence relation on  $ \pf {L(D) }$ defined as follows: $\alpha \sim_{\mt D}  \alpha'$ if and only if:
 \begin{itemize}
     \item $\delta(s,\alpha)= \delta(s, \alpha')$ and
     \item there are  entangled convex sets $ C_1, \dots,C_n\subseteq \pf {L( \mt D)}$  such that:
     \begin{description}
         \item[-] $[\alpha, \alpha']^\pm \subseteq \bigcup_{i=1}^{n} C_i$;
         \item[-]  $C_i\cap I_{\delta(s,\alpha)}\neq \emptyset$, for all $i\in \{1, \dots,n\}$.
     \end{description} 
 \end{itemize}
\end{definition}

Note that $ \sim_\mathcal{D} $ is indeed an equivalence
relation (in particular, it is transitive). When the DFA $\mt D$ is clear from the context, we shall drop the subscript $\mt D$ in $\sim_{\mt D}$. 

\medskip

In the following lemma we prove that the equivalence $\sim$ has finite index and $\mt L(\mt D)$ is equal to the union of some of its classes. 


\begin{lemma}\label{lem:sim_finite}
Let $ \mathcal D$ be a DFA. Then, $\sim$ has a finite number of classes on  $\pf {L(\mt D)}$ and 
$\mt L(\mt D)$ is equal to the union of some $\sim$-classes.
\end{lemma}

\begin{proof} Consider an  e.c. decomposition $\mt V=\{V_1, \dots, V_m\}$ of $\mt D$, whose existence is guaranteed by Theorem \ref{thm:fdt}. Since all $V_i$'s are    entangled convex sets in $\pf {L( \mt D)}$,   two words belonging to the same $V_i$ and ending in the same state belong to the same $\sim$-class.  Hence,  the number of $\sim$-classes is at most $m\times |Q| $.
  Moreover,   $\alpha\sim \beta$ implies $\delta(s,\alpha)=\delta(s,\beta)$. Hence,    $\alpha\in \mt {L(\mt D)}$ and $\alpha\sim \beta$  imply  $\beta\in \mt {L(\mt D)}$, proving  that  $\mt L(\mt D)$ is equal to the union of some $\sim $-classes. \qed
\end{proof}

\begin{lemma}\label{lem:simequivalence} Let $\mt D$ be a DFA. Then, the equivalence relation $\sim$ is right-invariant. 
\end{lemma}
\begin{proof} Let $\mt L=\mt L(\mt D)$, assume $\alpha \sim \alpha' $ and  let $a\in \Sigma$ be such that $\alpha a \in \pf{L }$. We must prove that $\alpha' a\in  \pf{L }$ and  $\alpha a\sim \alpha' a$. Since $\alpha \sim \alpha' $ we know that $\delta(s,\alpha)= \delta(s, \alpha')$ and there exist entangled convex sets $ C_1, \dots, C_n $ such that $[\alpha, \alpha']^\pm \subseteq C_1 \cup \dots \cup C_n $ and $ C_i\cap I_{\delta(s,\alpha)}\neq \emptyset$ for all $i=1, \dots,n$. We must prove that $\alpha' a\in  \pf{L }$, $\delta(s,\alpha a)= \delta(s, \alpha' a)$, and there exist entangled convex sets $ C'_1, \dots, C'_{n'} $ such that $[\alpha a, \alpha' a]^\pm \subseteq C'_1 \cup \dots \cup C'_{n'} $ and $ C'_i\cap I_{\delta(s,\alpha a)}\neq \emptyset$ for all $i=1, \dots,n'$.

From $\delta(s,\alpha)= \delta(s, \alpha')$ and $\alpha a \in \pf{L }$ we immediately obtain $\alpha' a \in \pf{L }$ and $\delta(s,\alpha a)= \delta(s, \alpha' a)$. Moreover, from $[\alpha, \alpha']^\pm \subseteq C_1 \cup \dots \cup C_n $ we obtain $[\alpha a, \alpha' a]^\pm = [\alpha, \alpha']^\pm a \subseteq C_1a \cup \dots \cup C_na $, and from $ C_i\cap I_{\delta(s,\alpha)}\neq \emptyset$ we obtain $ C_i a \cap I_{\delta(s,\alpha a)}\neq \emptyset $. We are only left with showing that every $ C_i a =\{\gamma a \; | \: \gamma \in C_i\}$ is an entangled convex set. The fact that they are convex follows directly from the definition of co-lex ordering. Let us prove that the $ C_i a $'s are entangled. Fix $ i $ and consider a monotone sequence $ (\alpha_j)_{j \in \mathbb{N}} $ witnessing  that $ C_i $ is entangled. Then $ (\alpha_j a)_{j \in \mathbb{N}} $ is a monotone sequence witnessing that $ C_i a $ is entangled. \qed
\end{proof}

We are now ready to complete the construction of the automaton $\mt D'$ using  $ \sim $.

\begin{definition}\label{def:D'}
Let  $\mt D$ be  a DFA and let $\mt L=\mt L(\mt D)$. 
Define $ \mathcal D' =(Q', s', \delta', F') $ by:
\begin{itemize}
	\item $ Q' = \{  [\alpha]_{\sim}: \alpha \in \pf L\}$;
	\item $ \delta'([\alpha]_{\sim},a) = [\alpha a]_{\sim} $ for every $ \alpha \in \pf L $ and for every $ a \in \Sigma $ such that $ \alpha a \in \pf L $;
	\item $ s' = [ \varepsilon ]_\sim $;
	\item $ F'= \{ [\alpha]_\sim:\alpha \in \mathcal L\} $.
\end{itemize}
\end{definition}

The equivalence relation $ \sim $ is right-invariant (Lemma \ref{lem:simequivalence}),   has finite index,  and $ \mathcal{L} $ is the union of some $ \sim $-classes (Lemma  \ref{lem:sim_finite}). Hence, $ \mathcal{D'} $ is a well-defined DFA, and $ \alpha \in [\beta]_\sim \iff \delta' (s', \alpha) = [\beta]_\sim $, which implies that for every $ \alpha \in \pf L $ it holds:
\begin{equation}\label{eq:sim}
  I_{[\alpha]_\sim} = [\alpha]_\sim
\end{equation}
and so $ \mathcal{L}(\mathcal{D}') = \mt L $.
 
In the following lemma we prove that it is safe to replace $C_1, \dots , C_n$ in Definition \ref{def:easy_sim} by the elements of a \emph{minimum-size} e.c. decomposition, that is, an e.c. decomposition with minimum cardinality.

\begin{lemma}\label{lem:equivalent_sim}
 Let $\mt D$ be a DFA and let $ \mathcal{V} = \{V_1, \dots, V_r \} $, with $ V_1 \prec \dots \prec V_r $, be a minimum-size e.c. decomposition of $\mt D$. 
Then,  $\alpha\sim  \alpha'$ holds if and only if:
\begin{itemize}
\item $\delta(s,\alpha)= \delta(s, \alpha')$ and 
 \item there exist integers $i \le j $ such that:
    \begin{description}
      \item[-] $[\alpha, \alpha']^{\pm} \subseteq \bigcup_{h=i}^j V_h$;
      \item[-] $V_h\cap I_{\delta(s,\alpha)}\neq \emptyset$, for all $h \in \{i, \dots , j \} $.
    \end{description}
\end{itemize}
\end{lemma}
\begin{proof}
 To prove   that $\alpha \sim \alpha'$ holds under the above  hypotheses, it is sufficient to recall  that $V_i , \dots , V_j$ are entangled convex sets and apply Definition \ref{def:easy_sim}.

Let us prove the reverse implication. Pick $ \alpha \sim \alpha' \in \text{Pref}(\mathcal{L(D)}) $ and let  $C_1, \dots,C_n$ be entangled convex sets such that $[\alpha, \alpha']^{\pm} \subseteq \bigcup_{i=1}^n C_i$ and $C_i\cap I_u\neq \emptyset$, for every $i=1, \dots,n$, where $ u = \delta(s,\alpha)= \delta(s, \alpha') $. By Lemma \ref{lem:convex_in_order} of Appendix \ref{app:minimal_convex} we can assume that $C_1\prec C_2 \prec \dots \prec C_n$. Let $i\leq j$ be such that $[\alpha, \alpha']^{\pm} \subseteq \bigcup_{h=i}^j V_h$ and $ V_h\cap  [\alpha, \alpha']^{\pm} \neq \emptyset$ for all $h\in \{i, \dots, j\} $. We just have to prove that $V_h\cap I_{u}\neq \emptyset$, for all $h\in \{i, \dots, j \} $.  From $ V_h\cap  [\alpha, \alpha']^{\pm} \neq \emptyset$ it follows that either  $V_h$ contains $\alpha$ or $\alpha'$,    or $V_h\subseteq [\alpha, \alpha']^{\pm} \subseteq   \bigcup_{i=1}^n C_i$. In the first case we have $V_h\cap I_{u}\neq \emptyset$ because $\alpha, \alpha'\in I_u$, while in the second case $V_h\cap I_{u}\neq \emptyset$  follows from  Lemma \ref{lem:noioso} of   Appendix \ref{app:minimal_convex}.\qed 
\end{proof}
Let $\mt D'$ be the automaton of Definition \ref{def:D'}. The following corollary  allows us to consider  an e.c. decomposition of $\mt D$ of minimum  size  as an e.c. decomposition of  $\mt D'$. 
\begin{corollary} \label{cor:D'} Any    e.c.  decomposition  of minimum size  $ V_1\prec \dots \prec V_r$  of $\mt D$ is also    an e.c. decomposition  of $\mt D'$. Moreover, 
for all $u' \in Q'$, there exist $i \le j$ such that 
$I_{u'} \subseteq \bigcup_{h=i}^j V_{h}$
and $V_{h}\cap I_{u'}\neq \emptyset$, for $h=i, \dots , j$. 
\end{corollary}
\begin{proof} 
  In order to prove that an e.c. decomposition of minimum size $ V_1\prec \dots \prec V_r$  of $\mt D$ is also    an e.c. decomposition  of $\mt D'$  we just have to check that   $V_h$ is entangled in $\mt D'$, for all $h=1, \dots,r$.  
  Let $u_1', \dots, u'_k$ be the pairwise distinct  $\mt D'$-states occurring in $V_h$. Notice that by the definition of $ \delta' $ we have $ \{u_1', \dots, u'_k\}=\{\delta'(s',\alpha) \;|\; \alpha \in V_h\}=
  \{[\alpha]_\sim \;|\; \alpha \in V_h\} $. Hence, for every $ j = 1, \dots, k $ there exists  $\alpha_j\in V_h$ such that $ u_j' = [\alpha_j]_\sim $. Then  the $\mt D$-states  
  $u_j=\delta(s, \alpha_j)$, for $j=1, \dots,k$,   occur in $V_h$. Notice that $u_1, \dots, u_k$ are pairwise distinct as well: if $u_i$ were equal to $u_j$  for $i\neq j$,  then from Lemma \ref{lem:equivalent_sim} we would have $\alpha_i\sim \alpha_j$   and $u_i'=[\alpha_i]_\sim=[\alpha_j]_\sim=u_j'$ would follow.  Since $V_h$ is entangled in $\mt D$,   there exists a monotone sequence $(\beta_i)_{i\in \mathbb N}$ in $V_h$ reaching each $u_j$ infinitely many times. Fix $j\in \{1, \dots,k\}$.  If $\delta(s,\beta_i)=u_j$ then from 
  $\delta(s,\alpha_j)=u_j$ and $\beta_i, \alpha_j\in V_h$ it follows $\alpha_j\sim \beta_i$ again by Lemma \ref{lem:equivalent_sim}, so that   $\delta'(s',\beta_i)=[\beta_i]_\sim=[\alpha_j]_\sim=u_j'$ in $\mt D'$. It follows   that  the   sequence  $(\beta_i)_{i\in \mathbb N}$  reaches   $u_j'$  infinitely many times. Hence, $V_h$ is entangled in $\mt D'$. 
  
As for the second part of the Corollary, if $u'=[\alpha]_\sim  \in Q'$ then $I_{u'}=[\alpha]_\sim$ (see Equation \ref{eq:sim} above). Let $i$   ($j$, respectively) be  the minimum (maximum) index $h$ with $[\alpha]_\sim\cap V_h\neq \emptyset$; then  $[\alpha]_\sim\subseteq V_i \cup \dots  \cup V_j $.  Fix $h\in \{i, \dots, j \} $ and consider  $u=\delta(s,\alpha)$. Since there exist $ \alpha', \alpha'' \in \text{Pref}(\mathcal{L(D)}) $ such that $ \alpha' \sim \alpha \sim \alpha'' $, $ \alpha' \in V_i $ and $ \alpha'' \in V_j $, then,     Lemma \ref{lem:equivalent_sim} implies $ \delta(s, \alpha') = \delta (s, \alpha'') = \delta (s, \alpha) = u $ and $V_h\cap I_{u}  \neq \emptyset$. Pick  $\beta \in V_h\cap I_{u}$. Then,  the same   lemma implies  $\beta \sim \alpha$ so that $\beta \in  I_{u'}$ and   $V_h\cap I_{u'}   \neq \emptyset$ as well.  \qed
\end{proof}

We can now prove that $\mt D'$ has width equal  (to its entanglement and) to the entanglement of $\mt D$. 
  \begin{theorem}\label{thm:equivalentholeproof}
If   $\mt D$ is  a DFA,   then $\text{ent}(\mt D)=\text{ent}(\mt D')= \text{width}(\mt D')$.
 \end{theorem}
 \begin{proof} 
Let us prove that $\text{ent}(\mt D')\leq \text{ent}(\mt D)$. Consider an entangled collection $ \{[\alpha_1]_\sim, \dots, [\alpha_h]_\sim \} $ of $h$ states in $ \mathcal{D'} $. Then, there is a monotone sequence $(\gamma_i)_{i\in \mathbb N}$ such that, for each $j\in \{1,\dots,h\}$ we have 
$\delta'(s', \gamma_i)= [\alpha_j]_\sim$ for infinitely many $i$'s.   Let $ \mathcal{V} $ be a minimum-size finite e.c. decomposition of $\mt D$. Since $\mt V$ is a finite partition and all the elements of $ \mathcal{V} $ are convex,  there exists $ V \in \mathcal{V} $ and $n_0$ such that   $\gamma_i\in V$ for all $i\geq n_0$.
In particular,  there are words $\beta_1, \dots,\beta_h$ in $V$ such that   $\delta'(s', \beta_{k}) = [\alpha_k]_\sim $, for every $ k = 1, \dots, h$. Define $ u_k = \delta (s, \beta_{k}) $ and notice that the states $ u_1, \dots, u_{h} $ are pairwise distinct. In fact, if $ u_r = u_s $  for $r\neq s$, then  by Lemma \ref{lem:equivalent_sim} we would have  $ \beta_{r} \sim \beta_{s} $ and $ [\alpha_r]_\sim=\delta'(s', \beta_{r})=[\beta_r]_\sim= [\beta_s]_\sim=\delta'(s', \beta_{s})= [\alpha_s]_\sim $,  
a contradiction.   Moreover, $ \{u_1, \dots, u_{h}\} $ is an entangled set in $ \mathcal{D} $, because all these states occur  in $ V $ (as witnessed by $ \beta_{ 1}, \dots, \beta_{ h} $) and $ V $ is an element of an e.c. decomposition.   Since this holds for any collection of   entangled states in $\mt D'$, it  follows   that $\text{ent}(\mt D')\leq \text{ent}(\mt D)$.

Let us now prove that $\text{ent}(\mt D)\leq \text{ent}(\mt D')$. Let $ \{u_1, \dots, u_{h} \} $ be an entangled set of $h$  states in $ \mathcal{D} $, witnessed by some monotone sequence $ (\alpha_i)_{i \in \mathbb{N}} $. Every $ I_{u_k} $ is equal to a  finite  union of some $ I_{u'} $'s, with $ u' \in Q' $,  and 
  $ (\alpha_i)_{i \in \mathbb{N}} $ goes through any $ u_k $ infinitely many times. Therefore, for all $k=1, \dots h$  there exist $ u'_k \in Q' $ such that $ I_{u'_k} \subseteq I_{u_k} $ and $ (\alpha_i)_{i \in \mathbb{N}} $ goes through $ u'_k $ infinitely many times. We conclude that $ u'_1, \dots, u'_h $ are pairwise distinct and $ \{u'_1, \dots, u'_h \} $ is an entangled set of states in $ \mathcal{D'} $, which implies $\text{ent}(\mt D)\leq \text{ent}(\mt D')$.

Finally, we prove that $ \text{width}(\mt D')= \text{ent}(\mt D')$. If $\mt V$ is  an e.c. decomposition of $\mt D$ of minimum size,  then   Corollary \ref{cor:D'} implies that   $\mt V$ is  an   e.c. decomposition of $\mt D'$ satisfying the hypothesis of  Lemma \ref{lem:width=ent} so that   $ \text{width}(\mt D')= \text{ent}(\mt D')$ follows from this lemma.   \qed
\end{proof}

If we start from  the  minimum DFA $\mt D_{\mt L}$ of a regular language $\mt L$, then,  as we shall see in Theorem \ref{thm:hasse},  the  automaton  $\mt D_{\mt L}'$ acquires a special role because it realizes the deterministic width of the language $\mt L$.

\begin{definition}\label{def:Hasse}\index{Hasse automaton} If $\mt D_{\mt L}$ is the minimum DFA of a regular language $\mt L$,  the DFA $\mt D_{\mt L}' $
is called the \emph{Hasse automaton} \index{Hasse automaton} for $\mt L$ and it is denoted by $\mt H_{\mt L}$.
\end{definition}

The above definition is motivated by the fact that   the width of the language can be  ``visualized" by the Hasse diagram of the partial order $\leq_{\mt H_{\mt L}}$.

 \begin{theorem}\label{thm:hasse} 
If $\mt D_{\mt L}$ is the minimum DFA of the regular language $\mt L$, then:
\[\text{width}^D(\mt L)= \text{width}(\mt H_{\mt L})= \text{ent}(\mt D_{\mt L})=\text{ent}(\mt L).\]
 \end{theorem}
\begin{proof}
     By  Lemma \ref{lem:minent} we have     
   $\text{ent}(\mt D_{\mt L})= \text{ent}(\mt L)$. Since $\text{ent}(\mt D)\leq \text{width}(\mt D)$ for all DFAs (Lemma \ref{lem:ent_leq_width}), we obtain $\text{ent}(\mt L)\leq \text{width}^D(\mt L), $ while from   Theorem \ref{thm:equivalentholeproof} we know that    $ \text{width}(\mt H_{\mt L})= \text{ent}(\mt D_{\mt L}) $. Hence, we have:
     \[\text{width}(\mt H_{\mt L})= \text{ent}(\mt D_{\mt L})= \text{ent}(\mt L) \leq \text{width}^D(\mt L)\leq \text{width}(\mt H_{\mt L}) \]
    and the conclusion follows. \qed
\end{proof}

The previous theorem allow us to give a first answer to Problem \ref{pr:aut_free}, that is, we provide  an automata-free characterization of the deterministic width of a regular language. Recall that a property is \emph{eventually} true for a sequence if it holds true for all but finitely many elements of the sequence.

\begin{corollary}\label{cor:monotone_sequences} Let $\mt L$ be a regular language. Then 
$\text{width}^D(\mt L)\leq p$ iff 
every (co-lexicographically) monotone sequence in $\pf L$ is eventually included in at most $p$ classes of the Myhill-Nerode equivalence $ \equiv_{\mathcal{L}}. $
\end{corollary}
 \begin{proof} Let $\mt D_{\mt L}$ be the minimum DFA for $\mt L$. 
 By definition, $k$   states $u_1,\dots,u_k$ are entangled in $ \mt D_{\mt L} $  iff there exists a monotone sequence  $(\alpha_j)_{j\in \mathbb N}$ such that,  for each $i=1, \dots, k$, we have    $\delta(s,\alpha_j)=u_i$ for infinitely many $j$'s.   Moreover, since  $\mt D_{\mt L}$ is minimum, if  $i\neq i'$  a word  arriving in $u_i$ and a word  arriving in $u_{i'}$ belong  to   different $\equiv_{\mt L}$-classes. Hence, 
$ \text{ent}(\mt D_{\mt L})> p$  iff 
there exists a  monotone sequence in $\pf L$ which eventually reaches more than $p$ classes of the Myhill-Nerode equivalence $ \equiv_{\mathcal{L}} $ infinitely often, and the corollary follows from the previous theorem. 
 \qed
 \end{proof}
 
Summarizing, the Hasse automaton $ \mt H_{\mt L} $ captures the deterministic width of a language. An interesting open question is whether it is possible to devise an effective procedure to build the Hasse automaton. More generally, also in view of the indexing and compression applications in Section \ref{subsec:ind_dfa}, we have two conflicting objectives: minimizing the width and minimizing the number of states. We will explore the latter objective in Section \ref{sec:minimal}.

 \subsection{Computing the Deterministic Width of a Regular Language}\label{sec:computing}
 
  In this section we shall use Theorem \ref{thm:hasse} --- stating that the deterministic width of $\mt L$ is equal to the entanglement of the minimum DFA for $\mt L$ --- to study  the complexity of Problem \ref{problem: language width} --- the problem of finding  the deterministic width of a language recognized by a given automaton. We show that if we are given a regular language $ \mathcal{L} $ by means of a DFA $\mathcal D$ accepting $ \mathcal{L} $ and a positive integer $p$, then  the problem \[\text{width}^D(\mathcal{L}) \stackrel{?}{\leq} p \] 
  is solvable in polynomial time for constant values of $p$. More precisely, we show that the problem of computing $ \text{width}^D(\mathcal{L}) $ is in the class \texttt{XP} with parameter $p$.
  This result is achieved by exhibiting a dynamic programming algorithm 
  that extends the ideas introduced in \cite{Alanko2021Wheeler} when solving the corresponding problem for Wheeler languages.

Theorem \ref{thm:hasse} suggests that the minimum DFA   contains all ``topological'' information required to compute the width of a language.
In the next theorem we clarify this intuition by providing a graph-theoretical characterization of the deterministic width of a language based on the minimum DFA recognizing the language.

\begin{theorem}\label{thm:conditions width}
    Let $ \mathcal{L} $ be a regular language and let 
    $ \mt D_{\mt L} $ be the minimum DFA of $ \mathcal{L} $, with set of states $ Q $. Let $ k \ge 2 $ be an integer. Then, $ \text{width}^D(\mathcal{L}) \ge k $ if and only if there exist strings $ \mu_1, \dots, \mu_k,  \gamma $ and  pairwise distinct states $ u_1, \dots, u_k \in Q $, such that   for every $ j = 1, \dots, k $:
    \begin{enumerate}
        \item $ \mu_j $ labels a path from the initial state $ s $ to $ u_j $;
        \item $ \gamma $ labels a cycle starting (and ending) at $ u_j $;
        \item either  $ \mu_1 , \dots , \mu_k \prec \gamma $  or $ \gamma \prec \mu_1 , \dots , \mu_k  $;
        \item $ \gamma $ is not a suffix of $ \mu_j $.
    \end{enumerate}
\end{theorem}
\begin{proof}
By Theorem \ref{thm:hasse} we have $ \text{width}^D(\mathcal{L}) =\text{ent}(   \mt D_{\mt L}  ) $.
We begin by proving that,  if the stated conditions hold true, then $\text{ent}(   \mt D_{\mt L}  ) \ge k $. Notice that for every integer $ i $ we have $ \mu_j \gamma^i \in I_{u_j } $. Moreover, the $ \mu_j $'s are pairwise distinct because the $ u_j $'s are pairwise distinct, so without loss of generality we can assume $ \mu_1 \prec \dots \prec \mu_k $.
\begin{enumerate}
    \item If $ \mu_1 \prec \dots \prec \mu_k \prec \gamma $, consider the increasing sequence:
    \begin{equation*}
        \mu_1 \prec \dots \prec \mu_k \prec \mu_1 \gamma \prec \dots \prec \mu_k \gamma \prec \mu_1 \gamma^2 \prec \dots \prec \mu_k \gamma^2 \prec \mu_1 \gamma^3 \prec \dots \prec \mu_k \gamma^3 \prec \dots
    \end{equation*}
    \item If $ \gamma \prec \mu_1 \prec \dots \prec \mu_k $, consider the decreasing sequence:
        \begin{equation*}
        \mu_k \succ \dots \succ \mu_1 \succ \mu_k \gamma \succ \dots \succ \mu_1 \gamma \succ \mu_k \gamma^2 \succ \dots \succ \mu_1 \gamma^2 \succ \mu_k \gamma^3 \succ \dots \succ  \mu_1 \gamma^3 \succ \dots
    \end{equation*}
    where   $\mu_1 \gamma^i \succ  \mu_k \gamma^{i+1}  $ holds  because $\mu_1 \succ \gamma  $ and $ \gamma $ is not a suffix of $ \mu_1 $.
\end{enumerate}
In both cases  the sequence witnesses that $ \{u_1, \dots, u_k \} $ is an entangled set of distinct states, so $\text{ent}(\mt D_{\mt L}) \ge k $.

Conversely, assume that $ \text{ent}(\mt D_{\mt L} ) \ge k $. This means that there exist distinct states $ v_1, \dots, v_k $ and  a monotone sequence  $ (\alpha_i)_{i \in \mathbb{N}} $ that reaches each of the $v_j$'s   infinitely many times. Let us show that, up to taking subsequences, we can assume not only that  $ (\alpha_i)_{i \in \mathbb{N}} $ reaches each of the $v_j$'s   infinitely many times, but it also satisfies additional properties.
\begin{itemize}
    \item Since $ |\Sigma| $ and $ |Q| $ are finite and $ (\alpha_i)_{i \in \mathbb{N}} $ is monotone, then up to removing a finite number of initial elements we can assume that all $ \alpha_i $'s end with the same $ m = |Q|^k $ characters, and we can write $ \alpha_i = \alpha'_i \theta $, for some $ \theta \in \Sigma^m $. Notice that such a new monotone sequence $ (\alpha_i)_{i \in \mathbb{N}} $ still reaches each of the $v_j$'s   infinitely many times.
    \item Up to taking a subsequence of the new $ (\alpha_i)_{i \in \mathbb{N}} $, we can assume that $ \alpha_i$ reaches $v_j$ if and only if $ i - j $ is a multiple of $ k $, that is, $\alpha_j, \alpha_{k + j},  \alpha_{2k + j} ,  \dots \in I_{v_j} $. Notice that such such a new monotone sequence $ (\alpha_i)_{i \in \mathbb{N}} $ still satifies $ \alpha_i = \alpha'_i \theta $ for every $ i $.
  \item Since $ |Q| $ is finite, up to taking a subsequence of the new $ (\alpha_i)_{i \in \mathbb{N}} $ we can assume that all   $\alpha_i$'s reaching the same $v_j$ spell the suffix $\theta$ visiting the same $m+1$ states  $x^j_0, x^j_1, \dots, x^j_m = v_j$.
\end{itemize} 
   


Consider the $k$-tuples $ (x^1_s, \dots, x^k_s) $, for $s \in \{0, \dots , m\} $ (corresponding to the states in column in Figure \ref{fig:th4}).
There are $ m + 1 = |Q|^k + 1 $ such $k$-tuples and therefore two of them must be equal. That is, there exist $ h, \ell $, with $ 0 \le h < \ell \le m $, such that   $ (x^1_h, \dots, x^k_h)=(x^1_\ell, \dots, x^k_\ell) $. Hence,  for all $ j \in \{1, \dots, k\} $  there is a cycle     $x^j_h, x^j_{h+1}, \dots, x^j_\ell  $,  all  these cycles are labelled by the same string $\gamma'$, and we can write  $ \theta = \phi \gamma' \psi $ for some  $ \phi $ and $ \psi $. 

\begin{figure}[h!] 
 \begin{center}
\begin{tikzpicture}[->,>=stealth', semithick, auto, scale=1]
        \node[state ]   (s) at (-10,-1)     {$s$};
        \node[]    at (-7,-1. )     {$\vdots$};
	    \node[state ]   (x10) at (-5,0)     {$x^1_{0}$};
	    \node[](x10+) at (-4,0){};
		\node[state] (u1) at (-2 ,0){\tiny $\stackrel{x^1_{h}=x^1_{\ell}}{\raisebox{-10pt}{\scriptsize$ u_1 $}} $};
		\node[]         (u1-) at (-3.4,0)     {};
		\node[]         (u1+) at (-0.6,0)     {};
	    \node[state ]   (v1) at (2,0)     {$v_1$};
	    \node[]   (v1-)       at (1,0)     {};
	    \node[]          at (2,-1)     {$\vdots$};
	    \node[]          at (-5,-1)     {$\vdots$};
	    \node[state ]   (xk0) at (-5,-2.5)     {$x^k_{0}$};
	    \node[] (xk0+) at (-4,-2.5){};
		\node[state ]   (uk) at (-2 ,-2.5) {\tiny$\stackrel{x^k_{h}=x^k_{\ell}}{\raisebox{-10pt}{\scriptsize$ u_k $}}$};
		 \node[]          at (-2,-1)     {$\vdots$};
		\node[]         (uk-) at (-3.4,-2.5)     {};
		\node[]         (uk+) at (-0.6,-2.5)     {};
 	    \node[state ]   (vk) at (2,-2.5)     {$v_k$};
 	    \node[]   (vk-)       at (1,-2.5)     {};
 	    \node[] (dot1) at (-3.7, 0 ){$\dots $};
		\node[] (phi1) at (-3.7, 0.5 ){$\phi$};
		\node[]   at (-3.7, -2.5 ){$\dots $};
		\node[]   at (-3.7, -1 ){$\vdots $};
		\node[] (phi2) at (-3.7, -2 ){$\phi$};
    	\node[] (psi1) at (0, 0.5 ){$\psi $};
    	\node[] at (0, 0  ){$\dots $};
    	\node[]   at (0, -1 ){$\vdots $};
    	\node[]  (psi2) at (0, -2 ){$\psi $};
    	\node[] at (0, -2.5 ){$\dots $};
    	\draw (v1-) edge [] node [] {} (v1);
    	\draw (vk-) edge [] node [] {} (vk);
    	\draw (u1-) edge [] node [] {} (u1);
    	\draw (u1) edge [] node [] {} (u1+);
		\draw (u1) edge  [dashed, loop above] node {$\gamma'$} (u1);
		\draw (uk-) edge [] node [] {} (uk);	
		\draw (x10) edge [] node [] {} (x10+);
		\draw (uk) edge [] node [] {} (uk+);
		\draw (uk) edge  [dashed, loop below] node {$\gamma'$} (uk);
		\draw (xk0) edge [] node [] {} (xk0+);
    	\draw (s) edge  [above] node [above]   {$\alpha_1', \alpha'_{k+1},\dots$} (x10);
    	\draw (s) edge [above] node [below]   {$\alpha_k', \alpha'_{2k},\dots$} (xk0);
\end{tikzpicture}
\end{center}
\caption{}\label{fig:th4}
\end{figure}
Let $u_1, u_2, \dots, u_k$ be the pairwise distinct states $x^1_h, x^2_h, \dots, x^k_h$ (they are distinct, because if $u_i=u_j$ for some $i\neq j$  we would have $v_i=v_j$). Hence, we have $k$ pairwise distinct states and $k$ equally labelled cycles. In order to fulfill the remaining conditions of the theorem, we proceed as follows.  Considering  the
 monotone sequence  $ (\alpha_i'\phi)_{i \in \mathbb{N}} $, reaching  each of the $u_i$'s infinitely many times,    we may  suppose without loss of generality (possibly eliminating a finite number of  initial elements)  that all $ \alpha_i'\phi $'s are co-lexicographically larger than $\gamma'$ or they are all co-lexicographically smaller than  $ \gamma'$.
 
 If $\gamma'$ is not a suffix of  any  $ \alpha_i'\phi $ we can choose $\gamma=\gamma'$ and,  considering $k$ words $\mu_1, \dots, \mu_k$ of the sequence $ (\alpha_i'\phi)_{i \in \mathbb{N}} $ arriving in $u_1, \dots, u_k$, respectively, we are done. 
 
 Otherwise, if $\gamma'$ is  a suffix of  some $ \alpha_i'\phi $, 
  pick $ 2k - 1 $ strings  $ \delta_1, \dots, \delta_{2k - 1} $ in  the sequence  $ (\alpha_i'\phi)_{i \in \mathbb{N}} $ such that $\delta_k$  ends in $u_k$, while   $ \delta_i $ and $ \delta_{k + i} $ end in  $ u_i$ for $ i = 1, \dots, k - 1 $,      and  \[ \delta_1 \prec \dots \prec \delta_k \prec\dots\prec \delta_{2k - 1}.\]

Let $ r $ be an integer such that $ |(\gamma')^r| > |\delta_i| $ for every $ i = 1, \dots, 2k - 1 $. Then   $ \gamma = (\gamma')^r $ is  the label of  a cycle from   $ u_i $, for every $ i = 1, \dots, k $, and $ \gamma $ is not a suffix of $ \delta_i $, for every $ i = 1, \dots, 2k - 1 $. We distinguish two cases:
\begin{enumerate}
    \item $ \delta_{k} \prec \gamma $. In this case, let $ \mu_1, \dots, \mu_k $ be equal to $ \delta_1, \dots, \delta_k $, respectively.
    \item $ \gamma \prec \delta_{k} $. In this case, let $ \mu_1, \dots, \mu_k $ be equal to $ \delta_k, \dots, \delta_{2k - 1} $, respectively. 
\end{enumerate}
In both cases, we have  either  $ \mu_1 , \dots , \mu_k \prec \gamma $  or $ \gamma \prec \mu_1 , \dots , \mu_k  $ and the conclusion follows. \qed
 
\end{proof}

\begin{example}\label{ex:lang_depend}
Let $ \mathcal{L} $ be a regular language. Let us prove that, in general, $ \text{width}^D(\mathcal{L}) $ and $ \text{width}^N(\mathcal{L}) $ may depend on the total order $ \preceq $ on the alphabet. Let $ \mathcal{D} $ be the DFA in Figure \ref{dependent}, and let $ \mathcal{L} $ be the language recognized by $ \mathcal{D} $. Notice that $ \mathcal{D} $ is the minimum DFA recognizing $ \mathcal{L} $.

First, assume that $ \preceq $ is the standard alphabetical order such that $ a \prec b \prec c \prec d $. Let us prove that $ \text{width}^D(\mathcal{L}) = 2 $. From Example \ref{ex:depedent}, we obtain $ \text{width}^D(\mathcal{L}) \le 2 $, and from Theorem \ref{thm:conditions width} we obtain $ \text{width}^D(\mathcal{L}) \ge 2 $ by choosing $ u_1 = q_2 $, $ u_2 = q_3 $, $ \mu_1 = a $, $ \mu_2 = b $, $ \gamma = c $. Notice that Corollary \ref{lem:widthupperbound} implies that $ \text{width}^N(\mathcal{L}) = \text{width}^D(\mathcal{L}) = 2 $.

Next, let $ \preceq $ be the total order such that $ a \prec c \prec b \prec d $. From Example \ref{ex:depedent} we immediately obtain $ \text{width}^N(\mathcal{L}) = \text{width}^D(\mathcal{L}) = 1 $.
\end{example}

The strings $ \mu_1, \dots, \mu_k ,$ and $ \gamma $ of  Theorem \ref{thm:conditions width} can be  determined by a dynamic programming algorithm whose running time can be computed using the following lemma.

  \begin{lemma}  \label{lem:bound}
Let $ \mathcal{D} $ be a DFA  with set of states $ Q $, and let $s_1, q_1,\dots, s_h,q_h \in Q$. Suppose there are strings  $ \nu_1 \preceq \dots \preceq \nu_h $  such that $\delta(s_i, \nu_i)=q_i$, for every $ i = 1, \dots, h $.
Then, there exist strings $ \nu'_1 \preceq \dots \preceq \nu'_h $ such that, for every $ i, j\in \{ 1, \dots, h\} $,  it holds: 
\begin{itemize}
   \item [-] 
$\delta(s_i ,\nu_i')=q_i$; 
 \item [-]$\nu_i= \nu_j $ iff  $\nu_i' =\nu_j'$;
  \item [-]$\nu_i \dashv \nu_j$ iff $\nu_i' \dashv \nu_j'$;
  \item [-] $|\nu'_i |   \le   h - 2 + \sum_{t = 1}^h |Q|^t$. 
\end{itemize}
\end{lemma}

\begin{proof}
We will prove the lemma for $ h = 3 $ (the extension to the general case is straightforward).
Given $ \varphi \in \Sigma^* $, we denote by $ \varphi(k)$ the $k$-th letter of $ \varphi $ from the right (if $ |\varphi| < k $ we write $ \varphi (k) = \varepsilon$, where  $\varepsilon$ is the empty string); therefore, if $\varphi\neq \varepsilon$, then  $\varphi(1)$ is the last letter of $\varphi$. 

Let $\nu_1\preceq\nu_2\preceq \nu_3$ be strings with $\delta(s_i,\nu_i)=q_i$, for $i=1,2,3$.  Let  $d_{3,2}$ be the first position from the right where $\nu_3$ and $\nu_2$ differ (if $\nu_3=\nu_2$, let $d_{3,2}=|\nu_3|$).    Since $ \nu_2 \preceq\nu_3$, we have $d_{3,2}\leq |\nu_3|$.
Defining  $d_{2,1}$ similarly, we have $d_{2,1}\leq |\nu_2|$.

We distinguish three cases.
\begin{enumerate}
\item $ d_{3, 2} = d_{2, 1} $.
\item $d_{3,2} < d_{2,1}$ (see Fig. \ref{figura-i}, assuming that $ \nu_1 \prec \nu_2 \prec \nu_3 $).
\item $ d_{2,1}<d_{3,2}$ (see Fig. \ref{figura-ii}, assuming that $ \nu_1 \prec \nu_2 \prec \nu_3 $).
  \end{enumerate}
Case 3  is analogous to case 2 and will not be considered. In cases 1 and 2,  $|\nu_3|\geq  d_{3,2}$ and $|\nu_2| \geq  d_{2,1}\geq d_{3,2}$, so that  $\nu_3, \nu_2$, and $ \nu_1$ end with  the same (possibly empty) word $\xi$ with $|\xi|= d_{3,2}-1$. 
Summing up:
\begin{equation*}
\nu_1=\theta_1 \nu_1(d_{3,2}) \xi\preceq \nu_2=\theta_2  \nu_2(d_{3,2}) \xi \preceq  \nu_3=\theta_3 \nu_3(d_{3,2})\xi
\end{equation*}
for some (possibly empty) strings $ \theta_1, \theta_2, \theta_3 $.

 \begin{figure}[H]
 \begin{subfigure}[b]{.5\textwidth}
    \begin{tabular}{lc|c|c|c|c|}
 \cline{2-6}
 $ \nu_3 \equiv $ & \multicolumn{3}{c|}{$ \dots \  \theta_3 \ \dots $}  & $ \nu_3(d_{3,2}) $ & $ \dots \  \xi \  \dots $ \\ \cline{2-6}
 \multicolumn{6}{c}{}\\ 
 $ \curlyvee $ &\multicolumn{3}{c}{ }& \multicolumn{1}{c}{$ \curlyvee $} & \multicolumn{1}{c}{}\\ 
 \multicolumn{6}{c}{}\\ \cline{2-6}
 \multirow{2}{*}{$ \nu_2 \equiv $} & \multicolumn{3}{c|}{$\dots \  \theta_2 \ \dots  $} & $ \nu_2(d_{3,2}) $ & $ \dots \  \xi \  \dots $ \\ \cline{2-6}
  & $ \dots \  \theta_{2}'\ \dots $ & $ \nu_2(d_{2,1}) $ & $ \dots \ \xi' \  \dots $ & $ \nu_2(d_{3,2}) $ & $ \dots \  \xi \  \dots $ \\ \cline{2-6}
 \multicolumn{6}{c}{}\\ 
 $ \curlyvee $ &\multicolumn{3}{c}{$ \curlyvee $ }& \multicolumn{1}{c}{$ \shortparallel$} & \multicolumn{1}{c}{}\\ 
 \multicolumn{6}{c}{}\\ \cline{2-6}
 \multirow{2}{*}{$ \nu_1 \equiv $} & \multicolumn{3}{c|}{$\dots \  \theta_1 \ \dots  $} & $ \nu_1(d_{3,2}) $ & $ \dots \  \xi \  \dots $ \\ \cline{2-6}
 & $ \dots \  \theta'_{1}\ \dots $ & $ \nu_1(d_{2,1}) $ & $ \dots \ \xi' \  \dots $ & $ \nu_1(d_{3,2}) $ & $ \dots \  \xi \  \dots $ \\ \cline{2-6}
 \end{tabular}
 \caption{Case $\nu_1\prec \nu_2 \prec \nu_3$ and $ d_{3,2} < d_{2,1} $.  }\label{figura-i}
 \end{subfigure}
 \begin{subfigure}[b]{.4\textwidth}
 \begin{tabular}{lc|c|c|c|c|}
 \cline{2-6}
 \multirow{2}{*}{$ \nu_3 \equiv $} & \multicolumn{3}{c|}{$\dots \  \theta_3 \ \dots  $} & $ \nu_3(d_{2,1}) $ & $ \dots \  \xi \  \dots $ \\ \cline{2-6}
 & $ \dots \  \theta'_3\ \dots $ & $  \nu_3(d_{3,2}) $ & $ \dots \ \xi' \  \dots $ & $ \nu_3(d_{2,1}) $ & $ \dots \  \xi \  \dots $ \\ \cline{2-6}
 \multicolumn{6}{c}{}\\ 
 $ \curlyvee $ &\multicolumn{3}{c}{$ \curlyvee $ }& \multicolumn{1}{c}{$ \shortparallel $} & \multicolumn{1}{c}{}\\ 
 \multicolumn{6}{c}{}\\ \cline{2-6}
 \multirow{2}{*}{$ \nu_2 \equiv $} & \multicolumn{3}{c|}{$\dots \  \theta_2 \ \dots  $} & $ \nu_2(d_{2,1}) $ & $ \dots \  \xi \  \dots $ \\ \cline{2-6}
  & $ \dots \  \theta'_2\ \dots $ & $ \nu_2(d_{3,2}) $ & $ \dots \ \xi' \  \dots $ & $ \nu_2(d_{2,1}) $ & $ \dots \  \xi \  \dots $ \\ \cline{2-6}
 \multicolumn{6}{c}{}\\ 
 $ \curlyvee $ &\multicolumn{3}{c}{ }& \multicolumn{1}{c}{$ \curlyvee $} & \multicolumn{1}{c}{}\\
 \multicolumn{6}{c}{}\\ \cline{2-6}
 $ \nu_1 \equiv $ & \multicolumn{3}{c|}{$ \dots \  \theta_1 \ \dots $}  & $ \nu_1(d_{2,1}) $ & $ \dots \  \xi \  \dots $ \\ \cline{2-6}
 \multicolumn{6}{c}{}\\ 
 \end{tabular}
 \caption{Case $\nu_1\prec \nu_2 \prec \nu_3$ and $ d_{2,1}<d_{3,2}$. }\label{figura-ii}
 \end{subfigure}
 \caption{}
 \end{figure}
 
Without loss of generality, we may assume that   $|\xi|  \leq |  Q|^3$. Indeed, 
 if $|\xi|>|  Q|^3$  then when we consider the triples of states visited while reading the last  $| \xi |$ letters in computations from $s_i$ to $q_i$ following  $\nu_i$, for $i=1,2,3$,   we should have met a repetition. If this were the case, we could erase a common factor from $\xi$, obtaining a shorter word $\xi_1$ such that 
 $\theta_1 \nu_1(d_{3,2})\xi_1\preceq \theta_2 \nu_2(d_{3,2}) \xi_1\preceq \theta_3\nu_3(d_{3,2}) \xi_1$, with the three new  strings    starting in  $s_1,s_2,s_3$  and ending in $q_1,q_2,q_3$, respectively,    respecting equalities and suffixes.    Then we can repeat  the argument until we reach a  word not longer than $|Q|^3$. 
 
If  $d_{3,2}=d_{2,1}$,   the order between the $\nu_i's$  is settled  in position  $d_{3,2}$.  Let $ r_1, r_2, r_3 $ be the states reached from $ s_1,s_2,s_3 $ by reading $ \theta_1, \theta_2, \theta_3 $, respectively.  For $i=1,2,3$, let $\bar \theta_i$ be the  label of a simple path from $s_i$ to $r_i$  and let    $\nu_i'=\bar \theta_i \nu_i(d_{3,2})\xi $.  Then    $\delta(s_i,\nu_i')= q_i$,   $\nu_1'\preceq \nu_2'\preceq \nu_3'$,    $|\nu_1'|, |\nu_2'|, |\nu_3'|\leq | Q| +| Q|^3$.

If  $d_{3,2}< d_{2,1}$ we have $ \nu_1(d_{3, 2}) = \nu_2 (d_{3, 2}) $. Moreover, $\theta_1 $ and $ \theta_2$ end  with  the same word $\xi'$   with $|\xi'|= d_{2,1}-d_{3,2}-1$ (see Fig. \ref{figura-i}), and we can write
\[ \theta_1=\theta_1' \nu_1(d_{2,1})\xi' \preceq  \theta_2=\theta'_2\nu_2(d_{2,1})\xi'. \]
Arguing as before we can assume, without loss of generality, that   $|\xi'| \leq | Q|^2 $. Moreover, we have $ \nu_1 (d_{2, 1}) \preceq \nu_2 (d_{2, 1})$ and,  as before,  we can assume that $ \theta'_1, \theta'_2 $ and $ \theta_3  $ label  simple paths.  Therefore,  $|\theta'_{1}|,  |\theta'_{2}|, |\theta_{3} |\leq |  Q| - 1 $.  Hence, in this case we can find $ \nu'_1, \nu'_2, \nu'_3 $ such that $\delta(s_i,\nu_i')= q_i$,   $\nu_1'\preceq \nu_2'\preceq \nu_3'$,    and  $|\nu_1'|, |\nu_2'|, |\nu_3'|\leq 1 + |  Q| +| {Q}|^2 + |  Q|^3$. 

Finally, the construction implies that $\nu_i= \nu_j $ iff  $\nu_i' =\nu_j'$, and $\nu_i \dashv \nu_j $ iff $ \nu_i' \dashv \nu_j'$. \qed
\end{proof}

We are now ready for a computational variant of Theorem \ref{thm:conditions width}.

\begin{corollary}\label{cor:conditions width}
     Let $ \mathcal{L} $ be a regular language and let $ \mt D_{\mt L} $ be the minimum DFA of $ \mathcal{L} $, with set of states $ Q $. Let $ k \ge 2 $ be an integer. Then, $ \text{width}^D(\mathcal{L}) \ge k $ if and only if there exist strings $ \mu_1, \dots, \mu_k, $ and $ \gamma $ and there exist pairwise distinct $ u_1, \dots, u_k \in Q $ such that, for every $ j = 1, \dots, k $:
     \begin{enumerate}
         \item $ \mu_j $ labels a path from the initial state $ s $ to $ u_j $;
         \item $ \gamma $ labels a cycle starting (and ending) at $ u_j $;
         \item either  $ \mu_1 , \dots , \mu_k \prec \gamma $  or $ \gamma \prec \mu_1 , \dots , \mu_k  $;
         \item $ |\mu_1|, \dots, |\mu_k| <|\gamma| \le  2( 2k - 2   + \sum_{t = 1}^{2k  } |Q|^t )$.
     \end{enumerate}
 \end{corollary}

 \begin{proof}
 $ (\Leftarrow) $ Since condition 4.  implies  that $\gamma$ is not a suffix of any of the $\mu_j$, $ \text{width}^D(\mathcal{L}) \ge k $  follows from Theorem \ref{thm:conditions width}. \\
  $ (\Rightarrow)$
 If  $ \text{width}^D(\mathcal{L}) \ge k $,    we use  Theorem \ref{thm:conditions width} and find words  $ \mu_1', \dots, \mu_k',  \gamma' $ and   pairwise distinct states  $ u_1, \dots, u_k \in Q $ such that   for every $ j = 1, \dots, k $:
    \begin{enumerate}
        \item $ \mu_j '$ labels a path from the initial state $ s $ to $ u_j $;
        \item $ \gamma' $ labels a cycle starting (and ending) at $ u_j $;
        \item either  $ \mu_1', \dots , \mu_k' \prec \gamma' $  or $ \gamma' \prec \mu_1' , \dots , \mu_k'  $;
        \item $ \gamma' $ is not a suffix of $ \mu_j' $.
    \end{enumerate}
   We only consider the case $\gamma' \prec \mu_1', \dots, \mu_k'$, since in  the case $\mu_1', \dots, \mu_k'\prec \gamma'$ the proof is similar. Up to an index permutation, we may suppose without loss of generality that $\gamma' \prec \mu_1'\prec \dots \prec  \mu_k'$. 
 Consider  the $2k$-words $\nu_i$ and states $s_i,q_i$ defined, for $i=1, \dots 2k$,  as follows:
 \begin{itemize}
     \item [-] $\nu_1=\dots=\nu_k=  \gamma'$, $s_1=q_1=u_1,\dots,s_k=q_k=u_k$;
     \item [-]  $\nu_{k+i}=\mu_i'$,    $s_{k+i}= s$,  $q_{k+i}=u_i$ for $i=1, \dots,k$,    where $s$ is  the initial state of $\mt D_\mt L$.
 \end{itemize}      
If we  apply    Lemma \ref{lem:bound} to these $2k$-words, we  obtain words $\nu_1'= \dots = \nu_k' \prec \nu_{k+1}'\prec \dots \prec  \nu_{2k}'$  such that, for all $i=1,\dots,k$:
    \begin{enumerate}
     \item $\delta(u_i,\nu_1')=u_i $,   that is, $\nu_1'$ labels a cycle from every $u_i$; 
 \item $\delta(s,\nu_{k+i}')=u_i $;
 \item  $\nu_1'$ is not a suffix of $\nu_{k+i}'$;
\item $ |\nu_1'| , |\nu_{k+i}'|    \le 2k - 2   + \sum_{t = 1}^{2k  } |Q|^t $. 
    \end{enumerate}
Let $r $ be the smallest integer   such that $|(\nu_1')^r|> max\{|\nu_{k+i}' |   ~|~   i \in \{1,\dots, k\}\}$.    Let   $\mu_1=\nu_{k+1}', \dots, \mu_k=\nu_{2k}' , \gamma= (\nu_1')^r$.  Since  $\nu_1'$ is not a suffix  
 of $\mu_{i}$, for all $i=1,\dots, k$, from   $\nu_1'  \prec \mu_{ 1}\prec \dots \prec  \mu_{ k}$  it follows  $\gamma=(\nu_1' )^r \prec \mu_{ 1}\prec \dots \prec  \mu_{ k}$, Moreover:
\[ |\mu_1|, \dots, |\mu_k| < |\gamma|\leq  max\{|\nu_{k+i}'| ~|~ i \in \{1,\dots, k\}\} +|\nu_1'|\leq   2( 2k - 2   + \sum_{t = 1}^{2k  } |Q|^t )\]  and the conclusion follows.  \qed
 \end{proof}

 We can finally provide an algorithmic solution to Problem \ref{problem: language width} in the deterministic case.

  \begin{theorem}\label{thm:dyn prog}
 Let $\mathcal L$ be  a regular language, given as input by means of any DFA $\mathcal D = (Q, s, \delta,F)$ recognizing $ \mathcal{L}$. 
Then, for any integer $p \geq 1$ we can decide whether $\text{width}^D(\mathcal L) \leq p$ in time $|\delta|^{O(p)}$.  
  \end{theorem}
  \begin{proof}
 We exhibit a dynamic programming algorithm based on Corollary \ref{cor:conditions width}, plugging in the value $k = p+1$ and returning true if and only if  $\text{width}^D(\mathcal L) \geq k$ is false.
 
 First, note that the alphabet's size is never larger than the number of transitions: $\sigma \leq |\delta|$, and that $|Q| \leq |\delta|+1$ since we assume that each state can be reached from $s$. 
 Up to minimizing $ \mathcal{D} $ (with Hopcroft's algorithm, running in time $O(|Q|\sigma\log |Q|) \leq |\delta|^{O(1)}$) we can assume that $ \mt D=\mt D_{\mt L} $ is the minimum DFA recognizing $ \mathcal{L} $. Let $N'= 2( 2k - 2   + \sum_{t = 1}^{2k  } |Q|^t )$  be the upper bound to the lengths of the strings $\mu_i$ ($1\leq i \leq k$) and $\gamma$ that need to be considered, and let $N = N'+1$ be the number of states in a path labeled by a string of length $N'$. Asymptotically, note that $N \leq |Q|^{O(k)} \leq |\delta|^{O(k)} $.
 The high-level idea of the algorithm is as follows. First, in condition (3) of Corollary \ref{cor:conditions width}, we focus on finding paths $\mu_j$'s smaller than $\gamma$, as the other case (all $\mu_j$'s larger than $\gamma$) can be solved with a symmetric strategy. Then:
 \begin{enumerate}
     \item For each state $u$ and each length $2\leq \ell \leq N$, we compute the co-lexicographically smallest path of length (number of states) $\ell$ connecting $s$ with $u$.
     \item For each $k$-tuple $u_1,\dots, u_k$ and each length $\ell \leq N$, we compute the co-lexicographically largest string $\gamma$ labeling $k$ cycles of length (number of states) $\ell$ originating (respectively, ending) from (respectively, in) all the states $u_1,\dots, u_k$.
 \end{enumerate}
 
Steps (1) and (2) could be naively solved by enumerating the strings $\mu_1, \dots, \mu_k$, and $\gamma$ and trying all possible combinations of states $u_1, \dots, u_k$. Because of the string enumeration step, however, this strategy would be exponential in $N$, i.e. doubly-exponential in $k$. We show that a dynamic programming strategy is exponentially faster. 

\texttt{Step (1)}. This construction is identical to the one  used  in \cite{Alanko2021Wheeler} for the Wheeler case $ (p = 1) $. For completeness, we report it here. 
Let $\pi_{u,\ell}$, with $u\in Q$ and $2 \leq \ell \leq N$, denote the predecessor of $u$ such that the co-lexicographically smallest path of length (number of states) $\ell$ connecting the source $s$ to $u$ passes through $\pi_{u,\ell}$ as follows: $s \rightsquigarrow \pi_{u,\ell} \rightarrow u$. 
 The node $\pi_{u,\ell}$ coincides with $s$ if $\ell=2$ and $u$ is a successor of $s$; in this case, the path is simply $s \rightarrow u$.
 If there is no path of length $\ell$ connecting $s$ with $u$, then we write $\pi_{u,\ell} = \bot$.
 We show that the set 
 $\{\pi_{u,\ell}\ :\ 2\leq \ell \leq N,\ u\in Q\}$ 
 stores in just polynomial space all co-lexicographically smallest paths of any fixed length $2 \leq \ell \leq N$ from the source to any node $u$. We denote such a path --- to be intended as a sequence $u_1 \rightarrow \dots \rightarrow u_\ell$ of states --- with $\alpha_\ell(u)$. 
 The node sequence $\alpha_\ell(u)$ can be obtained recursively (in $O(\ell)$ steps) as $\alpha_\ell(u) = \alpha_{\ell-1}(\pi_{u,\ell}) \rightarrow u$, where $\alpha_{1}(s) = s$ by convention.
 Note also that $\alpha_\ell(u)$ does not fully specify the sequence of edges (and thus labels) connecting those $\ell$ states, since two states may be connected by multiple (differently labeled) edges. However, the corresponding co-lexicographically smallest sequence $\lambda^-(\alpha_\ell(u))$ of $\ell-1$ labels is uniquely defined as follows:
 $$
 \left\{
 \begin{array}{ll}
    \lambda^-(\alpha_\ell(u)) = \mathrm{min}\{a\in\Sigma\ |\ \delta(s,a)=u\} & \mathrm{if}\ \ell=2 ,\\
    \lambda^-(\alpha_\ell(u)) = \lambda^-(\alpha_{\ell-1}(\pi_{u,\ell}) \rightarrow u) = \lambda^-(\alpha_{\ell-1}(\pi_{u,\ell})) \cdot \mathrm{min}\{a\in\Sigma\ |\ \delta(\pi_{u,\ell},a)=u\} & \mathrm{if}\ \ell > 2.
 \end{array}\right.
 $$
 

 It is not hard to see that each $\pi_{u,\ell}$ can be computed in $|\delta|^{O(k)}$ time using dynamic programming. First, we set $\pi_{u,2} = s$ for all successors $u$ of $s$. Then, for $\ell = 3, \dots, N$:

$$
\pi_{u,\ell} = \underset{v\in \text{Pred}(u)}{\mathrm{argmin}} \Big(
  \lambda^-( \alpha_{\ell-1}(v) ) \cdot  \mathrm{min}\{a\in\Sigma\ |\ \delta(v,a)=u\} \Big)    
$$

where $\text{Pred}(u)$ is the set of all predecessors of $u$ and the $\mathrm{argmin}$ operator compares strings in co-lex  order. In the equation above, if none of the $\alpha_{\ell-1}(v)$ are well-defined (because there is no path of length $\ell-1$ from $s$ to $v$), then $\pi_{u,\ell} = \bot$. 
Note that computing any particular $\pi_{u,\ell}$ requires comparing co-lexicographically $|\mathrm{Pred(u)}| \leq |Q|$ strings of length at most $\ell \leq N \leq |\delta|^{O(k)}$, which overall amounts to $|\delta|^{O(k)}$ time.
Since there are $|Q|\times N = |\delta|^{O(k)}$ variables $\pi_{u,\ell}$ and each can be computed in time $|\delta|^{O(k)}$, overall Step (1) takes time $|\delta|^{O(k)}$.
This completes the description of Step (1).
 
\texttt{Step (2)}. Fix a $k$-tuple $u_1,\dots, u_k$ and a length $2\leq \ell \leq N$. Our goal is now to show how to compute the co-lexicographically largest string $\gamma$ of length $\ell -1$ labeling $k$ cycles of length (number of states) $\ell$  originating (respectively, ending) from (respectively, in) all the states $u_1,\dots, u_k$. Our final strategy will iterate over all such $k$-tuples of states (in time exponential in $k$) in order to find one satisfying the conditions of Corollary \ref{cor:conditions width}. 

Our goal can again be solved by dynamic programming. Let $u_1,\dots, u_k$ and $u'_1,\dots, u'_k$ be two $k$-tuples of states, and let $2\leq \ell \leq N$. Let moreover $\pi_{u_1,\dots, u_k, u'_1,\dots, u'_k, \ell}$ be the $k$-tuple $\langle u''_1,\dots, u''_k \rangle$ of states (if it exists) such that there exists a string $\gamma$ of length $\ell-1$ with the following properties:
\begin{itemize}
    \item For each $1\leq i \leq k$, there is a path $u_i \rightsquigarrow u''_i \rightarrow u'_i$ of length (number of nodes) $\ell$ labeled with $\gamma$, and
    \item $\gamma$ is the co-lexicographically largest string satisfying the above property. 
\end{itemize}

If such a string $\gamma$ does not exist, then we set $\pi_{u_1,\dots, u_k, u'_1,\dots, u'_k, \ell} = \bot$.

Remember that we fix $u_1,\dots, u_k$.
For $\ell = 2$ and each $k$-tuple $u'_1,\dots, u'_k$, it is easy to compute $\pi_{u_1,\dots, u_k, u'_1,\dots, u'_k, \ell}$:  this $k$-tuple is $\langle u_1,\dots, u_k \rangle$ (all paths have length 2) if and only if there exists $c\in\Sigma$ such that $u'_i = \delta(u_i,c)$ for all $1\leq i\leq k$ (otherwise it does not exist). Then, $\gamma$ is formed by one character: the largest such $c$.

For $\ell >2$, the $k$-tuple $\pi_{u_1,\dots, u_k, u'_1,\dots, u'_k, \ell}$ can be computed as follows. Assume we have computed those variables for all lengths $\ell'<\ell$.
Note that for each such $\ell'<\ell$ and $k$-tuple $u''_1, \dots, u''_k$, the variables $\pi_{u_1,\dots, u_k, u''_1,\dots, u''_k, \ell'}$ identify $k$ paths $u_i \rightsquigarrow u''_i$ of length (number of nodes) $\ell'$. Let us denote with $\alpha_{\ell'}(u''_i)$ such paths, for $1\leq i\leq k$.

Then, $\pi_{u_1,\dots, u_k, u'_1,\dots, u'_k, \ell}$ is equal to  $\langle u''_1,\dots, u''_k \rangle$ maximizing co-lexicographically the string $\gamma'\cdot c$ defined as follows:

\begin{enumerate}
    \item $u'_i = \delta(u''_i,c)$ for all $1\leq i\leq k$,  
    \item $\pi_{u_1,\dots, u_k, u''_1,\dots, u''_k, \ell-1} \neq \bot$, and
    \item $\gamma'$ is the co-lexicographically largest string labeling all the paths $\alpha_{\ell-1}(u''_i)$. Note that this string exists by condition (2), and it can be easily built by following those paths in parallel (choosing, at each step, the largest character labeling all the $k$ considered edges of the $k$ paths).
\end{enumerate}
 
If no $c\in \Sigma $ satisfies condition (1), or condition (2) cannot be met, then $\pi_{u_1,\dots, u_k, u'_1,\dots, u'_k, \ell} = \bot$.

Note that $\pi_{u_1,\dots, u_k, u_1,\dots, u_k, \ell}$ allows us to identify (if it exists) the largest string $\gamma$ of length $\ell-1$ labeling $k$ cycles originating and ending in each $u_i$, for $1\leq i \leq k$. 

Each tuple $\pi_{u_1,\dots, u_k, u'_1,\dots, u'_k, \ell}$ can be computed in $|\delta|^{O(k)}$ time by dynamic programming (in order of increasing $\ell$), and there are $|\delta|^{O(k)}$ such tuples to be computed (there are $|Q|^{O(k)} \leq |\delta|^{O(k)}$ ways of choosing $u_1,\dots, u_k, u'_1,\dots, u'_k$, and $N \le |\delta|^{O(k)}$). Overall, also Step (2) can therefore be solved in $|\delta|^{O(k)}$ time. 

To sum up, we can check if the conditions of Corollary \ref{cor:conditions width} hold as follows:

\begin{enumerate}
    \item We compute $\pi_{u,\ell}$ for each $u\in Q$ and $\ell \leq N$. This identifies a string $\mu_u^\ell$ for each such pair $u\in Q$ and $\ell \leq N$: the co-lexicographically smallest one, of length $\ell-1$, labeling a path connecting $s$ with $u$.
    \item For each $k$-tuple $u_1, \dots, u_k$ and each $\ell \leq N$, we compute $\pi_{u_1,\dots, u_k, u_1,\dots, u_k, \ell}$. This identifies a string $\gamma_{u_1,\dots, u_k}^\ell$ for each such tuple $u_1,\dots, u_k$ and $\ell \leq N$: the co-lexicographically largest one, of length $\ell-1$, labeling $k$ cycles  originating and ending in each $u_i$, for $1\leq i \leq k$. 
    \item We identify the $k$-tuple $u_1, \dots, u_k$ and the lengths $\ell_i < \ell \leq N$ (if they exist) such that $\mu_{u_i}^{\ell_i} \prec \gamma_{u_1,\dots, u_k}^\ell$ for all $1\leq i \leq k$.
\end{enumerate}
 
The conditions of Corollary \ref{cor:conditions width} hold if and only if step 3 above succeeds for at least one $k$-tuple $u_1, \dots, u_k$ and lengths $\ell_i < \ell \leq N$, for $1\leq i\leq k$. Overall, the algorithm terminates in $|\delta|^{O(k)} = |\delta|^{O(p)}$ time. \qed
 
\end{proof}

\subsection{Relation with Star-Free Languages}\label{sec: relation star free}

 Theorem \ref{thm:conditions width} allows us to describe the levels of the width hierarchy looking to cycles in the minimum automata for the languages. 
 This results resemble another very well known result  on a class of subregular languages, the star-free ones, which can also be described by inspecting the cycles in the minimum DFA for the language.  
 \begin{definition}
  A regular language is said to be star-free if it can be described by a regular expression constructed from the letters of the alphabet, the empty set symbol, all boolean operators (including complementation),  and concatenation (but no Kleene star). 
 \end{definition}
  A well-known    automata  characterization of star-free languages is given by using counters.   A counter in a DFA is a sequence of pairwise-distinct states $u_0,\dots,u_n$ (with $n\geq 1$) such that there exists a non-empty string $\alpha$ with $\delta(u_0, \alpha)=u_1, \dots, \delta(u_{n-1}, \alpha)=u_n,  \delta(u_n , \alpha)=u_0$.
 A language is star-free if and only if  its minimum DFA has no counters \cite{SCHUTZENBERGER1965190, mcnaughton}.  

We can  easily prove   that a Wheeler language, i.e.  a language $\mathcal L$ with $ \text{width}^N(\mathcal{L})=\text{width}^D(\mathcal{L})=1$   for a fixed order of the alphabet,  is always star-free. Indeed, if the minimum DFA for a language has a counter $u_0,\dots,u_n$ with string $\alpha$, and $ \gamma \in I_{u_0} $, then $(\gamma \alpha^n)_{n \in \mathbb N}$ is a monotone sequence (increasing or decreasing depending on which string between $ \gamma $ and $ \gamma \alpha $ is smaller)  which  is not ultimately included  in one  class of the Myhill-Nerode equivalence $ \equiv_{\mathcal{L}}$ (because in a minimum DFA the $ I_u$'s are exactly equal to Nerode classes). Hence, the language is not Wheeler by  Corollary  \ref{cor:monotone_sequences}.

 This implies that  the first level of the deterministic width hierarchy is included in the class of star-free languages. On the other hand, in the next example we prove that  there is   an infinite sequence of star-free languages  $(\mathcal L_n)_{n\in \mathbb N}$ over the two letter alphabet $\{a,b\}$  such that $\text{width}^D(\mathcal L_n)=n$, for both total orders $ \preceq $ on $\{a,b\}$.

\begin{example}
In Figure \ref{fig:starfree}  we depicted a DFA $\mathcal D_n$ with $3n$  states accepting the language $\mathcal L_n= \bigcup_{j=0}^{n-1} b^jab^*a^{j+1} $. Notice  that:
\begin{enumerate}
    \item for every state $ u $ and for every $ 1 \le j \le n $, we have that $ \delta(u, aba^j) $ is defined and final if and only if $ u=q_j $;
    \item for every state $ u $ in the second or third row and for every $ 1 \le j \le n $, we have that $ \delta(u,  ba^j) $ is defined and final if and only if $ u = r_j $;
    \item for every state $ u $ in the third row and for every $ 1 \le j \le n $, we have that $ \delta(u, a^{j-1})$ is defined and final if and only if $ u = s_j $.
\end{enumerate}
We conclude that $\mathcal D_n$ is the minimum DFA of $\mathcal L_n $.

Since $\mathcal D_n$ has no counters (because every cycle is a self-loop), the above mentioned characterization of star-free languages tells us that     $\mathcal L_n$ is star-free. Let us prove  that $ \text{ent}(\mathcal{D}_n)= n$, so that $\text{width}^D(\mathcal L_n)=n$ follows from Theorem \ref{thm:hasse}. Notice that (1) states in the first row are reached by only one string, (2) states in the second row are reached infinitely many times only by string ending with $ b $, and (3) states in the third row are reached only by strings ending with $ a $. This implies $ \text{ent}(\mathcal{D}_n)\le n $, because the words belonging to a monotone sequence witnessing an entanglement between states will definitely end by the same letter, so only states belonging to the same row may belong to an entangled set. Finally, the $n$ states in the second level are entangled , as it witnessed by the monotone sequence: 
\[a\prec ba\prec bba\prec \dots \prec b^{n-1}a \prec ab\prec bab\prec bbab\prec \dots \prec b^{n-1}ab\prec  \dots\prec ab^k\prec bab^k\prec bbab^k \prec    \dots\] if $a\prec b$, and by the monotone sequence: 
\[b^{n-1} a  \succ b^{n-2} a  \succ \dots \succ a  \succ  b^{n-1}ab \succ b^{n-2} ab  \succ \dots \succ ab  \succ
\dots \succ b^{n-1}ab^k \succ b^{n-2} ab^k  \succ \dots \succ ab^k  \succ
\dots \]
if $b\prec a$. 
Hence,  in both cases we have  $ \text{ent}(\mathcal{D}_n)= n$.

 \begin{figure}[h!] 
 \begin{center}
\begin{tikzpicture}[->,>=stealth', semithick, auto, scale=1]
 \tikzset{every state/.style={minimum size=10pt}}
\node[state, initial above](0) at (0,0)	{$\;\;q_1\;\;$};
\node[state, label=above:{}] (01)  at (0,-1.5)    	{$\;\;r_1\;\;$};
\node[state,accepting, label=above:{}] (02) at (0,-3)    	{ $\;\;s_1\;\;$};
\node[state, label=above:{}] (1)  at (2,0)    	{$\;\;q_2\;\;$};
\node[state, label=above:{}] (11)  at (2,-1.5)    	{$\;\;r_2\;\;$};
\node[state,  label=above:{}] (12) at (2,-3)    	{$\;\;s_2\;\;$};

\node[state, label=above:{}] (2) at (4,0)    	{$\;\;q_3\;\;$};
\node[state, label=above:{}] (21)  at (4,-1.5)     	{$\;\;r_3\;\;$};
\node[state, label=above:{}] (22)  at (4,-3)     	{$\;\;s_3\;\;$};

  \node[] (dot0) [right  of=2] {$\dots $};
   \node[] (dot1) [right  of=21] {$\dots $};
   \node[] (dot2) [right  of=22] {$\dots $};
 
  \node[state, label=above:{}] (n) at (7,0)     	{$q_{n-1}$};
  \node[state, label=above:{}] (n1)  at (7,-1.5)     	{$r_{n-1}$};
  \node[state, label=above:{}] (n2)  at (7,-3)     	{$s_{n-1}$}; 

  \node[state, label=above:{}] (m) at (9,0)   	{$\; \;q_n\; \;$};
  \node[state, label=above:{}] (m1)  at (9,-1.5)     	{$\; \;r_n\; \;$};
 \node[state, label=above:{}] (m2)  at (9,-3)   {$\; \;s_n\; \;$};

  \draw (0) edge  [left] node [] {$a$} (01);
 \draw (01) edge  [loop left] node {$b$} (01);
   \draw (01) edge  [left] node [] {$a$} (02);
  \draw (0) edge  [above] node [] {$b$} (1);
   \draw (1) edge  [left] node [] {$a$} (11);
 \draw (11) edge  [loop left] node {$b$} (11);
 \draw (11) edge  [left] node [] {$a$} (12);
  \draw (1) edge  [above] node [] {$b$} (2);
     \draw (2) edge  [left] node [] {$a$} (21);
 \draw (21) edge  [loop left] node {$b$} (21);
 \draw (21) edge  [left] node [] {$a$} (22);
  \draw (22) edge  [above] node [] {$a$} (12);
    \draw (12) edge  [above] node [] {$a$} (02);
 
  \draw (n) edge  [left] node [] {$a$} (n1);
   \draw (n1) edge  [loop left] node {$b$} (n1);
      \draw (n1) edge  [left] node [] {$a$} (n2);

      \draw (n) edge  [above] node [] {$b$} (m);
      
        \draw (m) edge  [left] node [] {$a$} (m1);
   \draw (m1) edge  [loop left] node {$b$} (m1);
      \draw (m1) edge  [left] node [] {$a$} (m2);
      
        \draw (m2) edge  [above] node [] {$a$} (n2);
    
\end{tikzpicture}
 \end{center}
 \caption{A minimum  DFA $\mt D_n$  recognizing a star-free language $\mathcal L_n$ with $\text{width}^D(\mt L_n)=n$ for the two possible orders on the alphabet $\{a,b\}$.}
 \label{fig:starfree}
\end{figure}
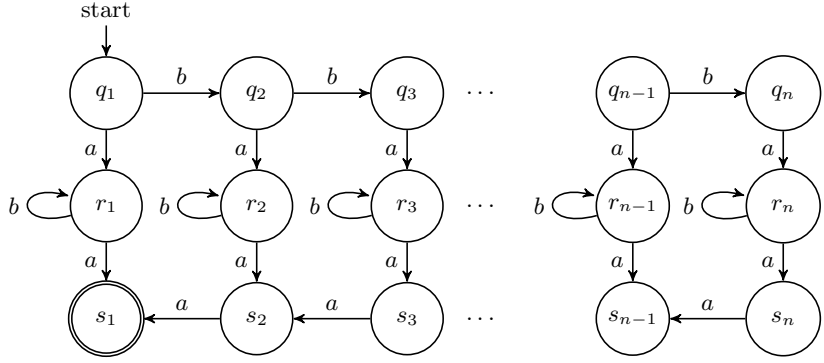 
\end{example}

\subsection{The Convex Myhill-Nerode Theorem}\label{sec:minimal}

In the previous  sections  we  described a hierarchy of regular languages by means of their deterministic  widths. A natural question is whether a corresponding Myhill-Nerode theorem can be provided for every level of the hierarchy: given a regular language $ \mathcal{L} $, if we consider all DFAs recognizing $ \mathcal{L} $ and having width equal to $ \text{width}^D (\mathcal{L}) $, is there a unique such DFA having the minimum number of states? In general, the answer is "no", as showed in Example \ref{ex:no_minimal}.

The non-uniqueness can be explained as follows. If a DFA of width $ p $ recognizes $ \mathcal{L} $, then $ \pf L $ can be partitioned into $ p $ sets, each of which consists of the (disjoint) union of some pairwise comparable $ I_q $'s.  However, in general the partition into $ p $ sets is not unique, so it may happen that two distinct partitions lead to two non-isomorphic minimal DFAs with the same number of states.  For example, in Figure \ref{fig:2minimumDFA}, we see two  non-isomorphic  DFAs (center and right) realizing the  width of the language and with the minimum number of states among all DFAs recognizing the same language and realizing the width of the language: the chain partition $ \{ \{0,1,4\}, \{2,3, 5, 3'\} \} $ of the DFA in the center induces the partition $ \{ac^*  \cup \{\varepsilon, e, h \}, bc^* \cup ac^*d \cup \{gd, ee, he,f,k,g \} \} $ of  $\pf L$, whereas the chain partition $ \{ \{0,1,3\}, \{2,4, 5, 4'\}\} $ of the DFA on the right induces the partition $\{ac^* \cup ac^*d \cup \{\varepsilon, gd,  ee,he, f,k \}, bc^* \cup \{e, h, g\}\}$ of   $\pf L$.

This example shows that  no uniqueness results can be ensured as long as partitions are not fixed. But what happens if we fix a partition? As we will prove in this section, once a partition is fixed, it is possible to prove a full Myhill-Nerode theorem, thereby providing a DFA-free characterization of languages of width equal to $p$   and a  minimum DFA for these languages.

More formally, let $ \mathcal{D} = (Q, s, \delta, F) $ be a DFA, and let $ \{Q_i\; | \;  1\leq i \leq p\} $ be a $ \leq_\mathcal{D} $-chain partition of $ Q $. For every $ i \in \{1, \dots, p \} $, define:
\begin{equation*}
    \text{Pref} (\mathcal{L(D)})^i = \{\alpha \in \text{Pref}(\mathcal{L(D)}) \; | \;\delta(s, \alpha) \in Q_i  \}.
\end{equation*}
Then $ \{\text{Pref}(\mathcal{L(D)}^i \; | \; 1\leq i \leq p\}$ is a partition of $ \text{Pref} (\mathcal{L(D)}) $, and from now on we will think of such a partition as fixed. We now consider the class of all DFAs accepting $ \mathcal{L} $ and inducing the considered partition.

\begin{definition}\label{def:psortableNFA}
Let $ \mathcal{D} = (Q, s, \delta, F) $ be a DFA, and let $ \mathcal{P} = \{U_1, \dots, U_p \} $ be a partition of $ \text{Pref}(\mathcal{L(D)}) $. We say that $ \mathcal{D} $ is $ \mathcal{P} $-sortable if there exists a $ \leq_\mathcal{D} $-chain partition $ \{Q_i\; | \;  1\leq i \leq p\} $ such that for every $ i \in \{1, \dots, p \} $:
\begin{equation*}
    \text{Pref}(\mathcal{L(D)})^i = U_i.
\end{equation*}
\end{definition}

We wish to give a DFA-free characterization of languages $ \mathcal{L} $ and  partitions $ \mathcal{P} $ of $ \pf L $ for which there exists a $ \mathcal{P} $-sortable DFA. As in the Myhill-Nerode theorem, we aim to determine which properties an equivalence relation $ \sim $ should satisfy to ensure that a canonical construction provides a $ \mathcal{P} $-sortable DFA. First, $ \mathcal{L} $ must be regular, so $ \sim $ is expected to be right-invariant. In order to develop some intuition on the required properties, let us consider an equivalence relation which plays a key role in the classical Myhill-Nerode theorem. Let $ \mathcal{D} = (Q, s, \delta, F) $ be a $ \mathcal{P} $-sortable DFA, and let $ \equiv_\mathcal{D} $ be the equivalence relation on $ \text{Pref}(\mathcal{L(D)})$ defined by 
\[\alpha \equiv_{\mt D} \beta \Leftrightarrow \delta(s,\alpha)=\delta(s, \beta). \]
   Notice that equivalent strings  end up in the same element of   $ \mathcal{P}$ (\emph{$ \mathcal{P} $-consistency}), and since all states in each $ \leq_\mathcal{D} $-chain $ Q_i $ are comparable, then each $ I_q $ must be convex in the corresponding element of $ \mathcal{P} $ (\emph{$ \mathcal{P} $-convexity}).  More formally  we consider the following definition, where, for every $ \alpha \in \pf L $, we denote by   $ U_\alpha $  the unique element $ U_i $ of $ \mathcal{P} $ such that $ \alpha \in U_i $.
 
\begin{definition}\label{def:PconsistentPconvex}
Let $ \mathcal{L} \subseteq \Sigma^* $ be a language, and let $ \sim $ be an equivalence relation on $ \pf L $. Let $ \mathcal{P} = \{U_1, \dots, U_p \} $ be a partition of $ \pf L $.
\begin{enumerate}
\item We say that $ \sim $ is \emph{$ \mathcal{P} $-consistent} if for every $ \alpha, \beta \in \pf L $, if $ \alpha \sim \beta $, then $ U_\alpha = U_\beta $.
\item Assume that $ \sim $ is $ \mathcal{P} $-consistent. We say that $ \sim $ is \emph{$ \mathcal{P} $-convex} if for every $ \alpha \in \pf L $ we have that $ [\alpha]_{\sim} $ is a convex in $ (U_\alpha, \preceq) $.
\end{enumerate}
\end{definition}

As we now  prove, these are exactly the required properties for a DFA-free characterization.

Let $\mt L \subseteq \Sigma^* $ be a language, and let $ \sim $ be an equivalence relation on $ \pf L $. We say that $ \sim $ \emph{respects} $ \pf L $ if:
\begin{equation*}
    (\forall \alpha, \beta \in \pf L)(\forall \phi \in \Sigma^*)(\alpha \sim \beta \land \alpha \phi \in \pf L \to \beta \phi \in \pf L).
\end{equation*}
Now, let us define the right-invariant, $ \mathcal{P} $-consistent and $ \mathcal{P} $-convex refinements of an equivalence relation $ \sim $.
\begin{enumerate}
    \item Assume that $ \sim $ respects $ \pf  L $. For every $ \alpha, \beta \in \pf  L $, define:
\begin{equation*}
    \alpha \sim^r \beta \iff (\forall \phi \in \Sigma^*)(\alpha \phi \in \pf L \to \alpha \phi \sim \beta \phi).
\end{equation*}
    We say that $ \sim^r $ is the \emph{right-invariant refinement} of $ \sim $.
    \item  Let $ \mathcal{P} = \{U_1, \dots, U_p \} $ be a partition of $ \pf  L $. For every $ \alpha, \beta \in  \pf  L$, define:
\begin{equation*}
    \alpha \sim^{cs} \beta \iff (\alpha \sim \beta) \land (U_\alpha = U_\beta)
\end{equation*}
    We say that $ \sim^{cs} $ is the \emph{$ \mathcal{P} $-consistent refinement} of $ \sim $.
    \item Let $ \mathcal{P} = \{U_1, \dots, U_p \} $ be a partition of $ \pf L $. Assume that $ \sim $ is $ \mathcal{P} $-consistent. For every $ \alpha, \gamma   \in \pf L $, define:
    \begin{equation*}
\begin{split}
    & \alpha \sim^{cv} \gamma \iff (\alpha \sim \gamma) \land \\
    & \land (\forall \beta \in \pf L)(((U_{\alpha} = U_{\beta}) \land (\min \{\alpha, \gamma \} \prec \beta \prec \max \{\alpha, \gamma \}) \to \alpha \sim \beta).
\end{split}
\end{equation*}
    We say that $ \sim^{cv} $ is the \emph{$ \mathcal{P} $-convex refinement} of $ \sim $.
\end{enumerate}
It is easy to check that $ \sim^r $ is the coarsest right-invariant equivalence relation refining $ \sim $, $ \sim^{cs} $ is the coarsest $ \mathcal{P} $-consistent equivalence relation refining $ \sim $ and $ \sim^{cv} $ is the coarsest $ \mathcal{P} $-convex equivalence relation refining $ \sim $.

We wish to prove that any equivalence relation that respects $ \text{Pref}(\mathcal{L}) $ admits a coarsest refinement being $ \mathcal{P} $-consistent, $ \mathcal{P} $-convex and right-invariant at once, because then we will be able to define an equivalence relation inducing the minimum ($ \mathcal{P} $-sortable) DFA. We first prove that if we  use the operators $cv$ and $r$, in this order,  over a $ \mathcal{P} $-consistent and right-invariant equivalence relation we do  not lose   $ \mathcal{P} $-consistency, nor  right-invariance, and we gain $ \mathcal{P} $-convexity. 
\begin{lemma}\label{lem:refinement}
Let $ \mathcal{L} \subseteq \Sigma^* $ be a language, and let $ \mathcal{P} $ be a partition of $ \text{Pref}(\mathcal{L}) $. If $\sim$ is a $ \mathcal{P} $-consistent and right-invariant equivalence relation on $ \text{Pref}(\mathcal{L}) $, then the relation $(\sim^{cv})^r$ 
is $ \mathcal{P} $-consistent, $ \mathcal{P} $-convex and right-invariant.
\end{lemma}
\begin{proof}
By definition $(\sim^{cv})^r$ is a right-invariant refinement. Moreover, $ \sim^{cv}$ and $(\sim^{cv})^r$ are $ \mathcal{P} $-consistent because they are refinements of the $ \mathcal{P} $-consistent equivalence relation $ \sim $. Let us prove that $ (\sim^{cv})^r $ is $\mathcal{P} $-convex. Assume that $\alpha, \beta, \gamma\in \pf L$ are such that  $ \alpha (\sim^{cv})^r \gamma $,   $ \alpha \prec \beta \prec \gamma $  and  $ U_\alpha  = U_\beta  $.  Being $(\sim^{cv})^r$ a $\mt P$-consistent relation, we have $ U_\alpha  = U_\beta =U_\gamma $. We must prove that $ \alpha (\sim^{cv})^r \beta $. Fix $ \phi \in \Sigma^* $ such that $ \alpha \phi \in \pf L $. We must prove that $ \alpha \phi \sim^{cv} \beta \phi $. Now, $ \alpha (\sim^{cv})^r \gamma $ implies $ \alpha \sim^{cv} \gamma $. Since $ \alpha \prec \beta \prec \gamma $ and $ U_\alpha = U_\beta = U_\gamma $, then the $ \mathcal{P} $-convexity of $ \sim^{cv} $ implies $ \alpha \sim^{cv} \beta $. In particular, $ \alpha \sim\beta $. Since $\sim$ is   right-invariant   we have  $ \alpha \phi \sim  \beta \phi $, and from the $\mt P$-consistency of $\sim$   we obtain    $ U_{\alpha \phi} = U_{\beta \phi} $. Moreover, $ \alpha (\sim ^{cv})^r \gamma $ implies $ \alpha \phi (\sim ^{cv})^r \gamma \phi $ by right-invariance, so $ \alpha \phi \sim ^{cv} \gamma \phi $. By $ \mathcal{P} $-convexity, from $ \alpha \phi \sim ^{cv} \gamma \phi $, $ U_{\alpha \phi} = U_{\beta \phi} $ and $ \alpha \phi \prec \beta \phi \prec \gamma \phi $ (since $ \alpha \prec \beta \prec \gamma) $ we conclude $ \alpha \phi \sim ^{cv} \beta \phi $. \qed
\end{proof}

\begin{corollary}\label{cor:coarsest}
Let $ \mathcal{L} \subseteq \Sigma^* $ be a nonempty language, and let $ \mathcal{P} $ be a partition of $ \text{Pref}(\mathcal{L}) $. Let $ \sim $ be an equivalence relation that respects $ \text{Pref}(\mathcal{L}) $. Then, there exists a (unique) coarsest $ \mathcal{P} $-consistent, $ \mathcal{P} $-convex and right-invariant equivalence relation refining $ \sim $.
\end{corollary}
\begin{proof}
The equivalence relation $ (\sim^{cs})^r $ is $ \mathcal{P} $-consistent (because it is a refinement of the $ \mathcal{P} $-consistent equivalence relation $ \sim^{cs} $) and right-invariant (by definition it is a right-invariant refinement), so by Lemma \ref{lem:refinement} the  equivalence relation $(((\sim^{cs})^r)^{cv})^r$ is $ \mathcal{P} $-consistent, $ \mathcal{P} $-convex and right-invariant. Moreover, every $ \mathcal{P} $-consistent, $ \mathcal{P} $-convex and right-invariant equivalence relation refining $ \sim $ must also refine $(((\sim^{cs})^r)^{cv})^r$, so $(((\sim^{cs})^r)^{cv})^r$ is the coarsest $ \mathcal{P} $-consistent, $ \mathcal{P} $-convex and right-invariant equivalence relation refining $ \sim $. \qed
\end{proof}

Corollary \ref{cor:coarsest} allows us to give the following definition.

\begin{definition}
Let $ \mt L \subseteq \Sigma^* $ be a language, and let $ \mathcal{P} = \{U_1, \dots, U_p \} $ be a partition of $ \pf L $. Denote by $ \equiv_\mathcal{L}^\mathcal{P} $ the coarsest $ \mathcal{P} $-consistent, $ \mathcal{P} $-convex and right-invariant equivalence relation refining the Myhill-Nerode equivalence $ \equiv_\mathcal{L} $.
\end{definition}

In particular, since $ \mathcal{L} $ is the union of some $ \equiv_\mathcal{L} $-classes, we also have that $ \mathcal{L} $ is the union of some $ \equiv_\mathcal{L}^\mathcal{P} $-classes.

Recall that, given a DFA $ \mathcal{D} = (Q, s, \delta, F) $, the equivalence relation $ \equiv_\mathcal{D} $ on $ \text{Pref}(\mathcal{L(D)}) $ is the one such that:
\begin{equation*}
    \alpha \equiv_\mathcal{D} \beta  \iff \delta (s, \alpha) = \delta (s, \beta).
\end{equation*}

Here are the key properties of $ \equiv_\mathcal{D} $, when $\mathcal D$ is a  $ \mathcal{P} $-sortable DFA.

\begin{lemma}\label{simA}
Let $ \mathcal{D} = (Q, s, \delta, F) $ be a $ \mathcal{P} $-sortable DFA, where $ \mathcal{P} = \{U_1, \dots, U_p \} $ is a partition of  $\pf L$ for  $ \mt L= \mathcal{L(D)} $. Then, $ \equiv_\mathcal{D} $  has finite index, it respects $ \pf L $, it is right-invariant,  $ \mathcal{P} $-consistent, $ \mathcal{P} $-convex, it refines $ \equiv_\mathcal{L}^\mathcal{P} $, and $ \mathcal{L} $ is the union of some $ \equiv_\mathcal{D} $-classes. In particular, $ \equiv_\mathcal{L}^\mathcal{P} $ has finite index.
\end{lemma}

\begin{proof}
The relation $ \equiv_\mathcal{D} $ has index equal to $ |Q| $. It respects $ \text{Pref}(\mathcal{L}) $ because if $\alpha \equiv_\mathcal{D}\beta $ and  $ \phi \in \Sigma^* $ satisfies $ \alpha \phi \in \text{Pref}(\mathcal{L}) $, then there exists $\gamma$ with $ \alpha \phi \gamma \in \mathcal{L}$ so $\delta(s,   \alpha \phi \gamma )\in F$. Since $\delta(s,   \alpha)=\delta(s,   \beta)$   we obtain $\delta(s,   \alpha \phi \gamma)=\delta(s,   \beta \phi \gamma)$ and so $ \beta \phi \gamma \in \mathcal{L}$ and 
$ \beta \phi \in \text{Pref}(\mathcal{L}) $ follows. Moreover, it is right-invariant because if $\alpha \equiv_\mathcal{D}\beta $ and $ \phi \in \Sigma^* $ is such that $ \alpha \phi \gamma \in \mathcal{L}$, then $ \beta \phi \in \text{Pref}(\mathcal{L}) $ and from $\delta(s,   \alpha)=\delta(s,   \beta)$ we obtain $\delta(s,   \alpha \phi)=\delta(s,   \beta \phi) $.

For every $ \alpha \in \pf L $ we have $ [\alpha]_{\equiv_\mathcal{D}} = I_{\delta(s, \alpha)} $, which implies that $ \equiv_\mathcal{D} $ is $ \mathcal{P} $-consistent. Moreover, $ \equiv_\mathcal{D} $ is $ \mathcal{P} $-convex, that is,   for every $ \alpha \in \text{Pref}(\mathcal{L}) $ we have that $[\alpha]_{\equiv_\mathcal{D}} = I_{\delta(s, \alpha)} $ is convex in $ U_\alpha $, because if $ u_1, \dots, u_k \in Q $ are such that $ U_\alpha = \bigcup_{i = 1}^k I_{u_i} $, then the $u_i$'s must be  pairwise $ \leq_\mathcal{D} $-comparable, being in the same $ \leq_\mathcal{D} $-chain. Moreover, $\equiv_\mathcal{D} $ refines  $ \equiv_\mathcal{L} $, because $\alpha \equiv_\mathcal{D} \beta $ implies that for every $ \phi \in \Sigma^* $ we have $ \delta(s, \alpha \phi) = \delta (s, \beta \phi) $ and so $ \alpha \phi \in \mathcal{L} $ iff $ \beta \phi \in \mathcal{L} $. Since $\equiv_\mathcal{L}^\mathcal{P} $ is the coarsest $ \mathcal{P} $-consistent, $ \mathcal{P} $-convex and right-invariant equivalence relation refining $ \equiv_\mathcal{L} $, and $\equiv_\mathcal{D} $ is a $ \mathcal{P} $-consistent, $ \mathcal{P} $-convex and right-invariant equivalence relation refining $ \equiv_\mathcal{L} $, we conclude that $ \equiv_\mathcal{D} $ also refines $ \equiv_\mathcal{L}^\mathcal{P} $, which in particular implies that $ \mathcal{L} $ is the union of some $ \equiv_\mathcal{D} $-classes. We know that $ \equiv_\mathcal{D} $ has finite index, so $ \equiv_\mathcal{L}^\mathcal{P} $ has finite index. \qed
\end{proof}

We can now explain how to canonically build a $ \mathcal{P} $-sortable DFA starting from an equivalence relation.

\begin{lemma}\label{from_equiv_to_DFA}

Let $ \mt L \subseteq \Sigma^* $ be a language, and let $ \mathcal{P} = \{U_1, \dots, U_p \} $ be a partition of $ \pf L $.  Assume that $ \mathcal{L} $ is the union of some classes of a $ \mathcal{P} $-consistent, $ \mathcal{P} $-convex, right-invariant equivalence relation $ \sim $ on $ \pf L $ of finite index. Then, $ \mathcal{L} $ is recognized by a $ \mathcal{P} $-sortable DFA $ \mathcal{D_\sim} = (Q_\sim, s_\sim, \delta_\sim, F_\sim) $ such that:
\begin{enumerate}
\item $ |Q_\sim| $ is equal to the index of $ \sim $;
\item $ \equiv_{D_\sim} $ and $ \sim $ are the same equivalence relation (in particular, $ |Q_\sim| $ is equal to the index of $ \equiv_{D_\sim} $).
\end{enumerate}
Moreover, if $ \mathcal{B} $ is a $ \mathcal{P} $-sortable DFA that recognizes $ \mathcal{L} $, then $ \mathcal{D_{\equiv_\mathcal{B}}} $ is isomorphic to $ \mathcal{B} $.
\end{lemma}

\begin{proof}
Define the DFA $ \mathcal{D_\sim} = (Q_\sim, s_\sim, \delta_\sim, F_\sim) $ as follows.
	\begin{itemize}
		\item $ Q_\sim = \{[\alpha]_\sim \; |\; \alpha \in \pf L\} $;
		\item $ s_\sim = [\varepsilon]_\sim $, where $ \varepsilon $ is the empty string;
		\item $ \delta_\sim ([\alpha]_\sim, a) = [\alpha a ]_\sim $, for every $ \alpha \in \Sigma^* $ and $ a \in \Sigma $ such that $ \alpha a \in \text{Pref}(\mathcal{L}) $.
		\item $ F_\sim = \{[\alpha]_\sim \;|\; \alpha \in \mathcal{L} \} $.
	\end{itemize}
Since $ \sim $ is right-invariant, it has finite index and $ \mathcal{L} $ is the union of some $ \sim $-classes, then $ \mathcal{D_\sim} $ is a well-defined DFA   and:
\begin{equation}\label{eq13}
\alpha \in [\beta]_\sim \iff \delta_\sim (s_\sim, \alpha) = [\beta]_\sim.
\end{equation}
which implies that for every $ \alpha \in \pf L $ it holds $ I_{[\alpha]_\sim} = [\alpha]_\sim $, and so $ \mathcal{L}(\mathcal{D_\sim}) = \mathcal{L} $.

For every $ i \in \{1, \dots, p \} $, define:
\begin{equation*}
    Q_i = \{[\alpha]_\sim \; |\; U_\alpha = U_i \}.
\end{equation*}

Notice that each $ Q_i $ is well-defined because $ \sim $ is $ \mathcal{P} $-consistent, and each $ Q_i $ is a $ \leq_{\mathcal{D_\sim}} $-chain because $ \sim $ is $ \mathcal{P} $-convex. It follows  that $ \{Q_i \; |\; 1\leq i \leq p\} $ is a $ \leq_{\mathcal{D_\sim}} $-chain partition of $ Q_\sim $.

From Equation \ref{eq13} we obtain:
\begin{equation*}
\begin{split}
    \text{Pref}(\mathcal{L(D_\sim)})^i & = \{\alpha \in \text{Pref}(\mathcal{L(D_\sim)})~ |~ \delta_\sim (s_\sim, \alpha) \in Q_i \}   \\
    & = \{\alpha \in \text{Pref}(\mathcal{L(D_\sim)})~ |~ (\exists [\beta]_\sim \in Q_i  ~ \alpha \in [\beta]_\sim) \}   \\
    & = \{\alpha \in \text{Pref}(\mathcal{L(D_\sim)})~ |~ U_\alpha = U_i \} = U_i.
\end{split}
\end{equation*}
In other words, $ \mathcal{D}_\sim $ witnesses that $ \mathcal{L} $ is recognized by a $ \mathcal{P} $-sortable DFA. Moreover:
\begin{enumerate}
\item The number of states of $ \mathcal{D}_\sim $ is clearly equal to the index of $ \sim $.
\item By Equation \ref{eq13}:
\begin{equation*}
    \alpha \equiv_{D_\sim} \beta \iff \delta_\sim (s_\sim, \alpha) = \delta_\sim (s_\sim, \beta) \iff [\alpha]_\sim = [\beta]_\sim \iff \alpha \sim \beta
\end{equation*}
so $ \equiv_{D_\sim} $ and $ \sim $ are the same equivalence relation.
\end{enumerate}
Finally, suppose $ \mathcal{B} $ is a $ \mathcal{P} $-sortable DFA that recognizes $ \mathcal{L} $. Notice that by Lemma \ref{simA} we have that $ \equiv_{\mathcal{B}} $ is a $ \mathcal{P} $-consistent, $ \mathcal{P} $-convex, right-invariant equivalence relation on $ \pf L $ of finite index such that $ \mathcal{L} $ is the union of some $ \equiv_\mathcal{B} $-classes, so $ \mathcal{D_{\equiv_\mathcal{B}}} $ is well-defined. Call $ Q_\mathcal{B} $ the set of states of $ \mathcal{B} $, and let $ \phi: Q_{\equiv_\mathcal{B}} \to Q_\mathcal{B} $ be the function sending $ [\alpha]_{\equiv_\mathcal{B}} $ into the state in $ Q_\mathcal{B} $ reached by reading $ \alpha $. Notice that $ \phi $ is well-defined because by the definition of $ \equiv_{\mathcal{B}} $ we obtain that all strings in $ [\alpha]_{\equiv_\mathcal{B}} $ reach the same state of $ \mathcal{B} $. It is easy to check that $ \phi $ determines an isomorphism between $ \mathcal{D_{\equiv_\mathcal{B}}} $ and $ \mathcal{B} $. \qed
   \end{proof}
   
     We now have all the required definitions to state our Myhill-Nerode theorem, which generalizes the one for Wheeler languages \cite{DBLP:conf/soda/AlankoDPP20}.

\begin{theorem}[Convex Myhill-Nerode Theorem]\label{thm:MN}
    Let $ \mathcal{L} $ be a language. Let $ \mathcal{P} $ be a partition of $ \pf L $. The following are equivalent:
    \begin{enumerate}
        \item $ \mathcal{L} $ is recognized by a $ \mathcal{P} $-sortable DFA.
        \item $ \equiv_\mathcal{L}^\mathcal{P} $ has finite index.
        \item $ \mathcal{L} $ is the union of some classes of a $ \mathcal{P} $-consistent, $ \mathcal{P} $-convex, right-invariant equivalence relation on $ \pf L $ of finite index.
    \end{enumerate}
Moreover, if one of the above statements is true (and so all the above statements are true), then there exists a unique minimum $ \mathcal{P} $-sortable DFA recognizing $ \mathcal{L} $ (that is, two $ \mathcal{P} $-sortable DFAs recognizing $ \mathcal{L} $ having the minimum number of states must be isomorphic).
\end{theorem}

\begin{proof}
$ (1) \to (2) $ It follows from Lemma \ref{simA}.

$ (2) \to (3) $ The desired equivalence relation is simply $ \equiv_\mathcal{L}^\mathcal{P} $. 

$ (3) \to (1) $ It follows from Lemma \ref{from_equiv_to_DFA}.

Now, let us prove that the minimum DFA is $ \mathcal{D_{\equiv_\mathcal{L}^\mathcal{P}}} $ as defined in Lemma \ref{from_equiv_to_DFA}. First, $ \mathcal{D_{\equiv_\mathcal{L}^\mathcal{P}}} $ is well-defined because $ \equiv_\mathcal{L}^\mathcal{P} $ is $ \mathcal{P} $-consistent, $ \mathcal{P} $-convex and right-invariant by definition; moreover,  it has finite index and $ \mathcal{L} $ is the union of some $ \equiv_\mathcal{L}^\mathcal{P} $-equivalence classes by Lemma \ref{simA}. Now, the number of states of $ \mathcal{D_{\equiv_\mathcal{L}^\mathcal{P}}} $ is equal to the index of $ \equiv_\mathcal{L}^\mathcal{P} $, or equivalently, of $ \equiv_{\mathcal{D_{\equiv_\mathcal{L}^\mathcal{P}}}} $. On the other hand, let $ \mathcal{B} $ be any $ \mathcal{P} $-sortable DFA recognizing $ \mathcal{L} $ non-isomorphic to $ \mathcal{D_{\equiv_\mathcal{L}^\mathcal{P}}} $. Then $ \equiv_\mathcal{B} $ is a refinement of $ \equiv_\mathcal{L}^\mathcal{P} $ by Lemma \ref{simA}, and it must be a strict refinement of $ \equiv_\mathcal{L}^\mathcal{P} $, otherwise $ \mathcal{D_{\equiv_\mathcal{L}^\mathcal{P}}} $ would be equal to $ \mathcal{D_{\equiv_\mathcal{B}}} $, which by Lemma \ref{from_equiv_to_DFA} is isomorphic to $ \mathcal{B} $, a contradiction. We conclude that the index of $ \equiv_\mathcal{L}^\mathcal{P} $ is smaller than the index of $ \equiv_\mathcal{B} $, so again by Lemma \ref{from_equiv_to_DFA} the number of states of $ \mathcal{D_{\equiv_\mathcal{L}^\mathcal{P}}} $ is smaller than the number of states of $ \mathcal{D_{\equiv_\mathcal{B}}} $ and so of $ \mathcal{B} $. \qed
\end{proof}

Notice that for a language $ \mathcal{L}$ Definition \ref{def:psortableNFA} implies that $ \text{width}^D(\mathcal{L}) = p $ if and only if (i) there exists a partition $ \mathcal{P} $ of size $ p $ such that $ \mathcal{L} $ is recognized by a $ \mathcal{P} $-sortable DFA and (ii) for every partition $ \mathcal{P'} $ of size less than $ p $ it holds that $ \mathcal{L} $ is not recognized by a $ \mathcal{P'} $-sortable DFA. As a consequence, $ \text{width}^D(\mathcal{L}) = p $ if and only if the minimum cardinality of a partition $ \mathcal{P} $ of $ \pf L $ that satisfies any of the statements in Theorem \ref{thm:MN} is equal to $ p $.
Given a $ \mathcal{P} $-sortable DFA recognizing $ \mathcal{L} $, it can be shown that the minimum $ \mathcal{P} $-sortable DFA recognizing $ \mathcal{L} $ can be built in polynomial time by generalizing the algorithm in \cite{alankocotpreDCC2022} (we do not provide the algorithmic details here because they would take us away   from the main ideas that we want to convey).

\section{Width-aware Encodings and Indexes for Regular Languages}\label{subsec:ind_dfa}

In this section we present compressed data structures for automata solving the \emph{compression} and the \emph{indexing} problems, that is  Problems \ref{problem:encoding}  and \ref{problem:indexing} of Section \ref{sec:definition indexing and compression}. 


When presenting our data structures in detail, we will assume to be working    with \emph{integer} alphabets of the form $\Sigma = [0,\sigma-1]$, that is, alphabets formed by all integers $\{0,1, \dots, \sigma-1\}$. Our data structure results hold in the word RAM model with words of size $w \in \Theta(\log u)$ bits, where $u$ is the size of the input under consideration (for example, $u$ may be the size of an automaton or the length of a string, depending on the input of the algorithm under consideration). When not specified otherwise, the space of our data structures is measured in words.  

Given an array $ S = S[1] S[2] \dots S[|S|] $, let $ S[l, r] = S[l] S[l + 1] \dots S[r - 1] S [r] $, if $ 1 \le l \le r \le |S| $, and $ S[l, r] = \emptyset $ if $ l > r $. 

Recall that the zero-order entropy of a sequence $S \in \Sigma^n$ of length $n$ over alphabet $\Sigma$ is \index{H(S)@ $H_0(S)$, zero-order entropy of $ S $} $H_0(S) = \sum_{c\in \Sigma} \frac{|S|_c}{n} \log_2 \frac{n}{|S|_c}$, where $|S|_c$ denotes the number of occurrences of character $c$ in $S$.  We will use some well-known properties of $H_0(S)$: the quantity $ n H_0(S) $ is a lower bound to the length of any encoding of $ S $ that encodes each character independently from the others via a prefix code of the alphabet $\Sigma$, and in particular $ H_0(S) \le \log_2 \sigma $.

\subsection{Path Coherence and Lower Bounds}\label{sec:path_coherence}

The reason why Wheeler automata admit an efficient indexing mechanism lies in two key observations: (i) on finite total orders a convex set can be expressed with $O(1)$ words by specifying its endpoints, and (ii) the set of states reached by a path labeled with a given string $\alpha$ forms a convex set (\emph{path-coherence}). We now show that the convex property holds true also for co-lex orders by generalizing the result in \cite{GAGIE201767}.

\begin{lemma}[Path-coherence]\label{lem:pathcoherence}
Let $ \mathcal{N} = (Q, s, \delta, F) $ be an NFA, $ \le $ be a co-lex order on $ \mathcal{N} $, $ \alpha \in \Sigma^*$, and $ U $ be a $ \le $-convex set of states. Then, the set $ U' $ of all states in $ Q $ that can be reached from $ U $ by following edges whose labels, when concatenated, yield $ \alpha $, is still a (possibly empty) $ \le $-convex set.
\end{lemma}

\begin{proof}
We proceed by induction on $ | \alpha | $. If $ |\alpha| = 0 $, then $ \alpha = \varepsilon $ and we are done. Now assume $ |\alpha| \ge 1 $. We can write $ \alpha = \alpha' a $, with $ \alpha' \in \Sigma^* $, $ a \in \Sigma $. Let $ u, v, z \in Q $ such that $ u < v < z $ and $ u, z \in U' $. We must prove that $ v \in U' $. By the inductive hypothesis, the set $ U'' $ of all states in $ Q $ that can be reached from some state in $ U $ by following edges whose labels, when concatenated, yield $ \alpha' $, is a $ \le $-convex set. In particular, there exist $ u', z' \in U'' $ such that $ u \in \delta(u', a) $ and $ z \in \delta(z', a) $. Since $ a \in \lambda (u) \cap \lambda (z) $ and $ u < v < z $, then $ \lambda (v) = \{a\} $ (otherwise by Axiom 1 we would obtain a contradiction), so there exists $ v' \in Q $ such that $ v \in \delta (v', a) $. From $ u < v < z $ and Axiom 2 we obtain $ u' \le v' \le z' $. Since $ u', z' \in U'' $ and $ U'' $ is a $ \le $-convex set, then $ v' \in U''$, and so $ v \in U' $. \qed
\end{proof}

 Note that Corollary \ref{cor:intervals} in Section \ref{sec:width of automaton} also follows from Lemma \ref{lem:pathcoherence} by picking $ U = \{s\} $, because  then $ U' = I_\alpha $.



As we will see, the above result implies that indexing mechanism can be extended to arbitrary finite automata by updating \emph{one} $\leq$-convex set for each character of the query pattern. This, however, does not mean that,in general, indexing can be performed as efficiently as on Wheeler automata: as we show next, in general it is not possible to represent a $\leq$-convex set in a partial order using constant space.

\begin{lemma}\label{lem:lower bound}
The following hold:
\begin{enumerate}
\item Any partial order $(V,\leq)$ of width $ p $  has at least $2^p$ distinct $\leq$-convex subsets.
\item  
For any $n$ and $p$ such that $1\leq p \leq n$, there exists a partial order $(V,\leq)$ of width $p$ and $|V|=n$ with at least $(n/p)^p$ distinct $\leq$-convex subsets.
\end{enumerate}
\end{lemma}

\begin{proof}
(1) Since $ V $ has width $ p $, there exists an antichain $ A $ of cardinality $ p $. It is easy to see that any subset $I \subseteq A$ is a distinct $\leq$-convex set. The bound $2^p$ follows. 
(2) Consider a partial order formed by $p$ mutually-incomparable total orders  $V_i$,  all having  $n/p$ elements. 
Since any total order of cardinality $n/p$ has $(n/p+1)(n/p)/2 + 1$ distinct convex sets  and   any combination of $\leq_{V_i}$-convex sets forms a distinct $\leq$-convex set, we obtain  at least \[\prod_{i=1}^p ((n/p+1)(n/p)/2+1) \geq \prod_{i=1}^p n/p = (n/p)^p\] distinct $\leq$-convex sets.   \qed
\end{proof}

\begin{remark}\label{rem:lower bounds bit}
Given an NFA $\mathcal N$ with $n$ states and a co-lex order $\leq$ of width $p$ on $\mathcal N$, Lemma \ref{lem:lower bound} implies an information-theoretic lower bound of $p$ bits for expressing a $\leq$-convex set, which increases to $\Omega(p\log(n/p))$ bits in the worst case. This means that, up to (possibly) a logarithmic factor, in the word RAM model $\Omega(p)$ time is needed to manipulate one $\leq$-convex set.
\end{remark}

\begin{remark} \label{rem:Dilworth}
If  $(V, \leq)$ is a partial order, $V'\subseteq V$ and $U$ is a convex subset of 
 $(V, \leq)$, then $U\cap V'$ is a convex set over the restricted partial order $(V', \leq_{V'})$. 
 In particular, if  $ \{V_i \; |\; 1\leq i \leq p\} $ is a partition of $ V $ then any  $ \le $-convex set $U$ is  the disjoint union of $ p $ (possibly empty) sets $ U_1, \dots, U_p $, where $ U_i = U \cap V_i $ is a convex set over the restriction  $(V_i, \le_{V_i}) $.
\end{remark}

The above remarks motivate the following strategy. Letting $p$ be the width of a partial order $ \le $, by Dilworth's theorem \cite{dilworth} there exists a $ \le $-chain partition $ \{Q_i \; |\; 1\leq i \leq p\}$ of $Q$ into $p$ chains. Then, Remark \ref{rem:Dilworth} implies that a $ \le $-convex set can be encoded by at most $p$ convex sets, each contained in a distinct chain, using $O(p)$ words. This encoding is essentially optimal by Remark \ref{rem:lower bounds bit}. 

\medskip

Using the above mentioned strategy we can now refine Lemma \ref{lem:pathcoherence} (path-coherence) and its  corollary (Corollary \ref{cor:intervals}).
 
\begin{lemma}\label{lem:lemmasimple}
Let $ \mathcal{N} = (Q, s, \delta, F) $ be an NFA, $ \le $ be a co-lex order on $ \mathcal{N} $, $ \{Q_i\}^p_{i=1} $ be   a $ \le $-chain partition of $ \mathcal{N} $, $ \alpha \in \Sigma^*$, and $ U $ be a $ \le $-convex set of states. Then, the set $ U' $ of all states in $ Q $ that can be reached from $ U $ by following edges whose labels, when concatenated, yield $ \alpha $, is the disjoint union of $ p $ (possibly empty) sets $ U'_1, \dots, U'_p $, where $ U'_i = U' \cap Q_i $ is $ \le_{Q_i} $-convex, for $ i = 1, \dots, p $.\\
In particular, if  $ \alpha \in \text{Pref}(\mathcal{L(A)}) $  then, $ I_\alpha $ is the disjoint union of $ p $ (possibly empty) sets $ I_\alpha^1, \dots, I_\alpha^p $, where $ I_\alpha^i = I_\alpha \cap Q_i $ is $ \le_{Q_i} $-convex, for $ i = 1, \dots, p $.
\end{lemma}

\subsection{Encoding DFAs and Languages: the Automaton BWT (aBWT)}\label{sec:aBWT}

Let us define a representation of   an automaton that is a generalization of the well-known \emph{Burrows-Wheeler transform} (BWT) of a string \cite{burrows1994block}. We call this generalization the \emph{automaton Burrows-Wheeler transform} \index{aBWT} (aBWT) and, just like the BWT of a string is an encoding of the string (that is, distinct strings have distinct BWTs), we will show that the aBWT of a DFA is an encoding of the DFA. We will also see that, on  NFAs,  the aBWT  allow us  to reconstruct the accepted language   and to   efficiently solve the string matching problem (Problem \ref{problem:indexing}), but in general it is not  an encoding since it is not sufficient to reconstruct the NFA's topology. A variant, using slightly more space and encoding NFAs, will be presented in Section \ref{sec:aaBWT}.

\definecolor{gr1}{rgb}{0.78, 0.78, 0.78}
\definecolor{gr2}{rgb}{0.9, 0.9, 0.9}

\newcolumntype{P}[1]{>{\centering\arraybackslash}p{#1}}
\newcolumntype{A}[1]{>{\centering\arraybackslash\columncolor{gr1}}p{#1}}
\newcolumntype{B}[1]{>{\centering\arraybackslash\columncolor{gr2}}p{#1}}


\begin{figure}[h!]
\centering
\begin{subfigure}{\textwidth}
	\begin{tikzpicture}[shorten >=1pt,node distance=1.5cm,on grid,auto]
	\footnotesize 
	\tikzstyle{every state}=[fill={rgb:black,1;white,10}]
	
	\node[state,initial,minimum size=5pt]   (q_0)                         {$v_1$};
	\node[state,minimum size=5pt]           (q_1)  [right of=q_0]    {$v_2$};
	\node[state,minimum size=5pt]           (q_3)  [right of=q_1]    {$v_3$};
	\node[state,accepting,minimum size=5pt] (q_2)  [right of=q_3]    {$v_6$};
	\node[state,minimum size=5pt]           (q_5)  [right of=q_2]    {$v_7$};

	\node[state,accepting,minimum size=5pt] (q_4)  [below of=q_3]    {$v_5$};

	\node[state,accepting,minimum size=5pt] (q_6)  [below of=q_5]    {$v_4$};
	
  	
	\path[->]

 	(q_0) edge [bend left=40] node {a}    (q_1)
 	(q_1) edge [bend left] node {b}    (q_2)
 	(q_2) edge [bend left, pos=0.51] node {a}    (q_3)
 	(q_2) edge [bend left=40] node {b}    (q_5)
 	(q_4) edge [pos=0.5,below] node {b}   (q_5)
 	(q_4) edge [bend left, pos=0.6] node {a}    (q_3)
 	(q_3) edge [bend left, left,pos=0.6] node {a}    (q_4)
 	(q_5) edge [bend left] node {b,c}    (q_6)
 	(q_6) edge [bend left, right] node {b}    (q_5);
 	
 	\node[state,color=white,text=black,inner sep=1pt,minimum size=0pt] (1)    at (10,-3)	{$v_1$};
	\node[state,color=white,text=black,inner sep=1pt,minimum size=0pt] (6)    at (10,-2)               {$v_2$};
	\node[state,color=white,text=black,inner sep=1pt,minimum size=0pt] (3)    at (10,-1)               {$v_3$};
	\node[state,color=white,text=black,inner sep=1pt,minimum size=0pt] (5)    at (11,-1)               {$v_5$};	
	\node[state,color=white,text=black,inner sep=1pt,minimum size=0pt] (2)    at (10,0)               {$v_6$};
	\node[state,color=white,text=black,inner sep=1pt,minimum size=0pt] (7)    at (10,1)               {$v_7$};
	\node[state,color=white,text=black,inner sep=1pt,minimum size=0pt] (4)    at (11,1)               {$v_4$};
 
 	\path[-]
	(1) edge node {}    (6)
	(6) edge node {}    (3)
	(3) edge node {}    (2)
	(2) edge node {}    (7)
	(6) edge node {}    (5)
	(5) edge node {}    (2)
	(2) edge node {}    (4)
	;
 \end{tikzpicture}
\end{subfigure}
\hfill
\begin{subfigure}{\textwidth}
      \centering
   \begin{tabular}{crP{25pt}|A{25pt}|A{25pt}|A{25pt}|A{25pt}|B{25pt}|B{25pt}|B{25pt}|}
    & \multicolumn{1}{c}{} & \multicolumn{1}{c}{{$\tt{CHAIN}$}} & \multicolumn{1}{c}{{1}} & \multicolumn{1}{c}{{0}} & \multicolumn{1}{c}{{0}} & \multicolumn{1}{c}{{0}} & \multicolumn{1}{c}{{1}} & \multicolumn{1}{c}{{0}} & \multicolumn{1}{c}{{0}}\\
    & \multicolumn{1}{c}{} & \multicolumn{1}{c}{{$\tt{FINAL}$}} & \multicolumn{1}{c}{{0}} & \multicolumn{1}{c}{{0}} & \multicolumn{1}{c}{{0}} & \multicolumn{1}{c}{{1}} & \multicolumn{1}{c}{{1}} & \multicolumn{1}{c}{{1}} & \multicolumn{1}{c}{{0}}\\
    & \multicolumn{1}{c}{} & \multicolumn{1}{c}{{$\tt{IN\_DEG}$}} & \multicolumn{1}{c}{{1}} & \multicolumn{1}{c}{{01}} & \multicolumn{1}{c}{{001}} & \multicolumn{1}{c}{{001}} & \multicolumn{1}{c}{{01}} & \multicolumn{1}{c}{{01}} & \multicolumn{1}{c}{{0001}}\\\cline{4-10}\\[-3.72mm]
    $\tt{OUT\_DEG}$ & \multicolumn{1}{c}{{$\tt{OUT}$}} & \multicolumn{1}{c|}{} &  1 &  2 &  3 &  4 & 5 &  6 &  7 \\\cline{3-10}\\[-3.72mm]
    01 & \multicolumn{1}{c|}{{(1,a)}} &  1 &  &  (1,a) &  &  &  &  &  \\\cline{3-10}\\[-3.72mm]
    01 & \multicolumn{1}{c|}{{(2,b)}} &  2 &  &  &  &  &  & (2,b)  &  \\\cline{3-10}\\[-3.72mm]
    01 & \multicolumn{1}{c|}{{(2,a)}} & 3 &  &  &  &  &  (2,a) &  &  \\\cline{3-10}\\[-3.72mm]
    01 & \multicolumn{1}{c|}{{(2,b)}} &  4 &  &  &  &  &  &  &  (2,b) \\\cline{3-10}\\[-3.72mm]
    001 & \multicolumn{1}{c|}{{(1,a),(2,b)}} &  5 &  &  &  (1,a) &  &  &  &  (2,b)\\\cline{3-10}\\[-3.72mm]
    001 & \multicolumn{1}{c|}{{(1,a),(2,b)}} &  6 &  &  &  (1,a) &  &  &  &  (2,b) \\\cline{3-10}\\[-3.72mm]
    001 & \multicolumn{1}{c|}{{(1,b),(1,c)}} &  7 &  &  &  &  (1,b) (1,c) &  &  &  \\\cline{3-10}\\[-3.72mm]
  \end{tabular}
  \vspace{5pt}
\end{subfigure}
\caption{A DFA $\mathcal D$ accepting $\mathcal L = ab(aa)^*(b(b+c))^*$, together with the Hasse diagram of its maximum co-lex order $ \le $ and the adjacency matrix of $\mathcal D$. In the following examples we consider the $ \le $-chain partition given by $ Q_1 = \{v_1, v_2, v_3, v_4 \} $, $ Q_2 = \{v_5, v_6, v_7 \} $. The adjacency matrix is sorted according to the total order $Q = \{v_1,v_2,v_3,v_4,v_5,v_6,v_7\}$. The two different shades of gray divide the edges by destination chain (either 1 or 2).
 Each edge is represented in this matrix as the pair $(i,c)$, where $i$ is the destination chain and $c\in\Sigma$ is the edge's label.
 This way of visualizing the adjacency matrix can be viewed as a two-dimensional representation of the automaton Burrows-Wheeler transform (aBWT, Definition \ref{def:aBWT}).
 The aBWT can be linearized in five sequences, as shown here and in Example \ref{ex:aBWT}.}
 \label{fig:abwtdfa}
\end{figure}
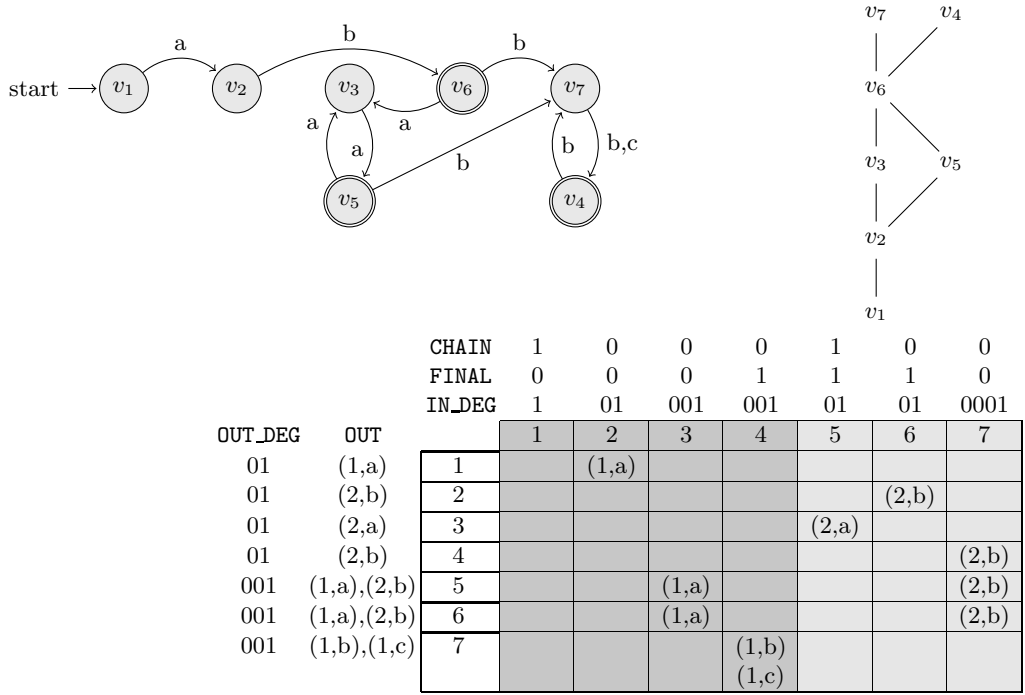

The aBWT is given for an automaton   $ \mathcal N =  (Q, s, \delta, F) $ and it depends on a co-lex order $ \le $ endowed with  a  fixed $ \le $-chain partition  $  \{Q_i \; |\; 1\leq i \leq p\} $ of $Q$  (we assume $s\in Q_1$, so $ s $ is the first element of $ Q_1 $). An intuition behind the aBWT is provided in Figure \ref{fig:abwtdfa}: after sorting the states in a total order which agrees with the  co-lex order $\le$ on  pairs whose elements belong  to the same class of the partition  $  \{Q_i \; |\; 1\leq i \leq p\}$ and drawing the transition function's adjacency matrix in this order, we build five sequences collecting the chain borders ($\tt{CHAIN}$), a boolean flag per state marking final states ($\tt{FINAL}$), the states' in-degrees ($\tt{IN\_DEG}$), the states' out-degrees ($\tt{OUT\_DEG}$), and the states' labels and destination chains ($\tt{OUT}$).

\begin{definition}[aBWT of an automaton]\label{def:aBWT}
Let $ \mathcal N =  (Q, s, \delta, F) $ be an NFA and let    $e = |\delta|$ be the number of $\mathcal N$-transitions.  Let $ \le $ be a co-lex  order on $ \mathcal N $, and let $  \{Q_i \; |\; 1\leq i \leq p\} $ be a $ \le $-chain partition of $ Q $, where w.l.o.g. $s \in Q_1$. 
Let $ \pi(v) $ be the unique map such that  $ v \in Q_{\pi(v)} $ and consider the total state order $Q=\{v_1, \dots, v_n\}$ such that, for every $ 1 \le i < j \le n $, it holds \footnote{Notice the overload on symbol $\le$, also used to indicate the co-lex order among states.} $ \pi(v_i) < \pi(v_j) \lor (\pi(v_i) = \pi(v_j) \land v_i < v_j) $.
The \emph{automaton Burrows-Wheeler transform} $\mathtt{aBWT}(\mathcal N, \le, \{Q_i \; |\; 1\leq i \leq p\}) $  of $(\mathcal N, \le, \{Q_i \; |\; 1\leq i \leq p\})$   consists of the following sequences. 
\begin{itemize}
    \item   $\mathtt{CHAIN} \in \{0, 1 \}^n $ is such that the $ i $-th bit is equal to 1 if and only if $ v_i $ is the first state of some chain $ Q_j $.
    \item  $\mathtt{FINAL} \in \{0, 1 \}^n $ is such that the $ i $-th bit is equal to 1 if and only if $ v_i \in F $.
    \item   $ \mathtt{IN\_DEG} \in \{0, 1\}^{e + n} $ stores the nodes' in-degrees in unary. More precisely, (1) $ \mathtt{IN\_DEG} $ contains exactly $ n $ characters equal to $ 1 $, (2) $ \mathtt{IN\_DEG} $ contains exactly $ e $ characters equal to $ 0 $, and (3) the number of zeros between the $(i-1)$-th character equal to one (or the beginning of the sequence if $i=1$) and  the $ i $-th character equal to $ 1 $  yields the in-degree of $v_i$.
    \item   $ \mathtt{OUT\_DEG} \in \{0, 1\}^{e + n} $ stores the nodes' out-degrees in unary. More precisely, (1) $ \mathtt{OUT\_DEG} $ contains exactly $ n $ characters equal to $ 1 $, (2) $ \mathtt{OUT\_DEG} $ contains exactly $ e $ characters equal to $ 0 $, and (3) the number of zeros between the $(i-1)$-th character equal to one (or the beginning of the sequence if $i=1$) and  the $ i $-th character equal to $ 1 $  yields the out-degree of $v_i$.
    \item   $ \mathtt{OUT} $ stores the edges' labels and destination chains, as follows.
    Sort all edges $(v_j,v_i,c)$ by their starting state $v_j$ according to their index $j$. Edges originating from the same state are further sorted by their label $c$. Edges sharing the starting state and label are further sorted by destination node $v_i$. 
    Then, $ \mathtt{OUT} $ is obtained by concatenating the pairs $(\pi(v_i),c)$ for all edges $(v_j,v_i,c)$ sorted in this order. 
\end{itemize}
\end{definition}

\begin{example}\label{ex:aBWT}
The aBWT of $(\mathcal D, \le,  \{Q_i \; |\; 1\leq i \leq 2\}) $ in Figure \ref{fig:abwtdfa} consists of the following sequences:
\begin{itemize}
    \item $\mathtt{CHAIN} = 1000100 $.
    \item $\mathtt{FINAL} = 0001110$.
    \item $ \mathtt{IN\_DEG} = 10100100101010001 $.
    \item $ \mathtt{OUT\_DEG} = 01010101001001001 $.
    \item $ \mathtt{OUT} = (1,a) (2,b) (2,a) (2,b) (1,a) (2,b) (1,a) (2,b) (1,b) (1,c) $.
\end{itemize}
\end{example}

It is not hard to show that the aBWT generalizes all existing approaches \cite{burrows1994block,Ferragina2005,mantaci2007extension,BOSS,GCSA,GAGIE201767}, for which $p=1$ always holds (and so sequences $\mathtt{CHAIN} $ and the first components of the pairs in $ \mathtt{OUT} $ are uninformative). For example, on strings also $ \mathtt{OUT\_DEG} $ and $ \mathtt{IN\_DEG} $ are uninformative ($\mathtt{FINAL} $ does not apply); the only sequence left is the concatenation of the second components of the pairs in $ \mathtt{OUT} $, that is, the classic Burrows-Wheeler transform (to be precise, its co-lexicographic variant).

  
In this section we will prove that if we only know the   aBWT  of an automaton we can reconstruct all the sets $ I_\alpha^i $ of Lemma \ref{lem:lemmasimple} (we recall that $ I_\alpha^i $ is the set of all states in the $i$-th chain being connected with the source by a path labeled $\alpha$), and in particular we can retrieve the language of the automaton. 
To this end, we first define   some   auxiliary sets of states of an NFA --- $S(\alpha)$ and $L(\alpha)$ --- and we prove that, for any $1\leq i\leq p$, on the $i$-th chain the convex set  corresponding to $I_\alpha^i$ lays between (the convex sets) $S(\alpha)\cap Q_i $ and $ L(\alpha)\cap Q_i$.
Intuitively, $S(\alpha)$ (respectively, $L(\alpha)$) is the set of all states $u$ whose associated regular language $I_u$ contains only strings co-lexicographically strictly smaller (respectively, larger) than $\alpha$.


\begin{definition}\label{def:S_i}
Let $ \mathcal{N} = (Q, s, \delta, F) $ be an NFA,  $ \le $ be a co-lex order on $ \mathcal{N} $,  and  $ \{Q_i\; | \;  1\leq i \leq p\} $ be a $ \le $-chain partition of $ Q $. Let $\alpha \in \Sigma^*$. Define:
	\begin{equation*}
	    \begin{split}
	        S(\alpha) & = \{  u \in Q\ |\ (\forall \beta\in I_{u})(\beta \prec \alpha)\} \\
	        L(\alpha) & = \{  u \in Q\ |\ (\forall \beta\in I_{u})(\alpha \prec \beta)\}.
	    \end{split}
	\end{equation*}
	Moreover, for every $ i = 1, \dots, p $ define $ S_i (\alpha) = S (\alpha) \cap Q_i $ and $ L_i (\alpha) = L (\alpha) \cap Q_i $.
\end{definition}

In the following, we see a $ \le $-chain $ Q_i $ as an array of sorted elements, so $Q_i[j]$ and $Q_i[1,k]$ denote the $j$-th smallest state in $Q_i$ and the $k$ smallest states in  $Q_i$, respectively.

In Lemma   \ref{lem:smallerlarger} we show    that in order to compute $ I_\alpha $ it will be sufficient to compute $ S(\alpha) $ and $ L(\alpha) $.

\begin{lemma}\label{lem:smallerlarger}
Let $ \mathcal{N} = (Q, s, \delta, F) $ be an NFA,  $ \le $ be a co-lex order on $ \mathcal{N} $,    $ \{Q_i\; | \;  1\leq i \leq p\} $ be a $ \le $-chain partition of $ Q $, and $\alpha \in \Sigma^* $. 
\begin{enumerate}
    \item If $ u, v \in Q $ are such that $ u \le v $ and $ v \in S(\alpha) $, then $ u \in S(\alpha) $. In particular, for every $ i = 1, \dots, p $ there exists $0 \leq l_i \leq |Q_i| $ such that $S_i(\alpha) = Q_i[1, l_i]$ (namely, $ l_i = |S_i(\alpha)| $). 
    \item If $ u, v \in Q $ are such that $ u \le v $ and $ u \in L(\alpha) $, then $ v \in L(\alpha) $. In particular, for every $ i = 1, \dots, p $ there exists $1 \leq r_i \leq |Q_i| + 1 $ such that $L_i(\alpha) = Q_i[r_i,|Q_i|]$ (namely, $ r_i = |Q_i| - |L_i(\alpha)| + 1 $).
    \item $ I_\alpha $, $ S(\alpha) $, and $ L(\alpha) $ are pairwise disjoint. In particular, it always holds that $ l_i < r_i $.
    \item Let $ 1 \le i \le p $. If $ I_\alpha^i \not = \emptyset $, then $ I_\alpha^i = Q_i [l_i + 1, r_i - 1] $, that is, $\{S_i(\alpha), I^i_\alpha, L_i(\alpha)\}$ is an ordered partition of $Q_i$.
\end{enumerate}
\end{lemma}

\begin{proof}
\begin{enumerate}
    \item Let $ \beta \in I_u $. We must prove that $ \beta \prec \alpha $. Now, if $ \beta \in I_v $, from $ v \in S(\alpha) $ we obtain $ \beta \prec \alpha $. If $ \beta \not \in I_v $, then for any $ \gamma \in I_v $ we have $ \beta \prec \gamma $ by Lemma \ref{lem:monot}. Again, we have $ \gamma \prec \alpha $, so we conclude $ \beta \prec \alpha $.
    \item Analogous to the previous point.
    \item We have $ I_\alpha \cap S(\alpha) = \emptyset $ because if $ u \in I_\alpha $, then $ \alpha \in I_u $, so $ u \not \in S(\alpha) $. Similarly, $ I_\alpha \cap L(\alpha) = \emptyset $. Finally, we have $ S(\alpha) \cap L(\alpha)=\emptyset $ because if there existed $ u \in S (\alpha) \cap L(\alpha) $, then  for any $ \beta \in I_u $ (there exists at least one such $\beta$ since   $I_u\neq \emptyset$) we would obtain $ \beta \prec \alpha \prec \beta $, a contradiction.
    \item To begin with, let us prove that, for every $ v \in I_\alpha, $ (1) if $ u < v $, then either $ u \in I_\alpha $ or $ u \in S(\alpha) $, and (2) if $ v < z $, then either $ z \in I_\alpha $ or $ z \in L(\alpha) $. We only prove (1),  the proof of (2)  being analogous. Assume that $ u \not \in I_\alpha $, and let $ \beta \in I_u $. We must prove that $ \beta \prec \alpha $ and, since $ \alpha \in I_v \setminus I_u $, this  follows from Lemma \ref{lem:monot}.
    
    Now, let $ 1 \le i \le p $ be such that $ I_\alpha^i \not = \emptyset $, and let us prove that $\{S_i(\alpha), I^i_\alpha, L_i(\alpha)\}$ is an ordered partition of $Q_i$. Consider $u\in I_\alpha^i$. Then, if $v\in Q_i\setminus I_\alpha^i$ we have either $v<u $ or $u<v$, hence what we have proved above implies that either $v\in S_i(\alpha)$ or $v\in L_i(\alpha)$. Therefore, if $ I_\alpha^i \not = \emptyset $ then $\{S_i(\alpha),  I_\alpha^i, L_i(\alpha) \}$ is an ordered partition of $Q_i$ and point 4 follows. \qed
\end{enumerate}
\end{proof}
 
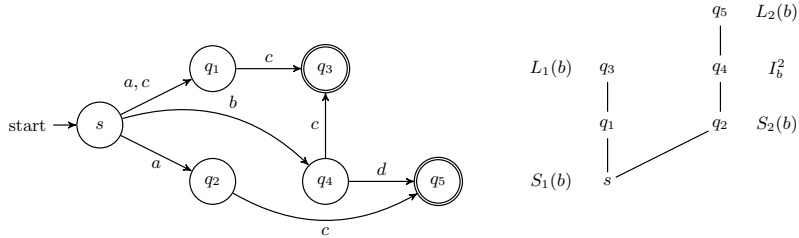
\begin{figure}[h!] 
 \begin{center}
 \scalebox{0.75}{ 
\begin{tikzpicture}[->,>=stealth', semithick, auto, scale=1]

 \node[state, initial] (0)    at (0,0)	{$s$};
\node[state] (1)    at ( 2,1)	{$q_1$};
 \node[state] (2)    at (2,-1)	{$q_2$};
\node[state,accepting] (3)    at (4,1)	{$q_3$};
 \node[state] (4)    at (4,-1)	{$q_4$};
  \node[state,accepting] (5)    at (6,-1)	{$q_5$};
 
\draw (0) edge [] node [] {$a,c$} (1); 
\draw (0) edge [] node [below] {$a$} (2);
\draw (1) edge [] node [] {$c$} (3);
 \draw (0) edge  [bend left] node {$b$} (4);
\draw (2) edge [bend right] node [below] {$c$} (5);
\draw (4) edge [] node [] {$c$} (3); 
 \draw (4) edge [] node [] {$d$} (5);

 \node[state,color=white,text=black,inner sep=1pt,minimum size=0pt] (10)    at (8,-1)	{$S_1(b)$};
 
  \node[state,color=white,text=black,inner sep=1pt,minimum size=0pt] (10)    at (8,1)	{$L_1(b)$};
  
  \node[state,color=white,text=black,inner sep=1pt,minimum size=0pt] (10)    at (12, 0)	{$S_2(b)$};
 
  \node[state,color=white,text=black,inner sep=1pt,minimum size=0pt] (10)    at (12,1)	{$I_b^2$};
    \node[state,color=white,text=black,inner sep=1pt,minimum size=0pt] (10)    at (12,2)	{$L_2(b)$};

\node[state,color=white,text=black,inner sep=1pt,minimum size=0pt] (10)    at (9,-1)	{$s$};
	\node[state,color=white,text=black,inner sep=1pt,minimum size=0pt] (11)    at (9,0)               {$q_1$};
		\node[state,color=white,text=black,inner sep=1pt,minimum size=0pt] (13)    at (9,1)               {$q_3$};
		 	\node[state,color=white,text=black,inner sep=1pt,minimum size=0pt] (12)    at (11,0)               {$q_2$};
			\node[state,color=white,text=black,inner sep=1pt,minimum size=0pt] (14)    at (11,1)               {$q_4$};	
				\node[state,color=white,text=black,inner sep=1pt,minimum size=0pt] (15)    at (11,2)               {$q_5$};
 
 	\path[-]
	(10) edge node {}    (11)
		(10) edge node {}    (12)
	(11) edge node {}    (13)
	(12) edge node {}    (14)
	(14) edge node {}    (15);
\end{tikzpicture}
}
\end{center}

 \caption{Consider the   NFA  in the figure and the  Hasse diagram of a co-lex order  $ \le  $ with  chain partition   $ Q_1 = \{s,q_1,q_3 \} $ and $ Q_2=\{q_2,q_4,q_5\} $.
 If we consider the word $ b$, then, on the one hand,  $I^1_b=\emptyset$ and 
 $\{S_1(b)=\{s\}, L_1(b)=\{q_3\}\}$ is not a partition of $Q_1$. On the other hand, since $I^2_b=\{q_4\}\neq \emptyset$, then  $\{S_2(b)=\{q_2\},I^2_b,  L_1(b)=\{q_5\}\}$ is   an ordered $Q_2$-partition. }
 \label{fig:not_part}
\end{figure}

\begin{remark}    Notice that if $ I_\alpha^i   = \emptyset $ then $\{S_i(\alpha),    L_i(\alpha) \}$ is not, in general,  an ordered partition of $Q_i$, as shown in  Fig.   \ref{fig:not_part}.
\end{remark}  

Our next step is to show how to recursively compute the sets $ S(\alpha) $ and $ L(\alpha) $ defined above. We begin with the following two lemmas.

\begin{lemma}\label{lem:Si alpha}
Let $ \mathcal{N} = (Q, s, \delta, F) $ be an NFA,  $ \le $ be a co-lex order on $ \mathcal{N} $,    $ \{Q_i\; | \;  1\leq i \leq p\} $ be a $ \le $-chain partition of $ Q $, $\alpha'\in \Sigma^*$, $ a \in \Sigma $,  and $ u \in Q $.
\begin{enumerate}
	\item  $u \in S(\alpha' a)$ if and only if (1) $ \text{max}_{\lambda(u)} \preceq a $ and (2) if $ u' \in Q $ is such that  $ u \in \delta(u', a) $, then $ u' \in S(\alpha')$.
	\item  $u \in L(\alpha' a)$ if and only if (1) $ a \preceq \text{min}_{\lambda(u)} $ and (2) if $ u' \in Q $ is such that $ u \in \delta(u', a) $, then $ u' \in L(\alpha')$.
\end{enumerate}
\end{lemma}

\begin{proof}
Let us prove the first statement.

$ (\Rightarrow) $  Let $ c \in \Sigma $ such that $ c \in \lambda (u) \setminus \{\#\} $. Let $ u' \in Q $ such that $ u \in \delta (u', c) $, and let $ \beta' \in I_{u'} $. Then $ \beta' c \in I_u $, so from $u \in S(\alpha' a)$ we obtain $ \beta' c \prec \alpha' a $, which implies $ c \preceq a $. Now assume that $ c = a $. Suppose for sake of contradiction that $ u' \not \in S(\alpha') $. This means that there exists $ \gamma' \in I_{u'} $ such that $ \alpha' \preceq \gamma' $. This implies $ \alpha' a \preceq \gamma' a $ and, since $ \gamma' a \in I_u $, we obtain $u \not \in S(\alpha' a) $, a contradiction.

$ (\Leftarrow) $ Let $ \beta \in I_u $. We must prove that $ \beta \prec \alpha' a $. If $ \beta = \varepsilon $ we are done, because $ \varepsilon \prec \alpha' a $. Now assume that $ \beta = \beta' b $. This means that there exists $ u' \in Q $ such that $ u \in \delta (u', b) $ and $ \beta' \in I_{u'} $. We know that $ b \preceq a $. If $ b \prec a $, then  $\beta \prec \alpha' a$  and we are done. If $ b = a $, then $ u' \in S(\alpha') $, so $ \beta' \prec \alpha' $, which implies $ \beta \prec \alpha' a $.

The proof of the second statement is analogus (the only difference being that in $ (\Leftarrow) $ it must necessarily be $ \beta \not = \varepsilon $, because $ a \preceq \text{min}_{\lambda(u)} $). \qed
\end{proof}

\begin{lemma}\label{lem:wheresgoes}
Let $ \mathcal{N} = (Q, s, \delta, F) $ be an NFA,  $ \le $ be a co-lex order on $ \mathcal{N} $,    $ \{Q_i\; | \;  1\leq i \leq p\} $ be a $ \le $-chain partition of $ Q $, $\alpha'\in \Sigma^*$, and $ a \in \Sigma $. Fix $ 1 \le i \le p $, and let $ S_i(\alpha' a) = Q_i[1, l_i] $ and $ L_i(\alpha' a) = Q_i[r_i, |Q_i|] $.
\begin{enumerate}
    \item If $ u' \in S(\alpha') $ and $ u \in Q_i $ are such that $ u \in \delta (u', a) $, then $ u \in Q_i[1, \min\{l_i + 1, |Q_i|\}] $.
    \item If $ u' \in L(\alpha') $ and $ u \in Q_i $ are such that $ u \in \delta (u', a) $, then $ u \in Q_i[\max\{r_i - 1, 1\}, |Q_i|] $.    
\end{enumerate}
\end{lemma}

\begin{proof}
We only prove the first statement, the proof of the second statement being entirely analogous. We can assume $ l_i < |Q_i| - 1 $, otherwise the conclusion is trivial. If $ a \prec \max(\lambda(Q_i[l_i + 1])) $ the conclusion is immediate by Axiom 1, so we can assume $ \max(\lambda(Q_i[l_i + 1])) \preceq a $. We know that $ Q_i[l_i + 1] \not \in S(\alpha' a) $, so by Lemma \ref{lem:Si alpha} there exists $ v' \in Q $ such that $ Q_i[l_i + 1] \in \delta (v', a) $ and $ v' \not \in S(\alpha') $. Suppose for sake of contradiction that $ Q_i[l_i + 1] < u $. By Axiom 2 we obtain $ v' \le u' $. From $ u' \in S(\alpha') $ and Lemma \ref{lem:smallerlarger} we conclude $ v' \in S(\alpha') $, a contradiction. \qed
\end{proof}

The following definition is instrumental in giving an operative variant  of Lemma \ref{lem:Si alpha} (i.e. Lemma \ref{lem:Si alpha2}) to be used in our algorithms. 

\begin{definition}\label{def: inout}
Let $ \mathcal{N} = (Q, s, \delta, F) $ be an NFA, $ \le $ be a co-lex order on $ \mathcal{N} $,  and  $ \{Q_i\; | \;  1\leq i \leq p\} $ be a $ \le $-chain partition of $ Q $. Let $ U \subseteq Q $.
We denote by $\mathtt{in}(U,a)$ the number of edges labeled with character $a$ that enter states in $U$:
	$$
	\mathtt{in}(U,a) = |\{(u', u)\ |\ u' \in Q, u \in U, u \in \delta(u', a) \}|.
	$$
We denote by $\mathtt{out}(U,i,a)$ the number of edges labeled with character $a$ that leave states in $U$ and enter the $i$-th chain:
	$$
	\mathtt{out}(U,i,a) = |\{(u', u)\ |\ u' \in U, u \in Q_i, u \in \delta (u', a) \}|.
	$$
\end{definition}

In the following lemma  we show how to compute the convex sets corresponding to $S_i(\alpha'a)$ and $L_i(\alpha'a)$, for every $i=1, \dots, p$, using the above definitions.   

\begin{lemma}\label{lem:Si alpha2}
Let $ \mathcal{N} = (Q, s, \delta, F) $ be an NFA,  $ \le $ be a co-lex order on $ \mathcal{N} $,    $ \{Q_i\; | \;  1\leq i \leq p\} $ be a $ \le $-chain partition of $ Q $, $\alpha'\in \Sigma^*$, $ a \in \Sigma $, and $ \alpha = \alpha' a $. For every $ j = 1, \dots, p $, let $ S_j(\alpha') = Q_j[1, l'_j] $ and $ L_j(\alpha') = Q_j[r'_j, |Q_j|] $. Fix $ 1 \le i \le p $, and let $ S_i(\alpha) = Q_i[1, l_i] $ and $ L_i(\alpha) = Q_i[r_i, |Q_i|] $.
\begin{enumerate}
    \item Let $ x = \mathtt{out}(S(\alpha), i, a) = \sum_{j=1}^p \mathtt{out}(Q_j[1,l'_j], i, a)$. Then, $ l_i $ is the largest integer $ 0 \le k \le |Q_i| $ such that (i) $\mathtt{in}(Q_i[1,k],a) \leq x $, and (ii) if $ k \ge 1 $, then $\max(\lambda(Q_i[k])) \preceq a$.
    \item Let $ y = \mathtt{out}(L(\alpha), i, a) = \sum_{j=1}^p \mathtt{out}(Q_j[r'_j,|Q_j|], i, a)$. Then, $ r_i $ is the smallest integer $ 1 \le k \le |Q_i| + 1 $ such that (i) $\mathtt{in}(Q_i[k,|Q_i|],a) \leq y $, and (ii) if $ k \le |Q_i | $, then $ a \preceq \min(\lambda(Q_i[k])) $.
\end{enumerate}
\end{lemma}

\begin{proof}
Again, we just prove the first statement since the proof of the second one is analogous.

Let $ z_i $ be the largest integer $ 0 \le k \le |Q_i| $ such that (i) $\mathtt{in}(Q_i[1,k],a) \leq x $, and (ii) if $ k \ge 1 $, then $\max(\lambda(Q_i[k])) \preceq a$. We want to prove that $ l_i = z_i $.

$ (\le) $ The conclusion is immediate if $ l_i = 0 $, so we can assume $ l_i \ge 1 $. It will suffice to prove that $\mathtt{in}(Q_i[1,l_i],a) \leq x $ and $\max(\lambda(Q_i[l_i])) \preceq a $. This follows from Lemma \ref{lem:Si alpha} and the definition of $ x $.

$ (\ge) $ The conclusion is immediate if $ l_i = |Q_i| $, so we can assume $ l_i < |Q_i| $. We only have to prove that if $ l_i + 1 \le k \le |Q_i| $, then either $\mathtt{in}(Q_i[1,k],a) > x $ or $\max(\lambda(Q_i[k])) \succ a$. By Axiom 1, it will suffice to prove that we have $\mathtt{in}(Q_i[1,l_i + 1],a) > x $ or $\max(\lambda(Q_i[l_i + 1])) \succ a$. Assume that $\max(\lambda(Q_i[l_i + 1])) \preceq a $. Since $ Q_i[l_i + 1] \not \in S(\alpha) $, by Lemma \ref{lem:Si alpha} there exists $ v' \in Q $ such that $ Q_i[l_i + 1] \in \delta (v', a) $ and $ v' \not \in S(\alpha') $. We will conclude that $\mathtt{in}(Q_i[1,l_i + 1],a) > x $ if we show that for every $ j = 1, \dots, p $, if $ u' \in Q_j $ and $ u \in Q_i $ are such that $ u' \in Q_j[1, l'_j] $ (and so $ u' \in S(\alpha') $) and $ u \in \delta (u', a) $, then it must be $ u \in Q_i[1, l_i + 1] $. This follows from Lemma \ref{lem:wheresgoes}. \qed
\end{proof}

We now use Lemma \ref{lem:Si alpha2} to retrieve the language of the automaton starting from the aBWT.

\begin{lemma}\label{lem:retrieve}
Let $ \mathcal N =  (Q, s, \delta, F) $ be an NFA, $ \le $ be a co-lex  order on $ \mathcal N $, and $\{Q_i \; |\; 1\leq i \leq p\} $ be a $ \le $-chain partition of $ Q $, with $s\in Q_1$. Let $v_1, \dots, v_n $ be the ordering of $ Q $ defined in Definition \ref{def:aBWT}. Assume that we do not know $ \mathcal{N} $, but we only know $\mathtt{aBWT}(\mathcal N, \le, \{Q_i \; |\; 1\leq i \leq p\} ) $. Then, for every $ \alpha \in \Sigma^* $ we can retrieve the set $ \{ i \in \{1, \dots, n\}\; | \; \alpha \in I_{v_i} \} $, which yields $ \delta (s, \alpha) $. 
\end{lemma}

\begin{proof}
First, let us prove that for every $ k = 1, \dots, n $, we can retrieve the labels --- with multiplicities --- of all edges entering $ v_k $. By scanning $\mathtt{CHAIN} $ we can retrieve the integers $ k_1 $ and $ k_2 $ such that the states in chain $Q_i $ are $ v_{k_1}, v_{k_1 + 1}, \dots, v_{k_2 - 1}, v_{k_2} $. By scanning $ \mathtt{OUT} $, which stores the label and the destinational chain of each edge, we can retrieve how many edges enter chain $Q_i $, and we can retrieve the labels - with multiplicities - of all edges entering chain $Q_i $. By considering the substring of $ \mathtt{IN\_DEG} $ between  the $ (k_1 - 1) $-th one and the $ k_2 $-th one we can retrieve the in-degrees of all states in chain $Q_i $. Since we know the labels - with multiplicities - of all edges entering chain $Q_i $ and the in-degrees of all states in chain $Q_i $, by Axiom 1 we can retrieve the labels - with multiplicities - of all edges entering each node in chain $Q_i $: order the multiset of incoming edge labels, scan the nodes in $ Q_i $ in order, and assign the labels to each node in $ Q_i $ in agreement with their in-degrees.

Let us prove that for every $ i = 1, \dots, p $ we can retrieve the integers $ l_i $ and $ r_i $ such that $ S_i(\alpha) = Q_i[1, l_i] $ and $L_i(\alpha) = Q_i[r_i,|Q_i|]$. We proceed by induction on $ |\alpha| $. If $ |\alpha| = 0 $, then $ \alpha = \varepsilon $, so for every $ i = 1, \dots, p $ we have $ l_i = 0 $, for every $ i = 2, \dots, p $ we have $ r_i = 1 $, and $ r_1 = 2 $. Now, assume $ |\alpha | > 0 $. We can write $ \alpha = \alpha' a $, with $ \alpha' \in \Sigma^* $ and $ a \in \Sigma $. By the inductive hypothesis, for $ j = 1, \dots, p $ we know the integers $ l'_j $ and $ r'_j $ such that $ S_j(\alpha') = Q_j[1, l'_j] $ and $L_j(\alpha') = Q_j[r'_j,|Q_j|]$. Notice that by using $ \mathtt{OUT\_DEG} $ and $ \mathtt{OUT} $ we can compute $ \mathtt{out}(Q_j[1,l'_j], i, a) $  for every $ j = 1, \dots, p $ (see Definition \ref{def: inout}). Since we know the labels - with multiplicities - of all edges entering each state, we can also compute $\mathtt{in}(Q_i[c, d],a) $   and $ \lambda(Q_i[k]))$ for every $ i = 1, \dots, p $,  $ 1 \le c \le d \le |Q_i| $, and $ 1 \le k \le |Q_i| $. By Lemma \ref{lem:Si alpha2} we conclude that we can compute $ l_i $ and $ r_i $ for every $ i = 1, \dots, p $, and we are done.
    
Now, let us prove that for every $ \alpha \in \Sigma^* $ we can retrieve the set $ \{ i \in \{1, \dots, n\} \; | \; \alpha \in I_{v_i} \} $. We proceed by induction on $ |\alpha| $. If $ |\alpha| = 0 $, then $ \alpha = \varepsilon $ and $ \{  i \in \{1, \dots, n\} \; | \; \varepsilon \in I_{v_i} \} = \{1 \} $. Now, assume $ |\alpha | > 0 $. We can write $ \alpha = \alpha' a $, with $ \alpha' \in \Sigma^* $ and $ a \in \Sigma $. By the inductive hypothesis, we know $ \{  i  \in \{1, \dots, n\} \; | \; \alpha' \in I_{v_i} \} $. For every $ i = 1, \dots, p $ we decide whether $ I^i_\alpha \not = \emptyset $ by using $ \{  i \in \{1, \dots, n\} \; | \; \alpha' \in I_{v_i} \} $, $ \mathtt{OUT\_DEG} $ and $ \mathtt{OUT} $. If $ I_\alpha^i \not = \emptyset $, then by Lemma \ref{lem:smallerlarger} we know that $ I_\alpha^i = Q_i [l_i + 1, r_i - 1] $, and we know how to determine $ l_i $ and $ r_i $. Hence, we can easily compute $ \{ i\in \{1, \dots, n\} \; | \; \alpha \in I_{v_i} \} $. \qed
\end{proof}

\begin{corollary}\label{cor:aBWT language} If $\mathtt{aBWT}(\mathcal N, \le, \{Q_i \; |\; 1\leq i \leq p\} ) = \mathtt{aBWT}(\mathcal N', \le', \{Q'_i \; |\; 1\leq i \leq p'\} ) $, then:
\begin{enumerate}
    \item 
  $ p = p' $;
  \item  for every $ 1 \le i \le p $ we have $ |Q_i| = |Q'_i | $;
   \item  $\mathcal L(\mathcal N)=\mathcal L(\mathcal N')$.
\end{enumerate}
\end{corollary}

\begin{proof}
Since $ \mathcal{N} $ and $ \mathcal{N}' $ share the sequence $ \mathtt{CHAIN} $, it must be $ p = p' $ and $ |Q_i| = |Q'_i | $ for every $ i $ . Fix a string $ \alpha \in \Sigma^* $. Then, by Lemma \ref{lem:retrieve} we conclude that the set $ \{ i \in \{1, \dots, n\}\; | \; \alpha \in I_{v_i} \} $ is the same for both $ \mathcal{N} $ and $ \mathcal{N}'$. Since $ \mathcal{N} $ and $ \mathcal{N}' $ share also the sequence $ \mathtt{FINAL} $, we conclude that $ \alpha $ is accepted by $ \mathcal{N} $ if and only if it is accepted by $ \mathcal{N}'$. \qed
\end{proof}

Corollary \ref{cor:aBWT language} ensures that $\mathtt{aBWT}(\mathcal N, \le, \{Q_i \; |\; 1\leq i \leq p\} ) $ 
is enough to reconstruct the language $\mathcal L(\mathcal N)$ of an NFA. 
Similarly to the string case, however (where the BWT is augmented with light data structures in order to achieve efficient indexing with the \emph{FM-index}  \cite{ferragina2000opportunistic}), we will need additional data structures built on top of the aBWT in order to solve efficiently string matching queries. 
In Section \ref{subsubsec:index} we will show how to extend the FM-index to automata by augmenting the aBWT with light data structures. 
 
While Corollary \ref{cor:aBWT language} establishes that the aBWT preserves the automaton's language, it does not state anything about whether it preserves the automaton's topology. In fact we now show that this is not, in general, the case.

\begin{definition}
Let $ \mathcal N =  (Q, s, \delta, F) $ be an NFA. We say that $ \mathcal{N} $ is \emph{distinguished by its paths} if for every $ v \in Q $ there exists $ \alpha \in \Sigma^* $ such that $ I_\alpha = \{v \} $. 
\end{definition}
 
\begin{remark}
If an  NFA is not  distinguished by its paths, then in general  we cannot retrieve its topology from its aBWT,  because there exist two non-isomorphic NFAs having the same aBWT: see Figure \ref{fig:notNFAs} for an example.

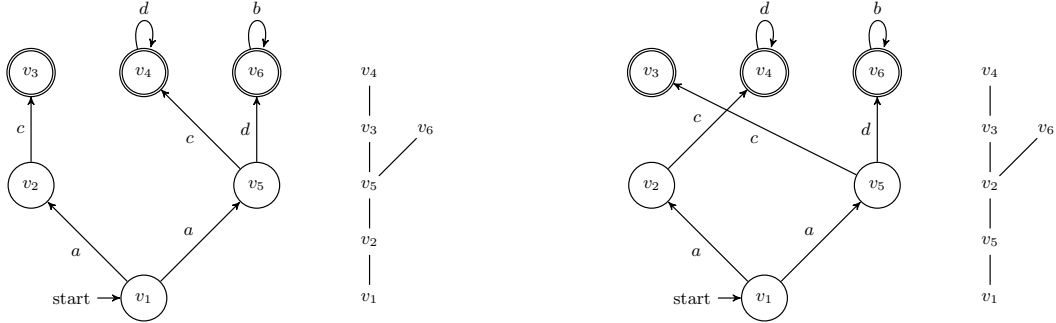
\begin{figure}[h!] 
 \begin{center}
 \scalebox{0.75}{ 
\begin{tikzpicture}[->,>=stealth', semithick, auto, scale=1]

\begin{scope}[shift={(-7,0)}]
\node[state, initial] (0)    at (0,0)	{$ v_1$};
\node[state] (1)    at (-2,2)	{$v_2$};
\node[state,accepting] (2)    at (-2,4)	{$v_3$};
\node[state,accepting] (3)    at (0,4)	{$v_4$};
\node[state] (4)    at (2,2)	{$v_5$};
\node[state,accepting] (5)    at (2,4)	{$v_6$};

\draw (0) edge [] node [] {$a $} (1); 
\draw (0) edge [] node [] {$a $} (4);
\draw (4) edge [] node [] {$d $} (5);
 \draw (5) edge  [loop above] node {$b$} (5);
\draw (1) edge [] node [] {$c $} (2);
\draw (4) edge [] node [] {$c $} (3); 
\draw (3) edge  [loop above] node {$d$} (3);

\begin{scope}[shift={(-4,0)}]
\node[state,color=white,text=black,inner sep=1pt,minimum size=0pt] (10)    at (8,0)	{$v_1$};
	\node[state,color=white,text=black,inner sep=1pt,minimum size=0pt] (11)    at (8,1)               {$v_2$};
		\node[state,color=white,text=black,inner sep=1pt,minimum size=0pt] (14)    at (8,2)               {$v_5$};
		\node[state,color=white,text=black,inner sep=1pt,minimum size=0pt] (12)    at (8,3)               {$v_3$};
			\node[state,color=white,text=black,inner sep=1pt,minimum size=0pt] (13)    at (8,4)               {$v_4$};
			\node[state,color=white,text=black,inner sep=1pt,minimum size=0pt] (15)    at (9,3)               {$v_6$};		
 
 	\path[-]
	(10) edge node {}    (11)
	(11) edge node {}    (14)
	(14) edge node {}    (12)
	(14) edge node {}    (15)
	(12) edge node {}    (13);
\end{scope}	
\end{scope}

\begin{scope}[shift={(4,6)}]
 \node[state, initial] (0')    at (0,-6)	{$ v_1$};	
	\node[state] (1')    at (-2,-4)	{$v_2$};
\node[state,accepting] (2')    at (-2,-2)	{$v_3$};
\node[state,accepting] (3')    at (0,-2)	{$v_4$};
\node[state] (4')    at (2,-4)	{$v_5$};
\node[state,accepting] (5')    at (2,-2)	{$v_6$};

\draw (0') edge [] node [] {$a $} (1'); 
\draw (0') edge [] node [] {$a $} (4');
\draw (4') edge [] node [] {$d $} (5');
 \draw (5') edge  [loop above] node {$b$} (5');
\draw (1') edge [] node [] {$c $} (3');
\draw (4') edge [] node [] {$c $} (2'); 
\draw (3') edge  [loop above] node {$d$} (3');

\begin{scope}[shift={(-4,0)}]
\node[state,color=white,text=black,inner sep=1pt,minimum size=0pt] (10')    at (8,-6)	{$v_1$};
	\node[state,color=white,text=black,inner sep=1pt,minimum size=0pt] (14')    at (8,-5)               {$v_5$};
		\node[state,color=white,text=black,inner sep=1pt,minimum size=0pt] (11')    at (8,-4)               {$v_2$};
		\node[state,color=white,text=black,inner sep=1pt,minimum size=0pt] (12')    at (8,-3)               {$v_3$};
			\node[state,color=white,text=black,inner sep=1pt,minimum size=0pt] (13')    at (8,-2)               {$v_4$};
			\node[state,color=white,text=black,inner sep=1pt,minimum size=0pt] (15')    at (9,-3)               {$v_6$};		
 
 	\path[-]
	(10') edge node {}    (14')
 	(14') edge node {}    (11')
 	(11') edge node {}    (12')
 	(11') edge node {}    (15')
 	(12') edge node {}    (13');
  \end{scope}
\end{scope}	
\end{tikzpicture}
}
 
 \end{center}
 \caption{Consider the two non-isomorphic NFAs $ \mathcal{N}_1 $ and $ \mathcal{N}_2 $ in the figure, both with set of states $ Q = \{v_1, v_2, v_3, v_4, v_5, v_6 \} $. Let $ \le_1 $ and $ \le_2 $ be the maximal co-lex orders given by the Hasse diagrams shown in the figure, and notice that in both cases if we consider $ Q_1 = \{v_1,v_2,v_3,v_4\} $ and $ Q_2=\{v_5,v_6\} $ we obtain a minimum-size chain partition $\mathcal Q = \{Q_1,Q_2\}$. It is easy to check that $ \mathtt{aBWT}(\mathcal N_1, \le_1, \mathcal Q) = \mathtt{aBWT}(\mathcal N_2, \le_2, \mathcal Q) $ because in both cases we have $\mathtt{CHAIN} = 100010 $, $\mathtt{FINAL} = 001101 $, $ \mathtt{OUT\_DEG} = 0010110100101 $, $ \mathtt{OUT} = (1,a) (2,a) (1,c) (1,d) (1,c) (2,d) (2,b)$, and $ \mathtt{IN\_DEG} = 1010100101001 $. Consistently with Theorem \ref{th:store DFA}, we have that $ \mathcal{N}_1 $ and $ \mathcal{N}_2 $ are not distinguished by their   paths.}
 \label{fig:notNFAs}
\end{figure}
\end{remark}

Let us prove that   $\mathtt{aBWT}(\mathcal N, \le, \{Q_i \; |\; 1\leq i \leq p\} ) $ is a one-to-one encoding for the  class of automata which are distinguished by   paths.

\begin{theorem}\label{th:store DFA}
Let $ \mathcal N =  (Q, s, \delta, F) $ be an NFA, $ \le $ be a co-lex  order on $ \mathcal N $, and $ \{Q_i \; |\; 1\leq i \leq p\} $ be a $ \le $-chain partition of $ Q $, with $s\in Q_1$. Assume that we do not know $ \mathcal{N} $, but we only know $\mathtt{aBWT}(\mathcal N, \le, \{Q_i \; |\; 1\leq i \leq p\}) $. Then, we can decide whether $ \mathcal{N} $ is distinguished by its paths and, if so, we can retrieve $ \mathcal{N} $.
\end{theorem}
 
\begin{proof}
Let $v_1, \dots, v_n $ be the ordering of $ Q $ in Definition \ref{def:aBWT}. We know that $ v_1 $ is the initial state and for every $ j = 1, \dots, n $ we can decide whether $ v_j $ is final by using $\mathtt{FINAL} $. Now, for every $ \alpha \in \Sigma^* $, let $ C_\alpha = \{ i \in \{1, \dots,  n\} \; | \; \alpha \in I_{v_i} \}  $. Notice that we can compute $ C_\alpha $ for every $ \alpha \in \Sigma^* $ by Lemma \ref{lem:retrieve}. Consider a list that contains pairs of the form $ (\alpha, C_\alpha) $. Initially, the list contains only $ (\varepsilon, \{1\}) $. Remove recursively an element $ (\alpha, C_\alpha) $ and for every $ a \in \Sigma $ add $ (\alpha a, C_{\alpha a}) $ to the list if and only if $ C_{\alpha a} $ is nonempty and it is not the second element of a pair which is or has already been in the list. This implies that after at most $ |\Sigma | \cdot 2^{n} $ steps the list is empty, and any non-empty  $C_\alpha$  has been the second element of some pair in the list. Then, we conclude that $ \mathcal{N} $ is distinguished by its paths if and only for every $ k = 1, \dots, n $ the set $ \{v_k \} $ has been the second element of some pair in the list. In particular, if $ \mathcal{N} $ is distinguished by its paths, then for every $ k = 1, \dots, n $ we know a string $ \alpha' \in \Sigma^* $ such that $ C_{\alpha'} = \{k\} $. We are only left with showing that we can use $ \alpha' $ to retrieve all edges leaving $ v_k $. Fix a character $ a \in \Sigma $. Then, compute $ C_{\alpha' a} $ using again Lemma \ref{lem:retrieve}. Then, $ v_k $ has $ |C_{\alpha' a}| $ outgoing edges labeled $ a $, whose indexes are given by $ C_{\alpha' a} $. \qed
\end{proof}
Since any DFA is distinguished by its paths, we obtain the following corollary: 

\begin{corollary}\label{cor:one-to-one DFA}
The aBWT is a one-to-one encoding over DFAs. 
\end{corollary}

By counting the number of bits required by the aBWT, we can determine the size of our encoding for NFAs being distinguished by their paths:

\begin{corollary}\label{cor:encoding DFA}
Let $\mathcal N$ be an NFA being distinguished by its paths (for example, a DFA), and let $p=\text{width}(\mt N)$. Then, we can store $\mathcal N$ using $\log(p\sigma) + O(1)$ bits per transition. If $\mathcal N$ is a DFA, this space can also be expressed as (at most) $\sigma\log(p\sigma) + O(\sigma)$ bits per state. If $\mathcal N$ is an NFA, this space can also be expressed as (at most) $2p\sigma \log(p\sigma) + O(p\sigma)$ bits per state.
\end{corollary}
\begin{proof}
The bound of $\log(p\sigma) + O(1)$ bits per transition follows directly from Definition \ref{def:aBWT} and Theorem \ref{th:store DFA}. Letting $|\delta|$ denote the number of transitions and $n$ denote the number of states, on DFAs the naive bound $|\delta| \leq n\sigma$ holds; this allows us to derive the bound of $\sigma\log(p\sigma) + O(\sigma)$ bits per state on DFAs. On arbitrary NFAs, we can use the bound $|\delta| \leq 2p\sigma n$ implied by Lemma \ref{lem: p sparsity}, yielding the bound of $2p\sigma \log(p\sigma) + O(p\sigma)$ bits per state on NFAs.
\end{proof}

We stress   that, while not being an encoding of the NFA, the aBWT still allows to reconstruct the language of the automaton and --- as we will show in the next subsection --- to solve subpath queries (Problem \ref{problem:indexing}) by returning the convex set of all states reached by a path labeled with a given input query string. 
In Section \ref{sec:aaBWT} we will augment the aBWT and obtain an injective encoding of arbitrary NFAs.

\subsection{An Index for NFAs and Languages}\label{subsubsec:index}

We now show how to  support subpath queries by augmenting the aBWT with light data structures and turning it into an index. In fact, our structure is a generalization of the FM-index \cite{ferragina2000opportunistic} to arbitrary automata. This contribution will solve Problem \ref{problem:indexing}.
Our index can be built in polynomial time for DFAs and exponential time for NFAs. In our companion paper \cite{parttwo} we will present an index for NFAs that can be built in polynomial time by circumventing the NP-hardness of computing a co-lex order of minimum width: the solution will be to switch to co-lex \emph{relations} (see also \cite{cotumaccio2022}).

Solving subpath queries on an NFA requires finding the subset $T(\alpha)$ of its states reached by some path labeled by the query string $\alpha$. In turn, note that there is a path labeled $ \alpha $ ending in state $ u $ if and only if $ I_u $ contains a string suffixed by $ \alpha $. This motivates the following definition.

\begin{definition}\label{def:R_i}
Let $ \mathcal{N} = (Q, s, \delta, F) $ be an NFA,  $ \le $ be a co-lex order on $ \mathcal{N} $,    $ \{Q_i\; | \;  1\leq i \leq p\} $ be a $ \le $-chain partition of $ Q $, and $\alpha \in \Sigma^*$. Define:
	\begin{align*}
	    T(\alpha) & = \{  u \in Q\ |\ (\exists \beta\in I_{u})( \alpha \dashv \beta)\}, \\
        R(\alpha) & = S(\alpha) \cup T(\alpha)=\{  u \in Q\ |\ (\forall \beta\in I_{u})(\beta \prec \alpha) \vee (\exists \beta\in I_{u})( \alpha \dashv \beta)\}.
	\end{align*}
Moreover, for every $ i = 1, \dots, p $ define $ T_i (\alpha) = T(\alpha) \cap Q_i $ and $ R_i (\alpha) = R (\alpha) \cap Q_i $.
\end{definition}

Intuitively, $T(\alpha)$ contains all states reached by a path labeled with $\alpha$, while $R(\alpha)$ contains all the states that are either reached by a string suffixed by $ \alpha $, or only reached by strings co-lexicographically smaller than $\alpha$.
Note that the goal of an index solving subpath queries (Problem \ref{problem:indexing}) is to compute the (cardinality of the) set $T(\alpha)$. The aim of the next lemma is to show that, once a co-lex order is fixed, $T(\alpha)$ always forms a range (a convex set). 
Indeed, we now prove a counterpart of Lemma \ref{lem:smallerlarger}, where for any $\alpha \in \Sigma^*$ we showed that $ S_i(\alpha) = Q_i[1, l_i] $, for some  $ 0 \le l_i \le |Q_i| $.

\begin{lemma}\label{lem:st}
Let $ \mathcal{N} = (Q, s, \delta, F) $ be an NFA,  $ \le $ be a co-lex order on $ \mathcal{N} $,  and  $ \{Q_i\; | \;  1\leq i \leq p\} $ be a $ \le $-chain partition of $ Q $. Let $\alpha \in \Sigma^*$. Then:
\begin{enumerate}
    \item $ S(\alpha) \cap T(\alpha) = \emptyset $.
    \item $ T(\alpha) $ is $ \le $-convex.
    \item If $ u, v \in Q $ are such that $ u \le v $ and $ v \in R(\alpha) $, then $ u \in R(\alpha) $. In particular, for every $ i = 1, \dots, p $ there exists $ 0 \le t_i \le |Q_i| $ such that $ T_i (\alpha) = Q_i[|S_i(\alpha)| + 1, t_i] $ (namely, $ t_i = |R_i (\alpha)| $).
\end{enumerate}
\end{lemma}

\begin{proof}
\begin{enumerate}
    \item If $ u \in T(\alpha) $, then there exists $ \beta \in I_u $ such that $ \alpha \dashv \beta $. In particular, $ \alpha \preceq \beta $, so $ u~\not \in~S(\alpha)=\{  v \in Q\ |\ (\forall \beta\in I_{v})(\beta \prec \alpha) \} $.
    \item It follows from Lemma \ref{lem:pathcoherence} by picking $ U = Q $.
    \item If $ v \in S(\alpha) $, then $ u \in S(\alpha) $ by Lemma \ref{lem:smallerlarger} and so $ u \in R(\alpha) $. Now, assume that $ v \in T(\alpha) $. If $u \in T(\alpha)$ we are done. If $ u \not \in T(\alpha) $ (and therefore $ u \not = v $), we want to prove that $ u \in S(\alpha) $, which implies $ u \in R(\alpha) $. Fix $ \beta \in I_u $; we must prove that $ \beta \prec \alpha $. Since $ v \in T(\alpha) $, then there exists $ \gamma \in \Sigma^* $ such that $ \gamma \alpha \in I_v $. Moreover, $ \gamma \alpha \not \in I_u $ because $ u \not \in T(\alpha) $. Since $ u < v $, by Lemma \ref{lem:monot} we conclude $ \beta \prec \gamma \alpha $. Since $ u \not \in T(\alpha) $ implies $\alpha \not \dashv \beta $, from  $ \beta \prec \gamma \alpha $ we conclude $\beta \prec \alpha$. \qed
\end{enumerate}
\end{proof}

We now show how to recursively compute the range on each chain $Q_i$ corresponding to $ R_i(\alpha) $.
Note that, by Lemma \ref{lem:Si alpha2}, we can assume to be able to recursively compute the range on each chain $Q_i$ corresponding to $ S_i(\alpha) $. A computational variant of Lemma \ref{lem:st} will allow us to compute $ T_i(\alpha) $ on each chain $1\leq i \leq p$. Each recursive step of this procedure --- dubbed here \emph{forward search} --- will stand at the core of our index.

\begin{lemma}[Forward search]\label{lem:forward search}
Let $ \mathcal{N} = (Q, s, \delta, F) $ be an NFA,  $ \le $ be a co-lex order on $ \mathcal{N} $,  and  $ \{Q_i\; | \;  1\leq i \leq p\} $ be a $ \le $-chain partition of $ Q $. Let $\alpha'\in \Sigma^*$, $ a \in \Sigma $ and $ \alpha = \alpha' a $. For every $ 1 \le i,j \le p $, let:
\begin{itemize}
    \item $ S_j(\alpha') = Q_j[1, l'_j] $;
    \item $ R_j(\alpha') = Q_j[1, t'_j] $;
    \item $ S_i(\alpha) = Q_i[1, l_i] $;
    \item $ R_i(\alpha) = Q_i[1, t_i] $.
\end{itemize}
Fix $ 1 \le i \le p $, and define $ c = \sum_{j=1}^p \mathtt{out}(Q_j[1,l'_j], i, a)$ and $ d = \sum_{j=1}^p \mathtt{out}(Q_j[1,t'_j], i, a) $. Then $ d \ge c $, and:
\begin{enumerate}
    \item If $ d = c $, then $ T_i(\alpha) = \emptyset $ and so $ t_i = l_i $.
    \item If $ d > c $, then $ T_i(\alpha) \not = \emptyset $ and $t_i$, with $ 1 \le t_i \le |Q_i| $, is the smallest integer such that $\mathtt{in}(Q_i[1, t_i], a) \ge d $.
\end{enumerate}
In particular, $ l_i $ can be computed by means of Lemma \ref{lem:Si alpha2}, and:

$$
T_i(\alpha) = Q_i[l_i+1,t_i].
$$
\end{lemma}

\begin{proof}
Since $ S(\alpha) \subseteq R(\alpha) $,  we have $ d \ge c $. Now, notice that $ T_i (\alpha) \not = \emptyset $ if and only there exists an edge labeled $ a $ leaving a state in $ T (\alpha') $ and reaching chain $ Q_i $, if and only if  $ d > c $. Hence, in the following we can assume $ T_i (\alpha) \not = \emptyset $. In particular, this implies $ l_i < |Q_i| $ and $ t_i \ge l_i + 1 $. By Lemma \ref{lem:wheresgoes} all   edges labeled $ a $, leaving a state in $ S(\alpha') $, and reaching chain $ Q_i $ must end in $ Q_i[1, l_i + 1] $. At the same time, since $ T_i (\alpha) \not = \emptyset $, the definition of $ t_i $ implies that there exists $ v' \in T(\alpha') $ (and so $ v' \in R(\alpha') $) such that $ Q_i[t_i] \in \delta (v', a) $. Hence, the conclusion will follow if we prove that if $ u', u \in Q $ are such that $ u \in Q_i[1, t_i - 1] $ and $ u \in \delta (u', a) $, then $ u' \in R(\alpha') $.   Since $ u < Q_i[t_i] $, from Axiom 2 we obtain $ u' \le v' $ and since $ v' \in R(\alpha') $, from Lemma \ref{lem:st} we conclude $ u' \in R(\alpha') $. \qed
\end{proof}

The next step is to show how to implement the forward search procedure of Lemma \ref{lem:forward search} using fast and small data structures, thereby obtaining an index. 
Before presenting the main result of this section (the aBWT-index of an automaton, Theorem \ref{thm:aBWT index}), we report a few results on data structures that will be the building blocks of our index. In the following lemma, $H_0(S) $ is the zero-order entropy of $ S\in \Sigma^*$.

\begin{lemma}[Succinct string \cite{BNtalg14}, Thm 5.2 and \cite{navarro2016compact}, Sec. 6.3]\label{lem:strings}
Let  $S \in \Sigma^n$ be a string over an integer
alphabet $\Sigma = [0,\sigma-1]$ of size $\sigma \leq n$. Then, there exists a data structure of $n H_0(S) (1+o(1)) + O(n)$ bits
supporting the following operations in time $O(\log\log \sigma)$:
\begin{itemize}
    \item \emph{Access: } $S[i]$, for any $1\leq i \leq n$.
    \item \emph{Rank: } $S.rank(i,c) = |\{j\in \{1,\dots, i\} \ |\ S[j]=c\}|$, for any $1\leq i \leq n$ and $c\in \Sigma$.
    \item \emph{Select: } $S.select(i,c) $ equals the integer $ j $ such that $S[j] = c$ and $S.rank(j,c) = i$, for any $1\leq i \leq S.rank(n,c)$ and $c\in \Sigma$.
\end{itemize}
Given $S$, the data structure can be built in $O(n\log\log \sigma)$ worst-case time. 
\end{lemma}
In other words, the operation $ S[i] $ simply returns the $ i $-th character  appearing in $ S $, the operation $ S.rank(i,c) $ returns the number of occurrences of character $ c $ among the first $ i $ characters of $ S $, and the operation $ S.select(i,c) $ returns the position of the $ i $-th occurrences of character $ c $ in $ S $ (if it exists). In the following, it will be expedient to assume $S.rank(0,c) = S.select(0, c) = 0 $, for $ c \in \Sigma $, and $S.select(i,c) = n + 1 $, for $ c \in \Sigma $ and $ i > S.rank(n,c) $.

Note that Lemma \ref{lem:strings} requires the cardinality $\sigma$ of the alphabet to be no larger than the length of the string. However, this will turn out to be too restrictive, for two reasons: (1) we would like to be able to handle also automata labeled with larger alphabets and, most importantly, (2) 
in our data structures (see the proof of Theorem \ref{thm:aBWT index}) we will also need to manage \emph{rank} and \emph{select} queries over strings defined not on $ \Sigma $, but $ [1, p] \times \Sigma $ (where $ p \le n $ is an integer specifying the width of the underlying automaton), so even if $ \sigma \le n $ it may still be $ p \cdot \sigma > n $. With the following lemma we cover this more general case. The requirement $|\Sigma| \leq n^{O(1)}$ ensures that characters fit in a constant number of computer memory words and thus they can be manipulated in constant time (as it is customary in the data compression field, we recall that in this paper we assume a computer memory word to be formed by $\Theta(\log n)$ bits --- see Section \ref{sec:basics}). Note that we lose fast access functionality (which however will not be required in our application of this data structure).

\begin{lemma}[Succinct string over large alphabet]\label{lem:strings2}
Let  $S \in \Sigma^n$ be a string over an integer
alphabet $\Sigma = [0,\sigma-1]$ of size $\sigma = |\Sigma| \leq n^{O(1)}$. Then, there exists a data structure of $n H_0(S) (1+o(1)) + O(n)$ bits, where $H_0(S) $ is the zero-order entropy of $ S $, supporting the following operations in time $O(\log\log \sigma)$:
\begin{itemize}
    \item \emph{Rank: } $S.rank(i,c) = |\{j\in \{1, \dots, i\} \ |\ S[j]=c\}|$, for  $1\leq i \leq n$ and   $c\in \Sigma$ that occurs in $ S $. 
    \item \emph{Select: } $S.select(i,c) $ equals the integer $ j $ such that $S[j] = c$ and $S.rank(j,c) = i$, for any $1\leq i \leq S.rank(n,c)$ and for any character $c\in \Sigma$ that occurs in $ S $.
\end{itemize}
Given $S$, the data structure can be built in expected $O(n\log\log \sigma)$ time. 
\end{lemma}

\begin{proof}
If $\sigma \leq n$, then we simply use the structure of Lemma \ref{lem:strings}. Otherwise ($\sigma > n$), 
let $\Sigma' = \{S[i]\ |\ 1\leq i \leq n\}$ be the \emph{effective alphabet} of $S$.
We build a minimal perfect hash function $h:\Sigma \rightarrow [0,|\Sigma'|-1]$ mapping (injectively) $\Sigma'$ to the numbers in the range $[0,|\Sigma'|-1]$ and mapping arbitrarily  $\Sigma \setminus \Sigma'$ to the range $[0,|\Sigma'|-1]$. 
We store $h$ using the structure described in \cite{hagerup2001efficient}. This structure can be built in $O(n)$ expected time, uses $O(n)$ bits of space, and answers queries of the form $h(x)$ in $O(1)$ worst-case time. 
Note that $|\Sigma'|\leq n$, so we can build the structure of Lemma \ref{lem:strings} starting from the string $S'\in [0,|\Sigma'| - 1]^n$ defined as $S'[i] = h(S[i])$. Then, \emph{rank} and \emph{select} operations on $S$ can be answered as $S.rank(i,c) = S'.rank(i,h(c))$ and $S.select(i,c) = S'.select(i,h(c))$, provided that $c\in \Sigma'$.
Notice that the zero-order entropies of $S$ and $S'$ coincide, since the character's frequencies remain the same after applying $h$ to the characters of $S$. We conclude that the overall space of the data structure is at most $nH_0(S)(1+o(1)) + O(n)$ bits. \qed
\end{proof}

Finally, we need a fully-indexable dictionary data structure. Such a data structure encodes a set of integers and supports efficiently  a variant of \emph{rank} and \emph{select} queries as defined below: 

\begin{lemma}[Fully-indexable dictionary \cite{Guy2016}, Theorem 4.1]\label{lem:dictionary}
A set $A = \{x_1, \dots, x_n\} \subseteq [1,u]$ of cardinality $n$ 
can be represented with a data structure of $n \log (u/n) + O(n)$ bits so that the following operations can be implemented in $O(\log\log (u/n))$ time:
\begin{itemize}
    \item \emph{Rank: } $A.rank(x) = |\{y \in A\ |\ y\leq x\}|$, for any $1\leq x \leq u$.
    \item \emph{Select: } $A.select(i) = x$ such that $x \in A$ and $A.rank(x) = i$, for any $1\leq i \leq |A|$.
\end{itemize}
Given $A$ as input, the data structure can be built in $O(n)$ worst-case time. \end{lemma}

\begin{remark}\label{rem:predecessor}
The queries of Lemma  \ref{lem:dictionary} can be used to solve in $O(\log\log (u/n))$ time also:
\begin{itemize}
    \item \emph{Predecessor: } the largest element of $ A $ smaller than or equal to $ x $, if it exists. For any $1\leq x \leq u$, $A.pred(x) = A.select(A.rank(x))$ if  $A.rank(x) > 0$, and $A.pred(x) = \bot$ otherwise.  
    \item \emph{Strict-Successor: } the smallest element of $ A $ strictly greater than   $ x $, if it exists. For any $1\leq x \leq u$, $A.succ(x) = A.select(A.rank(x)+1)$ if  $A.rank(x) < |A|$, and $A.succ(x) = \bot$ otherwise. 
    \item \emph{Membership: } For any $1\leq x \leq u$, $x \in A$ if  and only if $x = A.pred(x)$.
\end{itemize}
\end{remark}

We are ready to present the main result of this section (Theorem \ref{thm:aBWT index}): a linear-space index supporting subpath queries on any automaton (Problem \ref{problem:indexing}) in time proportional to $ p^2 \cdot \log\log (p\sigma) $ per query character ($p$ being the automaton's width).


\begin{theorem}[aBWT-index of a finite-state automaton]\label{thm:aBWT index}
Let $\mathcal N = (Q, s, \delta, F)$ be an NFA on alphabet $\Sigma$ of size $\sigma=|\Sigma| \leq e^{O(1)}$, where $e = |\delta|$ is the number of $\mathcal N$-transitions. 
Assume that we are given a $ \le $-chain partition $ \{Q_i\; | \;  1\leq i \leq p\} $, for some  co-lex order $\le$ of width $ p $.
Then, in expected time $ O(e \log \log \sigma)$,
we can build a data structure using 
$e\log(p\sigma)(1+o(1)) + O(e)$  bits 
that, given a query string $\alpha\in \Sigma^m$, answers the following queries in $O(m\cdot p^2 \cdot \log\log (p\sigma))$ time:
\begin{enumerate}
    \item compute the set $ T(\alpha) $ of all states reached by a path on $ \mathcal{N} $ labeled $\alpha$, represented by means of $p$ ranges on the chains in $ \{Q_i\; | \;  1\leq i \leq p\} $;
    \item compute  the set $ I_\alpha $ of all states reached by a path labeled with $\alpha$ originating in the source, represented by means of $p$ ranges on the chains in $ \{Q_i\; | \;  1\leq i \leq p\} $ and, in particular, decide whether $ \alpha \in \mathcal{L(\mathcal{N})} $.
\end{enumerate}
\end{theorem}

\begin{proof}
Let $n = |Q|$. In this proof we assume that the states of $ Q $ have been sorted like in Definition \ref{def:aBWT}: if $ \pi(v) $, for $v\in Q$, is the unique integer such that  $ v \in Q_{\pi(v)} $, then we consider the ordering $v_1, \dots, v_n $ of $ Q $ such that for every $ 1 \le i < j \le n $ it holds  $ \pi(v_i) < \pi(v_j) \lor (\pi(v_i) = \pi(v_j) \land v_i < v_j) $. 
Moreover, we assume that $ s \in Q_1 $ (again like in Definition \ref{def:aBWT}), so $ s = v_1 $.
For every $ i = 1, \dots, p $, let $e_i = |\{(u,v,a)\ |\ \delta(u,a)=v, u\in Q, v\in Q_i, a\in \Sigma\}|$ be the number of edges entering the $i$-th chain, let $ \Sigma_i = (\bigcup_{u\in Q_i} \lambda(u)) \setminus \{\# \} $ be the set of characters labeling edges entering the $i$-th chain, and let $\sigma_i = |\Sigma_i|$.

We store the following data structures:

\begin{itemize}
    \item One fully-indexable succinct dictionary (Lemma \ref{lem:dictionary}) on each $\Sigma_i$ to map $\Sigma_i \subseteq [0,\sigma-1]$ to  $[0,\sigma_i-1]$. The total number of required bits is $\sum_{i=1}^p (\sigma_i \log(\sigma/\sigma_i) + O(\sigma_i)) \leq \sum_{i=1}^p (e_i \log(\sigma/\sigma_i) + O(e_i)) = e \log\sigma - \sum_{i=1}^p (e_i \log \sigma_i) + O(e)$. As a consequence, we can solve rank, select, predecessor, strinct-successor and membership queries on each dictionary in $ O(\log \log (\sigma / \sigma_i)) \subseteq O(\log \log \sigma ) $ time.
    \item 
    The bitvector $ \texttt{CHAIN} \in \{0, 1\}^{n} $ of Definition \ref{def:aBWT}
    represented by the data structure of Lemma \ref{lem:strings}. 
    The number of required bits is $ n H_0(\texttt{CHAIN}) (1+o(1)) + O(n) = O(n) \subseteq O(e) $. As a consequence, we can solve rank and select queries on $ \texttt{CHAIN} $ in $ O(1) $ time. In particular, in $ O(1) $ time we can compute $ |Q_i |$, for $ i = 1, \dots, p $, because $ |Q_i| = \texttt{CHAIN}.select(i + 1, 1) - \texttt{CHAIN}.select(i, 1) $.  
    \item 
    The bitvector $ \texttt{FINAL} \in \{0, 1\}^{n} $  of Definition \ref{def:aBWT}
    represented by the data structure of Lemma \ref{lem:strings}.
    The number of required bits is again $ O(n) \subseteq O(e) $.
    \item 
    The bitvector $ \mathtt{OUT\_DEG} \in \{0, 1\}^{e + n} $ of Definition \ref{def:aBWT}
    represented by the data structure of Lemma \ref{lem:strings}.
    The number of required bits is $ (n + e) H_0(\mathtt{OUT\_DEG}) (1+o(1)) + O(n + e) \subseteq O(e) $. As a consequence, we can solve rank and select queries on $ \mathtt{OUT\_DEG} $ in $ O(1) $ time.
    \item 
    The string $ \mathtt{OUT} \in ([1, p] \times \Sigma)^e $ of Definition \ref{def:aBWT} 
    represented by the data structure of Lemma \ref{lem:strings2} 
    (the assumption on the size on the alphabet in Lemma \ref{lem:strings2}  is satisfied because $ |[1, p] \times \sigma| = p \cdot \sigma \le n \cdot \sigma \le (e + 1) \cdot \sigma = e^{O(1)} $). The number of required bits is $ e H_0(\mathtt{OUT}) (1+o(1)) + O(e) $. We will bound the quantity $ e H_0(\mathtt{OUT}) $ by exhibiting a prefix-free encoding of $ \mathtt{OUT} $. The key idea is that if $ (i, c) \in [1, p] \times \Sigma $ occurs in $ \mathtt{OUT} $, then it must be $ c \in \Sigma_i $, so we can encode $ (i, c) $ by using $\lceil \log(p+1) \rceil \leq \log p + 1 $ bits encoding  $i $, followed by $\lceil \log(\sigma_i+1) \rceil \leq \log\sigma_i + 1$ bits encoding $ c $ (note that this part depends on $i$). We clearly obtain a prefix code, so we conclude $ e H_0(\mathtt{OUT}) \le \sum_{i = 1}^p e_i (\log p + \log \sigma_i + O(1)) = e \log p + \sum_{i = 1}^p (e_i \log \sigma_i) + O(e) $ bits. 
    Observing that $\sum_{i = 1}^p (e_i \log \sigma_i) \leq e\log\sigma$, we conclude that the number of required bits for $\mathtt{OUT}$ is bounded by $e H_0(\mathtt{OUT}) (1+o(1)) + O(e) = \left(e \log p + \sum_{i = 1}^p (e_i \log \sigma_i) + O(e)\right)(1+o(1)) + O(e) \leq (1+o(1)) e \log p + \sum_{i = 1}^p (e_i \log \sigma_i) + o(e\log\sigma) + O(e) $ bits. Notice that in $O( \log\log (p\sigma)) $ time we can solve rank and select queries on $ \mathtt{OUT} $ (that is, queries $ \mathtt{OUT}.rank (j, (i, c)) $ and $ \mathtt{OUT}.select (j, (i, c)) $) for all $ 1 \le i \le p $ and for all $ c \in \Sigma $. Indeed, given $ i $ and $ c $, we first check whether $ c \in \Sigma_i $ by solving a membership query on the  dictionary for $\Sigma_i$ in $ O(\log \log \sigma) $ time. If $ c \not \in \Sigma_i $, then we immediately conclude that $ \mathtt{OUT}.rank (j, (i, c)) = 0 $ and $ \mathtt{OUT}.select (j, (i, c)) $ is undefined. If $ c \in \Sigma_i $, then $ (i, c)$ appears in $\mathtt{OUT} $, so the conclusion follows from Lemma \ref{lem:strings2}.    
    \item The bitvector 
    $ \mathtt{IN\_DEG} \in \{0, 1\}^{e + n} $ of Definition \ref{def:aBWT} represented by the data structure of Lemma \ref{lem:strings}.
    The number of required bits is again $ O(e) $. As a consequence, we can solve rank and select queries on $ \mathtt{IN\_DEG} $ in $ O(1) $ time.
    \item A bitvector $ \mathtt{IN}'$, represented by the data structure of Lemma \ref{lem:strings}, built as follows. We sort all edges $(v_j,v_i,c)$ by end state $v_i$ and, if the end state is the same, by label $c$.
    Then, we build a string $ \mathtt{IN} \in \Sigma^e $ by concatenating all labels of the sorted edges. Finally, $\mathtt{IN}' \in \{0, 1 \}^e $ is the bitvector such that $\mathtt{IN}' [k] = 1 $ if and only if $ k = 1 $ or $\mathtt{IN}[k] \neq \mathtt{IN}[k-1]$ or the $ k $-th edge and the $ (k - 1) $-th edge reach distinct chains. 
    The number of required bits is $ O(e) $.    
\end{itemize}

For an example, consider the automaton of Figure \ref{fig:abwtdfa}. All sequences except for bitvector $\mathtt{IN'}$ are reported in Example \ref{ex:aBWT}. To build bitvector $\mathtt{IN'}$, we first build the string $\mathtt{IN}$ of all incoming labels of the sorted edges: $\mathtt{IN} = aaabcabbbb$. Then, bitvector $\mathtt{IN'}$ marks with a bit '1' (i) the first character of each maximal unary substring in $\mathtt{IN}$, and (ii) the characters of $\mathtt{IN}$ labeling the first edge in each chain: $\mathtt{IN'} = 1001111000$.

By adding up the space of all components (note that the terms $- \sum_{i=1}^p (e_i \log \sigma_i)$ in the dictionaries $\Sigma_i$ and $\sum_{i=1}^p (e_i \log \sigma_i)$ in sequence $\mathtt{OUT}$  cancel out), we conclude that our data structures take at most $e\log(p\sigma)(1+o(1)) + O(e)$ bits.

We proceed by showing how to solve queries 1 (string matching) and 2 (membership).

(1) Let us prove that, given a query string $\alpha\in \Sigma^m$, we can use our data structure to compute, in time $O(m\cdot p^2 \cdot \log\log (p\sigma))$, the set $ T(\alpha) $ of all states reached by a $\alpha$-path on $ \mathcal{N} $, presented by  $p$ convex sets on the chains    $   \{Q_i\; | \;  1\leq i \leq p\} $. By Lemma \ref{lem:forward search}, it will suffice to show how to compute $ R(\alpha) $ and $ S(\alpha) $. We can recursively compute each $ R(\alpha) $ and $ S(\alpha) $ in time proportional to $ m $ by computing $ R(\alpha') $ and $ S(\alpha') $ for all prefixes $ \alpha' $ of $ \alpha $. Hence, we only have to show that we can update $ R (\alpha') $ and $ S(\alpha') $ with a new character, in $O(p^2 \cdot \log\log (p\sigma))$ time. We start with the empty prefix $\varepsilon$, whose corresponding sets are $ R (\varepsilon) = Q $ and $ S (\varepsilon) = \emptyset $. For the update we apply Lemmas \ref{lem:Si alpha2} and \ref{lem:forward search}. 
An inspection of the two lemmas reveals the we can update $ R (\alpha') $ and $ S(\alpha') $ with a new character by means of $O(p^2) $ calls to the following queries.

\begin{itemize}
    \item \texttt{(op1)} for any $1\leq i,j\leq p$, $1\leq k \leq |Q_j|$, and $a\in \Sigma$, compute $\mathtt{out}(Q_j[1,k],i,a)$; 
    \item \texttt{(op2)} for any $1\leq i \leq p$, $a\in\Sigma$, and $h\geq 0$, find the largest integer $0\leq k \leq |Q_i|$ such that 
	$\mathtt{in}(Q_i[1,k],a) \leq h$; 
	\item \texttt{(op3)} for any $1\leq i \leq p$, $a\in\Sigma$, and $z\geq 1$, find the smallest integer $1 \leq t \leq |Q_i|$ such that $\mathtt{in}(Q_i[1,t],a) \geq z$, if it exists, otherwise report that it does not exist.
	\item \texttt{(op4)} for any $1\leq i \leq p$ and $a\in\Sigma$, find the largest integer $0\leq h \leq |Q_i|$ such that, if $ h \ge 1 $, then $\max(\lambda(Q_i[h])) \preceq a$.
\end{itemize}

As a consequence, we are left to show that we can solve each query in $O( \log\log (p\sigma)) $ time.

(\texttt{op1}). The states in $ Q_j[1,k] $ correspond to the convex set of all states (in the total order $v_1, \dots, v_n$ of Definition \ref{def:aBWT}) whose endpoints are $v_l $ and $ v_r $, where $ l = \texttt{CHAIN}.select(j, 1) $ and $ r = l + k - 1 $.  
Considering the order of edges used to define $\mathtt{OUT}$, the set of all edges leaving a state in $ Q_j[1,k] $ forms a convex set in $\mathtt{OUT}$ and in $\mathtt{OUT\_DEG}$. Define:
\begin{itemize}
    \item $ x = \mathtt{OUT\_DEG}.rank (\mathtt{OUT\_DEG}.select(l - 1, 1), 0) $; 
    \item $ y = \mathtt{OUT\_DEG}.rank (\mathtt{OUT\_DEG}.select(r, 1), 0) $.
\end{itemize}  
Notice that $ x $ is equal to the number of edges leaving all states before $ v_l $, while $ y $ is the number of edges leaving all states up to $ v_r $ included. As a consequence, we have $ x \le y $. If $ x = y $, then there are no edges leaving states in $ Q_j[1,k] $ and we can immediately conclude $ \mathtt{out}(Q_j[1,k],i,a) = 0 $. Assuming $ x < y $,   $ x + 1 $ and $ y $ are the endpoints of the convex set of all edges leaving states in $ Q_j[1,k] $. Hence, we are left with counting the number of such edges labeled $ a $ and reaching chain $Q_i $. Notice that $ \mathtt{OUT}.rank(x, (i, a)) $ is the number of all edges labeled $ a $ and reaching chain $ i $ whose start state comes before $ v_l $, whereas $ \mathtt{OUT}.rank(y, (i, a)) $ is the number of all edges labeled $ a $ and reaching $ i $ whose start state comes before or is equal to $ v_r $. We can then conclude that $ \mathtt{out}(Q_j[1,k],i,a) = \mathtt{OUT}.rank(y, (i, a)) - \mathtt{OUT}.rank(x, (i, a)) $.

(\texttt{op2}). First, we check whether $ a \in \Sigma_i $ by a membership query on the dictionary for $\Sigma_i$. If $ a \not \in \Sigma_i $, we immediately conclude that the largest $ k $ with the desired properties is $ k = |Q_i| $. Assume $ a \in \Sigma_i $ and notice that the states in $ Q_i $ correspond to the convex set of all states whose endpoints are $ v_l $ and $ v_r $, where $ l = \texttt{CHAIN}.select(i, 1) $ and $ r = \texttt{CHAIN}.select(i + 1, 1) - 1 $. Considering the order of edges used to define $\mathtt{IN}$, the set of all edges entering a state in $ Q_j[1,k] $ forms a convex set in $\mathtt{IN}$ and in $\mathtt{IN\_DEG}$. Define 
\begin{itemize}
    \item $ x = \mathtt{IN\_DEG}.rank (\mathtt{IN\_DEG}.select(l - 1, 1), 0) $;
    \item $ y = \mathtt{IN\_DEG}.rank (\mathtt{IN\_DEG}.select(r, 1), 0) $.
\end{itemize}   
Notice that $ x $ is equal to the number of edges reaching all states before $ v_l $, while $ y $ is the number of edges reaching all states coming before or equal to $ v_r $. Since $ a \in \Sigma_i$, we have $ x < y $, and $ x + 1 $ and $ y $ are the endpoints of the convex set of all edges reaching a state in $ Q_i $. The next step is to determine the smallest edge labeled $ a $ reaching a state in $ Q_i $. First, notice that the number of characters smaller than or equal to $a$ in $\Sigma_i$ can be retrieved, by Lemma \ref{lem:dictionary}, as $ \Sigma_i.rank (a) $. Notice that $ f = \mathtt{IN'}.rank(x, 1) $ yields the number of 0-runs in $ \mathtt{IN'} $ pertaining to chains before chain $ Q_i $ so that, since we know that $ a \in \Sigma_i $, then $ g = \mathtt{IN'}.select(f + \Sigma_i.rank (a), 1) $ yields the smallest edge labeled $ a $ in the convex set of all edges reaching a state in $ Q_i $. We distinguish two cases for the parameter $h$ of \texttt{op2}:
\begin{itemize}
    \item $ h = 0 $. In this case, the largest $ k $ with the desired properties is equal to the position on chain $Q_i $ of the state reached by the $ g $-th edge minus one. The index of the state reached by the $ g $-th edge is given by $ p = \mathtt{IN\_DEG}.rank (\mathtt{IN\_DEG}.select(g, 0), 1) + 1 $, so the largest $ k $ with the desired properties is $k = p - l $.
    \item $ h > 0 $. The quantity $ h' = \mathtt{IN'}.select(f + \Sigma_i.rank (a) + 1, 1) -g = \mathtt{IN'}.select(f + \Sigma_i.rank (a) + 1, 1) - \mathtt{IN'}.select(f + \Sigma_i.rank (a), 1)  $ yields the number of edges labeled $ a $ in the convex set of all edges reaching a state in $ Q_i $. If $ h' \le h $, then we conclude that the largest $ k $ is $ |Q_i| $. Hence, assume that $ h' > h $. We immediately obtain that the $ (h + 1) $-th smallest edge labeled $a$ reaching a state in $ Q_i $ is the $ (g + h) $-th edge, and the largest $ k $ with the desired properties is equal to the position on chain $Q_i $ of the state reached by the $ (g + h) $-th edge minus one. Analogously to case 1, the index of the state reached by this edge is given by $ p = \mathtt{IN\_DEG}.rank (\mathtt{IN\_DEG}.select(g + h, 0), 1) + 1 $, so the largest $ k $ with the desired properties is $ k = p - l $.
\end{itemize}

(\texttt{op3}). We simply use operation (\texttt{op2}) to compute the largest integer $0\leq k \leq |Q_i| $ such that  $\mathtt{in}(Q_i[1,k],a) \leq z - 1$. If $ k = |Q_i| $, then the desired integer does not exist, otherwise it is equal to $ k + 1 $.

(\texttt{op4}). We first decide whether $ \Sigma_i.succ(a) $ is defined. If it is not defined, then the largest integer with the desired property is $ |Q_i| $. Now assume that $ \Sigma_i.succ(a) $ is defined. Then the largest integer with the desired property is simply the largest integer $ 0 \le k \le |Q_i| - 1 $ such that $\mathtt{in}(Q_i[1,k],\Sigma_i.succ(a)) \leq 0 $, which can be computed using (\texttt{op2}).

(2) Let us prove that, given a query string $\alpha\in \Sigma^m$, we can use our data structure to compute $I_\alpha $ in $O(m\cdot p^2 \cdot \log\log (p\sigma))$ time, represented as $p$ ranges on the chains in $\{Q_i\; | \;  1\leq i \leq p\}$. We claim that it will suffice to run the same algorithm used in the previous point, starting with $ R = \{v_1 \} $ and $ S = \emptyset $. Indeed, consider the automaton $\mathcal N' $ obtained from $\mathcal N $ by adding a new initial state $ v_0 $ and adding exactly one edge from $ v_0 $ to $ v_1 $ (the old initial state) labeled with $ \# $, a character smaller than every character in the alphabet $\Sigma$. Let $ \le' $ the co-lex order on $\mathcal N' $ obtained from $ \le $ by adding the pair $ \{(v_0, v_1) \} $, and consider  the $ \le' $-chain partition obtained from $  \{Q_i\; | \;  1\leq i \leq p\}  $ by adding $ v_0 $ to $ Q_1 $. It is immediate to notice that for every $ k = 1, \dots, n $ and for every string $ \alpha \in \Sigma^* $ we have that $ v_k \in I_\alpha $ on $ \mathcal{N} $ if and only if $ v_k \in T(\#\alpha) $ on $ \mathcal{N'} $. Since $ v_0 $ has no
incoming edges and on $ \mathcal{N}'$ we have $ R(\#) = \{v_0, v_1 \} $ and $ S (\#) = \{v_0 \} $, the conclusion follows. Now, given $ I_\alpha $, we can easily check whether $ \alpha \in \mathcal{L(N)} $. Indeed, for every $ i = 1, \dots, p $ we know the integers $ l_i $ and $ t_i $ such that $ I_\alpha^i = Q_i[l_i + 1, t_i] $, and we decide whether some of these states are final by computing $ f = \texttt{CHAIN}.select (i, 1) $, and then checking whether $ \texttt{FINAL}.rank(f + l_i - 1, 1) - \texttt{FINAL}.rank(f + t_i - 1, 1) $ is larger than zero.
\qed
\end{proof}

Notice that, in order to apply Theorem \ref{thm:aBWT index},   we need a chain partition of a co-lex order over the automaton. In the case of a  DFA,  Corollary \ref{cor:complexity width dfa} allows us to compute in polynomial time (with high probability) a $ \le $-chain partition of  optimal width, so from Theorem \ref{thm:aBWT index} we obtain the following result.

\begin{corollary}\label{cor:indexDFA}
Let $\mathcal D = (Q, s, \delta, F)$ be a DFA. Then, the data structure of Theorem \ref{thm:aBWT index} can be built on $\mathcal D$ with parameter  $ p = \text{width}({\mathcal D}) $ in expected $\tilde O(|\delta|^2)$ time.
\end{corollary}

Note that the data structure of Corollary \ref{cor:indexDFA} implicitly supports the navigation of the labeled graph underlying $\mathcal D$ starting from the initial state: a node is represented as the $p$ ranges (on the $p$ chains) obtained when computing $ I_\alpha $. This enables deciding whether $\alpha \in \mathcal{L(\mathcal{D})}$ in $O(p^2\log\log(p\sigma))$ time per character of $\alpha$. 
Even if testing membership with our structure is inefficient on DFAs for large $p$ (on a classic DFA representation, this operation can be easily implemented in $O(|\alpha|)$ time), for small values of $p$ on NFAs our membership procedure can be much faster than the classical ones (determinization or dynamic programming), provided our index has been built on the NFA.

\subsection{Encoding NFAs}\label{sec:aaBWT}

We now show   a simple extension of the aBWT of Definition \ref{def:aBWT}, yielding an injective encoding of NFAs. 
It turns out that in order to achieve these goals it is sufficient to collect the components of the aBWT and, in addition, the origin chain of every edge (see Figure \ref{tab:aaBWT} for an example): 

\begin{definition}\label{def:aaBWT}
Let $ \mathcal N =  (Q, s, \delta, F) $ be an NFA and let $e = |\delta|$ be the number of $\mathcal N$-transitions.  Let $ \le $ be a co-lex  order on $ \mathcal N $, and let $  \{Q_i \; |\; 1\leq i \leq p\} $ be a $ \le $-chain partition of $ Q $, where w.l.o.g. $s \in Q_1$. 
Let $ \pi(v) $ and $Q=\{v_1, \dots, v_n\}$ 
be the map and the total state order defined in  Definition \ref{def:aBWT}.
Define a new sequence
$ \mathtt{IN\_CHAIN} \in [1,p]^e$, storing the edges' origin chains, as follows.
Sort all edges $(v_j,v_i,c)$ by increasing destination index $i$, breaking ties by label $c$ and then by origin index $j$. 
Then, $ \mathtt{IN\_CHAIN}$ is obtained by concatenating the elements $\pi(v_j)$ for all edges $(v_j,v_i,c)$ sorted in this order.
\end{definition}

\newcolumntype{?}{!{\vrule width 1pt}}

\begin{figure}
\begin{minipage}{6cm}
 \scalebox{0.75}{ 
\begin{tikzpicture}[->,>=stealth', semithick, auto, scale=1]

 \node[state, initial] (0)    at (0,0)	{$ v_1$};
\node[state] (1)    at (-2,2)	{$v_2$};
\node[state,accepting] (2)    at (-2,4)	{$v_3$};
\node[state,accepting] (3)    at (0,4)	{$v_4$};
\node[state] (4)    at (2,2)	{$v_5$};
\node[state,accepting] (5)    at (2,4)	{$v_6$};

\draw (0) edge [] node [] {$a $} (1); 
\draw (0) edge [] node [] {$a $} (4);
\draw (4) edge [] node [] {$d $} (5);
 \draw (5) edge  [loop above] node {$b$} (5);
\draw (1) edge [] node [] {$c $} (2);
\draw (4) edge [] node [] {$c $} (3); 
\draw (3) edge  [loop above] node {$d$} (3);
\end{tikzpicture}
}
\end{minipage}
\begin{minipage}{5cm}
  \centering
   \begin{tabular}{crc|P{25pt}|P{25pt}|P{25pt}|P{25pt}?P{25pt}|P{25pt}|}
    & \multicolumn{1}{c}{} & \multicolumn{1}{c}{{$\tt{CHAIN}$}} & \multicolumn{1}{c}{{1}} & \multicolumn{1}{c}{{0}} & \multicolumn{1}{c}{{0}} & \multicolumn{1}{c}{{0}} & \multicolumn{1}{c}{{1}} & \multicolumn{1}{c}{{0}}\\
    & \multicolumn{1}{c}{} & \multicolumn{1}{c}{{$\tt{FINAL}$}} & \multicolumn{1}{c}{{0}} & \multicolumn{1}{c}{{0}} & \multicolumn{1}{c}{{1}} & \multicolumn{1}{c}{{1}} & \multicolumn{1}{c}{{0}} & \multicolumn{1}{c}{{1}}\\
    & \multicolumn{1}{c}{} & \multicolumn{1}{c}{{$\tt{IN\_DEG}$}} & \multicolumn{1}{c}{{1}} & \multicolumn{1}{c}{{01}} & \multicolumn{1}{c}{{01}} & \multicolumn{1}{c}{{001}} & \multicolumn{1}{c}{{01}} & \multicolumn{1}{c}{{001}}\\
    & \multicolumn{1}{c}{} & \multicolumn{1}{c}{{$\tt{IN\_CHAIN}$}} & \multicolumn{1}{c}{} & \multicolumn{1}{c}{{1}} & \multicolumn{1}{c}{{1}} & \multicolumn{1}{c}{{21}} & \multicolumn{1}{c}{{1}} & \multicolumn{1}{c}{{22}} \\\cline{4-9}\\[-3.72mm]
    $\tt{OUT\_DEG}$ & \multicolumn{1}{c}{{$\tt{OUT}$}} & \multicolumn{1}{c|}{} &  1 &  2 &  3 &  4 & 5 &  6 \\\cline{3-9}\\[-3.72mm]
    001 & \multicolumn{1}{c|}{{(1,a),(2,a)}} &  1 &  &  (1,1,a) &  &  & (1,2,a)  &  \\\cline{3-9}\\[-3.72mm]
    01 & \multicolumn{1}{c|}{{(1,c)}} & 2 &  &  & (1,1,c) &  &  &    \\\cline{3-9}\\[-3.72mm]
    1 & \multicolumn{1}{c|}{} & 3 &  &  &  &  &  &   \\\cline{3-9}\\[-3.72mm]
    01 & \multicolumn{1}{c|}{{(1,d)}} & 4 &  &  &  & (1,1,d) &  & \\\cline{3-9}\\[-7.6mm]\\\cline{3-9}\\[-3.72mm]
    001 & \multicolumn{1}{c|}{{(1,c),(2,d)}} & 5 &  &  &  & (2,1,c) &  & (2,2,d)\\\cline{3-9}\\[-3.72mm]
    01 & \multicolumn{1}{c|}{{(2,b)}} & 6 &  &  &  &  &  & (2,2,b) \\\cline{3-9}\\[-3.72mm]
  \end{tabular}
  \end{minipage}
  \vspace{5pt}
  \caption{Augmented aBWT of an NFA (the one in Figure \ref{fig:notNFAs} on the left), using the chain partition $\{\{v_1,v_2,v_3,v_4\}, \{v_5,v_6\}\}$. In addition to the aBWT of Definition \ref{def:aBWT}, we add a sequence \texttt{IN\_CHAIN} collecting the origin chain of every edge. For each labeled edge $(u,v,a)$, in the adjacency matrix we show the triple $(\pi(u), \pi(v), a)$, that is, the origin chain, destination chain, and label of the edge (the matrix is visually divided in 4 sectors, corresponding to all combinations of origin and destination chains). Vector \texttt{IN\_CHAIN} collects vertically the incoming chains, that is, the first component of each triple in the corresponding column of the adjacency matrix.}\label{tab:aaBWT}
\end{figure}

The following lemma presents a function that will ultimately allow us to show that our augmented aBWT is indeed an injective encoding on NFAs.

\begin{lemma}\label{lem:aaBWT LF}
    Let $ 1 \le i, j \le p $ and $ a \in \Sigma $. Let $ w $ be the number of edges labeled with character $ a $ leaving any state in $ Q_j $ and entering any state in $ Q_i $. Let $ B_{j, i, a} = (f_1, f_2, \dots, f_w) $ be state indices such that (i) $ 1 \le f_1 \le f_2 \le \dots \le f_w \le n $, (ii) if $ 1 \le k \le n $ occurs in $ B_{j, i, a} $, then $ \pi(v_k) = j $ and (iii) if $ 1 \le k \le n $ occurs $ t \ge 1 $ times in $ B_{j, i, a} $, then there exist exactly $ t $ edges labeled with character $ a $ leaving $ v_k $ and entering a state in $ Q_i $. Let $ C_{j, i, a} = (g_1, g_2, \dots, g_w) $ be state indices such that (i) $ 1 \le g_1 \le g_2 \le \dots \le g_w \le n $, (ii) if $ 1 \le h \le n $ occurs in $ C_{j, i, a} $, then $ \pi(v_h) = i $ and (iii) if $ 1 \le h \le n $ occurs $ t \ge 1 $ times in $ C_{j, i, a} $, then there exist exactly $ t $ edges labeled with character $ a $ entering $ v_h $ and leaving a state in $ Q_j $. Then, $ \{(v_{f_\ell}, v_{g_\ell}, a ) \;|\; 1 \le \ell \le w \} $ is the set of all edges labeled with character $ a $, leaving a state in $ Q_j $ and entering a state in $ Q_i $.
\end{lemma}


\begin{proof}
    Consider the set of all edges labeled with characater $ a $, leaving a state in $ Q_j $ and entering a state in $ Q_i $. Then, $ B_{j, i, a} $ is obtained by picking and sorting all start states of these edges, and $ C_{j, i, a} $ is obtained by picking and sorting all end states of these edges. The conclusion follows from Axiom 2. \qed
\end{proof}

Lemma \ref{lem:aaBWT LF} allows to reconstruct the topology of a NFA starting from our augmented aBWT, as we show in the next lemma.


\begin{lemma}\label{lem:faithful NFA encoding}
    The aBWT of Definition \ref{def:aBWT}, in addition to sequence ${\tt IN\_CHAIN}$ of Definition \ref{def:aaBWT}, is a one-to-one encoding over the NFAs.
\end{lemma}

\begin{proof}
    From $\mathtt{CHAIN} $ and $ \mathtt{FINAL} $ we can retrieve the chain of each state, and we can decide which states are final. We only have to show how to retrieve the set $\{(v_f,v_g,a)\ |\ v_g\in \delta(v_f,a),\ v_f,v_g\in Q, a\in \Sigma\}$ of all NFA's transitions. By Lemma \ref{lem:aaBWT LF}, we only have to prove that for every $ 1 \le i, j \le p $ and for every $ a \in \Sigma $ we can retrieve $ B_{j, i, a} $ and $ C_{j, i, a} $. We will use ideas similar to those employed in the proof of Lemma \ref{lem:retrieve}.

    Let us show how to retrieve $ B_{j, i, a} $ for every $ 1 \le i, j \le p $ and for every $ a \in \Sigma $. Fix $ i $, $ j $ and $ a $. From $ \mathtt{OUT\_DEG} $ we can retrieve the number of edges leaving each state in the $ j $-th chain. Then, the definition of $ \mathtt{OUT} $ implies that, for every state in the $ j $-th chain, we can retrieve the label and the destination chain of each edge leaving the state, so we can decide how many times a state in the $ j $-th chain occurs in $ B_{j, i, a} $.

    Let us show how to retrieve $ C_{j, i, a} $ for every $ 1 \le i, j \le p $ and for every $ a \in \Sigma $. Fix $ i $, $ j $ and $ a $. From $ \mathtt{IN\_DEG} $ we can retrieve the number of edges entering each state in the $ i $-th chain. From $ \mathtt{OUT} $ we can retrieve all characters (with multiplicities) labeling some edge entering the $ i $th-chain. As a consequence, Axiom 1 implies that, for every state in the $ i $-th chain, we can retrieve the label of each edge entering the state. Moreover, the definition of $ \mathtt{IN\_CHAIN}$ implies that, for every state in the $ i $-th cain, we can retrieve the the start chain of each edge entering the state, so we can decide how many times a state in the $ i $-th chain occurs in $ C_{j, i, a} $. \qed
\end{proof}

By analyzing the space required by our extension of the aBWT and applying Lemma \ref{lem:faithful NFA encoding}, we obtain:

\begin{corollary}\label{cor:encoding NFA}
Let $\mathcal N$ be an NFA, and let $p=\text{width}(\mt N)$. Then, we can store $\mathcal N$ using $\log(p^2\sigma) + O(1)$ bits per transition. This space can also be expressed as (at most) $2p\sigma \log(p^2\sigma) + O(p\sigma)$ bits per state.
\end{corollary}
\begin{proof}
    The bound of $\log(p^2\sigma) + O(1)$ bits per transition follows easily from the definitions of aBWT (Definition \ref{def:aBWT}) and $\tt IN\_CHAIN$ (Definition \ref{def:aaBWT}). In order to bound this space as a function of the number of states, we use the bound
    $|\delta| \leq 2p\sigma n$ implied by Lemma \ref{lem: p sparsity}, where $|\delta|$ is the number of transitions. \qed
\end{proof}

In Theorem \ref{thm:aBWT index} we provided an aBWT-index supporting pattern matching queries on any NFA. Being a superset of the aBWT, our (indexed) augmented aBWT can clearly support the same operations (in the same running times) of Theorem \ref{thm:aBWT index}, albeit using additional $\log p$ bits per edge. Actually, these operations turn out to be much simpler on the augmented aBWT than on the aBWT (thanks to the new sequence \texttt{IN\_CHAIN}). This is possible because by means of our augmented aBWT, given \emph{any} convex set $ U $ of states (represented by means of $ p $ ranges on the chains) and a character $ c $, one can compute the convex set of all states that can be reached from $ U $ by following edges labeled $ c $ (see Lemma \ref{lem:pathcoherence}). However, the queries' running times remain the same as in Theorem \ref{thm:aBWT index} so we will not describe them here.

\section{Conclusions}\label{sec:conc}

In this paper we considered the theoretical and practical implications of studying a specific \emph{partial} order on the states of a finite automaton. The considered partial order is a particularly natural one: the one obtained lifting to sets the \emph{co-lex} order of strings reaching every given state of the underlying automaton. 

In this work we extensively argued that our proposed point of  view allows us to sensibly classify regular languages and their complexities, from a both \emph{practically} and \emph{theoretically} interesting perspective. The central measure for the classification we put forward is the (minimum) \emph{width} of the above mentioned partial order. In this paper and in \cite{parttwo} we show that such measure induces a proper hierarchy of regular languages and make a number of observations on the levels of this hierarchy. An interesting feature of the classification obtained is that languages at higher levels of the hierarchy have a larger \emph{entangled} collection of states that makes them less prone to index-ability --- and, consequently, to any subsequent algorithmic analysis. 

From a theoretical perspective, we showed that a canonical automaton minimizing the width to its entanglement size (the \emph{Hasse automaton} of a language), can be built at any given level.

From a practical perspective, in the last part of the paper we introduced a technique to build very efficient indexes, exploiting the partial order considered. As it turned out, the indexes proposed can be used in both the deterministic and non-deterministic case, but can be built in polynomial time only in the former case. In the companion paper \cite{parttwo} we show how to overcome this issue.

As a matter of fact, our full proposal can be divided in two macro steps: in the first one, the one presented in this paper, we illustrated and proved results which mainly apply to the deterministic-width hierarchy and to indexing and encoding regular languages (a task that we have shown to be computationally easy on DFAs). In a subsequent paper, we will illustrate the extensions of our classification results to the non-deterministic-width hierarchy, as well as polynomial-time algorithms for sorting non-deterministic automata by means of \emph{co-lex relations} (a more general concept than co-lex orders, enabling indexing and avoiding NP-hardness).

\subsection*{A Glimpse into the Non-Deterministic Case}

\label{glimpse nondeterministic}
What happens if we consider the notion of width applied to non-deterministic automata instead of deterministic ones?   In this paragraph, we sketch the main results that will be presented in the  companion paper \cite{parttwo}.

\begin{enumerate}
    \item There exist NFAs without a maximum co-lex order,  suggesting  that  Problem \ref{pr:width_complexity} (Automata-width problem) could become more complex when  applied  to  NFAs.  Indeed, in \cite{gibney2022complexity} it is proved that  even deciding whether $\text{width}(\mt N)= 1$   (i.e., deciding whether $\mt N$ is Wheeler)    is an NP-complete problem. For this reason, a smallest-width NFA index cannot be built in polynomial time by means of co-lex orders (assuming $ P \not = NP $). We will show that the solution relies on switching to \emph{co-lex relations} (see \cite{cotumaccio2022}): the smallest-width relation can be computed in polynomial time, it enables indexing, and its width is never larger than that of a smallest-width co-lex order. 
    \item Another way to bypass the previous obstacle to indexing NFAs is to consider the   class   consisting of all  NFAs admitting  a maximum co-lex order. We will prove that this class --- dubbed the $\mathcal {MAX}$ class --- is a polynomial time  decidable class of automata, strictly between the DFA and NFA classes,   for which  the maximum co-lex order is always   computable  in polynomial time.  This implies that the NP-completeness of determining the width of an NFA   is due to automata outside the $ \mathcal {MAX} $ class. 
    
 
    \item The language-width problem becomes more complex: starting from an NFA $\mt N$,  already deciding whether $\text{width}^D(\mt L(\mt N))=1$ (equivalently, $\text{width}^N(\mt L(\mt N))=1$) is a PSPACE complete problem (see \cite{dagostino2023complexity}).
    \item The language  hierarchies based on the two (deterministic/non-deterministic) notions of width do not coincide, except for level 1 (Wheeler languages). More precisely, we will prove that there exist regular languages $ \mathcal{L} $, whose deterministic width $p>1$ can be chosen arbitrarily large, such that $p = \text{width}^{D}(\mathcal L) = 2^{ \text{width}^{N}(\mathcal L)} -1$, matching the upper-bound given in Corollary \ref{lem:widthupperbound}. 
    \end{enumerate}

\section{Acknowledgements and Fundings}

We wish to thank Gonzalo Navarro for providing us correct references for the succinct string data structures used in Lemma \ref{lem:strings2} and Manuel C\'aceres for pointing out the work \cite{KoganParterICALP}.\\

\emph{Nicola Prezza}: funded by the European Union (ERC, REGINDEX, 101039208). Views and opinions expressed are however those of the author(s) only and do not necessarily reflect those of the European Union or the European Research Council. Neither the European Union nor the granting authority can be held responsible for them.\\

\emph{Alberto Policriti}: Paritally supported by project funded under the National Recovery and Resilience Plan (NRRP), Mission 4 Component 2 Investment 1.4 - Call for tender No. 3138 of 16 December 2021, rectified by Decree n.3175 of 18 December 2021 of Italian Ministry of University and Research funded by the European Union – NextGenerationEU;
Project code CN\_00000033, Concession Decree No. 1034 of 17 June 2022 adopted by the Italian Ministry of University and Research, CUP G23C22001110007, Project title “National Biodiversity Future Center - NBFC”.

\bibliographystyle{alpha}
\bibliography{main}

\printindex\label{sec:index}

 \appendix 
 \section{Partitions and Orders }\label{app:minimal_convex}

This section is devoted to the proof of Theorem \ref{thm:fdt} and to other useful properties of entangled convex sets. All these results    follow from  general results valid for    arbitrary total orders and   partitions. From now on, we fix a total order $ (Z, \le) $ and a finite partition $ \mathcal P = \{ P_{1}, \dots , P_{m}\} $ of $ Z $. 
We first   give a  notion of entanglement, with respect to  $\mt P$,   for   subsets $X\subseteq Z$.  The main result of this section, Theorem \ref{thm:generalfdt},  states that there   always exists a finite, \emph{ordered} partition $\mt V$ of $Z$ composed of       entangled convex sets.    

It is convenient to think of the elements of  $\mathcal P$ as letters of an alphabet, forming finite or infinite strings while labelling element of $Z$. A finite string $P_{ 1}\dots P_{ k} \in \mathcal P^{*}$  is said to be {\em generated} by  $X \subseteq Z $, if there exists a  sequence $x_{1}\leq\dots  \leq x_{k}$  of elements in $X$   such that 
$x_{j}\in P_{j}$, for all $j=1, \dots,k$.
We also say that $P_{ 1}\dots P_{ k} $ {\em occurs in $ X $  at  $x_{1}, \dots, x_{k}$}.
Similarly,  an infinite string  $P_{1}\dots P_{k} \dots   \in \mathcal P^{\omega}$ is  generated by  $X \subseteq Z $ if there exists  a monotone  sequence  $(x_i)_{i\in \mathbb N}$  of elements in $X$   such that 
$x_{j}\in P_{j}$, for all  $j\in \mathbb N$.
Notice that if $ X $ is a finite set, then there exists an index $ i_0 $ such that for every $ i \ge i_0 $ it holds $ P_i = P_{i_0} $.
We can now re-state the notion of entanglement in this, more general, context.

\begin{definition}\label{def:entangled}
Let $ (Z, \le) $ be a total order, let $ \mathcal{P} $ be a   partition of $ Z $, and let $ X \subseteq Z $.
\begin{enumerate}
    \item We define $\mathcal P_{X}=\{P\in \mathcal P : P \cap X\neq\emptyset\}$.
    \item If $\mathcal P'=\{P_1, \dots, P_m\}\subseteq \mathcal P$, we say that $\mathcal P'$ is  {\em entangled} in $X$   if the infinite string 
    $(P_1  \dots  P_m)^\omega$ is generated by $X$.
    \item We say that $ X $ is {\em entangled} if $ \mathcal P_{X}$ is entangled in $ X $.
\end{enumerate}
\end{definition}

The property of $X$ being entangled is captured by the occurrence of an infinite string $(P_1 \dots P_m)^\omega$. In fact, as proved in the following lemma, finding $(P_1 \dots P_m)^k$ for arbitrarily big $k$ is sufficient to guarantee the existence of $(P_1 \dots P_m)^\omega$.

\begin{lemma}\label{lem:m_implies_wm}
Let $ (Z, \le) $ be a total order, let $ \mathcal{P} $ be a partition of $ Z $, let $ X \subseteq Z $, and let $ \mathcal P'=\{P_1, \dots, P_m\}\subseteq \mathcal P$.  The following are equivalent:
\begin{enumerate}
    \item For every $k\in \mathbb N$, the string $ (P_1 \dots P_m)^k$ is generated by $X$.
    \item $ (P_1 \dots P_m)^\omega $ is generated by $ X $.
\end{enumerate}
\end{lemma}
\begin{proof}
 The nontrivial implication is $ (1) \Rightarrow (2) $. If $ m = 1 $, then by choosing $ k = 1 $ we obtain that there exists $ x \in X $ such that $ x \in P_1 $, so  $ (P_1)^\omega $ occurs in $ X $, as witnessed by the monotone sequence $ (x_i)_{i \in \mathbb{N}} $ such that $ x_i = x $ for every $ i \in \mathbb{N} $. Thus, in the following we can assume $ m \ge 2 $. This implies that for every $ k $, if $ (P_1 \dots P_m)^k$ occurs in $ X $ at $ x_1, x_2, \dots, x_{m_k} $, then $ x_1 <  x_2 < \dots <x_{m_k} $, that is, the inequalities are strict. If $(1)$ holds, we prove that we can find an infinite  family $ (Y_i)_{i \ge 1} $ of   pairwise disjoint subsets of $X$, each containing an occurrence of $(P_1 \dots P_m)$, such that for every pair of distinct integers $ i, j $ it holds either $ Y_i < Y_j $ (that is, each element in $Y_i$ is smaller than each  element in $Y_j$)  or $ Y_j < Y_i $. This will imply $ (2) $, because if the set $ \{i \ge 1 | (\forall j > i)(Y_i < Y_j) \} $ is infinite, then $ (2) $ is witnessed by an increasing sequence, and if $ \{i \ge 1 | (\forall j > i)(Y_i < Y_j) \} $ is finite, then $ (2) $ is witnessed by a decreasing sequence. \\
Let us show a recursive construction of $ (Y_i)_{i \ge 1} $.
We say that $ \langle X_{1}, X_{2} \rangle $ is a \emph{split} of $ X $ if $ \{X_{1}, X_{2}\} $ is a partition of $ X $ and  
$ X_1 < X_2 $. 
Given a split $ \langle X_{1}, X_{2} \rangle $ of $ X $, we claim that (1) must hold for either $ X_{1} $ or $ X_{2}$ (or both). 
In fact, reasoning by  contradiction, assume there exists $ \bar{k} $ such that   the string $ (P_1 \dots P_m)^{\bar k}$  is neither generated by $ X_1 $ nor by $ X_2 $. 
This promptly leads to a contradiction, since $ (P_1 \dots P_m)^{2\bar{k}}$ is   generated by $ X $ and hence,if $ (P_1 \dots P_m)^{\bar{k}}$ is not generated by $X_1$, then it must be generated by $ X_2$.\\
Now consider an occurrence of  $ (P_1 \dots P_m)^2$ generated by $ X $ and a split   $ \langle X_{1}, X_{2}\rangle $ such that $(P_1 \dots P_m)$ is generated by $ X_{1} $ and $(P_1 \dots P_m)$ is generated by $ X_{2} $. Now, if (1) holds for $ X_{1} $, then define $ Y_1 = X_2 $ and repeat the construction using $ X_1 $ instead of $ X $. If (1) holds for $ X_{2} $, then define $ Y_1 = X_1 $ and repeat the construction using $ X_2 $ instead of $ X $. We can then recursively define a family $ (Y_i)_{i \ge 1} $ with the desired properties.
\qed 
\end{proof}

We now introduce the notion of an \emph{entangled convex  decomposition}, whose aim is to identify  entangled regions of $(Z,\leq)$ with respect to a partition $\mathcal P$.  

\begin{definition}\label{def:ec decomposition}
Let $ (Z, \le) $ be a total order and let $ \mathcal{P} $ be  a partition of $ Z $. We say that a partition  $ \mathcal V  $ of $Z$ is an \emph{entangled, convex  decomposition  of $ \mathcal P$ in $(Z,\leq)$} (\emph{e.c. decomposition}, for short) if
all the elements of $ \mathcal V  $ are entangled (w.r.t. the partition $\mt P$)  convex sets in $ (Z, \le) $.
\end{definition}

 \begin{example}
 Consider the total order $ (\mathbb{Z}, \le) $, where $\mathbb Z$ is the set of all integers and $ \le $ is the usual order on $\mathbb Z$. Let $\mathcal P=\{P_1,P_2,P_3\}$ be the partition of $ \mathbb{Z} $ defined as follows:
   \begin{align*}
& P_1=\{n \leq 0 : n \text{~ is odd}\} \cup \{n >  0 : n \equiv 1 ~\text{mod } 3\}&\\
&  P_2=\{n\leq 0 : n \text{~ is even}\}  \cup \{n >  0 : n \equiv 2 ~\text{mod } 3\} , &\\
 &P_3=\{n >  0 : n \equiv 0 ~\text{mod } 3\} &
\end{align*}

The partition $\mathcal P$ generates the following \emph{trace}  over $\mathbb Z$:
\[\dots P_1P_2P_1P_2 \dots P_1P_2 P_1P_2P_3P_1P_2P_3\dots\]

Now define $\mt V=\{V_1,V_2\}$, where 
$V_1=\{n\in \mathbb Z : n\leq 0\}$, $V_2=\{n\in \mathbb Z : n>0\}$. It is immediate to check that $ \mathcal{V} $ is an e.c. decomposition of $ \mathcal{P} $ in $ (\mathbb{Z}, \le) $. More trivially, even $\mt V'=\{\mathbb Z\}$ is an e.c. decomposition of $\mathcal P$ in $ (\mathbb{Z}, \le) $.
 \end{example}

Below we prove that if $ \mathcal{P} $ is a finite partition, then there always exists a \emph{finite} e.c. decomposition of $ \mathcal{P} $.

\begin{theorem} \label{thm:generalfdt}
    Let $ (Z, \le) $ be a total order, and let $ \mathcal{P} = \{P_1, \dots, P_m \} $ be a finite partition of $ Z $. Then, $ \mathcal{P} $ admits a finite e.c. decomposition in $ (Z, \le) $.
\end{theorem}

\begin{proof} We proceed by induction on $ m = |\mathcal{P} | $. If $ m = 1 $, then $ \mathcal{P} = \{Z \} $, so $ \{Z \} $ is an e.c. decomposition of $ \mathcal{P} $ in $ (Z, \le) $. Assume $ m \ge 2 $ and notice that  we may also  assume that the sequence $ (P_1 \dots P_m)^\omega $ is \emph{not}  generated by $Z$,  otherwise    the partition $\mt V=\{Z\}$ is a finite e.c. decomposition of $\mathcal P$ in $(Z,\leq)$ and we are done.  
Since $ (P_1 \dots P_m)^\omega $ is not  generated by $Z$, for any permutation   $ \pi $  of the set $ \{1, \dots, m \}$ the sequence 
$ (P_{\pi (1)}, \dots, P_{\pi (m)})^\omega $ is not generated by $ Z $ and therefore, by Lemma \ref{lem:m_implies_wm},  for any $\pi$  there exists an integer $ s_\pi $ such that $ (P_{\pi (1)}, \dots, P_{\pi (m)})^{s_\pi}$ is not generated by $ Z $. 
Using this property we    prove that there exists    a finite partition $ \mathcal{V} $ of $ Z $ into convex sets such that for  every $ V \in \mathcal{V} $ and  every  $ \pi $, the string $ (P_{\pi (1)}, \dots, P_{\pi (m)})^2 $ does not occur in $ V $.

Consider the following procedure:

\begin{algorithmic}[1]
    \STATE{$\mathcal V \leftarrow \{Z \}$;}
    \COMMENT{initialise the partition}
    \FOR{$ \pi $ permutation of $ \{1, \dots, m \}$}
          \WHILE{exists an element in $ \mathcal{V} $ generating $ (P_{\pi(1)}\dots P_{\pi(m)})^{2} $}
            \STATE{let $ V \in \mathcal{V} $ generating $ (P_{\pi(1)}\dots P_{\pi(m)})^{2} $;}
            \STATE{$ \mathcal{V} \leftarrow \mathcal{V} \setminus V $;}
            \STATE{let $ \alpha_{1}  < \dots < \alpha_{m}  < \alpha_{1}' < \dots < \alpha_{m}' $ in $ V $ be such that  $\alpha_{j}, \alpha_{j}' \in P_{\pi(j)}$, for $j=1,\dots, m$;}
            \STATE{$ \mathcal V \leftarrow \mathcal{V} \cup \{\{ \alpha \in V \ | \ \alpha \leq \alpha_{m}\}, \{ \alpha \in V \ | \ \alpha > \alpha_{m}\} \} $;}
          \ENDWHILE
    \ENDFOR
            \RETURN{$\mathcal V$}
\end{algorithmic}

The procedure starts with $ \mathcal{V} = \{Z \} $ and recursively partitions a $V$ in $ \mathcal{V} $ into two nonempty convex sets as long as $ (P_{\pi (1)}, \dots, P_{\pi (m)})^2 $ occurs in $V$. Notice that the procedure ends after at most $ \sum_{\pi} s_\pi $ iterations, returning a finite partition $ \mathcal{V} $ of $ Z $ into convex sets. 

Now   fix   $V\in \mt V$ and consider the partition $ \mt P_{|V} = \{P \cap V \;|\; P \in \mathcal{P} \wedge P \cap V \neq \emptyset \} =  \{P \cap V \;|\; P \in \mathcal{P}_V \} $ of $ V $, where $\mathcal{P}_V$ is as in Def. \ref{def:entangled} and $ |\mt P_{|_V}| \leq m $. 
  To complete the proof  it will be enough to prove that $\mt P _{|V} $ admits a finite e.c. decomposition in $(V, \leq)$ because then a finite e.c. decomposition of $\mt P$ in $(Z, \leq)$ can be obtained by merging all the decompositions of $\mt P_{|V} $, for  $V\in \mt V$.
 
 If $|\mt P_{|V}|<m$, then $\mt P_{|V}$  admits an e.c. decomposition by inductive hypothesis. Otherwise, $|\mt P_{|V}|=m $, say $\mt P_{|V}=\{P'_{1}, \dots, P'_{m}\} $. 
By construction,  we know that for any permutation $\pi$ of $\{1, \dots,m \}$ the   string $ (P'_{\pi (1)}, \dots, P'_{\pi (m )})^2 $ does not occur in $ V $.  Example \ref{ex:e.c-decomposition} below may help with an intuition for the rest of the argument.  
 Let $ k \ge 1 $ be the number of distinct permutations $ \pi $ of $ \{1, \dots, m  \} $ such that $ (P'_{\pi (1)}, \dots, P'_{\pi (m )}) $ occurs in $ V $. We proceed by induction on $ k $. If $ k = 1 $, then each set $   P'_j $ is convex  and trivially entangled, so $ \{  P'_j \;|\;  1 \le j \le m  \} $ is an e.c. decomposition of $ \mt P_{|V} $ in $ (V, \le) $. Now assume $ k \ge 2 $ and fix a permutation  $ \pi$ such that $ (P'_{\pi (1)}, \dots, P'_{\pi (m )}) $ occurs in $ V $. Define:
\begin{equation*}
    V_1 = \{\alpha \in V \;|\; \text{$ \exists~ \alpha_1, \dots, \alpha_m $ with $ \alpha \le \alpha_1 < \dots <\alpha_{m} $ and $ \alpha_i \in P'_{\pi  (i)}$ \}}
\end{equation*}
and $ V_2 =V \setminus V_1 $.   Let us prove that  $V_1$ and $ V_2 $ are nonempty.  Just observe that if $ \alpha_1, \dots,   \alpha_{m } $ is a witness for $ (P'_{\pi (1)}, \dots, P'_{\pi (m )}) $ in $ V $, then $ \alpha_1 \in V_1 $. Moreover,     $ \alpha_{m }   \in V\setminus V_1=V_2 $, otherwise,    since $m>1$ and   $P'_{\pi (1 )}\neq P'_{\pi (m )}$, the sequence  $ (P'_{\pi (1)}, \dots, P'_{\pi (m )})^2 $ would occur in $ V $. The previous observation   implies also  that  $ (P'_{\pi (1)}, \dots, P'_{\pi (m )})$ does not occur in $ V_1 $, nor in $ V_2 $.  Moreover $ V_1 $ and $ V_2 $ are clearly convex. To conclude it will  suffice to prove that the $V_i$-partition $ \mt P_{|V_i}= \{P \cap V_i \;|\; P \in \mathcal{P}_{|V}, P \cap V_i\neq \emptyset \}  $  admits a finite e.c. decomposition in $ (V_i, \le) $, for $ i = 1, 2 $. For any given $i$, if $ |\mt P_{|V_i}| <   m $, we conclude by the inductive hypothesis on $ m $. If, instead, $ |\mt P_{|V_i}| = m $,  say $ \mt P_{|V_i} = \{P''_1, \dots, P''_{m} \}$,
we use the inductive hypothesis on $ k $:  the number of distinct permutations $ \pi' $ of $ \{1, \dots, m \} $ such that $ (P''_{\pi' (1)}, \dots, P''_{\pi' (m)}) $ occurs in $ V_i $ is less than $k$, since $ (P'_{\pi (1)}, \dots, P'_{\pi(m)})$ occurs in $ V $ while $ (P''_{\pi  (1)}, \dots, P''_{\pi  (m)}) $ does not occur   in $ V_i $.
 \qed
\end{proof}

\begin{example}\label{ex:e.c-decomposition}
We give an example of the final part of the construction described in Theorem \ref{thm:generalfdt}. Suppose $\mt P_{|V}=\{P_1', P_2',P_3'\}$  leaves the following trace over $(V,\leq)$: 
\[ (P'_1P'_2)^\omega (P'_3P'_2)^\omega(P'_1)^\omega(P'_2)^\omega \]
Notice that, as assumed  in the last part of the above proof,  for any permutation $\pi$ of $\{1,2,3\}$   the sequence 
$(P'_{\pi(1)}P'_{\pi(2)}P'_{\pi(3)})^2$ does not appear in $V$; however, the sequence    $(P'_{\pi(1)}P'_{\pi(2)}P'_{\pi(3)})$ appears in $V$   for $\pi=id$. 
If we fix $\pi=id$ and consider the sets $V_1,V_2$  as in the above proof, then the partition $\mt P_{|V}$  leaves the following traces on  the sets  $V_1,V_2$:
\[ (P'_1P'_2)^\omega \qquad \text{and} \qquad  (P'_3P'_2)^\omega(P'_1)^\omega(P'_2)^\omega,\]
respectively. Notice that $P'_3$ does not appear in $V_1$, while the sequence $(P'_1P'_2P'_3)$ does not appear in $V_2$, so that the inductive hypothesis can be applied.\qed
\end{example}

We say that an e.c. decomposition of $ \mathcal{P} $ in $ (Z, \le) $ is a \emph{minimum-size} e.c. decomposition if it has minimum cardinality. 
As  shown in the following remark, minimum-size e.c. decompositions  ensure  additional interesting properties.

 \begin{remark}\label{rem: minimality}
Let $ (Z, \le) $ be a total order, let $ \mathcal{P}  $ be a finite partition of $ Z $, and let $ \mathcal V= \{V_{1}, \dots , V_{r} \}$ be a minimum-size e.c. decomposition of $ \mathcal{P} $ in $ (Z, \le)$, where $V_{1}< \dots < V_{r}$. Then, for every $1\leq i<r$, we have $\mathcal P_{V_i}\not \subseteq  \mathcal P_{V_{i+1}}$, where $ \mathcal{P}_{V_i}=\{P\in \mathcal P  \;|\; P\cap V_i\neq \emptyset \} $ (see Definition \ref{def:entangled}). In fact,  if this were not the case, $ \mathcal V'= \{V_{1}, \dots , V_{i-1}, V_i\cup V_{i+1}, \dots V_{r} \}$ would be a smaller size e.c. decomposition of $ \mathcal{P} $ in $ (Z, \le) $. Similarly, for every $ 1< i\leq r $, it must be $\mathcal P_{V_i}\not \subseteq  \mathcal P_{V_{i-1}}$. 
In conclusion, for every $ i = 1, \dots, r $, there exist $ R_i \in \mathcal{P}_{V_i} \setminus \mathcal{P}_{V_{i + 1}}$ and $ L_i \in \mathcal{P}_{V_i} \setminus \mathcal{P}_{V_{i - 1}} $, where we assume $ V_0 = V_{r + 1} = \emptyset $.
\end{remark}

In general, a minimum-size e.c. decomposition is not unique. 


\begin{example}
Let us show that even in the special case when   $(Z,\leq)=( \text{Pref}(\mathcal{L(D)}), \preceq)$ and $\mt P=\{I_u \;|\; u\in Q\}$ we can have more than one   minimum-size e.c. decomposition.
 
Consider the DFA $ \mathcal{D} $ in Figure \ref{fig:2minent}. Notice that in every e.c. decomposition of $ \mathcal{D} $ one element is $ \{\varepsilon \} $, because $ I_0 = \{\varepsilon \} $. Moreover, every e.c. decomposition of $ \mathcal{D} $ must have cardinality at least three, because, since  $ 1 <_{\mathcal{D}} 3 $, states $1$ and $3$ are not entangled.  It is easy to check that:
\begin{equation*}
    \mathcal{V} = \{\{\varepsilon \}, \{ac^*\cup bc^*\}, \{[b(c+d)^*\setminus bc^*]\cup f(c+d)^*\cup gd^*\}   \}
\end{equation*}
and:
\begin{equation*}
    \mathcal{V}' = \{\{\varepsilon \}, \{ac^*\cup b(c+d)^*\cup [f(c+d)^*\setminus fd^*]\}, \{fd^*\cup gd^*\}\}
\end{equation*}
are two distinct minimum-size e.c. decompositions of $ \mathcal{D} $.

\begin{figure}[h!] 
 \begin{center}
\begin{tikzpicture}[->,>=stealth', semithick, auto, scale=1]
\node[state, initial below] (0) at (0,0)	{$0$};
\node[state, accepting,label=above:{}] (1)    at (0,1.5)	{$1$};
\node[state,accepting, label=above:{}] (2)    at (-2,1)	{$2$};
\node[state, accepting,label=above:{}] (3)    at (2,1)	{$3$};
\draw (0) edge [] node [] {$a$} (1);
\draw (0) edge [] node [above] {$b,f$} (2);
\draw (0) edge [] node [above] {$g$} (3);
 \draw (1) edge  [loop above] node {$c$} (1);
 \draw (2) edge  [loop above] node {$c,d$} (2);
  \draw (3) edge  [loop above] node {$d$} (3);
\end{tikzpicture}
 \end{center}
 	\caption{An automaton $\mt D$ admitting two distinct minimum-size e.c. decompositions.}\label{fig:2minent}
\end{figure}
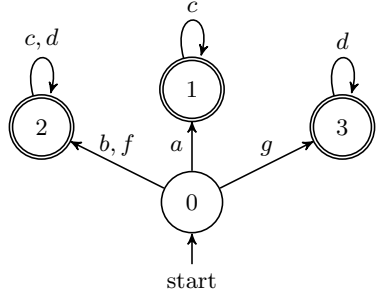
\end{example}

We need the following lemmas on entangled convex sets in  Section \ref{sec:Hasse automaton of L}. Assume that $ (Z, \le) $ is a total order and  $ \mathcal{P} $ is a partition of $ Z $.
\begin{lemma}\label{lem:noioso}
  Let $ \mathcal V$ be a minimum-size e.c. decomposition  of $ \mathcal{P} $. Assume that $V\in \mt V $ is such that there exist $C_1< \dots < C_n$
  entangled convex sets with $V\subseteq \bigcup_{i=1}^n C_i$. Then, for all  $P\in \mt P$ it holds:
\begin{eqnarray*}
\forall i \in \{1, \dots , n\} (C_i\cap P\neq \emptyset) & \rightarrow & V\cap P\neq \emptyset.
\end{eqnarray*}
\end{lemma}

\begin{proof}

 If $ C_j \subseteq V $ for some $ j $, then $V\cap P\neq \emptyset$ since $V\cap P \supseteq C_j\cap P\neq \emptyset$. Otherwise, consider the smallest $i$ such that $C_i\cap V\neq \emptyset$. Since $C_1< \dots < C_n$, $ V $ is convex and $V$  does not contain any $C_j$,  it must be $ V \subseteq C_i \cup C_{i + 1} $, (where we assume  $C_{i + 1}=\emptyset$ if $i=n$).   Let $ \mathcal V= \{V_{1}, \dots , V_{r}\}$ with $V_{1}< \dots < V_{r}$. For all $j=1, \dots, r$,  consider the elements $R_j\in \mathcal{P}_{V_j} \setminus \mathcal{P}_{V_{j + 1}}$ and $  L_j\in \mathcal{P}_{V_j} \setminus \mathcal{P}_{V_{j - 1}} $ (where  we assume $ V_0 = V_{r + 1} = \emptyset $), as in Remark \ref{rem: minimality} .  Let $s$ be such that $V=V_s$. We distinguish three cases. 
\begin{enumerate}
    \item   $ V_s \cap  C_{i+1}=\emptyset $. In this case, it must be $ V_s \subseteq C_i $. Let $ V_{s - h}, V_{s - h +1},  \dots, V_s, \dots, V_{s+k -1}, V_{s+k} $ ($ h, k \ge 0 $) be all the elements of $ \mathcal{V} $ contained in $ C_i $. Since $ V_1 < \dots < V_r $ , we conclude:
    \begin{equation*}
        V_{s - h }\cup \dots  \cup V_{s}\cup \dots \cup V_{s+k}\subseteq C_i  \subseteq V_{s-h-1 }\cup V_{s-h}\cup \dots\cup V_{s}\cup \dots \cup V_{s+k}\cup V_{s+k+1 }.
    \end{equation*}
    We know that $L_{s - h}, \dots,  L_{s},R_{s}, \dots R_{s+k}$ occur in $V_{s-h}\cup \dots \cup V_{s+k}$, so they also occur in $C_i$. Moreover, we also know that $ P $ occurs in $ C_i $. Since $ C_i $ is entangled,   there exists a sequence $ \alpha_{s - h} \leq \dots \leq \alpha_{s} \leq \beta \leq \gamma_{s} \leq \dots \leq \gamma_{s + k} $ of elements in $C_i$   witnessing that the ordered sequence $ L_{s - h}, \dots,  L_{s}, P, R_{s}, \dots R_{s+k} $ occurs in $ C_i $; it follows that $ L_{s - h}, \dots,  L_{s}, P, R_{s}, \dots R_{s+k} $ occurs  in $ V_{s-h -1 }\cup V_{s-h}\cup \dots\cup V_{s}\cup \dots \cup V_{s+k}\cup V_{s+k+1 } $ as well. Since $ L_{s - h} $ does not occur in $ V_{s - h - 1} $, then the ordered  sequence $ L_{s - h + 1}, \dots,  L_{s}, P, R_{s}, \dots R_{s+k} $ occurs  in $ V_{s - h}\cup \dots\cup V_{s}\cup \dots \cup V_{s+k}\cup V_{s+k+1 } $. Now, $ L_{s - h + 1} $ does not occur in $ V_{s - h} $, so the ordered sequence $ L_{s - h + 2}, \dots,  L_{s}, P, R_{s}, \dots R_{s+k} $ occurs    in $ V_{s - h + 1}\cup \dots\cup V_{s}\cup \dots \cup V_{s+k}\cup V_{s+k+1 } $. Proceeding in this way, we obtain that the sequence $ P, R_{s}, \dots R_{s+k} $ occurs in this order in $ V_{s} \cup \dots\cup V_{s+k}\cup V_{s+k+1} $. Now suppose for sake of a contradiction that $ P $ does not occur in $ V_s $. As before we obtain that $ R_{s}, \dots R_{s+k} $ occurs in this order in $ V_{s + 1}\cup \dots\cup V_{s+k}\cup V_{s+k+1 } $, $ R_{s+ 1}, \dots R_{s+k} $ occurs in this order in $ V_{s + 2}\cup \dots\cup V_{s+k}\cup V_{s+k+1 } $, and so on. We finally conclude that $ R_{s + k} $ occurs in $ V_{s+k+1 } $, a contradiction.
    \item $ V_s \cap  C_{i}=\emptyset $. In this case, it must be $ V_s \subseteq C_{i + 1} $ and one concludes as in the previous case.
    \item $ V_s \cap  C_{i} \not = \emptyset $ and $ V_s\cap  C_{i + 1} \not = \emptyset $. In this case, let $ V_{s - h}, \dots, V_{s - 1} $ ($ h \ge 0 $) be all elements of $ \mathcal{V} $ contained in $ C_i $, and let $ V_{s + 1}, \dots, V_{s + k} $ ($ k \ge 0 $) be all elements of $ \mathcal{V} $ contained in $ C_{i + 1} $. As before:
    \begin{equation*}
    V_{s - h} \cup \dots \cup V_{s - 1} \subseteq C_i \subseteq V_{s -h - 1} \cup V_{s - h} \cup \dots \cup V_{s - 1} \cup V_s
    \end{equation*}
    and:
    \begin{equation*}
    V_{s + 1} \cup \dots \cup V_{s + k} \subseteq C_{i+1} \subseteq V_s \cup V_{s + 1} \cup \dots \cup V_{s + k} \cup V_{s + k + 1}.
    \end{equation*}
    Now, assume by contradiction that $ P $ does not occur in $ V_s $. First, let us prove that $ L_s $ does not occur in $ C_i $. Suppose by contradiction that $ L_s $ occurs in $ C_i $. We know that $ L_{s - h}, \dots, L_{s - 1} $ occurs in $ C_i $, and we also know that $ P $ occurs in $ C_i $. Since $ C_i $ is entangled, then $ L_{s - h}, \dots, L_{s - 1}, L_s, P $ should occur in this order in $ C_i $ and so also in $ V_{s -h - 1} \cup V_{s - h} \cup \dots \cup V_{s - 1} \cup V_s $; however, reasoning  as in case 1, this  would imply that $ P $ occurs in $ V_s $, a contradiction. Analogously, one shows that $ R_s $ does not occur in $ C_{i + 1} $.
    
    Since $ R_s $ and $ L_s $ occur in $ V_s $, then there exists a monotone sequence in $ V_s $ whose trace consists of alternating values of $ R_s $ and $ L_s $. But $ V_s \subseteq C_i \cup C_{i + 1} $ and $ C_i \prec C_{i + 1} $, so the monotone sequence is definitely contained in $ C_i $ or $ C_{i + 1} $. In the first case we would obtain that $ L_s $ occurs in $ C_i $, and in the second case we would obtain that $ R_s $ occurs in $ C_{i + 1} $, so in both cases we reach a contradiction. \qed
\end{enumerate}

\end{proof}

\begin{lemma}\label{lem:union}
 Let $ C \subseteq Z$ be an   entangled convex set and consider any pair of convex sets $ C_1 , C_2 $ such that $C = C_1 \cup C_2 $. Then, there exists $ i \in \{1, 2 \} $ such that $ C_i $ is entangled and $ \mathcal{P}_{ C_i} = \mathcal{P}_{ C} $. 
\end{lemma}
\begin{proof}
Let $ (z_i)_{i \geq 1} $ be a monotone sequence witnessing the entanglement of $ C $. Then,   infinitely many $z_j$'s appear in either $C_1$ or $C_2$ (or both).  In the former case $ C_1 $ is entangled:   since  $C_1$ is convex, if $j_0\geq 1$ is such that $ z_{j_0} \in C_1 $, then  the subsequence $ (z_j)_{j \ge j_0} $ is  in $ C_1 $ and, clearly, we have $\mathcal{P}_{ C_1} = \mathcal{P}_{ C}$.  In the latter case, analogously, $C_2$ is entangled and $\mathcal{P}_{ C_2} = \mathcal{P}_{ C}$. \qed
\end{proof}

\begin{lemma}\label{lem:entangled}
 Let $ C_1, C_2\subseteq Z $ be entangled convex sets. Then, at least one the following holds true:
\begin{enumerate}
 \item $ C_1 \setminus C_2 $  is entangled and convex and $ \mathcal{P}_{ C_1 \setminus C_2} = \mathcal{P}_{ C_1} $;
  \item  $ C_2 \setminus C_1 $  is entangled and convex and  $ \mathcal{P}_{ C_2 \setminus C_1} = \mathcal{P}_{ C_2} $;
    \item $ C_1 \cup C_2 $ is entangled and convex and    $ \mathcal{P}_{ C_1 \cup C_2} = \mathcal{P}_{C_1}$ or $ \mathcal{P}_{ C_1 \cup C_2} = \mathcal{P}_{C_2}$.
\end{enumerate}
\end{lemma}
\begin{proof}
If $ C_1 \cap C_2 = \emptyset $,  (1) and  (2)  hold. If  $ C_2  \subseteq C_1 $ or $ C_1  \subseteq C_2 $, (3) holds. 
In the remaining cases observe that $ C_1 \setminus C_2 $, $ C_2 \setminus C_1 $, $ C_1 \cap C_2 $ and $ C_1 \cup C_2 $ are convex. Since $ C_1 = (C_1 \setminus C_2) \cup (C_1 \cap C_2) $ and $ C_2 = (C_2 \setminus C_1) \cup (C_1 \cap C_2) $, by Lemma \ref{lem:union} we conclude that at least one the following holds true:
\begin{enumerate}
    \item   $ C_1 \setminus C_2 $ is entangled and convex and $ \mathcal{P}_{C_1 \setminus C_2} = \mathcal{P}_{C_1} $, or $ C_2 \setminus C_1 $ is entangled and convex and $ \mathcal{P}_{C_2 \setminus C_1} = \mathcal{P}_{C_2} $;
    \item the intersection $ C_1 \cap C_2 $ is entangled and convex  and $ \mathcal{P}_{C_1 \cap C_2} = \mathcal{P}_{C_1} = \mathcal{P}_{C_2}$.
\end{enumerate}
In the first case we are done, while in the second case we have $ \mathcal{P}_{C_1 \cap C_2} = \mathcal{P}_{C_1} = \mathcal{P}_{C_2} = \mathcal{P}_{C_1} \cup \mathcal{P}_{C_2} = \mathcal{P}_{C_1 \cup C_2} $. Since $C_1\cap C_2 \subseteq C_1\cup C_2$  and $C_1\cap C_2$ is entangled,   we conclude that $ C_1 \cup C_2 $ is entangled. \qed
\end{proof}


\begin{lemma} \label{lem:convex_in_order}
 Let $ C_1, \dots, C_n \subseteq Z$ be    entangled convex sets. Then, there exist $ m \le n $ pairwise disjoint, entangled convex sets $ C'_1, \dots, C'_m \subseteq Z$,  such that: 
 \begin{itemize}
 \item $ C'_1 < \dots < C'_m $ and $ \bigcup_{i =1}^n C_i = \bigcup_{i = 1}^m C'_i $; 
 \item if  $P\in \mathcal P$ occurs in all $ C_i$'s, then it occurs in all $C_i'$'s as well. 
 \end{itemize}
\end{lemma}

\begin{proof}  We can suppose without loss of generality   that the $C_i$'s are non-empty and we proceed by induction on the number of intersections $ r = |\{(i, j)| i < j \wedge C_i \cap C_j \not = \emptyset \}| $. 

If $ r = 0 $, then the $ C_i $'s are pairwise disjoint  and, since they are convex, they are comparable. Hence, it is sufficient to take $ C'_1, \dots, C'_n $ as the permutation of the $ C_i $'s such that $ C'_1< \dots< C'_n $. 

Now assume $ r \ge 1 $ and let, without loss of generality,  $ C_1 \cap C_2 \not = \emptyset $. We now produce a new sequence of at most $n$ entangled convex sets to which we can apply the inductive hypothesis. By Lemma \ref{lem:entangled} at least one among $ C_1 \setminus C_2 $, $ C_2 \setminus C_1 $ and $ C_1 \cup C_2 $, is an entangled convex set. 
If $ C_1 \cup C_2 $ is  an entangled convex set, then let $ C_1 \cup C_2, C_3, \dots, C_n $ be the new sequence. Otherwise, if $ C_1 \setminus C_2 $ (the case $ C_2 \setminus C_1 $ analogous) is entangled, let the new sequence be $ C_1 \setminus C_2, C_2, C_3, \dots, C_n $. In both cases the number of intersections decreases: this is clear in the first case, while in the second case $ C_1 \setminus C_2 \subseteq C_1 $ and $ (C_1 \setminus C_2) \cap C_2 = \emptyset $. 

In all the above cases  Lemma \ref{lem:entangled} implies that if a $P \in \mathcal{P}$ occurs in all $C_i$'s, then  it occurs in all elements of the new family and  we can conclude by the inductive hypothesis. \qed
\end{proof}

Below we prove a fairly intuitive result on the intersection of convex sets. 

\begin{lemma}\label{lem:con-int}
Let $ (Z, \le) $ be a total order. If $ C_1, \dots , C_{n} \subseteq Z$ are non-empty, convex sets such that $ C_{i}\cap C_{j} \neq \emptyset $ for all  $i, j\in \{ 1 , \dots , n\} $, then $ \bigcap_{i=1}^{n} C_{i}\neq \emptyset $.     
\end{lemma}
\begin{proof}
We proceed by induction on $ n $. Cases $ n=1, 2 $ are trivial, so assume $ n \ge 3 $. For every $ i \in \{1, \dots , n\} $, the set:
\[\bigcap_{\substack{k \in \{1, \dots , n\} \\ k \neq i}} C_{k}
\]
is nonempty by the inductive hypothesis, so we can pick an element $ d_i $. If for some distinct $ i $ and $ j $ we have $ d_{i}=d_{j} $, then such an element witnesses that $ \bigcap_{i=1}^{n} C_{i}\neq \emptyset $. Otherwise, assume without loss of generality that $ d_{1} < \dots < d_{n} $. Fix any integer $ j $ such that  $ 1 < j < n $, and let us prove that $ d_j $ witnesses that $ \bigcap_{i=1}^{n} C_{i}\neq \emptyset $. We only have to prove that $ d_j \in C_j $. This follows from $ d_1, d_n \in C_j $ and the fact that  $ C_j $ is convex. \qed
\end{proof}


\end{document}